\documentclass[11pt,a4paper,openright,twoside]{book}

\usepackage{amsmath,amssymb,amsfonts}
\usepackage{color}
\usepackage{float}
\usepackage[bf, footnotesize]{caption}
\usepackage{graphicx,epsfig,url}
\usepackage{mathrsfs}
\usepackage{latexsym}
\usepackage{amsthm}
\usepackage{graphics}
\usepackage[latin1]{inputenc}
\usepackage[italian,english]{babel}
\usepackage{feynmf}
\usepackage{fancyhdr}
\usepackage{indentfirst}
\usepackage{layout}
\usepackage{calc}
\usepackage{pifont}
\usepackage{gensymb}
\usepackage{multicol}
\usepackage{mathcomp}
\usepackage{appendix}
\usepackage[Lenny]{fncychap}
\usepackage{epigraph}
\usepackage{wasysym}
\usepackage{soul}
\usepackage[colorlinks=true,linkcolor=red,anchorcolor=black,citecolor=blue,filecolor= cyan,menucolor=red,runcolor=filecolor,urlcolor=magenta,bookmarks=true,bookmarksnumbered=true]{hyperref}

\usepackage{foekfont}
\usepackage{emerald}

\usepackage{lettrine}

\usepackage{engord}

\usepackage{ifdraft}

\fancypagestyle{plain}{
\fancyhf{}}
\newenvironment{abstract}
{\clearpage\null \vfill\begin{center}
\bfseries \abstractname \end{center}}
{\vfill \null}

\newenvironment{citazione}
{\null \vfill\begin{center}
\bfseries \end{center}}
{\vfill \null}

\def\dst{\displaystyle}
\def\be{\begin{equation}}
\def\ee{\end{equation}}
\def\bes{\begin{subequations}}
\def\ees{\end{subequations}}
\def\bry{\begin{array}}
\def\ery{\end{array}}
\def\bit{\begin{itemize}}
\def\eit{\end{itemize}}
\def\ben{\begin{enumerate}}
\def\een{\end{enumerate}}
\def\bedes{\begin{description}}
\def\endes{\end{description}}
\def\ble{\begin{lemma}}
\def\ele{\end{lemma}}
\def\bte{\begin{teorema}}
\def\ete{\end{teorema}}
\def\bcor{\begin{corollario}}
\def\ecor{\end{corollario}}
\def\bedef{\begin{definizione}}
\def\endef{\end{definizione}}
\def\bnote{\begin{note}}
\def\enote{\end{note}}
\def\bex{\begin{esempio}}
\def\eex{\end{esempio}}
\def\bd{\begin{diagram}}
\def\ed{\end{diagram}}

\def\bfi{\begin{figure}[h]}
\def\efi{\end{figure}}

\def\l{\label}

\def\f{\frac}
\def\({\left(}
\def\){\right)}
\def\>{\right\rangle}
\def\<{\left\langle}
\def\ovl{\overline}

\def\a{\alpha}

\def\d{\delta}
\def\g{\gamma}
\def\G{\Gamma}

\def\D{\Delta}

\def\o{\omega}
\def\O{\Omega}

\def\s{\sigma}
\def\S{\Sigma}

\newcommand{\psl}{p\llap{/\kern-0.3pt}}
\newcommand{\vsl}{v\llap{/\kern-0.3pt}}
\newcommand{\ppsl}{p'\llap{/\kern+1.5pt}}
\newcommand{\ksl}{k\llap{/\kern-0.3pt}}
\newcommand{\Ksl}{K\llap{/\kern+2.5pt}}
\newcommand{\qsl}{q\llap{/\kern-0.3pt}}
\newcommand{\qpsl}{q'\llap{/\kern+1.5pt}}
\newcommand{\Qsl}{Q\llap{/\kern+2.5pt}}
\newcommand{\Psl}{P\llap{/\kern+2.5pt}}
\newcommand{\Qtsl}{\tilde{Q}\llap{/\kern+2.5pt}}
\newcommand{\desl}{\partial\llap{/\kern+0.3pt}}
\newcommand{\Dsl}{D\llap{/\kern+1.5pt}}
\newcommand{\Asl}{A\llap{/\kern+0.3pt}}
\newcommand{\etasl}{\eta\llap{/\kern+0.5pt}}
\newcommand{\lsl}{l\llap{/\kern-0.3pt}}
\newcommand{\hsl}{h\llap{/\kern-0.3pt}}
\newcommand{\MET}{E\llap{/\kern1.5pt}_T}

\def\La{\mathscr{L}}
\def\Mamp{\mathcal{A}}

\def\pvec{{\bf p}}

\def\kvec{{\bf k}}

\def\de{\partial}
\def\demub{\de_{\mu}}

\def\demua{\de^{\mu}}

\begin{document}
\selectlanguage{english}
\numberwithin{footnote}{chapter}
\numberwithin{equation}{section}
\newtheorem{note}{Note}[section]
\newtheorem{theorem}{Theorem}[section]
\newtheorem{lemma}{Lemma}[section]
\newtheorem{corollario}{Corollary}[section]
\newtheorem{esempio}{Example}[section]

\selectlanguage{english}
\begin{titlepage}
\begin{center} 
{\LARGE \textsc{GRADUATE COURSE IN PHYSICS \\\vspace{3mm} UNIVERSITY OF PISA}} \\
\vspace{5mm}
\Large \textsc{The School of Graduate Studies in Basic Sciences ``GALILEO GALILEI''}
\vspace{3.5cm}
\end{center} 
\begin{center} 
\huge
Ph.D. Thesis
\vspace{1cm}
\end{center} 
\begin{center}
{\bf \textsf{\Huge Signals of composite\\ particles at the LHC\\}}
\end{center}
\vspace{4cm}
\begin{center}
\LARGE July, 2011
\end{center}
\vspace{2cm}
\begin{center}
\begin{tabular}{ll}
{\bf Candidate} \hspace{2.9cm}& \hspace{2.9cm} {\bf Supervisor}\\
\textsc{Dr. Riccardo Torre} \hspace{2.9cm} & \hspace{2.9cm} \textsc{Prof. Riccardo Barbieri}\\\\
& \hspace{2.9cm} {\bf Thesis reader}\\
& \hspace{2.9cm} \textsc{Prof. Alessandro Strumia}\\
\end{tabular}
\end{center}
\end{titlepage}

\null\vspace{\stretch{1}}
\begin{flushright}
A mia madre
\end{flushright}
\vspace{\stretch{2}}\null 

\selectlanguage{english}
\begin{abstract}
\markboth{{\scshape Abstract}}{{\scshape Abstract}}
\addcontentsline{toc}{chapter}{Abstract}
\indent \lettrine{C}{omposite} particles generated by an unknown strong dynamics can be responsible for the ElectroWeak Symmetry Breaking (EWSB) and can substitute the Standard Model (SM) Higgs boson in keeping perturbative unitarity in the longitudinal $WW$ scattering up to a cut-off $\Lambda\approx 4\pi v$. These new states can be sufficiently light to be observed at the Large Hadron Collider (LHC) and they can even be the first manifestation of new physics at the LHC. Their couplings among themselves and with the SM particles, can be described using reasonable effective Lagrangians and Chiral Perturbation Theory.\\
In the first part of this thesis different possibilities for a strongly interacting EWSB are discussed in details with particular attention to the roles of unitarity in the longitudinal $WW$ scattering and of ElectroWeak Precision Tests (EWPT). Higgsless models with composite vectors and scalars, based on the $SU\(2\)_{L}\times SU\(2\)_{R}/SU\(2\)_{L+R}$ custodial symmetry, are discussed in the context of ElectroWeak Chiral Lagrangians and the phenomenology of the pair productions is studied for the high energy and high luminosity phase of the LHC.\\
In the second part of the thesis the possible signals of single particle production at the early LHC, with $7$ TeV of center of mass energy and $1-5$ fb$^{-1}$ of integrated luminosity, are treated with a phenomenological Lagrangian approach. The final states containing at least one photon emerge as the most promising channels for an early discovery already with tens of inverse picobarns of integrated luminosity.\\
Finally, in the last part of this work, the role of a composite iso-singlet vector in Dark Matter models is discussed and the related LHC phenomenology is studied, giving particular attention to the $Z\gamma$ final state.
\end{abstract}

\newpage
\begin{citazione}
\begin{center}
{\it Per chi viaggia in direzione ostinata e contraria\\
col suo marchio speciale di speciale disperazione,\\
e tra il vomito dei respinti muove gli ultimi passi\\
per consegnare alla morte una goccia di splendore,\\
di umanit\`a, di verit\`a.}
\end{center}
\qquad\qquad\qquad\,\,\qquad\qquad\qquad\qquad\qquad\qquad\qquad{\scshape Fabrizio De Andr\`e}
\end{citazione}

\tableofcontents
\markboth{{\scshape Contents}}{{\scshape Contents}}



\newpage
\chapter*{Introduction}\l{chap1}
\addcontentsline{toc}{chapter}{Introduction}
\markboth{{\scshape Introduction}}{{\scshape Introduction}}
\pagenumbering{arabic}
\lettrine{T}{he} LHC is finally on and in the next few years it will probably answer many of the open questions in the high energy particle physics. The high center of mass (c.o.m.) energy of the LHC makes it sensitive to energy scales well above the Fermi scale $v\approx 246$ GeV, making it possible to discover the mechanism that generates the ElectroWeak Symmetry Breaking (EWSB). The evidence for new phenomena at the Fermi scale is indirectly written in the impressive results of previous collider experiments at CERN (i.e. UA1 and UA2 at the SPS, Aleph, Delphi, Opal and L3 at the LEP) and at Fermilab (i.e. CDF and D0 at the Tevatron): the high precision measurement of all the SM parameters but the Higgs boson mass. In particular, the gauge sector of the SM (SM) has been tested with very high accuracy and the existence in nature of massive vector particles can be translated, in the language of quantum field theory, to the statement that they should be the gauge bosons of a spontaneously broken non-abelian gauge theory. Therefore, a mechanism which realizes the $SU\(2\)_{L}\times U\(1\)_{Y}\rightarrow U\(1\)_{Q}$ spontaneous symmetry breaking must be present. Moreover, the violation of perturbative unitarity in the scattering of longitudinally polarized gauge bosons makes the existence of new degrees of freedom universally accepted.

There are in principle two possibilities to realize the EWSB: we can have a weakly coupled dynamics, mediated by at least one fundamental scalar which acquires a Vacuum Expectation Value (VEV) breaking the ElectroWeak (EW) symmetry or a strong dynamics that becomes non-perturbative above the Fermi scale, realizing the breaking through some condensate. In the first case, a relatively light fundamental Higgs boson exists and the SM, perhaps with a proper supersymmetric extension at the weak scale, necessary to ameliorate the stability of the EW scale to radiative corrections, can perturbatively describe physics up to higher energies, perhaps even the GUT or the Planck scale, without significant changes. In the second case, in which an unknown strong dynamics is responsible for the breaking of the EW symmetry, the SM physics cannot be perturbatively extrapolated up to energies far above the Fermi scale, and new degrees of freedom become relevant at that energy. 

In this thesis we are interested in the strongly interacting hypothesis. The main goal of this work is to apply an effective approach, guided by general principles, but unbiased by explicit model building, to the description of new states at the Fermi scale in different contexts. Since we are going to describe three complementary but well separated implications of new composite particles at the LHC the thesis is consistently divided into three parts.

In Part~\ref{PARTI} we introduce the main results about Spontaneous Symmetry Breaking (SSB) discussing in some details the EWSB in the SM. We consider extensions of the SM only making the assumption, standard in this framework, that the new strong dynamics responsible for the EWSB is by itself invariant under a global chiral $SU\(2\)_{L} \times SU\(2\)_{R}$ symmetry, spontaneously broken to the diagonal $SU\(2\)_{L+R}$ by a non-linear $\s$-model. Making this `custodial symmetry hypothesis' we study the case in which both a vector triplet and a scalar singlet of $SU\(2\)_{L+R}$ are present in the low energy spectrum, i.e. have a mass below the cutoff $\Lambda \approx 4\pi v \approx 3$ TeV. In this case the role of unitarizing the different scattering amplitudes is shared by the scalar and the vector and the perturbative unitarity in the elastic longitudinal gauge boson scattering does not completely constrain the couplings of the scalar and the vector to the gauge and the Goldstone Bosons (GB), only implying a relation among them. Therefore, in this case there is a wider region in the parameter space that is allowed by perturbative unitarity in the elastic longitudinal gauge boson scattering. In this framework we study the phenomenology of the pair productions of the new states, which are sensitive to the most model dependent couplings, that cannot be tested in the single resonant production and that are essential in distinguishing among different explicit models. As we shall see, the pair productions at the LHC require a large statistics and can be accessed only in the nominal energy and luminosity run of the LHC, namely $\sqrt{s}=14$ TeV and $\La=10^{34}$ cm$^{-2}$ s$^{-1}$ $=10$ nb$^{-1}$ s$^{-1}$.

In Part~\ref{PARTII} we apply an effective approach to the discussion of some new physics signals accessible in the first run of the LHC, i.e. with $\sqrt{s}=7$ TeV and $L=\int \mathcal{L} dt\approx 1-5$ fb$^{-1}$. The first run of the LHC is the first possibility to explore the Fermi scale beyond the reach of the Tevatron ($\sqrt{s}=1.96$ TeV). It is therefore very important to understand what we can expect from this run. It was recently pointed out that with about $5$ fb$^{-1}$ of integrated luminosity, the combination of the data collected by both the ATLAS and the CMS experiments can suffice to exclude, or even to discover, the SM Higgs boson down to the LEP limit of $114$ GeV \cite{Collaboration:1323856,CMSCollaboration:2010p1118}. Even if this is certainly the most important task of the early LHC, we think that an open attitude in the expectation for possible signals of new physics is necessary at this stage. Model building prejudices normally play an important role in the determination of the experimental strategies. However, we think that such prejudices should be avoided as much as possible in order to develop model independent search strategies and to maximize the sensitivity to a large range of new physics models. For these reasons, we discuss some  possible signals of new physics that we can expect at the early LHC in a model independent way. In particular we point out that, for parton luminosity reasons, the most favorable states, in comparison with the Tevatron, are new resonances in the $gg$- and $qg$-channels. Moreover, always guided by our open attitude, we also discuss the possibility that a new charged vector state in the $q\bar{q}$-channel, an iso-singlet $W'$, being very weakly constrained by present bound, can be very light and with couplings large enough to appear as an early signal in the first run of the LHC.

In Part~\ref{PARTIII} we consider some possible relations between Dark Matter (DM) and new physics at the EW scale. The present constraints from direct detection experiments like CDMS and XENON100 do not rule out a composite DM candidate, either a scalar or a Dirac fermion, which can couple to the SM particles through a new EW neutral iso-singlet vector. We consider a DM candidate with a mass in the TeV range and a new EW neutral vector coupled to the hypercharge $B$ gauge boson through a kinetic mixing. The new vector state can be accessed at the LHC in its advanced phase giving rise to an interesting phenomenology in the final states with two gauge bosons.
\part{A Strong Sector for the ElectroWeak Symmetry Breaking}\l{PARTI}
\chapter{Unitarity and ElectroWeak Precision Tests}\l{chap2}
\setlength{\epigraphwidth}{.4\textwidth}     
\vspace{-2mm}\epigraph{Whenever I have found out that I have blundered, or that my work has been imperfect, and when I have been contemptuously criticized, and even when I have been overpraised, so that I have felt mortified, it has been my greatest comfort to say hundreds of time to myself that ``I have worked as hard and as well as I could, and no man can do more than this''.}%
              {\textsc{Charles Darwin}}
\vspace{3mm}     
\epigraph{I have not failed. I've just found 10000 ways that won't work. }
              {\textsc{Thomas Edison}}
\vspace{7mm}
\lettrine{I}{n} this chapter we introduce the basic concepts and results on SSB, focusing in particular on the EW theory in the SM. One of the main results is to show that the EWSB can be realized in the SM in two different ways:
\bit
\item {\bf linear realization}: the $SU\(2\)_{L}\times U\(1\)_{Y}\to U\(1\)_{Q}$ SSB is realized with a linear $\s$-model by introducing an $SU\(2\)_{L}$ complex doublet which acquires a VEV. Three of the four degrees of freedom in the complex doublets are the GBs eaten up by the massive gauge bosons while a new scalar degree of freedom, the Higgs boson, appears in the spectrum. The resulting theory turns out to be perturbative up to very high energies provided that the Higgs boson is sufficiently light ($m_{h}\lesssim 1$ TeV). The agreement with ElectroWeak Precision Tests (EWPT) requires the Higgs boson to be lighter than $\(145,149,194\)$ GeV at $\(90,95,99\)\%$ Confidence Level (CL) \cite{Nakamura:2010dr}.
\item {\bf non-linear realization}: a non-linear field is constructed with the GBs of the $SU\(2\)_{L}\times SU\(2\)_{R}\to SU\(2\)_{L+R}$ SSB. This field breaks spontaneously the gauge symmetry through a non-linear $\s$-model. In this case no additional degrees of freedom appear in the spectrum. The resulting theory violates perturbative unitarity at the TeV scale and cannot account for the EWPT. New degrees of freedom are needed to ameliorate the high energy behavior of the scattering amplitudes and to account for the EWPT.
\eit

In order to show explicitly the differences between these two approaches we compute, using the Equivalence Theorem, the amplitudes for the scattering process $W^{a}_{L}W^{b}_{L}\to W^{c}_{L}W^{d}_{L}$ and the one loop contributions to ElectroWeak Precision Observables (EWPO) $S$ and $T$ in different frameworks.

This chapter is organized as follows. In Section~\ref{SSBOAGaugeS} we introduce the concept of SSB of a global symmetry and we enunciate the Goldstone's Theorem making the example of the chiral symmetry in QCD with two flavors. In Section~\ref{SSBOAGlobalS} we introduce the Higgs phenomenon and the Equivalence Theorem and we study the SSB in the framework of gauge theories discussing the two possible realizations of the EWSB. Sections~\ref{TSMWAFHB}, \ref{TROTHB}, \ref{TCOAHLMWAHCV} and \ref{ASVS} are devoted to the study of the perturbative unitarity and the EWPT in four different frameworks, respectively: the SM without the Higgs boson, the SM plus a new composite scalar, the SM plus two new composite vectors of opposite parity and the SM with both a scalar and two composite vectors in the low energy spectrum.



\section{Spontaneous symmetry breaking of a global symmetry}\l{SSBOAGaugeS}
A continuous global symmetry is spontaneously broken when the ground state is not invariant under the action of some generators that are called the broken generators. A fundamental result on SSB is the Goldstone's Theorem \cite{Goldstone:1961p3591,Goldstone:1962p3592,Nambu:1961p3590,Nambu:1960p3588,Nambu:1961p3589}. In Quantum Field Theory (QFT) this theorem can be enunciated as follows\footnote{An example of a non-relativistic application of the theorem is the magnetization of a ferromagnet: above the Curie temperature the magnetization is zero and the system is invariant under simultaneous rotation of all the spins, while below the Curie temperature the system acquires a magnetization and the ground state with all the spins aligned is not invariant anymore.}:\\

\noindent {\bf Goldstone's Theorem}\\
{\it Consider a theory invariant under a continuous global symmetry with $n$ generators $T^{i}$, $i=1,\ldots,n$. If the ground state of the theory (the vacuum), is not invariant under the action of some of the generators $T^{i}$, $i=k,\ldots, n, $ then there must exist $n-k$ massless bosons, scalar or pseudo-scalar, associated to each broken generator (i.e. each generator that does not annihilate the vacuum) and having the same quantum numbers under the unbroken subgroup. These massless bosons are called the Nambu-Goldstone Bosons or simply Goldstone Bosons (GB) of the spontaneously broken symmetry.}

\subsection{Spontaneous symmetry breaking of chiral symmetry in QCD with two flavors}
A simple and important example of a spontaneously broken global symmetry is the chiral symmetry $SU\(2\)_{L}\times SU\(2\)_{R}$ in QCD with two flavors. This symmetry is defined by the transformations
\be\l{chsy}
\bry{ll}
\Psi_{L}\to \Psi_{L}'=U_{L}\Psi_{L}\qquad &U_{L}\in SU\(2\)_{L}\\
\Psi_{R}\to \Psi_{R}'=U_{R}\Psi_{R}\qquad &U_{R}\in SU\(2\)_{R}\,,
\ery
\ee
where $\Psi_{L,R}$ are the left- and right-handed doublets containing the $u$ and $d$ quarks. The matrices $U_{L,R}$ can be written in terms of the $SU\(2\)_{L,R}$ generators $T^{a}_{L,R}$ as
\be\l{chsy2}
U_{L}=e^{-i\a_{L}^{a}T_{L}^{a}}\,,\qquad\qquad U_{R}=e^{-i\a_{R}^{a}T_{R}^{a}}\,.
\ee
Now, the existence of a nonzero vacuum expectation value of a quark-antiquark pair
\be\l{chsy3}
\<0|\bar{\Psi}\Psi|0\>=\<0|\bar{\Psi}_{L}\Psi_{R}+\bar{\Psi}_{R}\Psi_{L} |0\>\neq 0\,,
\ee
which transforms non trivially under Eq.~\eqref{chsy} when $U_{L}\neq U_{R}$, is an indication that the chiral symmetry $SU\(2\)_{L}\times SU\(2\)_{R}$ is spontaneously broken down to the vectorial $SU\(2\)_{V}\equiv SU\(2\)_{L+R}$ ($U_{L}=U_{R}$). The Goldstone's Theorem therefore implies the existence of as many massless spin-$0$ particles as the number of broken generators, i.e. three, with the same quantum numbers. In particular we expect three massless spin-$0$ particles to be generated by the three axial vector currents. On the other hand, the real strong interactions do not contain any massless particles since the chiral symmetry $SU\(2\)_{L}\times SU\(2\)_{R}$ is only approximate. For these reasons we expect the corresponding pseudo-GBs to have a small mass. In fact we know experimentally that an isospin triplet of relatively light mesons exists, the pions, which are parity odd, as we would expect for a quark-antiquark bound state. Moreover, the smallness of the pion masses $\sim 130$ MeV with respect to the first resonance mass, i.e. the rho mass $\sim 780$ MeV, guarantees that the chiral symmetry $SU\(2\)_{L}\times SU\(2\)_{R}$ is a good approximate symmetry of the strong interactions.

\section{Spontaneous symmetry breaking of a gauge symmetry}\l{SSBOAGlobalS}
As we have seen, the Goldstone's Theorem implies the appearance of new physical degrees of freedom in the spectrum and holds only for theories with a spontaneously broken continuous global symmetry. In the case of gauge theories the appearance of the GB is associated to the appearance of the masses and longitudinal polarizations of the gauge bosons corresponding to the broken generators. In this case no new physical degrees of freedom appear in the spectrum and the Goldstone's Theorem is substituted by the Higgs phenomenon \cite{Anderson:1963p3593,Higgs:1964p3594,Englert:1964p3595,Higgs:1966p3596,Kibble:1967p3597}:\\

\noindent {\bf Higgs phenomenon}\\
{\it The GBs (in this case sometimes referred  to as ``would be'' GBs) associated to the global symmetry breaking corresponding to the spontaneous breaking of the gauge symmetry  do not manifest explicitly in the physical spectrum. They instead ``provide'' the longitudinal polarization to the new massive gauge bosons. For this reason they are said to be ``eaten up'' by the corresponding massive gauge bosons.}

\subsection{ElectroWeak Symmetry Breaking in the SM: the linear realization}
In this case the most relevant example is the SM of EW interactions. The minimal EW Lagrangian is given by
\be\l{EWLag}
\La_{\text{EW}}=-\f{1}{4}B_{\mu\nu}B^{\mu\nu}-\f{1}{4}W_{\mu\nu}^{a}W^{\mu\nu\,a}+i\ovl{\Psi} \Dsl \Psi\,,
\ee
where $B_{\mu\nu}$ and $W_{\mu\nu}^{a}$ are respectively the $U\(1\)_{Y}$ and $SU\(2\)_{L}$ gauge vectors, $\Psi$ is a vector containing all the SM fermion fields and $D_{\mu}$ is the covariant derivative acting on the fermion fields according to\footnote{We adopt a normalization for the hypercharge such that the electric charge is given by $Q=T^{3}_{L}+Y$.}
\be\l{covdevferm}
D_{\mu}\Psi=\(\demub-ig W_{\mu}^{a} T^{a}-ig'YB_{\mu}\)\,,\qquad T^{a}=\f{\s^{a}}{2}
\ee
with $\s^{a}$ the ordinary Pauli matrices.
All the particles in Eq.~\eqref{EWLag} are massless and $\La_{\text{EW}}$ is invariant under the gauge group $SU\(2\)_{L}\times U\(1\)_{Y}$. Since we know experimentally that the weak gauge bosons $W^{\pm}$ and $Z$ and the fermions have masses and that the only symmetry that is realized in nature is the Electromagnetic $U\(1\)_{Q}$, we need a mechanism to realize the SSB $SU\(2\)_{L}\times U\(1\)_{Y}/U\(1\)_{Q}$ and to generate the fermion masses. One way to do it is to add new ingredients to the minimal EW Lagrangian \eqref{EWLag}. The minimal choice is to add a complex scalar doublet of $SU\(2\)_{L}$ with hypercharge $Y=1/2$ defined by
\be\l{HiggsDoub}
H=\(\bry{c}H^{+}\(x\)\\ H^{0}\(x\)\ery\)=\f{1}{\sqrt{2}}\(\bry{c}\pi^{1}\(x\)-i\pi^{2}\(x\)\\ h\(x\)+i\pi^{0}\(x\)\ery\)\,,
\ee
where $h\(x\),\pi^{1}\(x\),\pi^{2}\(x\),\pi^{3}\(x\)$ are four real scalar functions of the space-time coordinates and the signs and normalizations have been chosen for later convenience. We can now write the EWSB Lagrangian
\be\l{SBLag}
\La_{\text{EWSB}}=\(D_{\mu}H\)^{\dag} \(D^{\mu}H\)-V\(H\)\,,\qquad V\(H\)=-\mu^{2}H^{\dag}H+\lambda \(H^{\dag}H\)^{2}\,,
\ee
where the SM covariant derivative acting on $H$ is given by
\be\l{CovDev}
D_{\mu}=\demub - \f{ig}{2}W^{\mu\,a}\s^{a}-\f{ig'}{2}B^{\mu}\,.
\ee
For $\mu^{2}>0$ the minimum of the potential $V\(H\)$ is at\footnote{We use the convention $v=\(\sqrt{2}G_{F}\)^{-1/2}$ so that $v\approx 246.22$ GeV.}
\be\l{HigPotMin}
H^{\dag}H=\f{2\mu^{2}}{\lambda}\equiv v^{2}
\ee
and therefore the field $H$ acquires a VEV and the $SU\(2\)_{L}\times U\(1\)_{Y}$ symmetry is spontaneously broken down to the electromagnetic $U\(1\)_{Q}$.
Note that the potential in Eq.~\eqref{SBLag}, only dependent upon the quantity
\be\l{HigSquar}
H^{\dag}H=\f{1}{2}\(h^{2}+\(\pi^{1}\)^{2}+\(\pi^{2}\)^{2}+\(\pi^{3}\)^{2}\)\,,
\ee
has an $SO\(4\)\sim SU\(2\)\times SU\(2\)$ global symmetry of which the gauged $SU\(2\)\times U\(1\)$ is a subgroup. We will see in the following that this $SO\(4\)$ symmetry has important consequences on the structure of the EWSB sector of the SM. Now, choosing the vacuum configuration 
\be\l{Vacuum}
\<H\>= \(\bry{c}0\\ \f{v}{\sqrt{2}}\ery\)\,,
\ee
substituting it into the Lagrangian \eqref{SBLag} and taking into account the definition of the covariant derivative \eqref{CovDev} we find the mass terms for the gauge bosons
\be\l{LMass}
\La_{M}=-\f{v^{2}}{4}\(g^{2}W^{+}_{\mu}W^{-\mu}+\f{g^{2}+g^{\prime\,2}}{2}Z_{\mu}Z^{\mu}\)\,,
\ee
where we have used the definition of the charge and mass eigenstates given by
\be\l{Eigen}
\bry{l}
W^{\pm}_{\mu}=\dst \f{1}{\sqrt{2}}\(W^{1}_{\mu}\mp iW^{2}_{\mu}\)\,,\\\\
A_{\mu}=\sin\theta_{W}W^{3}_{\mu}+\cos\theta_{W}B_{\mu}\,,\\\\
Z_{\mu}=\cos\theta_{W}W^{3}_{\mu}-\sin\theta_{W}B_{\mu}\,,
\ery
\ee
and we have defined the weak mixing angle
\be\l{thetaW}
\tan\theta_{W}=\f{g'}{g}\,.
\ee
The $W$ and the $Z$ bosons have now acquired the masses
\be\l{masses}
m_{W}^{2}=\f{g^{2}v^{2}}{4}\,,\qquad\qquad m_{Z}^{2}=\f{\(g^{2}+g^{\prime\,2}\)v^{2}}{4}\,.
\ee
These masses are related by the formula
\be\l{rho1}
\f{m_{W}^{2}}{m_{Z}^{2}\cos^{2}\theta_{W}}\equiv \rho=1\,.
\ee
and in the limit $g'\to 0$ they become degenerate. This is not a consequence of the gauge symmetry but rather of the global $SO\(4\)\sim SU\(2\)_{L}\times SU\(2\)_{R}$ custodial symmetry of the potential in Eq.~\eqref{SBLag}.
To prove this fact let us define the $2\times 2$ matrix
\be\l{HiggsMatr}
\mathcal{H}=\(i\s^{2}H^{*},H\)=\(\bry{cc}H^{0\,*}& H^{+}\\ -H^{-}&H^{0}\ery\)\,,
\ee
transforming under the chiral $SU\(2\)_{L}\times SU\(2\)_{R}$ as
\be\l{HiggsMatrTransf}
\mathcal{H}\to g_{L}^{\dag}\mathcal{H}g_{R}\,,\qquad\qquad g_{L,R}=e^{-i\f{\o_{L,R}^{a}\s^{a}}{2}}\in SU\(2\)_{L,R}\,.
\ee
We can now write the symmetry breaking Lagrangian \eqref{SBLag} in the equivalent form
\be\l{SBLag2}
\La_{\text{EWSB}}=\<D^{\mu}\mathcal{H}\(D_{\mu}\mathcal{H}\)^{\dag}\> -V\(\mathcal{H}\)\,,
\ee
with 
\be\l{CovDev2}
D_{\mu}\mathcal{H} =\demub \mathcal{H} -i\hat{W}_{\mu}\mathcal{H} +i\mathcal{H} \hat{B}_{\mu}\,,
\ee
and
\be\l{Gaugematrices}
\hat{W}_{\mu}=\f{g}{2}W_{\mu}^{a}\s^{a}\,,\qquad\qquad \hat{B}_{\mu}=\f{g'}{2}B_{\mu}\s^{3}\,,
\ee
and where $\<\,\>$ is the trace over $SU\(2\)$.
Now, since the $SU\(2\)_{L}$ of the global $SU\(2\)_{L}\times SU\(2\)_{R}$ symmetry coincides with the $SU\(2\)_{L}$ gauge group, the coupling $g$ does not break the $SU\(2\)_{L}\times SU\(2\)_{R}$ symmetry and the $W_{\mu}^{a}$ transforms as a triplet of $SU\(2\)_{L}$, whereas the coupling $g'$ of the hypercharge $U\(1\)_{Y}$ whose generator coincides with the $T_{3R}$ of $SU\(2\)_{R}$ breaks the full $SU\(2\)_{L}\times SU\(2\)_{R}$ symmetry. Moreover, since the vacuum configuration
\be\l{vacuum2}
\<\mathcal{H}\>=\(\bry{cc}v& 0\\ 0&v \ery\)
\ee
is left invariant under the diagonal $SU\(2\)_{V}\equiv SU\(2\)_{L+R}$ with $g_{L}=g_{R}$ in Eq.~\eqref{HiggsMatrTransf}, the neutral and charged vector boson masses are degenerate in the limit $g'\to 0$.\\
Finally, since $\mathcal{H}$ transforms linearly under $SU\(2\)_{L}$, the Lagrangian \eqref{SBLag2} is an example of a linear $\s$-model and for these reason the SSB in the EW SM is said to be linearly realized.

\subsection{ElectroWeak Symmetry Breaking in the SM: the non-linear realization}\l{sec2.2.2}
The $SU\(2\)_{L}\times U\(1\)_{Y}\to U\(1\)_{Q}$ SSB can also been realized through a non-linear $\s$-model without the need of introducing in the spectrum the physical Higgs boson. To prove this fact let us proceed as follows: we perform a polar parametrization of the field $\mathcal{H}$ according to
\be\l{PolarHiggs}
\mathcal{H}=\f{1}{\sqrt{2}}\(v+h\(x\)\)\S\(x\)\,,\qquad\qquad \S\(x\)=e^{i\f{\pi^{a}\s^{a}}{v}}\,.
\ee
Taking the limit $\lambda\gg 1$ (heavy Higgs), we can rewrite the EWSB Lagrangian in the form
\be\l{SBLag3}
\La_{\text{EWSB}}=\f{v^{2}}{4}\<D^{\mu}\S\(D_{\mu}\S\)^{\dag}\> +O\(\f{H}{v}\)\,,
\ee
where the covariant derivative is the same as in \eqref{CovDev2}.
Expanding the Lagrangian \eqref{SBLag3} to the zero order in $\pi$, i.e. setting $\S=1$ we find again the mass terms in Eq.~\eqref{LMass}. 

As we will see in some details in the following, the Lagrangian \eqref{SBLag3} describes the interactions of the GBs with the SM gauge bosons in the framework of
\be
SU\(2\)_{L}\times SU\(2\)_{R}/SU\(2\)_{L+R}
\ee
spontaneous symmetry breaking and is the starting point of the formalism of the ElectroWeak Chiral Lagrangian (EWChL). The EWSB generated by the Lagrangian \eqref{SBLag3} is realized non-linearly, in the sense that the $\hat{\pi}$ fields transform non linearly under the gauge symmetry.

\subsection{The Goldstone Boson Equivalence Theorem}\l{EQTEO}
So far we have shown that we can realize the EWSB with a linear $\s$-model, i.e. explicitly introducing a new scalar degree of freedom known as the Higgs boson or with a non-linear $\s$-model in the context of ElectroWeal Chiral Lagrangians. In the next section we show that the role of the Higgs boson goes well beyond the simple breaking of the EW symmetry being of fundamental importance in guaranteeing the perturbative unitarity of the SM up to scales much higher than the Fermi scale. We will also show that it plays a leading role in making the SM in agreement with EWPT.\\

Before taking these steps, it is useful to introduce another important result concerning the Higgs phenomenon known as the Equivalence Theorem. This Theorem can be enunciated as follows:\\

\noindent {\bf Equivalence Theorem}\\
{\it In any renormalizable $R_{\xi}$ gauge of a spontaneously broken gauge theory, the scattering amplitude involving $N$ longitudinally polarized gauge bosons is equal, up to corrections of $O\(M_{W}/\sqrt{s}\)$, to $\(-i\)^{N}$ times the same scattering amplitude computed substituting all the external massive gauge boson legs with the corresponding GBs:}
\be\l{eqteo1}
\bry{l}
\dst \Mamp\Big(W_{L}\(p_{1}\),\ldots, W_{L}\(p_{n}\)+X\to W_{L}\(k_{1}\),\ldots,W_{L}\(k_{m}\)+Y\Big)\\\\
\dst =\(-i\)^{n+m}\Mamp\Big(\pi\(p_{1}\),\ldots, \pi\(p_{n}\)+X\to \pi\(k_{1}\),\ldots,\pi\(k_{m}\)+Y\Big)\(1+O\(\f{M_{W}}{\sqrt{s}}\)\)\,.
\ery
\ee 
This theorem was firstly proved by Cornwall {\it et al.} \cite{Cornwall:1974kt} and by Vayonakis \cite{Vayonakis:1976vu} at tree-level and for only one boson leg and then it was extended at any order (in the Feynman-'t Hooft gauge) by Lee, Quigg and Thacker \cite{Lee:1977p1572}. The complete proof at any order in perturbation theory and for any number of legs was given by Chanowitz and Gaillard \cite{Chanowitz:1985hj} and Gounaris {\it et al.} \cite{Gounaris:1986cr}. Many other works analyzed the influence of renormalization scheme and of gauge choices on the Equivalence Theorem as e.g. Refs.~\cite{Yao:1988aj} and \cite{Veltman:1989ud}. A proof of the theorem in the Chiral Lagrangian formalism was given by Dobado and Pelaez \cite{Dobado:1994p61,Dobado:1994p84} and by He {\it et al.} \cite{He:1993qa,He:1994fe}.

\section{The Standard Model without the fundamental Higgs boson}\l{TSMWAFHB}
We consider the Higgsless SM described by the Lagrangian
\be\l{HlessSM}
\La_{\hsl-\text{SM}}=\La_{\text{EW}}+\La_{\hsl-\text{EWSB}}+\La_{\hsl-\text{Yuk}}\,,
\ee
where $\La_{\text{EW}}$ has been defined in Eq.~\eqref{EWLag} and we have defined the Higgsless EWSB Lagrangian
\be\l{HlessSB}
\La_{\hsl-\text{EWSB}}=\f{v^{2}}{4}\<D^{\mu}\S\(D_{\mu}\S\)^{\dag}\>\,.
\ee
and the Higgsless Yukawa interaction
\be\l{HlessYuk}
\La_{\hsl-\text{Yuk}}=-m_{i}\bar{\psi}_{L\,i}\S\psi_{R\,i}\,.
\ee
In the two following subsections we use this Lagrangian and the Equivalence Theorem to discuss respectively, the perturbative unitarity in the elastic longitudinal $WW$ scattering and the EWPO $S$ and $T$.

\subsection{Elastic unitarity}\l{ElasticUnitarity1}
In the $g'=0$ limit, using the $SU\(2\)_{V}$ and the Bose symmetries, we can decompose the four longitudinal gauge boson amplitudes as
\be\l{k2}
\Mamp\(W^{a}_{L}W^{b}_{L}\to W^{c}_{L}W^{d}_{L}\)=\Mamp\(s,t,u\)\d^{ab}\d^{cd}+\Mamp\(t,s,u\)\d^{ac}\d^{bd}+\Mamp\(u,t,s\)\d^{ad}\d^{bc}\,.
\ee
Moreover, using the Equivalence Theorem \eqref{eqteo1}, at high energy (i.e. $s\gg M_{W}^{2}$) we can write
\be\l{k1}
\Mamp\(W^{a}_{L}W^{b}_{L}\to W^{c}_{L}W^{d}_{L}\)\approx\Mamp\(\pi^{a}\pi^{b}\to \pi^{c}\pi^{d}\)\,.
\ee

The relevant Feynman diagrams\footnote{In this thesis the ``arrow of time'', from initial to final state for the Feynman diagrams, flows from bottom to top.} for the $\pi^{a}\pi^{b}\to \pi^{c}\pi^{d}$ process in the SM without the Higgs boson are given in Fig.~\ref{fig:fourpionsHiggsless}.
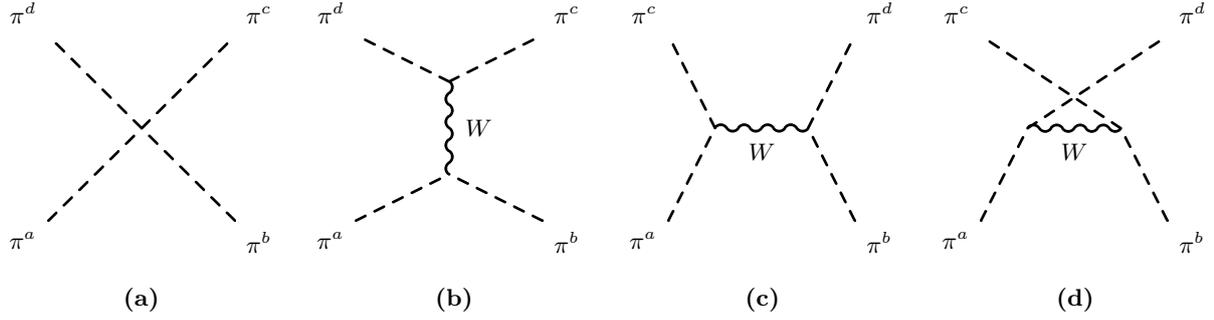
\begin{figure}[!htb]
\centering
\begin{minipage}[t]{3.3cm}
\begin{center}
\begin{fmffile}{pppp1}
  \fmfframe(1,7)(1,7){ 
          \begin{fmfgraph*}(70,70)
          \fmfstraight
          \fmflabel{\footnotesize{$\pi^{c}$}}{o2}
          \fmflabel{\footnotesize{$\pi^{a}$}}{i1}
          \fmflabel{\footnotesize{$\pi^{d}$}}{o1}
          \fmflabel{\footnotesize{$\pi^{b}$}}{i2}          
          \fmfbottom{i1,i2}
          \fmftop{o1,o2}
          \fmf{dashes}{i1,v1}
          \fmf{dashes}{i2,v1}
          \fmf{dashes}{v1,o1}
          \fmf{dashes}{v1,o2}
          \end{fmfgraph*}} \\\vspace{4mm}  \footnotesize{{\bf (a)}}\l{pppp1}\vspace{4mm}
\end{fmffile}
\end{center}
\end{minipage}
\ \hspace{3mm} \
\begin{minipage}[t]{3.3cm}
\begin{center}
        \begin{fmffile}{pppp2}
  \fmfframe(1,7)(1,7){ 
   \begin{fmfgraph*}(70,70)
          \fmfstraight
    \fmfbottom{i1,i2}
    \fmftop{o1,o2}
          \fmflabel{\footnotesize{$\pi^{c}$}}{o2}
          \fmflabel{\footnotesize{$\pi^{a}$}}{i1}
          \fmflabel{\footnotesize{$\pi^{d}$}}{o1}
          \fmflabel{\footnotesize{$\pi^{b}$}}{i2}   
    \fmf{dashes}{i1,v1}
    \fmf{dashes}{i2,v1}
    \fmf{dashes}{v2,o1}
    \fmf{dashes}{v2,o2}
    \fmf{photon,label={\footnotesize{$W$}}}{v1,v2}
    \end{fmfgraph*}
        }   \\\vspace{4mm}  \footnotesize{{\bf (b)}}\l{pppp2}%
      \end{fmffile}
\end{center}
\end{minipage}
\ \hspace{3mm} \
\begin{minipage}[t]{3.3cm}
\begin{center}
        \begin{fmffile}{pppp3}
  \fmfframe(1,7)(1,7){ 
   \begin{fmfgraph*}(70,70)
          \fmfstraight
    \fmfleft{i1,i2}
    \fmfright{o1,o2}
    \fmflabel{\footnotesize{$\pi^{a}$}}{i1}
    \fmflabel{\footnotesize{$\pi^{b}$}}{o1}
    \fmflabel{\footnotesize{$\pi^{c}$}}{i2}
    \fmflabel{\footnotesize{$\pi^{d}$}}{o2}
    \fmf{dashes}{i1,v1}
    \fmf{dashes}{v1,i2}
    \fmf{dashes}{o1,v2}
    \fmf{dashes}{v2,o2}
    \fmf{photon,l.side=right,label={\footnotesize{$W$}}}{v1,v2}
    \end{fmfgraph*}
        }   \\\vspace{4mm}  \footnotesize{{\bf (c)}} \l{pppp3}%
      \end{fmffile}
\end{center}
\end{minipage}
\ \hspace{3mm} \
\begin{minipage}[t]{3.3cm}
\begin{center}
        \begin{fmffile}{pppp4}
  \fmfframe(1,7)(1,7){ 
   \begin{fmfgraph*}(70,70)
          \fmfstraight
    \fmfleft{i1,i2}
    \fmfright{o1,o2}
    \fmflabel{\footnotesize{$\pi^{a}$}}{i1}
    \fmflabel{\footnotesize{$\pi^{b}$}}{o1}
    \fmflabel{\footnotesize{$\pi^{c}$}}{i2}
    \fmflabel{\footnotesize{$\pi^{d}$}}{o2}
    \fmf{dashes}{i1,v1}
    \fmf{phantom}{v1,i2}
    \fmf{dashes}{o1,v2}
    \fmf{phantom}{v2,o2}
    \fmf{photon,l.side=right,label={\footnotesize{$W$}}}{v1,v2}
    \fmffreeze
    \fmf{dashes}{v1,o2}
    \fmf{dashes}{v2,i2}
    \end{fmfgraph*}
        }  \\\vspace{4mm}  \footnotesize{{\bf (d)}} \l{pppp4}\vspace{4mm}%
      \end{fmffile}
\end{center}
\end{minipage}
\caption{The four Feynman diagrams contributing to the four-pion scattering in the SM without the Higgs boson: (a) represents the four-pion contact interaction; (b), (c) and (d) are respectively the $s$, $t$ and $u$ channels of the $W$ exchange contribution.}\l{fig:fourpionsHiggsless}
\end{figure}\\
A simple calculation shows that the contribution of the first Feynman diagram grows asymptotically with the c.o.m.~energy squared $s$, while the contributions of the last three Feynman diagrams, i.e. the diagrams with a $W$ boson exchange, have a constant asymptotic behavior and are therefore subleading with respect to the first one. The final result for the asymptotic amplitude is
\be\l{k3}
\Mamp\(s,t,u\)_{\hsl-\text{SM}}\approx \f{s}{v^{2}}\,.
\ee
This amplitude grows with the c.o.m.~energy squared $s$, leading to a violation of perturbative unitarity of the $S$-matrix.

One way to quantify the energy scale at which perturbative unitarity is violated is to study the perturbative unitarity of the fixed ``isospin'' amplitudes, where ``isospin'' refers to the generators of $SU\(2\)_{V}$. From Eq.~\eqref{k2} we can simply compute the fixed total isospin amplitudes $T\(I\)$ obtaining
\be\l{k3a}
\bry{l}
T\(0\)=3\Mamp\(s,t,u\)+\Mamp\(t,s,u\)+\Mamp\(u,t,s\)\,,\\\\
T\(1\)=\Mamp\(t,s,u\)-\Mamp\(u,t,s\)\,,\\\\
T\(2\)=\Mamp\(t,s,u\)+\Mamp\(u,t,s\)\,.
\ery
\ee
These amplitudes can be expanded in partial waves according to
\be\l{k3b}
T\(I\)=32\pi\sum_{l=0}^{\infty}a_{l}^{I}\(2l+1\)P_{l}\(\cos\theta\)\,,
\ee
where $P_{l}\(\cos\theta\)$ are the Legendre polynomials and the expansion coefficients $a_{l}^{I}$ are given by
\be\l{k3c}
a_{l}^{I}=\f{1}{64\pi}\int_{-1}^{1}d\(\cos\theta\)P_{l}\(\cos\theta\)T\(I\)\,.
\ee
Now, taking into account only the elastic channel, the Optical Theorem written in the form
\be\l{k3d}
\Im\(a_{l}^{I}\)=\sqrt{1-\f{4m_{W}^{2}}{s}} |a_{l}^{I}|^{2}\,,\qquad\mathop \to \limits_{s\gg M_{W}^{2}}\qquad \Im\(a_{l}^{I}\)= |a_{l}^{I}|^{2}\,,
\ee
implies the perturbative unitarity condition
\be\l{k3e}
\Im\(a_{l}^{I}\)=\Im\(|a_{l}^{I}|e^{i\d_{l}^{I}}\)=|a_{l}^{I}|\sin\(\d_{l}^{I}\)\geq |a_{l}^{I}|^{2}\qquad\Longrightarrow \qquad  |a_{l}^{I}|\leq \sin\(\d_{l}^{I}\)
\ee
ant thus
\be\l{k3f}
|a_{l}^{I}|\leq 1\,,\qquad \forall\, I=0,1,2\quad \text{and}\quad\forall \,l=0,1,\ldots \infty\,.
\ee
It is simple to prove that the strongest unitarity constraint is given by $a^{0}_{0}\leq 1$. In this case we have \cite{Lee:1977p1572}
\be\l{k4}
\lim_{s\gg M_{W}^{2}}  \(a_{0}^{0}\)_{\hsl-\text{SM}}=\f{1}{64\pi}\f{s}{v^{2}}\int_{-1}^{1}dx\Big[3-\f{1}{2}\(1-x\)-\f{1}{2}\(1+x\)\Big]=\f{s}{16\pi v^{2}}\,,
\ee
so that the strongest unitarity constraint gives the elastic unitarity bound
\be\l{k5}
|\lim_{s\gg M_{W}^{2}}\(a_{0}^{0}\)_{\hsl-\text{SM}}|\leq 1 \quad\Longrightarrow\quad \sqrt{s}\leq 4\sqrt{\pi} v =\Lambda_{\hsl-\text{SM}}\approx 1.7\, \text{TeV}\,.
\ee
on the Higgsless SM cutoff.
In other words, Eq.~\eqref{k5} says that the SM without the Higgs boson violates perturbative unitarity at $\Lambda_{\hsl-\text{SM}}\approx 1.7 \,\text{TeV}$\,.

\subsection{ElectroWeak Precision Tests}\l{EWPT1}
The importance of the oblique corrections to the EWPO in the SM and in its extensions has been historically introduced in the nineties by Peskin and Takeuchi \cite{Peskin:1990p1615,Peskin:1992p1662} and by Altarelli, Barbieri and collaborators \cite{Altarelli:1991p1619,Altarelli:1991fk,Altarelli:1993bh,Altarelli:1993p1620,Altarelli:1994iz,Altarelli:1997et}. These oblique corrections can be conveniently expressed in terms of the SM gauge boson vacuum polarization correlators through the effective Lagrangian
\be\l{oblcorr}
\La_{\text{vac-pol}}=-\f{1}{2}W_{\mu}^{3}\Pi_{33}^{\mu\nu}\(q^{2}\)W_{\nu}^{3}-\f{1}{2}B_{\mu}\Pi_{00}^{\mu\nu}\(q^{2}\)B_{\nu}-W_{\mu}^{3}\Pi_{30}^{\mu\nu}\(q^{2}\)B_{\nu}-W_{\mu}^{+}\Pi_{WW}^{\mu\nu}\(q^{2}\)W_{\nu}^{-}\,,
\ee
where the correlators $\Pi^{\mu\nu}_{ij}\(q^{2}\)$ are defined by
\be\l{corr}
\Pi^{\mu\nu}_{ij}\(q^{2}\)=\(\f{q^{\mu}q^{\nu}}{q^{2}}-g^{\mu\nu}\)\Pi_{ij}\(q^{2}\)\,.
\ee
In particular, we can define the two parameters $S$ and $T$ as
\bes
\begin{align}
& \hat{S}=\f{g}{g'}\Pi_{30}'\(M_{Z}^{2}\)=\f{g}{g'}\f{d\Pi_{30}\(q^{2}\)}{dq^{2}}\Bigg|_{q^{2}=M_{Z}^{2}}\,,\qquad &&S=\f{4\sin^{2}\theta_{W}}{\a}\hat{S}\,,\l{Shat}\\\notag\\
&\hat{T}=\f{\Pi_{33}\(0\)-\Pi_{WW}\(0\)}{m_{W}^{2}}\,,\qquad && T=\f{1}{\a}\hat{T}\,.\l{That}
\end{align}
\ees
From these definitions it immediately follows that $\hat{T}=\rho-1$, where $\rho$ is the function defined in Eq.~\eqref{rho1}. Therefore, the $T$ parameter is a measure of the breaking of the custodial symmetry. This fact can also be seen in terms of the GBs as follows. In the mass and charge eigenstate basis for the gauge bosons (see Eq.~\eqref{Eigen}) and for the GBs
\be\l{GBChargeEigen}
\pi^{\pm}=\dst \f{1}{\sqrt{2}}\(\pi^{1}\mp i\pi^{2}\)\,,\qquad\qquad \pi^{0}=\pi^{3}\,,
\ee
expanding the EWSB Lagrangian \eqref{HlessSM} to the first order in $\pi^{a}$ we find the Lagrangian
\be\l{P1}
\La\(\pi^{a}\)=Z^{\(+\)}\Bigg|\demub \pi^{+}+\f{gv}{2}W^{+}_{\mu}\Bigg|^{2}+\f{Z^{\(0\)}}{2}\Bigg|\demub \pi^{0}+\f{gv}{2\cos\theta_{W}}Z_{\mu}\Bigg|^{2}
\ee
where we have introduced the two field-strength renormalization constants
\be\l{P2}
Z^{\(+\)}\equiv1+\d Z^{\(+\)}\,,\qquad\qquad\qquad Z^{\(0\)}\equiv1+\d Z^{\(0\)}
\ee
of the charged and neutral GBs.
Expanding the mass terms in Eq.~\eqref{P1} we can write
\be\l{P3}
\La_{M}=Z^{\(+\)}\f{g^{2}v^{2}}{4}W^{+}_{\mu}W^{-\,\mu}+\f{Z^{\(0\)}}{2}\f{g^{2}v^{2}}{4\cos^{2}\theta_{W}}Z_{\mu}Z^{\mu}\,.
\ee
\begin{figure}[!htb]
\begin{minipage}[b]{3.2cm}
  \centering
   \begin{fmffile}{Shat1}
 \fmfframe(1,7)(1,7){
          \begin{fmfgraph*}(90,45)
          \fmfstraight
          \fmftop{t1,t2,t3}
          \fmfbottom{b1,b2,b3}
          \fmfleft{l1,l2,l3,l4,l5}
          \fmfright{r1,r2,r3,r4,r5}
          \fmf{phantom}{v3,v5}          
          \fmf{photon,tension=.6}{l3,v1}
          \fmf{dashes,left,label={\footnotesize{$\pi$}},tension=.2}{v1,v2}
          \fmf{dashes,left,label={\footnotesize{$\pi$}},tension=.2}{v2,v1}
          \fmf{photon,tension=.6}{v2,r3}
          \fmf{phantom}{t2,v3}
          \fmf{phantom,tension=.195}{v3,v4}
          \fmf{phantom}{v4,b2}
	 \fmffreeze
          \end{fmfgraph*}}\vspace{-5mm}
      \end{fmffile}
      \caption*{(a)}
       \end{minipage}
 \ \hspace{5mm} \
\begin{minipage}[b]{3.2cm}
  \centering
   \begin{fmffile}{Shat2}
 \fmfframe(1,7)(1,7){
          \begin{fmfgraph*}(90,45)
          \fmfstraight
          \fmftop{t1,t2,t3}
          \fmfbottom{b1,b2,b3}
          \fmfleft{l1,l2,l3,l4,l5}
          \fmfright{r1,r2,r3,r4,r5}
          \fmf{phantom}{v3,v5}          
          \fmf{photon,tension=.6}{l3,v1}
          \fmf{fermion,left,label={\footnotesize{$f_{i}$}},tension=.2}{v1,v2}
          \fmf{fermion,left,label={\footnotesize{$f_{j}$}},tension=.2}{v2,v1}
          \fmf{photon,tension=.6}{v2,r3}
          \fmf{phantom}{t2,v3}
          \fmf{phantom,tension=.195}{v3,v4}
          \fmf{phantom}{v4,b2}
	 \fmffreeze
          \end{fmfgraph*}}\vspace{-5mm}
      \end{fmffile}
      \caption*{(b)}
       \end{minipage}
 \ \hspace{5mm} \
\begin{minipage}[b]{3.2cm}
  \centering
   \begin{fmffile}{That1}
 \fmfframe(1,7)(1,7){
          \begin{fmfgraph*}(90,45)
          \fmfstraight
          \fmftop{t1,t2,t3}
          \fmfbottom{b1,b2,b3}
          \fmfleft{l1,l2,l3,l4,l5}
          \fmfright{r1,r2,r3,r4,r5}
          \fmf{phantom}{v3,v5}          
          \fmf{dashes,tension=.6}{l3,v1}
          \fmf{photon,left,label={\footnotesize{$B$}},tension=.2}{v1,v2}
          \fmf{dashes,left,label={\footnotesize{$\pi$}},tension=.2}{v2,v1}
          \fmf{dashes,tension=.6}{v2,r3}
          \fmf{phantom}{t2,v3}
          \fmf{phantom,tension=.195}{v3,v4}
          \fmf{phantom}{v4,b2}
	 \fmffreeze
          \end{fmfgraph*}}\vspace{-5mm}
      \end{fmffile}
      \caption*{(c)}
       \end{minipage}
 \ \hspace{5mm} \
\begin{minipage}[b]{3.2cm}
  \centering
   \begin{fmffile}{That2}
 \fmfframe(1,7)(1,7){
          \begin{fmfgraph*}(90,45)
          \fmfstraight
          \fmftop{t1,t2,t3}
          \fmfbottom{b1,b2,b3}
          \fmfleft{l1,l2,l3,l4,l5}
          \fmfright{r1,r2,r3,r4,r5}
          \fmf{phantom}{v3,v5}          
          \fmf{dashes,tension=.6}{l3,v1}
          \fmf{fermion,left,label={\footnotesize{$f_{i}$}},tension=.2}{v1,v2}
          \fmf{fermion,left,label={\footnotesize{$f_{j}$}},tension=.2}{v2,v1}
          \fmf{dashes,tension=.6}{v2,r3}
          \fmf{phantom}{t2,v3}
          \fmf{phantom,tension=.195}{v3,v4}
          \fmf{phantom}{v4,b2}
	 \fmffreeze
          \end{fmfgraph*}}\vspace{-5mm}
      \end{fmffile}
      \caption*{(d)}
       \end{minipage}
 \caption{The four Feynman diagrams giving the leading (1-loop) contributions to the $\hat{S}$ and $\hat{T}$ parameters in the Higgsless SM: (a) and (b) are the contributions to $\hat{S}$, while (c) and (d) are the contributions to $\hat{T}$.}\l{fig:STHatHiggsless}
\end{figure}
From this last equation we immediately derive the all order relation\footnote{The relations we are introducing among the radiative corrections to the mass of the gauge bosons and the wave function renormalizations of the GBs, are exact in $\lambda_{t}$ (top Yukawa coupling) and/or in $\lambda$ (Higgs quartic coupling), but are only approximate in the gauge couplings. They are therefore useful to compute the leading logarithms in $m_{h}$ (or analogously in $\Lambda$), i.e. in the limit $m_{h},\Lambda \gg m_{W}$ \cite{Barbieri:1993p1648}.}
\be\l{P4}
\rho\equiv \f{m_{W}^{2}}{m_{Z}^{2}\cos^{2}\theta_{W}}=\f{Z^{\(+\)}}{Z^{\(0\)}}\,.
\ee
Substituting the definitions \eqref{P2} into the relation \eqref{P4} and taking into account that $\hat{T}=\rho-1$ we can write
\be\l{P5}
\bry{lll}
\hat{T}&=&\dst \f{Z^{\(+\)}}{Z^{\(0\)}}-1=\f{1+\d Z^{\(+\)}}{1+\d Z^{\(0\)}}-1\\\\
&\approx&\dst  \d Z^{\(+\)}-\d Z^{\(0\)}+O\(\a^{2}\)\,.
\ery
\ee
Comparing Eqs.~\eqref{P5} with the definition of $\hat{T}$ given in Eq.~\eqref{That} we can derive the $O\(\a\)$ relations
\be\l{P6}
\d Z^{+}=-\f{\Pi_{WW}\(0\)}{m_{W}^{2}}\,,\qquad \d Z^{0}=-\f{\Pi_{33}\(0\)}{m_{W}^{2}}\,.
\ee 

In the Higgsless SM the leading one-loop contributions to $\hat{S}$ come from the GB and the top loops (Fig.~\ref{fig:STHatHiggsless} (a) and (b)). Moreover, since only the gauge coupling $g'$ breaks the custodial symmetry, the leading one-loop contributions to $\hat{T}$ come from the loop containing a $B$ boson and from the top loop (Fig.~\ref{fig:STHatHiggsless} (c) and (d)). The leading terms in $m_{t}$ and $\Lambda$ are given by
\be\l{P7}
\bry{lll}
\dst \hat{S}_{\hsl-\text{SM}}&=&\dst -\f{\a}{12\pi\sin^{2}\theta_{W}}\log\(\f{m_{t}}{m_{Z}}\)+\f{\a}{24\pi\sin^{2}\theta_{W}}\log\(\f{\Lambda}{m_{W}}\)\,,\\\\
\dst \hat{T}_{\hsl-\text{SM}}&=&\dst \f{3\,\a\, m_{t}^{2}}{16\pi m_{W}^{2}\sin^{2 }\theta_{W}}-\f{3\,\a}{8\pi\cos^{2}\theta_{W}}\log\(\f{\Lambda}{m_{W}}\)\,,
\ery
\ee
In Eqs.~\eqref{P7} we can identify the two well known infrared contributions
\be\l{P8}
\bry{lll}
\dst \hat{S}_{\text{IR}}&=&\dst \f{\a}{24\pi\sin^{2}\theta_{W}}\log\(\f{\Lambda}{m_{W}}\)\,,\\\\
\dst \hat{T}_{\text{IR}}&=&\dst -\f{3\a}{8\pi\cos^{2}\theta_{W}}\log\(\f{\Lambda}{m_{W}}\)\,.
\ery
\ee
From these relations we immediately see that in the SM without the Higgs boson the large logarithmic contributions to $\hat{S}$ and $\hat{T}$ must be cut-off by an energy scale $\Lambda$. 
\begin{figure}[htbp!]
\centering
\includegraphics[width=8cm]{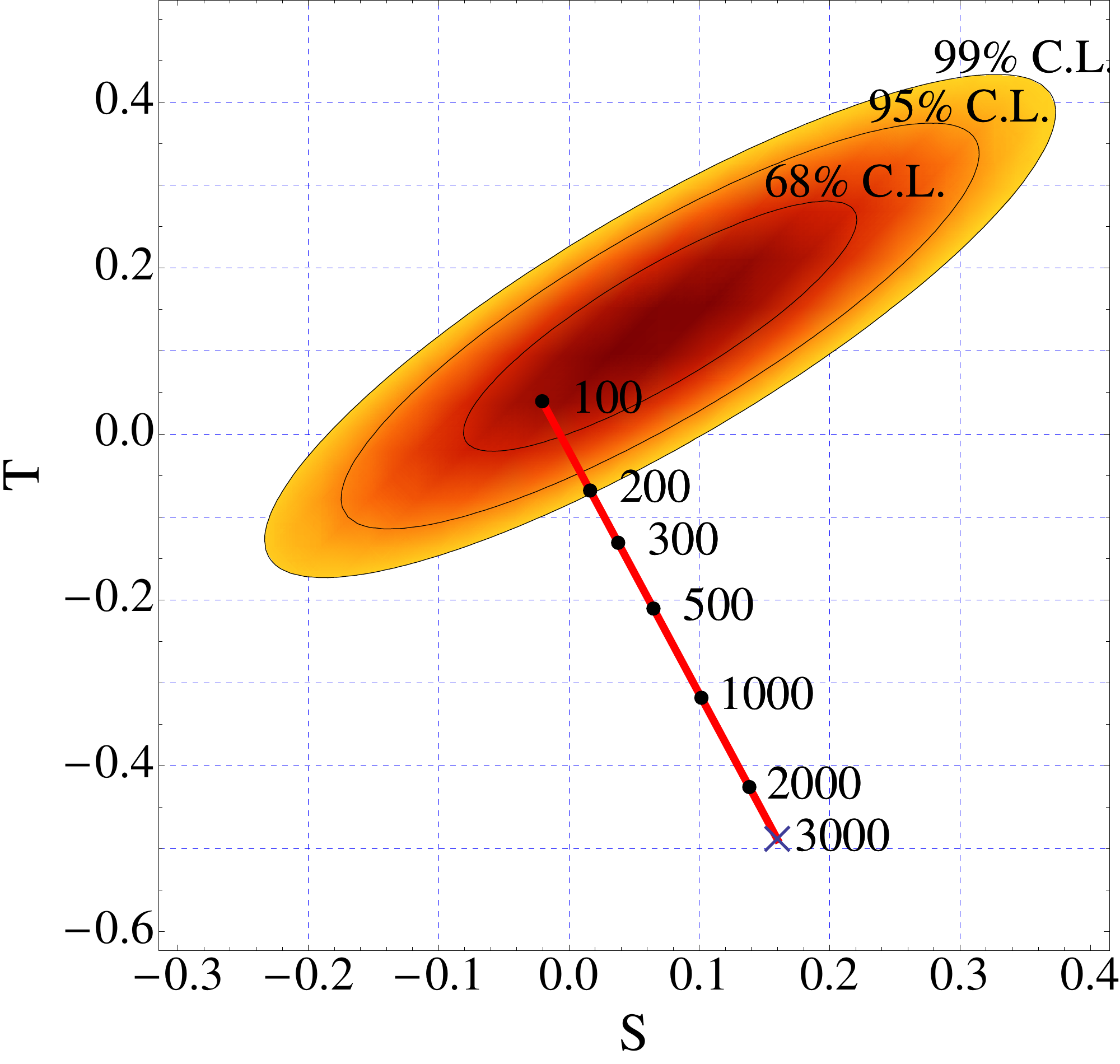}
\caption{Experimental allowed regions and theoretical predictions for the $S$ and $T$ parameters in the Higgsless SM for $100<\Lambda<3000$ GeV (Eqs.~\eqref{P7}). The experimental allowed regions are chosen as in Ref.~\cite{ALEPHCollaboration:2005p1834}. For the definition of the $\(0,0\)$ point see the footnote \ref{nota1}. For the theoretical prediction we have used the most updated value of the top mass $m_{t}=173.3$ GeV \cite{TevatronElectroWeakWorkingGroup:2010wu}.}
\l{fig:EWPTHiggsless}
\end{figure}
In fact, the effect of the SM Higgs boson on the $\hat{S}$ and $\hat{T}$ parameters of Eqs.~\eqref{P8} is exactly to cut-off the logarithms by substituting the scale $\Lambda$ with the Higgs boson mass $m_{h}$. 
In Fig.~\ref{fig:EWPTHiggsless} we have plotted the experimental allowed region in the $\(S,T\)$ plane\footnote{The origin of the axes in the $\(S,T\)$ plane is chosen in such a way that $\(S_{\text{SM}},T_{\text{SM}}\)\Big|_{m_{h}=150 \text{ GeV },\,m_{t}=175 \text{ GeV }}\equiv\(0,0\)$. All the plots represent deviations from these values.\l{nota1}} compared with the theoretical predictions for $\Lambda< 3$ TeV (or equivalently $m_{h}<3$ TeV). It is simple to see that the experimental bounds on $S$ and $T$ imply $\Lambda\lesssim 200$ GeV that fixes a cutoff for the Higgsless SM of the order of the EW scale. As we will see in the next section, the same bound can be read $m_{h}\lesssim 200$ GeV for the SM Higgs boson mass. The plot in Fig.~\ref{fig:EWPTHiggsless} only contains the logarithmic contributions of Eqs.~\eqref{P7}. It turns out that introducing also the finite terms that vanish in the limit $m_{h}\to 0$ the straight line in Fig.~\ref{fig:EWPTHiggsless} acquires a slight bending shape slightly changing the limit on the Higgs boson mass. However, a precise determination of the limits on the Higgs boson mass requires a global fit to all the EWPO. The result of the global fit is \cite{Nakamura:2010dr}
\be\l{Higgsmassglobfit}
m_{h}=90^{+27}_{-22} \text{ GeV}\,,\qquad m_{h}<\(145,149,194\) \text{ GeV at } \(90,95,99\)\% \text{ CL}\,.
\ee

\section{The role of a composite scalar and the fundamental Higgs boson limit}\l{TROTHB}
In this section we generalize the Higgsless SM discussed in the previous section adding a scalar field, coupled to the SM fields through a general effective Lagrangian. We will see that for a particular choice of the parameters the scalar coincides with the SM Higgs boson, i.e. can be embedded with the GBs into a linear doublet of $SU(2)_{L}$. In this case the Lagrangian will reduce exactly to the SM Lagrangian.\\
Adding a scalar field to the Higgsless SM Lagrangian \eqref{HlessSM} we obtain
\be\l{HSM}
\La_{h-\text{SM}}=\La_{\hsl-\text{SM}}+\La_{h-\text{kin}}+\La_{h-\text{pot}}+\La_{h-\text{int}}\,,
\ee
where $\La_{\hsl-\text{SM}}$ is the Higgsless SM Lagrangian \eqref{HlessSM},
\be\l{Hkin}
\La_{h-\text{kin}}=\f{1}{2}\demub h \demua h\,,
\ee
\be\l{Hpot}
\La_{h-\text{pot}}=-\f{m_{h}^{2}}{2}h^{2}-\f{d_{3}}{6}\(\f{3m_{h}^{2}}{v}\)h^{3}-\f{d_{4}}{24}\(\f{3m_{h}^{2}}{v}\)h^{4}+\ldots \,,
\ee
and\footnote{We are assuming that the constant $c$ is equal for all the fermions but it can in principle be different. In that case, a $c_{i}$ would appear in the Lagrangian \eqref{Hint} in place of $c$.}
\be\l{Hint}
\La_{h-\text{int}}=\f{v^{2}}{4}\<D^{\mu}\S\(D_{\mu}\S\)^{\dag}\>\(2a\f{h}{v}+b\f{h^{2}}{v^{2}}+\ldots\)-m_{i}\bar{\psi}_{L\,i}\S\(c\f{h}{v}+\ldots\)\psi_{R\,i}\,.
\ee
Here $a$, $b$, $c$, $d_{3}$ and $d_{4}$ are arbitrary real numbers. Note that we use the notation $\La_{h-\text{SM}}$ to indicate the SM Lagrangian plus a general scalar singlet $h$. When we will instead consider the SM Higgs boson we will simply use the notation $\La_{\text{SM}}$. Neglecting operators of dimension grater than $4$, for the special choice $a=b=c=d_{3}=d_{4}=1$ the scalar $h$ can be embedded into the linear doublet $\mathcal{H}$ of Eq.~\eqref{PolarHiggs} and the Lagrangian \eqref{HSM} reduces to the SM Lagrangian with the ``standard'' Higgs boson. \\
In the following subsections we see how the perturbative unitarity and the EWPT change in presence of the new scalar singlet $h$ discussing the SM limit of the couplings.

\subsection{Elastic unitarity}\l{ElasticUnitarity2}
In the $g'=0$ limit, the new scalar gives rise to the three new Feynman diagrams in Fig.~\ref{fig:fourpionsHiggs} contributing to the four-pion scattering.
\begin{figure}[!htb]
\centering
\begin{minipage}[t]{4cm}
\begin{center}
        \begin{fmffile}{ppppH1}
  \fmfframe(1,7)(1,7){ 
   \begin{fmfgraph*}(70,70)
          \fmfstraight
    \fmfbottom{i1,i2}
    \fmftop{o1,o2}
    \fmflabel{\footnotesize{$\pi^{a}$}}{i1}
    \fmflabel{\footnotesize{$\pi^{d}$}}{o1}
    \fmflabel{\footnotesize{$\pi^{b}$}}{i2}
    \fmflabel{\footnotesize{$\pi^{c}$}}{o2}
    \fmf{dashes}{i1,v1}
    \fmf{dashes}{i2,v1}
    \fmf{dashes}{v2,o1}
    \fmf{dashes}{v2,o2}
    \fmf{dashes,label={\footnotesize{$h$}}}{v1,v2}
    \end{fmfgraph*}
        }   \\\vspace{4mm}  \footnotesize{{\bf (a)}}\l{ppppH1}%
      \end{fmffile}
\end{center}
\end{minipage}
\ \hspace{5mm} \
\begin{minipage}[t]{4cm}
\begin{center}
        \begin{fmffile}{ppppH2}
  \fmfframe(1,7)(1,7){ 
   \begin{fmfgraph*}(70,70)
          \fmfstraight
    \fmfleft{i1,i2}
    \fmfright{o1,o2}
    \fmflabel{\footnotesize{$\pi^{a}$}}{i1}
    \fmflabel{\footnotesize{$\pi^{b}$}}{o1}
    \fmflabel{\footnotesize{$\pi^{c}$}}{i2}
    \fmflabel{\footnotesize{$\pi^{d}$}}{o2}
    \fmf{dashes}{i1,v1}
    \fmf{dashes}{v1,i2}
    \fmf{dashes}{o1,v2}
    \fmf{dashes}{v2,o2}
    \fmf{dashes,l.side=right,label={\footnotesize{$h$}}}{v1,v2}
    \end{fmfgraph*}
        }   \\\vspace{4mm}  \footnotesize{{\bf (b)}} \l{ppppH2}%
      \end{fmffile}
\end{center}
\end{minipage}
\ \hspace{5mm} \
\begin{minipage}[t]{4cm}
\begin{center}
        \begin{fmffile}{ppppH3}
  \fmfframe(1,7)(1,7){ 
   \begin{fmfgraph*}(70,70)
          \fmfstraight
    \fmfleft{i1,i2}
    \fmfright{o1,o2}
    \fmflabel{\footnotesize{$\pi^{a}$}}{i1}
    \fmflabel{\footnotesize{$\pi^{b}$}}{o1}
    \fmflabel{\footnotesize{$\pi^{c}$}}{i2}
    \fmflabel{\footnotesize{$\pi^{d}$}}{o2}
    \fmf{dashes}{i1,v1}
    \fmf{phantom}{v1,i2}
    \fmf{dashes}{o1,v2}
    \fmf{phantom}{v2,o2}
    \fmf{dashes,l.side=right,label={\footnotesize{$h$}}}{v1,v2}
    \fmffreeze
    \fmf{dashes}{v1,o2}
    \fmf{dashes}{v2,i2}    
    \end{fmfgraph*}
        }  \\\vspace{4mm}  \footnotesize{{\bf (c)}} \l{ppppH3}\vspace{4mm}%
      \end{fmffile}
\end{center}
\end{minipage}
\caption{The three Feynman diagrams contributing to the four-pion scattering through a scalar singlet exchange: (b), (c) and (d) are respectively the $s$, $t$ and $u$ channels of the contribution given by the exchange of the scalar singlet $h$.}\l{fig:fourpionsHiggs}
\end{figure}
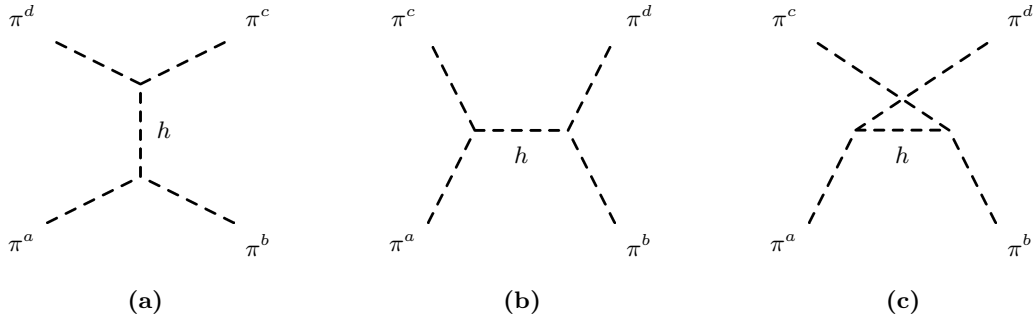\\
A simple calculation gives for the scalar contribution the result
\be\l{k6}
\Mamp\(s,t,u\)_{h}\approx -\f{a^{2}}{v^{2}}\f{s^{2}}{s-m_{h}^{2}}\,.
\ee
Summing up the two amplitudes \eqref{k6} and \eqref{k3} we find for the SM with a scalar singlet the amplitude
\be\l{k7}
\Mamp\(s,t,u\)_{h-\text{SM}}=\Mamp\(s,t,u\)_{\hsl-\text{SM}}+\Mamp\(s,t,u\)_{h}\approx \f{s}{v^{2}}-\f{a^{2}}{v^{2}}\f{s^{2}}{s-m_{h}^{2}}=\f{s}{v^{2}}\(1-\f{a^{2}s}{s-m_{h}^{2}}\)\,.
\ee
This amplitude grows again with the c.o.m.~energy squared $s$ and the perturbative unitarity is violated at an energy scale $\Lambda_{h-\text{SM}}$. To estimate this scale we should compute the $a_{0}^{0}$ coefficient corresponding to the amplitude \eqref{k7}. Substituting Eq.~\eqref{k7} into Eq.~\eqref{k3a} and the result into Eq.~\eqref{k3c} we find
\be\l{k7bis}
\bry{lll}
\dst \lim_{s\gg m_{h}^{2}}\(a^{0}_{0}\)_{h-\text{SM}}&=&\dst \f{1}{64\pi v^{2}}\int_{-1}^{1}dx \lim_{s\gg m_{h}^{2}}\bigg[3s\(1-\f{a^{2}s}{s-m_{h}^{2}}\)\\\\
&&\dst-\f{s\(1-x\)}{2}\(1-\f{a^{2}s\(1-x\)}{s\(1-x\)+2m_{h}^{2}}\)-\f{s\(1+x\)}{2}\(1-\f{a^{2}s\(1+x\)}{s\(1+x\)+2m_{h}^{2}}\)\bigg]\\\\
&=&\dst \f{s\(1-a^{2}\)}{16\pi v^{2}}-\f{5a^{2}m_{h}^{2}}{32\pi v^{2}}+\f{1}{v^{2}}O\(\f{m^{4}}{s}\)\,.
\ery
\ee
The $\(a^{0}_{0}\)_{h-\text{SM}}$ coefficient for the SM with a general scalar singlet grows with $s$ unless $a=1$. This is exactly what we expected since for $a=1$ the trilinear coupling of the scalar to the Goldstone and the gauge bosons is exactly that of the SM Higgs boson. For the cutoff of the SM with a scalar singlet we find, from the strongest unitarity constraint,
\be\l{k5bis}
|\lim_{s\gg m_{h}^{2}}\(a_{0}^{0}\)_{h-\text{SM}}|\leq 1 \quad\Longrightarrow\quad \Lambda_{h-\text{SM}}\leq \sqrt{\f{16\pi v^{2}}{\(1-a^{2}\)}+\f{5a^{2}m_{h}^{2}}{2}}\,.
\ee
It was proved that in general theories in which the Higgs boson is the pseudo-GB of an extended global symmetry one has $a<1$, while in the case of a dilaton\footnote{The dilaton is the Goldstone boson of the spontaneously broken conformal symmetry predicted in some extensions of the SM.} one can have either $a<1$ or $a>1$ \cite{Low:2009gl,Contino:2010vc}. Therefore in the case of a Higgs boson which is a pseudo-GB we have that the cutoff varies over the range\footnote{We set an upper bound at the Planck scale, where the effects of gravity are supposed to become important.}
\be\l{cutoff1}
4\sqrt{\pi} v\leq \Lambda_{h-\text{SM}}\leq M_{\text{Planck}}  \quad\Longrightarrow\quad 1.7\,\text{TeV}\lesssim \Lambda_{h-\text{SM}}\leq M_{\text{Planck}}\,.
\ee
This means that values of $a$ from $0$ to $1$ interpolates from Higgsless models (see the Higgsless SM cutoff \eqref{k5}) to the SM with the fundamental Higgs boson passing through theories in which the Higgs boson is different from the SM one, perhaps a composite object arising as a pseudo-GB of and extended global symmetry. In the case $a>1$ the cutoff is not bounded from below, but in order to keep it above the cutoff of the Higgsless SM one should have a large $a\gtrsim 4.5\,\(9.5\)$ for a scalar mass $m_{h}\approx 246\,\(114\)$ \text{GeV} of the order of the EW scale. Notice that this can give rise to a very different collider phenomenology with a highly fermio-phobic Higgs boson (i.e., with a suppressed $h\to b\bar{b}$ Branching Ratio and an enhanced $h\to \gamma\gamma$ Branching Ratio even for a light Higgs boson).\\
In the case $a=1$ we obtain the result corresponding to the SM Higgs. In particular we can write the amplitude \eqref{k7} in the form
\be\l{k7tris}
\Mamp\(s,t,u\)_{\text{SM}}=\lim_{a\to 1}\Mamp\(s,t,u\)_{h-\text{SM}}=\f{s}{v^{2}}-\f{1}{v^{2}}\f{s^{2}}{s-m_{h}^{2}}=-\f{m_{h}^{2}}{v^{2}}\(\f{s}{s-m_{h}^{2}}\)\,.
\ee
This amplitude is finite in the high energy limit where we obtain
\be\l{k8}
\Mamp\(s,t,u\)_{\text{SM}} \longrightarrow_{_{\hspace{-0.6cm}s\gg m_{h}^{2}}} -\f{m_{h}^{2}}{v^{2}}+O\(\f{m_{h}^{2}}{s}\)\,.
\ee
The first coefficient of the partial wave expansion of the fixed isospin amplitude $T\(0\)$  is now given only by the finite part in Eq.~\eqref{k7bis}
\be\l{a00sm}
\bry{lll}
\dst \(a_{0}^{0}\)_{\text{SM}}&=&\dst -\f{1}{64\pi}\f{m_{h}^{2}}{v^{2}}\int_{-1}^{1}dx \bigg[\f{3s}{s-m_{h}^{2}}+\f{s\(1-x\)}{s\(1-x\)+2m_{h}^{2}}+\f{s\(1+x\)}{s\(1+x\)+2m_{h}^{2}}\bigg]\\\\
&=&\dst -\f{1}{16\pi}\f{m_{h}^{2}}{v^{2}}\(\f{5}{2}+\f{3}{2\(x_{h}-1\)}-\f{\log\(x_{h}+1\)}{x_{h}}\)\,,
\ery
\ee
where we have defined $x_{h}=s/m_{h}^{2}$. In the high energy limit we simply obtain
\be\l{k9}
\lim_{s\gg m_{h}^{2}}\(a^{0}_{0}\)_{\text{SM}}= \bigg[-\f{1}{16\pi}\f{m_{h}^{2}}{v^{2}}\(\f{5}{2}+\f{3}{2\(x_{h}-1\)}-\f{\log\(x_{h}+1\)}{x_{h}}\)\bigg]=-\f{5}{32\pi}\f{m_{h}^{2}}{v^{2}}\,.
\ee
In this case the elastic unitarity constraint puts a bound on the Higgs boson mass
\be\l{k10}
|a^{0}_{0}|\leq 1\quad \Longrightarrow \quad m_{h}^{2}\leq \f{32\pi v^{2}}{5}\approx \(1\,\text{TeV}\)^{2}\,.
\ee
Therefore, provided that the Higgs boson is lighter than about $1$ TeV the SM is perturbatively predictive up to a scale much higher than the EW scale\footnote{We are ignoring the hierarchy problem of the SM which can be (partially) solved embedding the SM in a suitable supersymmetric extension.} (e.g. the GUT or the Planck scale). This is a significant result, since it cannot be obtained in strongly interacting models like Higgsless or composite Higgs models.

\subsection{ElectroWeak Precision Tests}\l{EWPT2}
The contribution of the scalar singlet to the $\hat{S}$ and $\hat{T}$ parameters are given by the Feynman diagrams of Fig.~\ref{fig:STHatHiggs}.
\begin{figure}[t]
\begin{minipage}[b]{8cm}
  \centering
   \begin{fmffile}{Shat3}
 \fmfframe(1,7)(1,7){
          \begin{fmfgraph*}(90,45)
          \fmfstraight
          \fmftop{t1,t2,t3}
          \fmfbottom{b1,b2,b3}
          \fmfleft{l1,l2,l3,l4,l5}
          \fmfright{r1,r2,r3,r4,r5}
          \fmf{phantom}{v3,v5}          
          \fmf{photon,tension=.6}{l3,v1}
          \fmf{dashes,left,label={\footnotesize{$\pi$}},tension=.2}{v1,v2}
          \fmf{dashes,left,label={\footnotesize{$h$}},tension=.2}{v2,v1}
          \fmf{photon,tension=.6}{v2,r3}
          \fmf{phantom}{t2,v3}
          \fmf{phantom,tension=.195}{v3,v4}
          \fmf{phantom}{v4,b2}
	 \fmffreeze
          \end{fmfgraph*}}
      \end{fmffile}
      \caption*{(a)}
       \end{minipage}
\begin{minipage}[b]{8cm}
  \centering
   \begin{fmffile}{That3}
 \fmfframe(1,7)(1,7){
          \begin{fmfgraph*}(90,45)
          \fmfstraight
          \fmftop{t1,t2,t3}
          \fmfbottom{b1,b2,b3}
          \fmfleft{l1,l2,l3,l4,l5}
          \fmfright{r1,r2,r3,r4,r5}
          \fmf{phantom}{v3,v5}          
          \fmf{dashes,tension=.6}{l3,v1}
          \fmf{photon,left,label={\footnotesize{$B$}},tension=.2}{v1,v2}
          \fmf{dashes,left,label={\footnotesize{$h$}},tension=.2}{v2,v1}
          \fmf{dashes,tension=.6}{v2,r3}
          \fmf{phantom}{t2,v3}
          \fmf{phantom,tension=.195}{v3,v4}
          \fmf{phantom}{v4,b2}
	 \fmffreeze
          \end{fmfgraph*}}
      \end{fmffile}
      \caption*{(b)}
       \end{minipage}
 \caption{The two Feynman diagrams giving the $h$ contributions to the $S$ and $T$ parameters: (a) is the contribution to $S$, while (b) is the contribution to $\hat{T}$.}\l{fig:STHatHiggs}
\end{figure}
From the calculation of these diagrams we find
\be\l{eq19}
\bry{lll}
\dst \D\hat{S}_{h}&=&\dst -\f{a^{2}\,\a}{24\pi\sin^{2}\theta_{W}}\log\(\f{\Lambda}{m_{h}}\)\,,\\\\
\dst \D\hat{T}_{h}&=&\dst \f{3\,a^{2}\,\a}{8\pi\cos^{2}\theta_{W}}\log\(\f{\Lambda}{m_{h}}\)\,.
\ery
\ee
As expected, in the case $a=1$, the contributions \eqref{eq19} are exactly the contributions of the SM Higgs boson and their effect is to cut-off the infrared contributions \eqref{P8} so that the sum of the two contributions becomes
\be\l{eq20}
\bry{lll}
\dst \hat{S}_{\text{SM}}\(m_{h},m_{t}\)&=&\dst -\f{\a}{12\pi\sin^{2}\theta_{W}}\log\(\f{m_{t}}{m_{Z}}\)+\f{\a}{24\pi\sin^{2}\theta_{W}}\log\(\f{m_{h}}{m_{W}}\)\,,\\\\
\dst \hat{T}_{\text{SM}}\(m_{h},m_{t}\)&=&\dst \f{3\,\a\, m_{t}^{2}}{16\pi m_{W}^{2}\sin^{2 }\theta_{W}}-\f{3\,\a}{8\pi\cos^{2}\theta_{W}}\log\(\f{m_{h}}{m_{W}}\)\,,
\ery
\ee
and is no longer dependent on the UV cutoff $\Lambda$. Fig.~\ref{fig:EWPTHiggsless} can now be read as the contribution of the SM Higgs boson of Eq.~\eqref{eq20} for a Higgs boson mass ranging between $100$ GeV and $3$ TeV.\\
On the contrary, for values $a\neq 1$, the sum of the infrared contributions \eqref{P8} and the scalar contributions \eqref{eq19} still grow logarithmically with the cutoff $\Lambda$ and we find
\be\l{eq21}
\bry{lll}
\dst \hat{S}_{h-\text{SM}}&=&\dst \hat{S}_{\text{SM}}\(m_{h},m_{t}\)+\(1-a^{2}\)\f{\a}{24\pi\sin^{2}\theta_{W}}\log\(\f{\Lambda}{m_{h}}\)\,,\\\\
\dst \hat{T}_{h-\text{SM}}&=&\dst \hat{T}_{\text{SM}}\(m_{h},m_{t}\)-\(1-a^{2}\)\f{3\a}{8\pi\cos^{2}\theta_{W}}\log\(\f{\Lambda}{m_{h}}\)\,,
\ery
\ee
where we have made manifest the logarithmic contributions proportional to the positive quantity\footnote{Here we are assuming $a<1$ since we have in mind a pseudo-GB Higgs.} $\(1-a^{2}\)$.
From the last relations we immediately see that a composite Higgs boson cannot avoid large contributions to $\hat{S}$ (positive) and $\hat{T}$ (negative). In Fig.~\ref{fig:EWPTCompHiggs} we have plotted the experimental allowed regions in the $\(S,T\)$ plane compared with the theoretical predictions for the SM with a composite scalar. The theoretical predictions are shown for $a=1/2$, $a=\sqrt{2/3}$ and $a=\sqrt{13}/4$ and for $100<m_{h}<3000$ GeV. The choice of these values of $a$ will be clear in the following, when we will add to the spectrum also a composite vector. In particular we will see that they correspond, through the relation \eqref{agvrelation}, to the values $G_{V}=v/2,\,v/3,\,v/4$ of the coupling of the heavy vector to the GBs and the gauge bosons.
\begin{figure}[h!]
\begin{minipage}[b]{4.9cm}
   \centering
   \includegraphics[width=4.8cm]{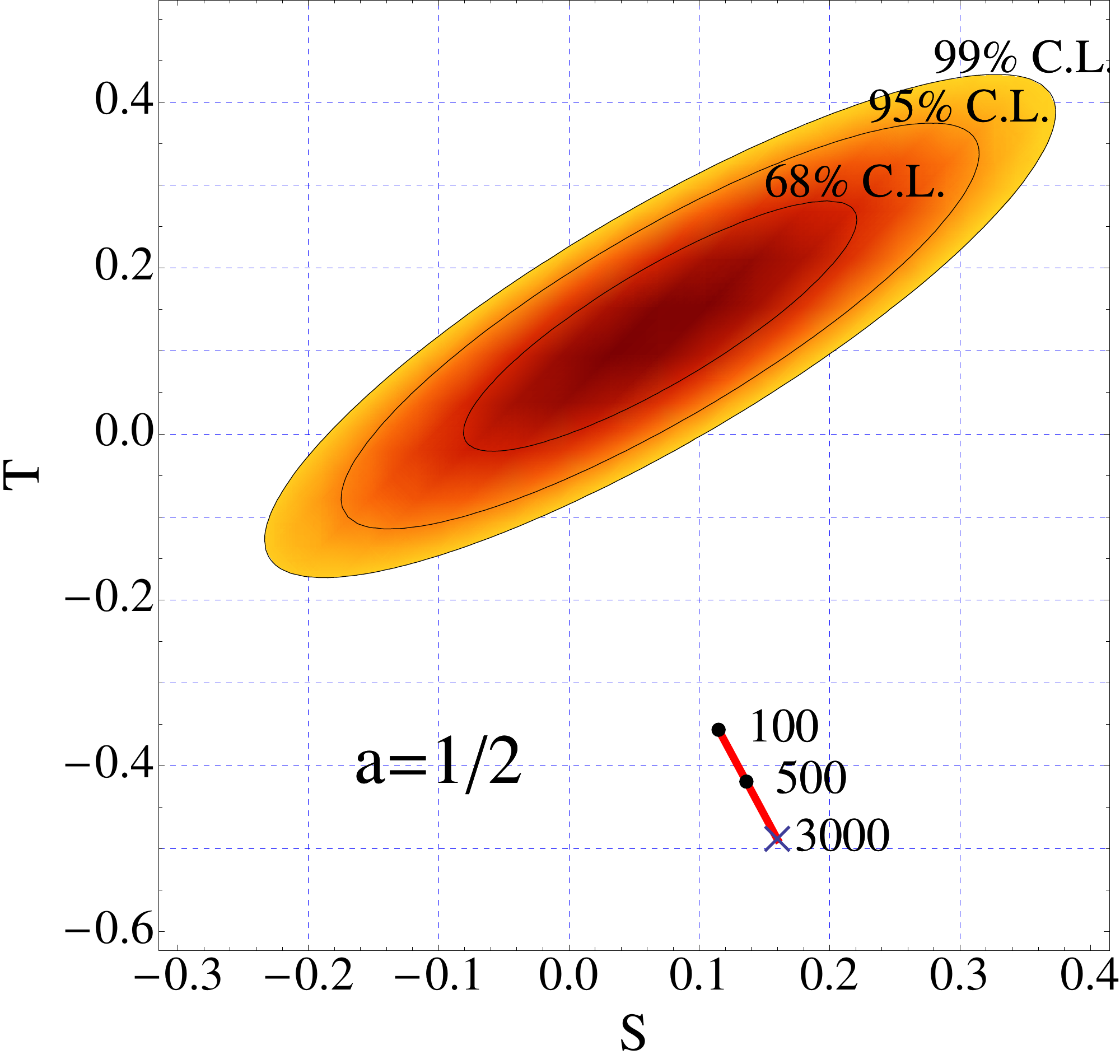}
 \end{minipage}
\ \hspace{3mm} \ 
\begin{minipage}[b]{4.9cm}
  \centering
   \includegraphics[width=4.8cm]{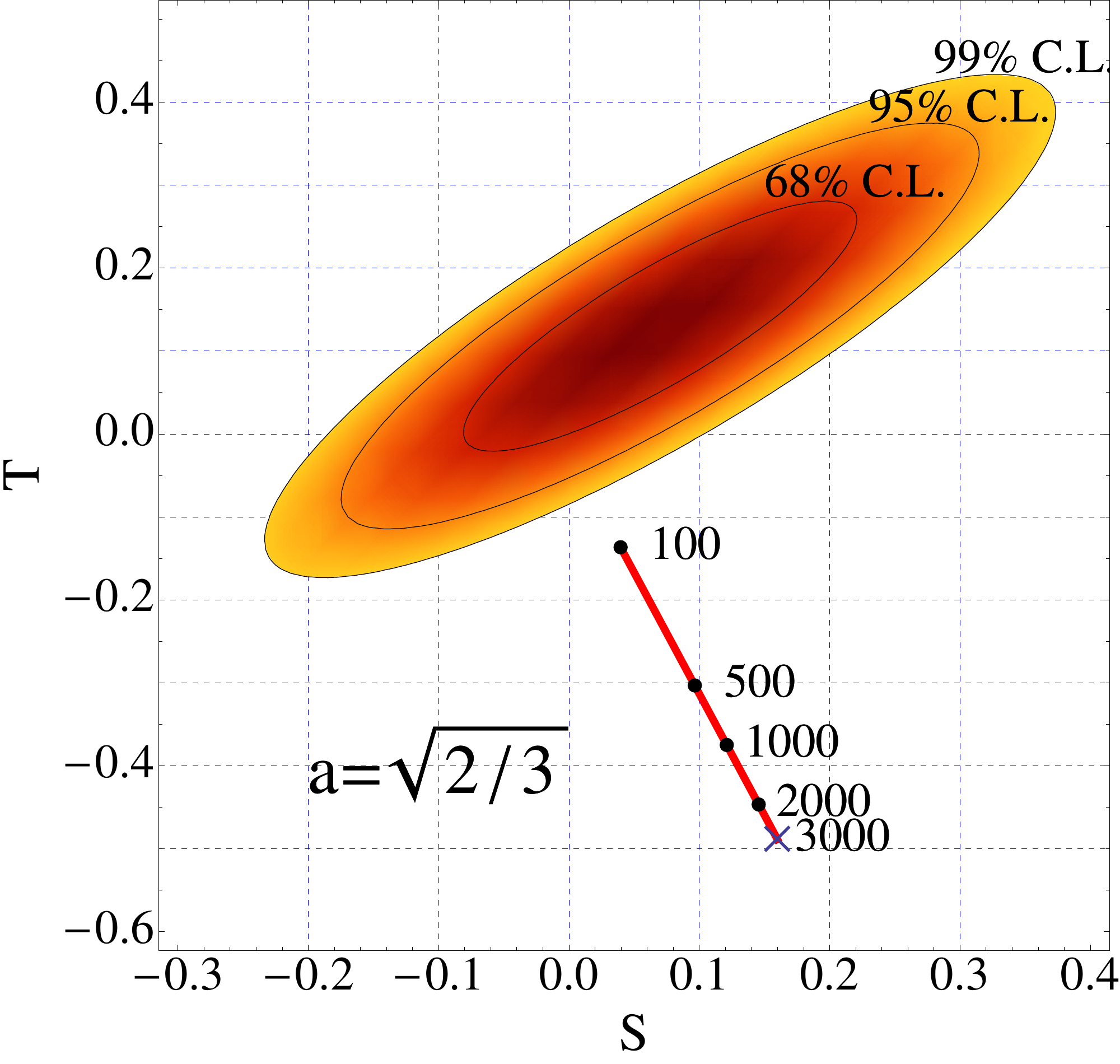}
 \end{minipage}
 \ \hspace{3mm} \ 
\begin{minipage}[b]{4.9cm}
  \centering
   \includegraphics[width=4.8cm]{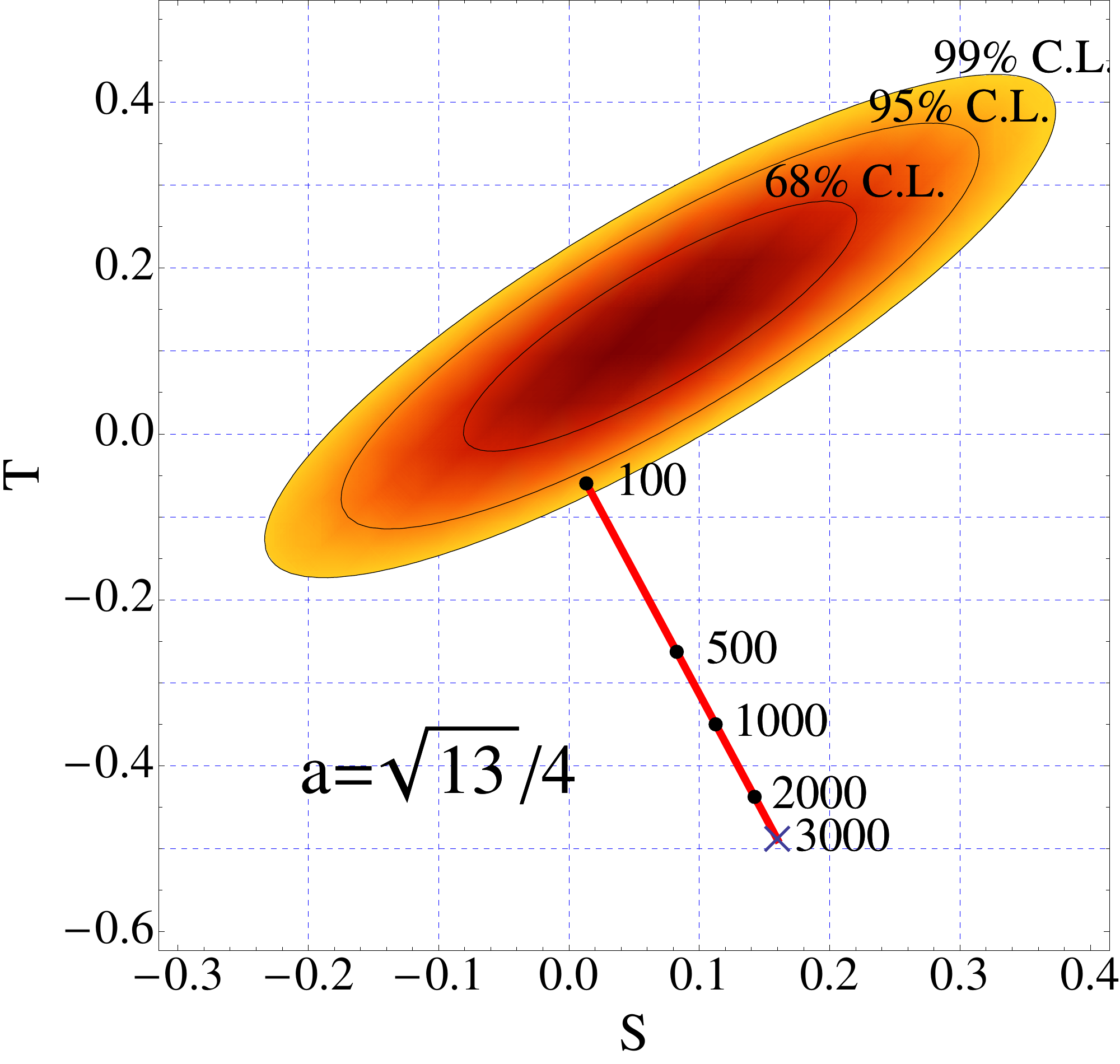}
 \end{minipage}
\caption{Experimental allowed regions and theoretical predictions for the $S$ and $T$ parameters in the SM with a composite Higgs boson (see Eq.~\eqref{eq21}) for $100<m_{h}<3000$ GeV and for three different values of $a$. The small cross at negative $T$ can be interpreted as the Higgsless SM with a cutoff $\Lambda = 3$ TeV.}
\l{fig:EWPTCompHiggs}
\end{figure}
From the plots of Fig.~\ref{fig:EWPTCompHiggs} we can see that small values of $a$ weaken the dependence of $S$ and $T$ on $m_{h}$. However, small values of $a$ give large contributions to $S$ (positive) and $T$ (negative) being strongly disfavored by the experimental bounds. In the limit $a=1$ we find again the SM with a fundamental Higgs boson (see Fig.~\ref{fig:EWPTHiggsless}). Therefore we can only expect small deviations from the $a=1$ relation. Finally, since in the case of a pseudo-GB Higgs only values of $a$ smaller than one are allowed, deviations from $a=1$ imply bounds on the Higgs mass that are more stringent than those required by the SM Higgs boson and usually require further assumptions on the nature of the SM particles (e.g. the top quark) or the introduction of other states \cite{Duccio:nuovo}.\\
As a concrete example, in the pseudo-GB Higgs models based on the coset $SO\(5\)/SO\(4\)$ we would have the relation \cite{Contino:2010vc,Agashe:2004ib,Contino:2006fd}
\be\l{eq22a}
a=\sqrt{1-\xi}\,,
\ee
where $\xi=v^{2}/f^{2}$ is the squared ratio of the EW scale over the symmetry breaking scale of the extended symmetry. From the considerations above and from Fig.~\ref{fig:EWPTCompHiggs}, in absence of new contributions coming from the fermion sector and in the case in which there are no vector resonances below the cutoff $4\pi f$, we can therefore expect in these models $\xi\lesssim 0.2$. These considerations show that it is non trivial to account for the EWPT only with a composite Higgs boson in the low energy spectrum, especially for a highly non SM-like Higgs boson ($a$ significantly different from $1$).

\section{The case of a Higgsless model with heavy composite vectors}\l{TCOAHLMWAHCV}
As we will see in details in the next chapter, alternatively to the introduction of a new scalar degree of freedom, the introduction of new heavy (gauge or composite) vectors is often considered in the literature (see e.g. Ref.~\cite{Barbieri:2008p1580}). Without entering in the details of the construction of the model, which will be extensively discussed in the next chapter, we are interested here in studying the unitarity and the EWPT of the SM plus a composite vector only assuming that it is a triplet under the custodial symmetry. The Lagrangian of this model is given by
\be\l{LagHigglsessVect}
\La_{V-\text{SM}}=\La_{\hsl-\text{SM}}+\La_{V}\,,
\ee
where the Higgsless SM Lagrangian $\La_{\hsl-\text{SM}}$ is given by Eq.~\eqref{HlessSM} and $\La_{V}$ is the Lagrangian describing the new vector field (see Section~\ref{sec3.1} and in particular Eq.~\eqref{Ltot}).

\subsection{Elastic unitarity}\l{ElasticUnitarity3}
In this case, the new Feynman diagrams of Fig.~\ref{fig:fourpionsVectors} contribute to the four-pion scattering.
\begin{figure}[t]
\centering
\begin{minipage}[t]{3.3cm}
\begin{center}
\begin{fmffile}{pppp1counter}
  \fmfframe(1,7)(1,7){ 
          \begin{fmfgraph*}(70,70)
          \fmfstraight
          \fmflabel{\footnotesize{$\pi^{c}$}}{o2}
          \fmflabel{\footnotesize{$\pi^{a}$}}{i1}
          \fmflabel{\footnotesize{$\pi^{d}$}}{o1}
          \fmflabel{\footnotesize{$\pi^{b}$}}{i2}          
          \fmfbottom{i1,i2}
          \fmftop{o1,o2}
          \fmf{phantom}{v1,v2}
          \fmf{dashes}{i1,v1}
          \fmf{dashes}{i2,v1}
          \fmf{dashes}{v1,o1}
          \fmf{dashes}{v1,o2}
          \fmfv{decor.shape=circle,decor.filled=empty}{v1}
          \fmfv{decor.shape=cross}{v2}
          \end{fmfgraph*}} \\\vspace{4mm}  \footnotesize{{\bf (a)}}\l{pppp1}\vspace{4mm}
\end{fmffile}
\end{center}
\end{minipage}
\ \hspace{3mm} \
\begin{minipage}[t]{3.3cm}
\begin{center}
        \begin{fmffile}{ppppV1}
  \fmfframe(1,7)(1,7){ 
   \begin{fmfgraph*}(70,70)
          \fmfstraight
    \fmfbottom{i1,i2}
    \fmftop{o1,o2}
    \fmflabel{\footnotesize{$\pi^{a}$}}{i1}
    \fmflabel{\footnotesize{$\pi^{d}$}}{o1}
    \fmflabel{\footnotesize{$\pi^{b}$}}{i2}
    \fmflabel{\footnotesize{$\pi^{c}$}}{o2}
    \fmf{dashes}{i1,v1}
    \fmf{dashes}{i2,v1}
    \fmf{dashes}{v2,o1}
    \fmf{dashes}{v2,o2}
    \fmf{dbl_plain,label={\footnotesize{$V$}}}{v1,v2}
    \end{fmfgraph*}
        }  \\\vspace{4mm}  \footnotesize{{\bf (b)}} \l{ppppV1}%
      \end{fmffile}
\end{center}
\end{minipage}
\ \hspace{3mm} \
\begin{minipage}[t]{3.3cm}
\begin{center}
        \begin{fmffile}{ppppV2}
  \fmfframe(1,7)(1,7){ 
   \begin{fmfgraph*}(70,70)
          \fmfstraight
    \fmfleft{i1,i2}
    \fmfright{o1,o2}
    \fmflabel{\footnotesize{$\pi^{a}$}}{i1}
    \fmflabel{\footnotesize{$\pi^{b}$}}{o1}
    \fmflabel{\footnotesize{$\pi^{c}$}}{i2}
    \fmflabel{\footnotesize{$\pi^{d}$}}{o2}
    \fmf{dashes}{i1,v1}
    \fmf{dashes}{v1,i2}
    \fmf{dashes}{o1,v2}
    \fmf{dashes}{v2,o2}
    \fmf{dbl_plain,l.side=right,label={\footnotesize{$V$}}}{v1,v2}
    \end{fmfgraph*}
        }  \\\vspace{4mm}  \footnotesize{{\bf (c)}}  \l{ppppV2}%
      \end{fmffile}
\end{center}
\end{minipage}
\ \hspace{3mm} \
\begin{minipage}[t]{3.3 cm}
\begin{center}
        \begin{fmffile}{ppppV3}
  \fmfframe(1,7)(1,7){ 
   \begin{fmfgraph*}(70,70)
          \fmfstraight
    \fmfleft{i1,i2}
    \fmfright{o1,o2}
    \fmflabel{\footnotesize{$\pi^{a}$}}{i1}
    \fmflabel{\footnotesize{$\pi^{b}$}}{o1}
    \fmflabel{\footnotesize{$\pi^{c}$}}{i2}
    \fmflabel{\footnotesize{$\pi^{d}$}}{o2}
    \fmf{dashes}{i1,v1}
    \fmf{phantom}{v1,i2}
    \fmf{dashes}{o1,v2}
    \fmf{phantom}{v2,o2}
    \fmf{dbl_plain,l.side=right,label={\footnotesize{$V$}}}{v1,v2}
    \fmffreeze
    \fmf{dashes}{v1,o2}
    \fmf{dashes}{v2,i2}
    \end{fmfgraph*}
        }    \\\vspace{4mm}  \footnotesize{{\bf (d)}}\l{ppppV3}%
      \end{fmffile}
\end{center}
\end{minipage}\vspace{-2mm}\caption{The four Feynman diagrams contributing to the four-pion scattering in the Higgsless SM with a new composite vector triplet: (a) is the contribution of the contact four-pion vertex given by the Lagrangian \eqref{Lcontact}; (b), (c) and (d) are the $s$, $t$ and $u$ channels of $V$ exchange contribution.}\l{fig:fourpionsVectors}
\end{figure}
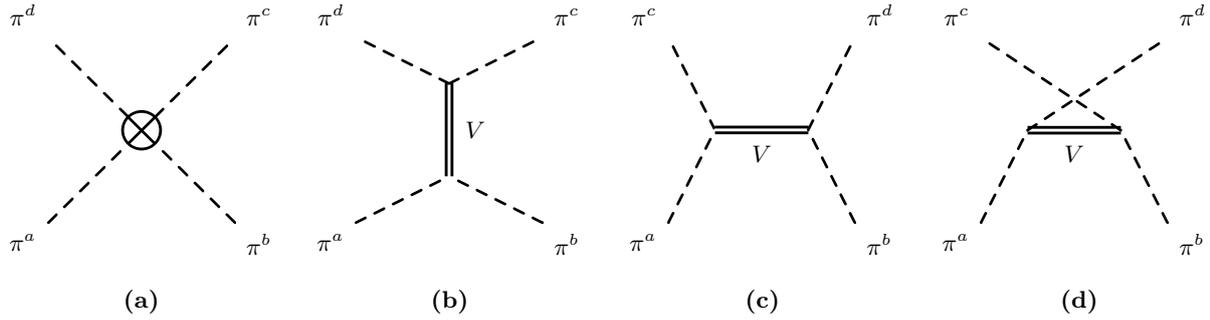\\
The amplitudes corresponding to these diagrams are
\be\l{k11}
\Mamp\(s,t,u\)_{\text{contact}}=-\f{8h_{2}}{v^{4}}\(s^{2}+2ut\)+\f{8h_{3}}{v^{4}}\(t^{2}+u^{2}\)
\ee
and 
\be\l{k12}
\Mamp\(s,t,u\)_{V}=\f{g_{V}^{2}}{v^{4}}\Bigg[s^{2}+2ut+M_{V}^{2}\(\f{s-u}{t-M_{V}^{2}}+\f{s-t}{u-M_{V}^{2}}\)\Bigg]\,.
\ee
These amplitudes grow with $s^{2}$ and have an asymptotic behavior that is worst than in the Higgsless SM (see Eq.~\eqref{k3}). On the other  hand there is a unique choice of the parameters $h_{1}$ and $h_{2}$ which cancels the growth proportional to $s^{2}$ from the sum of the amplitudes \eqref{k11} and \eqref{k12}: $h_{2}=g_{V}^{2}/8$ and $h_{3}=0$. As we will see in the next chapter, these values of the parameters are those predicted by a minimal gauge model (see Eq.~\eqref{gaugeparam}). It is worth noting that this cancellation holds for an arbitrary number of gauge vectors for $s\gg M_{V_{i}}^{2}$ . Indeed, in the case of more than one vector, the $g_{V}^{2}$ in Eq.~\eqref{k12} becomes a sum over the $\tilde{g}_{V_{i}}^{2}$ (we are using the same notation of Section~\ref{sec3.3.2} to indicate the couplings in the mass-eigenstate basis), while the relations \eqref{newcontactmass} provide the cancellation of the terms proportional to $s^{2}$.
This is an explicit example of a general result concerning massive vectors in field theory: the gauge invariance protects the scattering amplitudes from terms that grow more than $s/v^{2}$. We will see other examples of this fact in Sections~\ref{sec4.2} and \ref{sec5.1} where we study the asymptotic behavior of the $W_{L}W_{L}\to V_{\lambda}V_{\lambda'}$ and $W_{L}W_{L}\to V_{L}h$ scattering amplitudes.

Coming back to the longitudinal $WW$ scattering we can now sum up the contributions \eqref{k3}, \eqref{k11} and \eqref{k12} (assuming the values of $h_{2}$ and $h_{3}$ that cancel the terms proportional to $s^{2}$) finding the final result
\be\l{k13}
\bry{lll}
\dst \Mamp\(s,t,u\)_{V-\text{SM}}&=&\dst\Mamp\(s,t,u\)_{\hsl-\text{SM}}+\Mamp\(s,t,u\)_{V}+\Mamp\(s,t,u\)_{\text{contact}}\\\\
&=&\dst \f{s}{v^{2}}-\f{G_{V}^{2}}{v^{4}}\Bigg[3s+M_{V}^{2}\(\f{s-u}{t-M_{V}^{2}}+\f{s-t}{u-M_{V}^{2}}\)\Bigg]\,.
\ery
\ee
The cancellation of the linear growth with $s$ of this amplitude occurs for
\be\l{k14}
G_{V}=\f{v}{\sqrt{3}}\,.
\ee
In this case the $a_{0}^{0}$ coefficient is given by
\be\l{k15}
\lim_{s\gg M_{W}^{2}}\(a^{0}_{0}\)_{V-\text{SM}}=\f{M_{V}^{2}}{16\pi v^{2}}\bigg[\f{s}{M_{V}^{2}}\(1-\f{3G_{V}^{2}}{v^{2}}\)+\f{2G_{V}^{2}}{v^{2}}\Big[\(2+\f{M_{V}^{2}}{s}\)\log\(\f{s}{M_{V}^{2}}+1\)-1\Big]\bigg]\,.
\ee
Requiring perturbative unitarity up to a cutoff $\Lambda= 4\pi v\approx 3$ TeV we find, from the strongest unitarity bound $|a_{0}^{0}|\leq 1$, the allowed region in the plane $\(M_{V},G_{V}\)$ depicted in Fig.~\ref{fig:UnitarityHiggsless}. 
\begin{figure}[t]
\centering
\includegraphics[width=8cm]{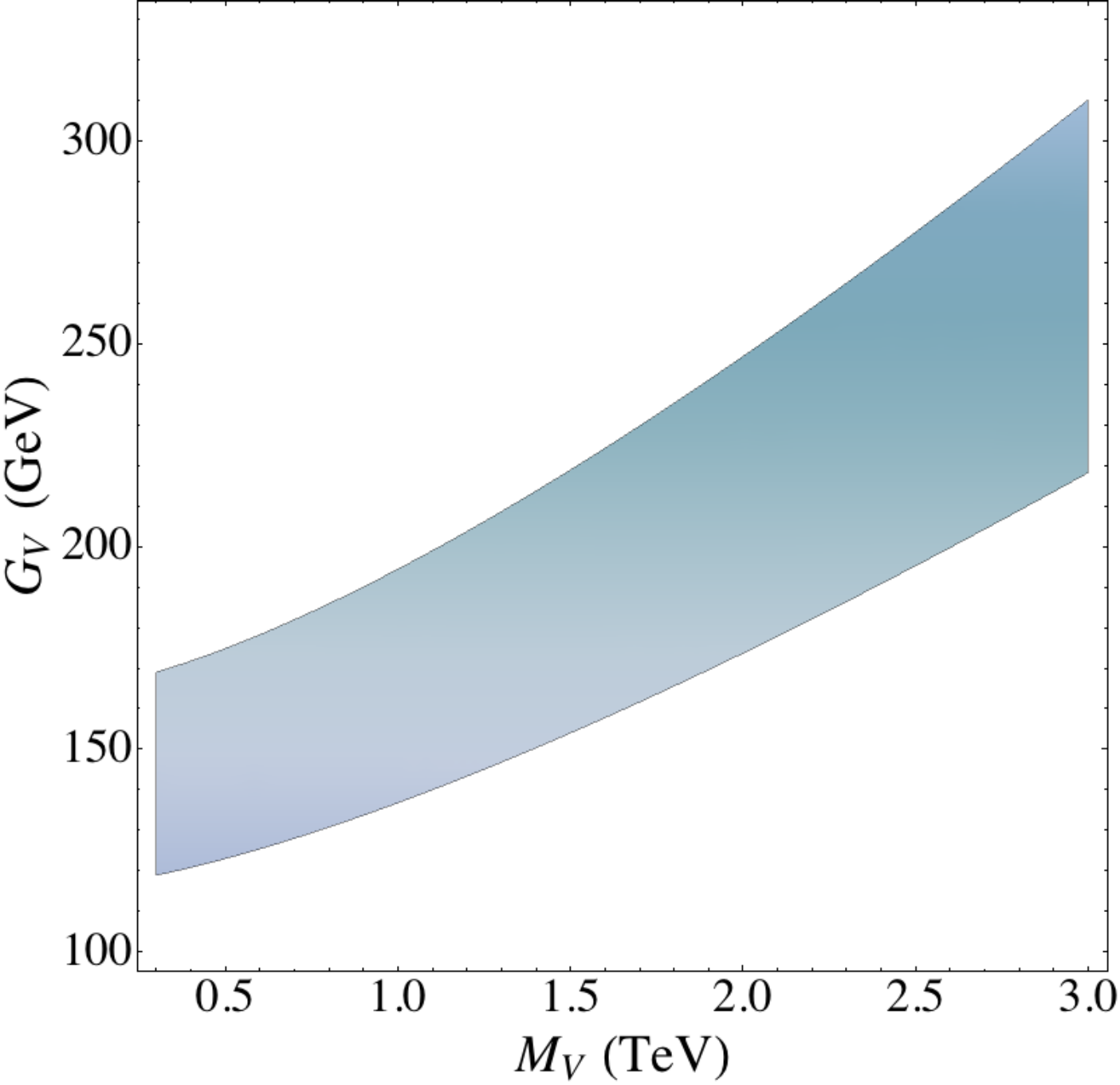}
\caption{Region in the parameter space $\(M_{V},G_{V}\)$ allowed by the strongest unitarity constraint in a Higgsless model with a heavy composite vector.}
\l{fig:UnitarityHiggsless}
\end{figure}
Note that the partial wave coefficient $a_{0}^{0}$ grows with the energy even for $G_{V}=v/\sqrt{3}$, albeit only logarithmically. This implies that perturbative unitarity is restored more efficiently for $G_{V}$ slightly above $v/\sqrt{3}$. It is also worth noting that the choice $G_{V}=v/\sqrt{3}$ does not coincide with the relation $G_{V}=v/2$ that one has in a minimal gauge model like the $3$-site model \cite{Matsuzaki:2007ft}. In fact it is known that in gauge models the linear growth with $s$ in the amplitude \eqref{k13} can be canceled only by the contribution of the full tower of resonances, i.e. by all the Kaluza Klein modes in 5D or all the heavy vectors in the 4D infinite sites limit, and even in this case the coefficient $a_{0}^{0}$ still grows logarithmically with $s$.
Finally, as can be read off Fig.~\ref{fig:UnitarityHiggsless}, the unitarity constraint does not set a strong upper bound on $M_{V}$ for suitable values of $G_{V}$: a value of $G_V$ between $150$ and $200$ GeV keeps the elastic longitudinal $WW$ scattering amplitude from saturating the unitarity bound below $\Lambda\approx 3$ TeV almost independently of $M_V \lesssim 1.5$ TeV \cite{Bagger:1993dz,Barbieri:2008p1580}.

\subsection{ElectroWeak Precision Tests}\l{EWPT3}
At tree-level the new vector contributes to the $S$ parameter through its mixing with the EW gauge bosons given by the diagram of Fig.~\ref{fig:STreeVector} while the tree-level contribution to the $T$ parameter is zero due to the custodial $SU\(2\)_{L+R}$ symmetry.
 \begin{figure}[t]
  \centering
   \begin{fmffile}{Shat4}
 \fmfframe(1,7)(1,7){
          \begin{fmfgraph*}(90,45)
          \fmfstraight
          \fmfleft{l1}
          \fmfright{r1}
          \fmflabel{\footnotesize{$B$}}{l1}
          \fmflabel{\footnotesize{$W$}}{r1}
          \fmf{photon,tension=.6}{l1,v1}
          \fmf{dbl_plain,tension=.6,label={\footnotesize{$V$}}}{v1,v2}
          \fmf{photon,tension=.6}{v2,r1}
          \end{fmfgraph*}}
      \end{fmffile}
 \caption{Feynman diagram representing the tree-level contribution of the composite vector $V_{\mu}^{a}$ to the $S$ parameter.}\l{fig:STreeVector}
\end{figure}
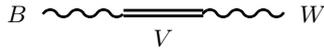
In formulae we have
\be\l{STtreeVector}
\bry{lll}
\dst \D\hat{S}_{V}^{\(\text{Tree}\)}&=&\dst \f{g^{2}}{4}\(\f{F_{V}^{2}}{M_{V}^{2}}-\f{F_{A}^{2}}{M_{A}^{2}}\)\,,\\\\
\dst \D\hat{T}_{V}^{\(\text{Tree}\)}&=&\dst 0\,,
\ery
\ee
where, for completeness, we have also considered the contribution of a heavy axial vector of mass $M_{A}$\footnote{Note that the presence of the axial vector does not modify the longitudinal elastic $WW$ scattering, so that the results of previous subsection remain unchanged.}. It was shown in the literature \cite{Barbieri:2008p1580,Cata:2010vn} that the one-loop contributions of composite vectors to the $S$ and $T$ parameters in the framework of Higgsless models can be sizable. Here we adopt the approach of Barbieri et al. \cite{Barbieri:2008p1580} considering only the one-loop contribution to $T$ that is the only parameter vanishing at tree-level\footnote{A more complete treatment of the one loop contributions to $S$ and $T$ in Higgsless models can be found in Ref.~\cite{Cata:2010vn}.}. In particular for the contributions given by the Feynman diagrams of Fig.~\ref{fig:ToneloopVector} we find
\begin{figure}[!htb]
\begin{minipage}[b]{4.5cm}
  \centering
   \begin{fmffile}{That4}
 \fmfframe(1,7)(1,7){
          \begin{fmfgraph*}(90,45)
          \fmfstraight
          \fmftop{t1,t2,t3}
          \fmfbottom{b1,b2,b3}
          \fmfleft{l1,l2,l3,l4,l5}
          \fmfright{r1,r2,r3,r4,r5}
          \fmf{phantom}{v3,v5}          
          \fmf{dashes,tension=.6}{l3,v1}
          \fmf{photon,left,label={\footnotesize{$B$}},tension=.2}{v1,v2}
          \fmf{dbl_plain,left,label={\footnotesize{$V,,A$}},tension=.2}{v2,v1}
          \fmf{dashes,tension=.6}{v2,r3}
          \fmf{phantom}{t2,v3}
          \fmf{phantom,tension=.195}{v3,v4}
          \fmf{phantom}{v4,b2}
	 \fmffreeze
          \end{fmfgraph*}}
      \end{fmffile}
       \end{minipage}
    \ \hspace{7mm} \
\begin{minipage}[b]{4.5cm}
   \centering
   \begin{fmffile}{That5}
 \fmfframe(1,7)(1,7){
          \begin{fmfgraph*}(90,45)
          \fmfstraight
          \fmftop{t1,t2,t3}
          \fmfbottom{b1,b2,b3}
          \fmfleft{l1,l2,l3,l4,l5}
          \fmfright{r1,r2,r3,r4,r5}
          \fmf{dashes,tension=.6}{l3,v1}
          \fmf{phantom,left,tension=.2}{v1,v2}
          \fmf{dashes,left,label={\footnotesize{$\pi$}},tension=.2}{v2,v1}
          \fmf{dashes,tension=.6}{v2,r3}
          \fmf{phantom}{t2,v3}
          \fmf{phantom,tension=.195}{v3,v4}
          \fmf{phantom}{v4,b2}
	 \fmffreeze
          \fmf{photon,left=.42,label={\footnotesize{$B$}}}{v1,v3}
          \fmf{dbl_plain,left=.42,label={\footnotesize{$V$}}}{v3,v2}
          \end{fmfgraph*}}
      \end{fmffile}
       \end{minipage}
 \ \hspace{7mm} \
\begin{minipage}[b]{4.5cm}
     \begin{fmffile}{That6}
 \fmfframe(1,7)(1,7){
          \begin{fmfgraph*}(90,45)
          \fmfstraight
          \fmftop{t1,t2,t3,t4,t5,t6,t7,t8,t9}
          \fmfbottom{b1,b2,b3,b4,b5,b6,b7,b8,b9}
          \fmfleft{l1,l2,l3,l4,l5}
          \fmfright{r1,r2,r3,r4,r5}
          \fmf{dashes,tension=.6}{l3,v1}
          \fmf{phantom,left,tension=.2}{v1,v2}
          \fmf{dashes,left,label={\footnotesize{$\pi$}},tension=.2}{v2,v1}
          \fmf{dashes,tension=.6}{v2,r3}
          \fmf{phantom}{t4,v3}
          \fmf{phantom,tension=.24}{v3,v4}
          \fmf{phantom}{v4,b4}
          \fmf{phantom}{t6,v5}
          \fmf{phantom,tension=.24}{v5,v6}
          \fmf{phantom}{v6,b6}
	 \fmffreeze
          \fmf{dbl_plain,left=.26,label={\footnotesize{$V$}}}{v1,v3}
          \fmf{photon,left=.26,label={\footnotesize{$B$}}}{v3,v5}
          \fmf{dbl_plain,left=.26,label={\footnotesize{$V$}}}{v5,v2}
          \end{fmfgraph*}}
      \end{fmffile}
       \end{minipage}
\caption{One loop contributions of the composite vector $V_{\mu}^{a}$ to the $\hat{T}$ parameter.}\l{fig:ToneloopVector}
\end{figure}
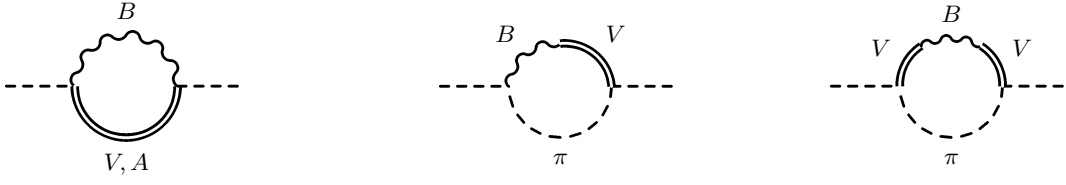
\be\l{eq22}
\bry{lll}
\dst \D\hat{T}_{V,A}^{\(\text{1-loop}\)}&=&\dst \f{3\pi \a}{4c_{W}^{2}}\Bigg[\f{\(F_{V}-2G_{V}\)^{2}}{M_{V}^{2}}+\f{F_{A}^{2}}{M_{A}^{2}}\Bigg]\f{\Lambda^{2}}{\(4\pi v\)^{2}}\\\\
&&\dst +\f{2\pi\a}{\(4\pi v\)^{2}c_{W}^{2}}\Bigg[\(4G_{V}^{2}+6F_{V}G_{V}-\f{3}{2}F_{V}^{2}\f{2G_{V}^{2}+v^{2}}{v^{2}}\)\log\(\f{\Lambda}{M_{V}}\)\\\\
&&\dst -G_{V}^{2}\log\(\f{\Lambda}{m_{W}}\)-\f{3}{2}F_{A}^{2}\log\(\f{\Lambda}{M_{A}}\)\Bigg]\,.\\\\
\ery
\ee
From these relations we immediately see that the quadratic contribution to $T$ is always positive. 
Note that in the case of a minimal gauge model in which the relations $F_{A}=0$ and $F_{V}=2 G_{V}$ hold, the quadratic contribution in Eq.~\eqref{eq22} vanishes and the remaining logarithmic contribution becomes
\be\l{eq23}
\bry{lll}
\dst \D\hat{T}_{V,A}^{\(\text{1-loop}\)}\Big|_{F_{V}=2G_{V}\,,F_{A}=0}
&=&\dst -\f{3\a}{8\pi c_{W}^{2}}\Bigg\{\log\(\f{M_{V}}{\Lambda}\)+\f{G_{V}^{2}}{3v^{2}}\log\(\f{M_{V}}{m_{W}}\)\\\\
&&\dst +\Bigg[\(1-2\f{G_{V}^{2}}{v^{2}}\)^{2}+\f{G_{V}^{2}}{v^{2}}\Bigg]\log\(\f{\Lambda}{M_{V}}\)\Bigg]\Bigg\}\,.\\\\
\ery
\ee
We immediately see that the first term in the curly brackets cuts-off the infrared contribution of Eq.~\eqref{P8} and only the term in the second line gives a logarithmic contribution to the $T$ parameter. Moreover, contrary to the quadratic contribution in Eq.~\eqref{eq22}, the logarithmic contribution in Eq.~\eqref{eq23} is always negative.\\
Going back to Eq.~\eqref{eq22} we see that the quadratic contribution is in general parametrically large, so that in absence of other contributions it is necessary to find a way to decrease its size. For example, the inclusion of a trilinear coupling of the form $\pi V A$ can in principle solve the problem. In fact this is exactly the case \cite{Barbieri:2008p1580}. It was shown that the introduction of vertices involving both the vector and the axial states can cut-off the quadratic divergence in Eq.~\eqref{eq22} substituting $\Lambda$ with the mass of the heavier resonances. In fact we can follow the argument of Barbieri et al. \cite{Barbieri:2008p1580} assuming that the cancellation involves heavy states of mass close to the cutoff, so that the estimate of the quadratic corrections in Eq.~\eqref{eq22} is reasonable with a cutoff $\Lambda\approx 3$ TeV. The contributions of the heavy vectors to the EWPT is shown in Fig.~\ref{fig:EWPTHVHM} for $\Lambda = 4\pi v$ and for different values of $M_{V}$. The other parameters are constrained by requiring the following conditions:
\ben
\item $\dst \f{F_{V}}{M_{V}}>\f{F_{A}}{M_{A}}$, which ensures that the tree-level contribution to $\hat{S}$ of Eq.~\eqref{STtreeVector} is always positive;
\item $\dst \lim_{s\gg M_{W}^{2}}\left|\(a^{0}_{0}\)_{V-\text{SM}}<1\right|$, where $\(a^{0}_{0}\)_{V-SM}$ is given by Eq.~\eqref{k15}. This constraint ensures that the strongest unitarity bound is always satisfied.
\een
\begin{figure}[h!]
\begin{minipage}[b]{4.9cm}
   \centering
   \includegraphics[width=4.8cm]{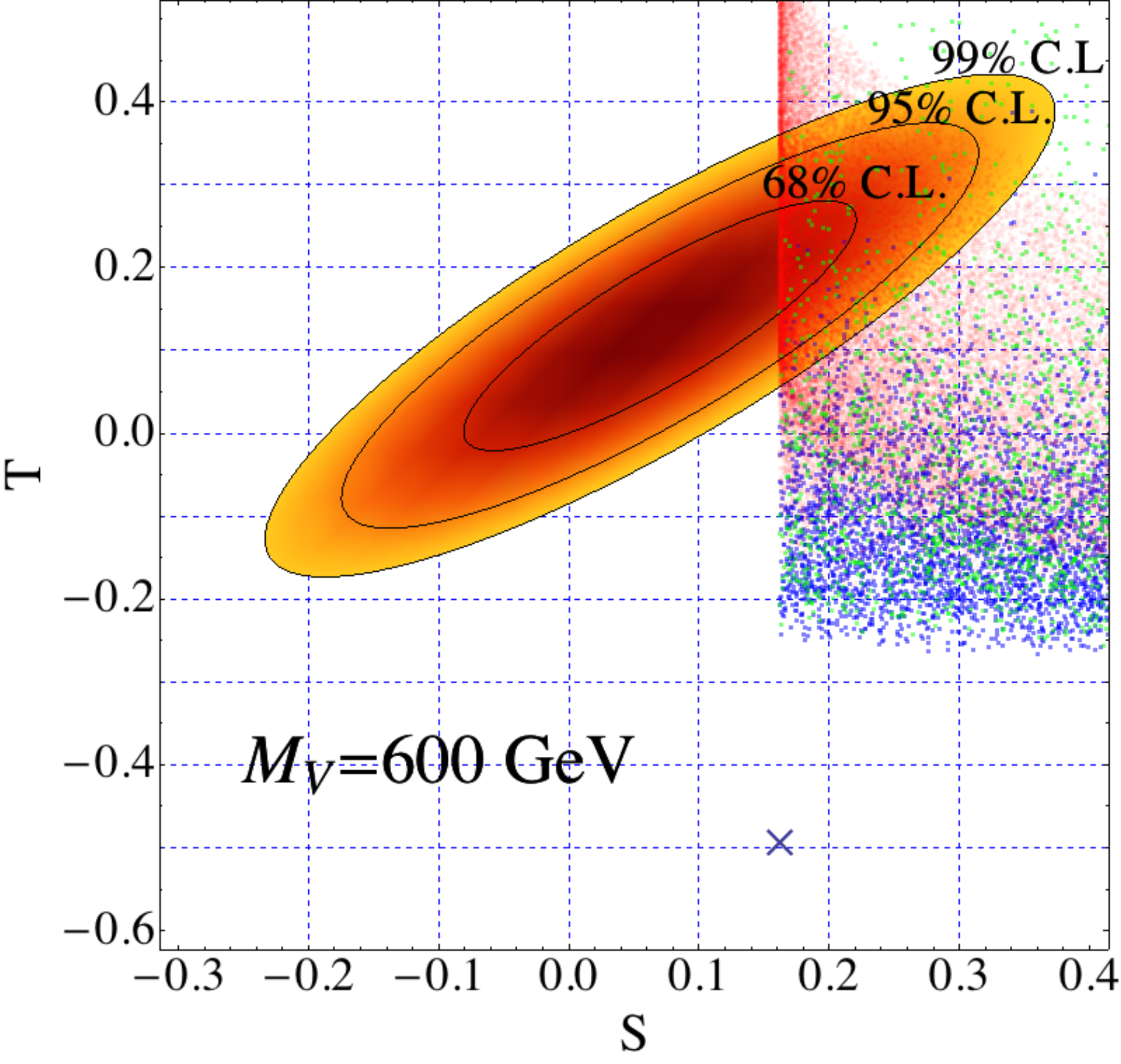}
 \end{minipage}
\ \hspace{3mm} \ 
\begin{minipage}[b]{4.9cm}
  \centering
   \includegraphics[width=4.8cm]{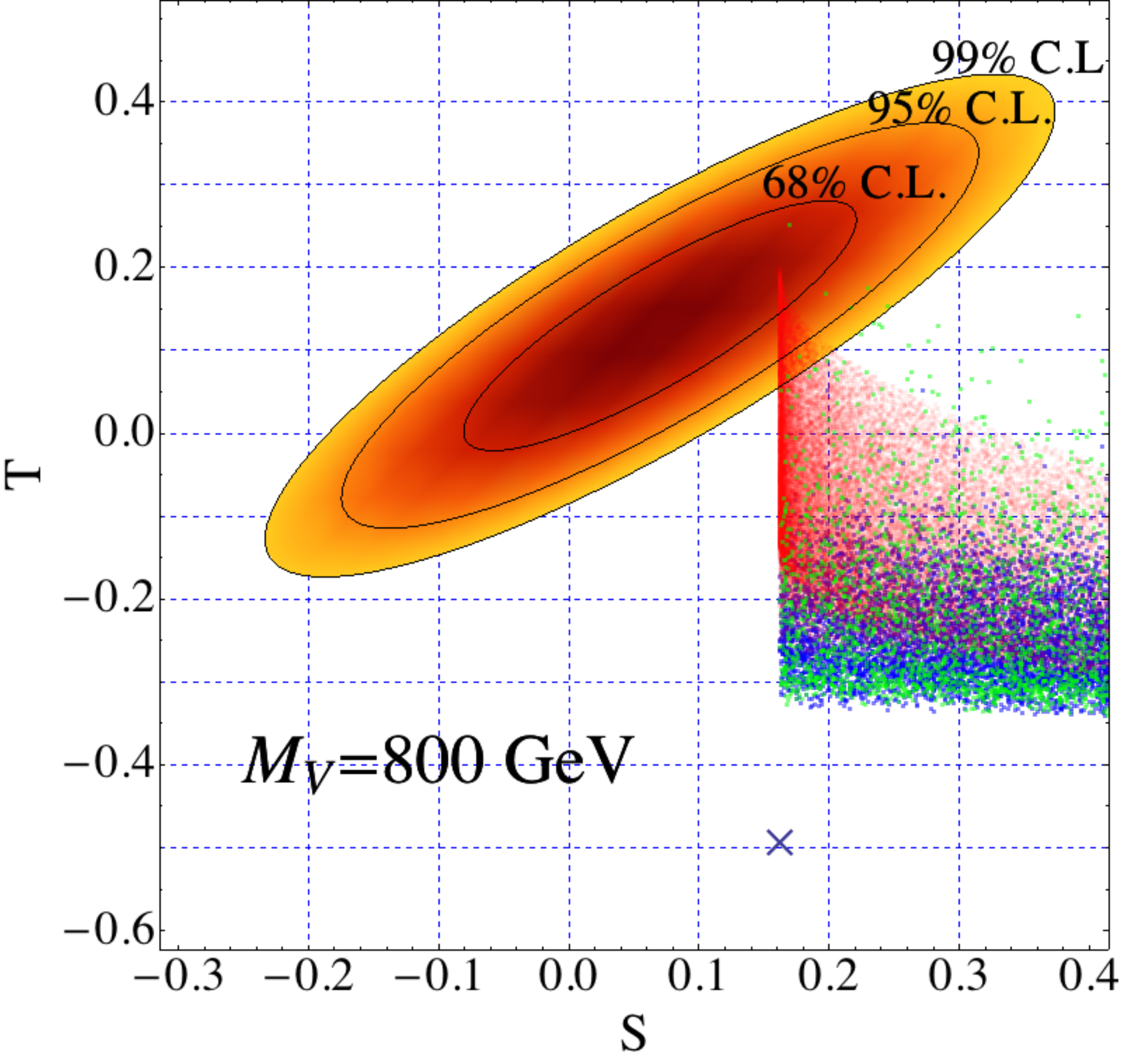}
 \end{minipage}
 \ \hspace{3mm} \ 
\begin{minipage}[b]{4.9cm}
  \centering
   \includegraphics[width=4.8cm]{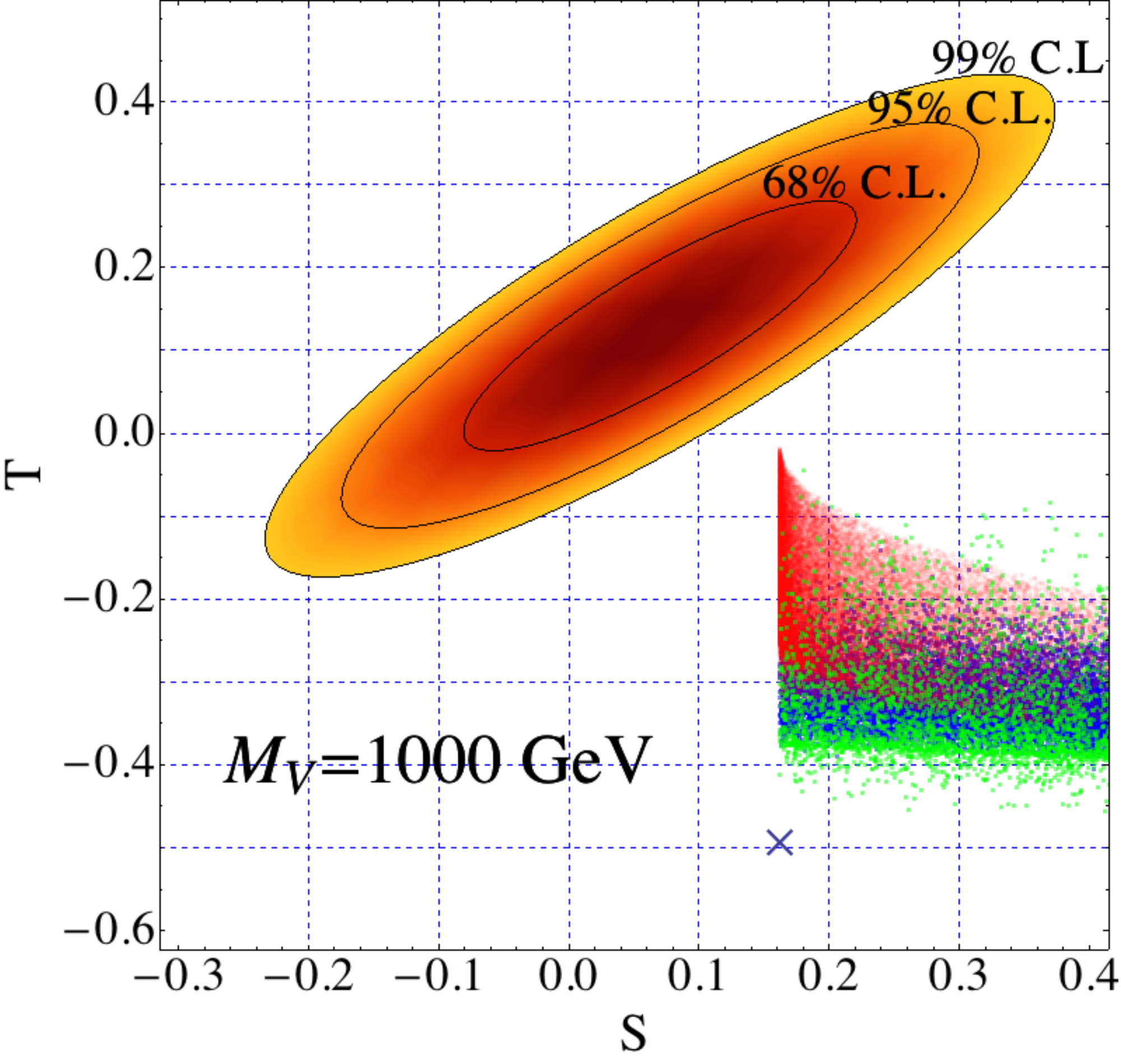}
 \end{minipage}
\caption{Experimental allowed regions and theoretical predictions for the $S$ and $T$ parameters in the SM with two composite vector bosons of opposite parity (see Eq.~\eqref{eq21}) for $M_{V}=600,\,800,\,1000$ GeV and for three different regions of the parameter space: $F_{V}< G_{V}$ (blue), $G_{V}<F_{V}<2 G_{V}$ (red) and $2G_{V}<F_{V}<2.5 G_{V}$ (green). The small cross at negative $T$ is the point $G_{V}=F_{V}=F_{A}=0$, i.e. the Higgsless SM with a cutoff $\Lambda=3$ TeV.}
\l{fig:EWPTHVHM}
\end{figure}
In the plots of Fig.~\ref{fig:EWPTHVHM}, the density of points gives a graphical representation of the population of the $\(S,T\)$ points as function of the parameters $G_{V}$, $F_{V}$, $M_{A}$, $F_{A}$ for fixed $M_{V}$ and constrained by the above two conditions. We can also quantify the level of fine-tuning of each point in the $\(S,T\)$ plane. To do this we follow Ref.~\cite{Barbieri:1988p1552} and define the level of fine-tuning in the definition of a set of observables $\mathcal{O}_{i}$ as functions of the parameters $a_{j}$ the quantity
\be\l{finetuning}
\Delta=\max\{\Delta_{ij}\}=\max\Bigg\{\left|\f{a_{j}}{\mathcal{O}_{i}}\f{\de \mathcal{O}_{i}\(a_{j}\)}{\partial a_{j}}\right|\Bigg\}\,.
\ee
By applying this definition to the set of observables $S$ and $T$ defined as functions of the parameters $M_{V}$, $G_{V}$, $F_{V}$, $M_{A}$ and $F_{A}$ we have studied the level of fine-tuning of any point in the parameter space. In Fig.~\ref{fig:EWPTHVHMFineTuning} the prediction for $S$ and $T$ obtained varying all the parameters is compared with the experimental bound using a different coloring to represent the level of fine-tuning, i.e. the value of $\Delta$ for each given point in the $\(S,T\)$ plane.
\begin{figure}[htbp!]
\centering
\includegraphics[width=8cm]{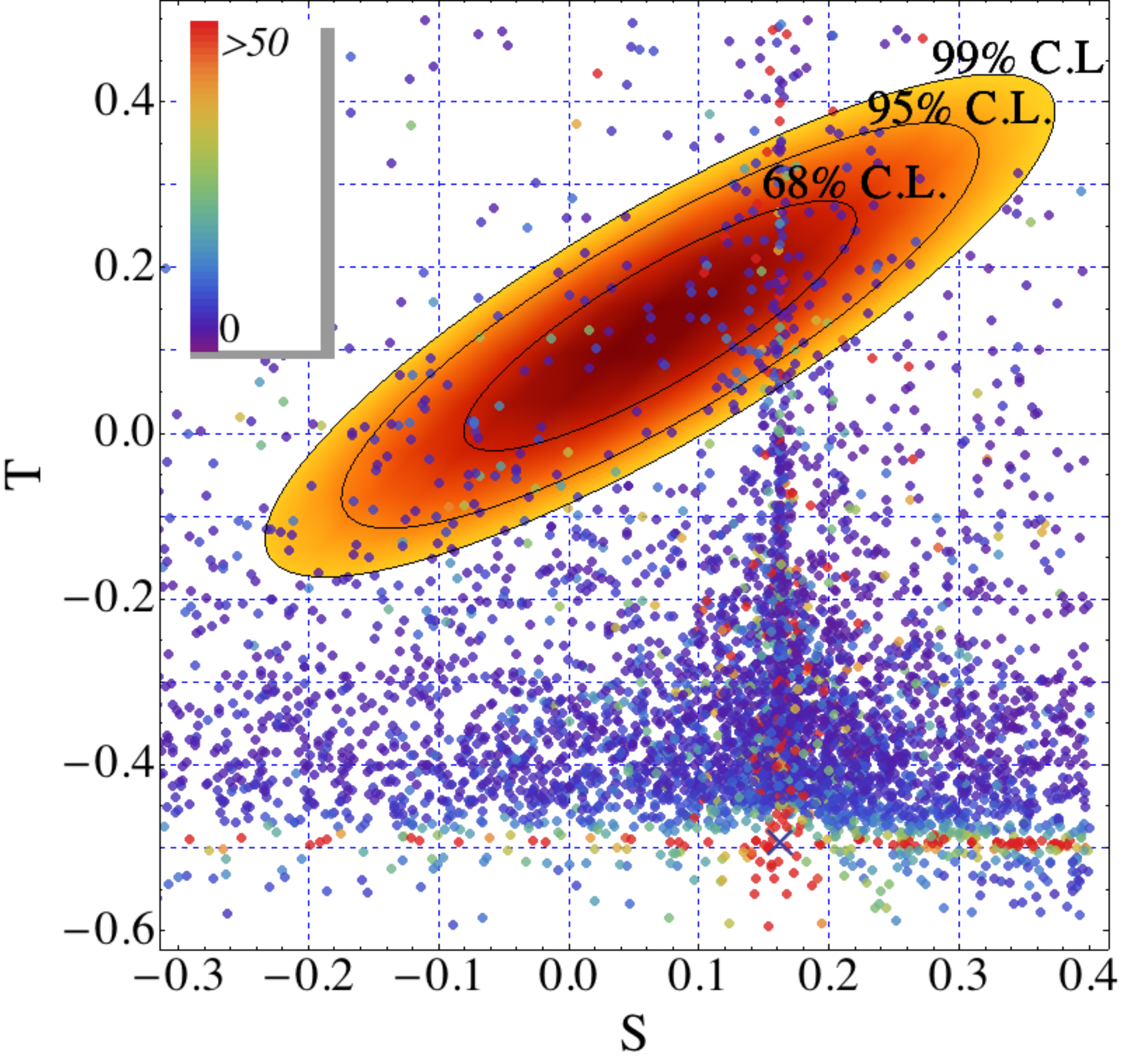}
\caption{Experimental allowed regions and theoretical predictions for the $S$ and $T$ parameters in the SM with two composite vector bosons of opposite parity obtained spanning over all the parameter space. The coloring of the points represents the level of fine-tuning $\Delta$ (see the text for details). The parameter space is always constrained to satisfy the strongest unitarity bound.}
\l{fig:EWPTHVHMFineTuning}
\end{figure}
From this figure we immediately see that even if the experimental ellipses are not the most populated regions, the level of fine-tuning necessary to satisfy the constraints is in the range $\Delta\lesssim 10\text{-}20$, i.e. a fine-tuning of $5$-$10\%$.

\section{A composite scalar--vector system}\l{ASVS}
We can consider now the case in which both a light scalar and a relatively light vector are present in the low energy spectrum. In this case the Lagrangian is given by the sum of the Higgsless SM Lagrangian, the scalar contribution, the vector contribution and a new interaction among the scalar and the vector, which is irrelevant for the present analysis but will be crucial in the study of the phenomenology (see Chapter~\ref{chap5})
\be\l{HVSM}
\bry{lll}
\La_{h-V-\text{SM}}=\La_{h-\text{SM}}+\La_{V}+\La_{h-V}\,,
\ery
\ee
where $\La_{h-\text{SM}}$, $\La^{V}$ and $\La_{h-V}$ are given by Eqs.~\eqref{HSM}, \eqref{Ltot} and \eqref{lvhinteraction} respectively.

\subsection{Elastic unitarity}\l{ElasticUnitarity4}
Since the term describing the trilinear interaction among the scalar and the heavy vectors is irrelevant for elastic $W_{L}W_{L}$ scattering, the amplitude for this process is given simply by the sum of the IR, the scalar and the vector contributions, i.e.
\be\l{mainamp}
\bry{lll}
\dst \Mamp\(s,t,u\)_{h-V-\text{SM}}&=&\dst\Mamp\(s,t,u\)_{\hsl-\text{SM}}+\Mamp\(s,t,u\)_{h}+\Mamp\(s,t,u\)_{V}+\Mamp\(s,t,u\)_{\text{contact}}\\\\
&=&\dst \f{s}{v^{2}}\(1-\f{a^{2} s}{s-m_{h}^{2}}-\f{3G_{V}^{2}}{v^{2}}\)-\f{G_{V}^{2}M_{V}^{2}}{v^{4}}\(\f{s-u}{t-M_{V}^{2}}+\f{s-t}{u-M_{V}^{2}}\)\,.
\ery
\ee
This amplitude grows with the c.o.m.~energy squared $s$ for generic values of the couplings $a$ and $G_{V}$. Let us study some limits of this relation. \\
First of all consider the limit $a\to 1$ and $G_{V}\to 0$. This correspond to the case in which $h$ is exactly the SM Higgs boson, i.e. can be embedded into a linear doublet of $SU\(2\)_{L}$.
In this case the amplitude \eqref{mainamp} has a constant asymptotic behavior and $a^{0}_{0}$ is given by the relation \eqref{a00sm}. We have shown in Section~\ref{ElasticUnitarity2} (Eqs.~\eqref{k9} and \eqref{k10}) that this relation has a finite limit for $s\gg m_{h}$ and that for $m_{h}\leq 32\pi v^{2}/5\approx 1$ TeV the perturbative unitarity is satisfied up to very high energy.\\
The second extreme case is the opposite one, the Higgsless limit $a=0$. In Section~\ref{ElasticUnitarity3} (Eqs.~\eqref{k14} and \eqref{k15}) we have shown that for $G_{V}=v/\sqrt{3}$ the term proportional to $s$ is again canceled in the amplitude \eqref{mainamp} even if a logarithmic dependence of $a_{0}^{0}$ on $s$ still remains.\\
Now let us consider the intermediate case, in which both the vector and the scalar contribute to the amplitude \eqref{mainamp}. We can again cancel the contribution proportional to $s$ in this amplitude by requiring
\be\l{agvrelation}
a=\sqrt{1-\f{3G_{V}^{2}}{v^{2}}}\,.
\ee
In this case the amplitude \eqref{mainamp} reads
\be\l{eq12e}
\bry{lll}
\dst\Mamp\(s,t,u\)_{h-V-\text{SM}}\Bigg|_{a=\sqrt{1-\f{3G_{V}^{2}}{v^{2}}}}
&=&\dst -\f{m_{h}^{2}}{v^{2}}\f{s}{s-m_{h}^{2}}\(1-\f{3G_{V}^{2}}{v^{2}}\)+\f{G_{V}^{2}M_{V}^{2}}{v^{4}}\(\f{u-s}{t-M_{V}^{2}}+\f{t-s}{u-M_{V}^{2}}\)\,,
\ery
\ee
and the $a_{0}^{0}$ coefficient in this case is easily computed to be
\be\l{eq12e}
\bry{lll}
\dst \lim_{s\gg m_{h}^{2}} \(a^{0}_{0}\)_{h-V-SM} &=&\dst -\f{1}{16\pi}\f{m_{h}^{2}}{v^{2}}\(1-\f{3G_{V}^{2}}{v^{2}}\)\(\f{5}{2}+\f{3}{2\(x_{h}-1\)}-\f{\log\(x_{h}+1\)}{x_{h}}\)\\\\
&&\dst +\f{G_{V}^{2}M_{V}^{2}}{8\pi v^{4}}\Big[\(2+\f{1}{x_{V}}\)\log\(x_{V}+1\)-1\Big]\,,
\ery
\ee
where $x_{h}=s/m_{h}^{2}$ and $x_{V}=s/M_{V}^{2}$
As in the case of the Higgsless model, the first coefficient of the partial wave expansion still grows logarithmically for large $s$. However, now the contribution of the scalar with opposite sign can partially compensate this growth for appropriate values of masses and couplings.
\begin{figure}[h!]
\begin{minipage}[b]{7.5cm}
   \centering
   \includegraphics[height=6cm]{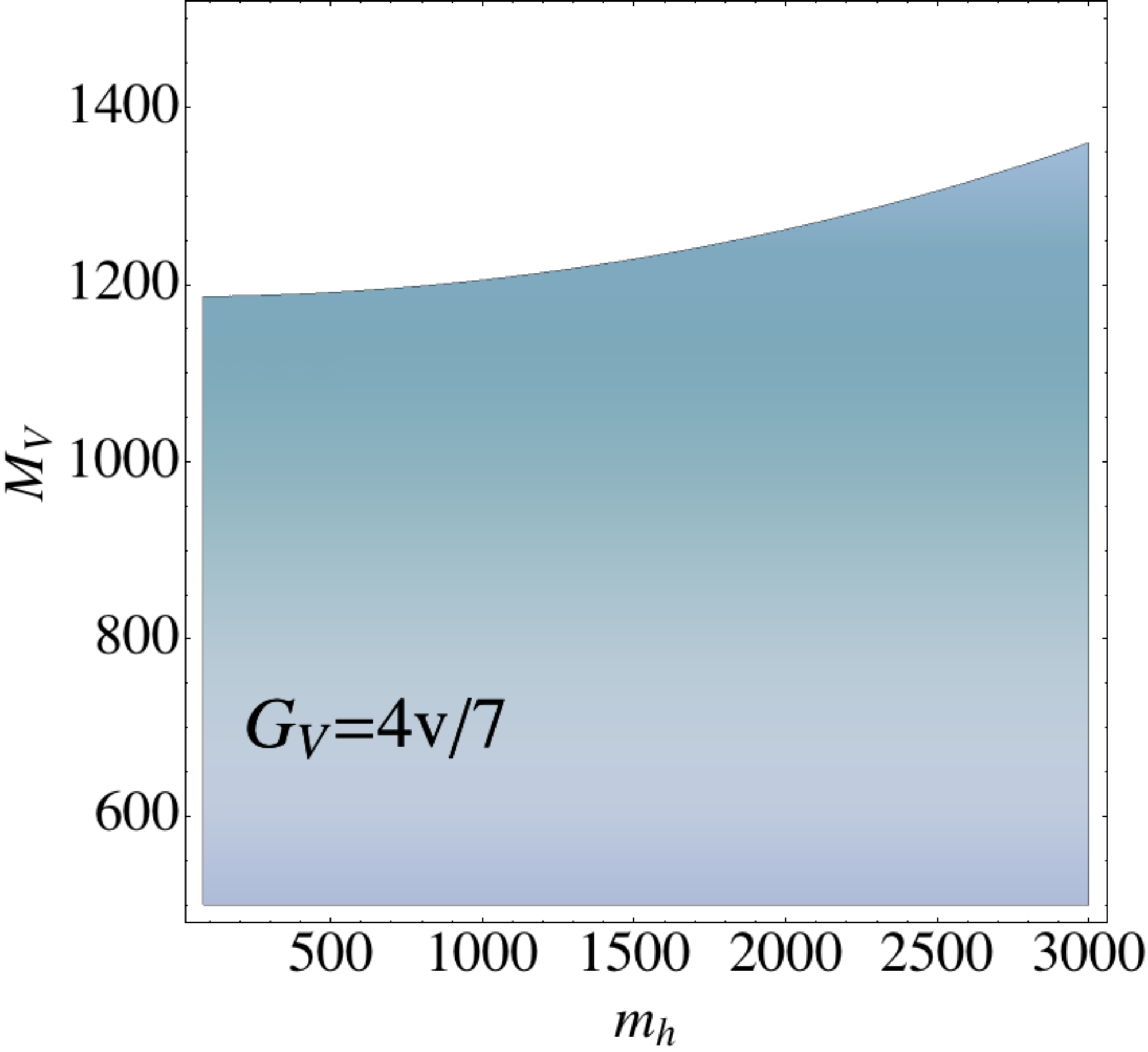}
 \end{minipage}
\ \hspace{-1mm} \ 
\begin{minipage}[b]{7.5cm}
  \centering
   \includegraphics[height=6cm]{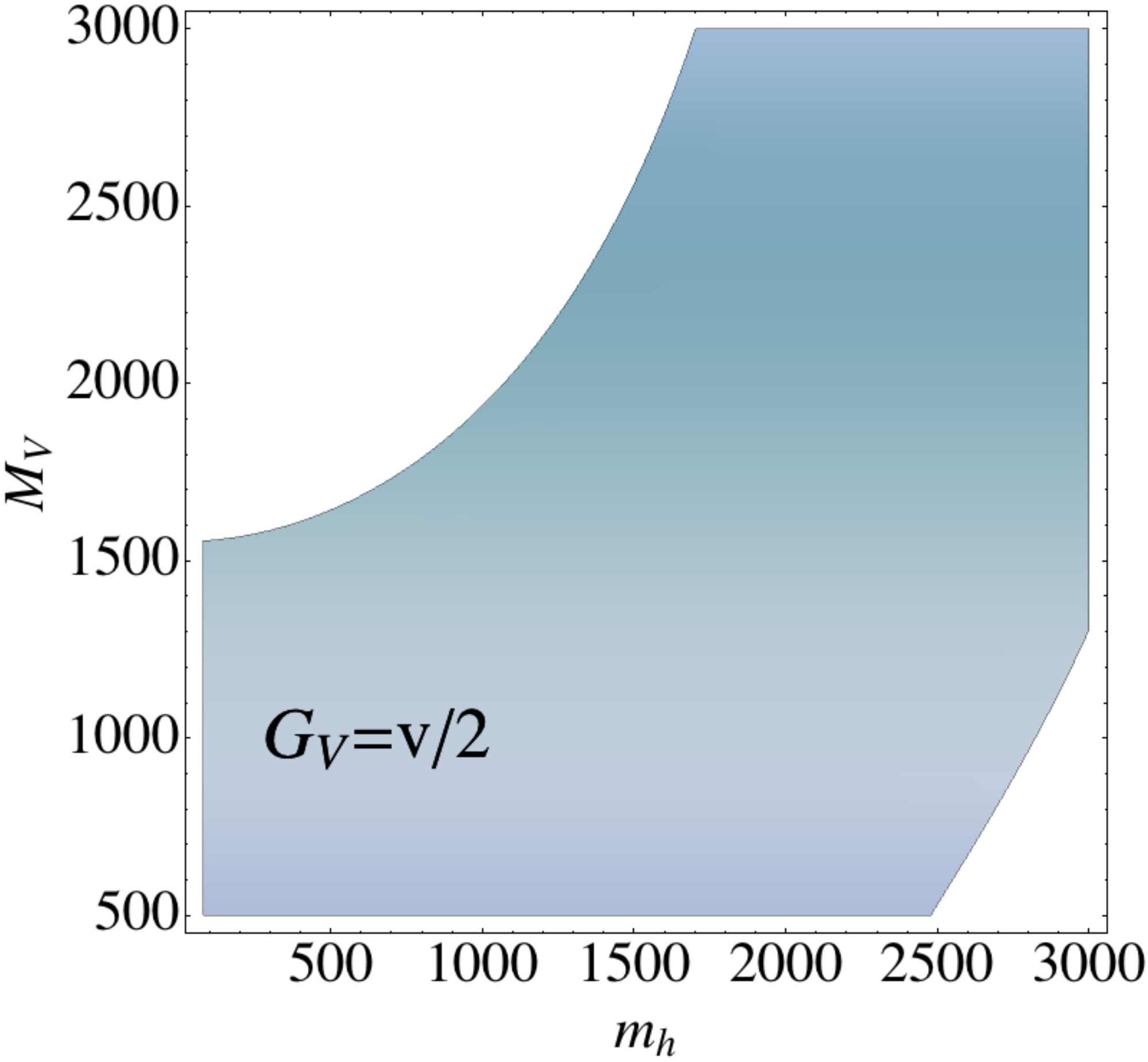}
 \end{minipage}
\ \hspace{-1mm} \\ \vspace{5mm}
 \begin{minipage}[b]{7.5cm}
  \centering
   \includegraphics[height=6cm]{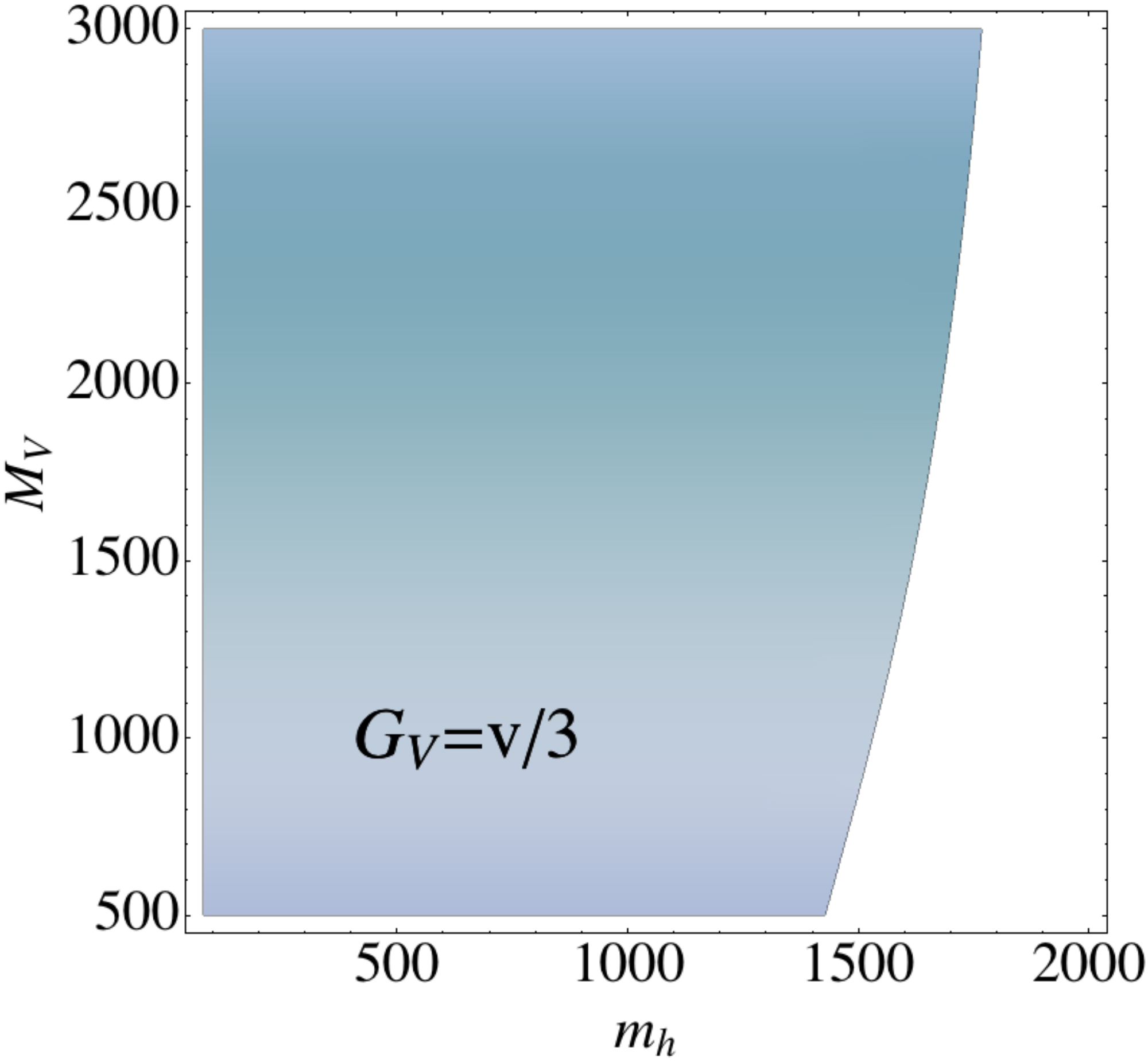}
 \end{minipage}
 \ \hspace{-1mm} \ 
 \begin{minipage}[b]{7.5cm}
  \centering
   \includegraphics[height=6cm]{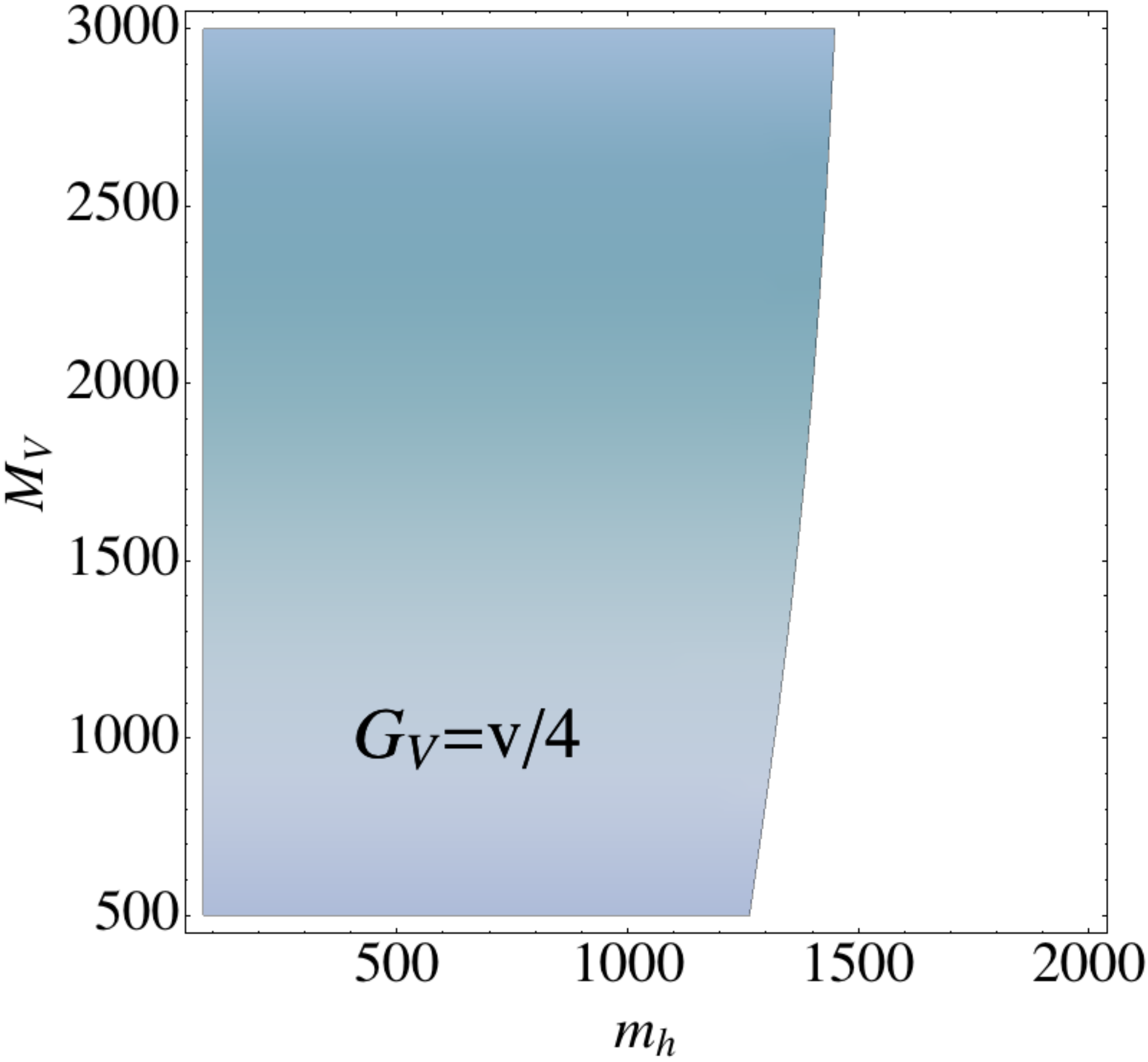}
 \end{minipage}
\caption{Allowed regions from the elastic unitarity constraint in the $\(m_{h},M_{V}\)$ plane for different values of the coupling $G_{V}$ in the range $80<G_{V}<160$ GeV.}
\l{Unitarityplots}
\end{figure}
We can use the strongest unitarity bound $|a_{0}^{0}|\leq 1$ to construct an allowed region in the $\(m_{h},M_{V}\)$ plane, for a fixed cutoff $\Lambda \approx 4\pi v$ and for some reference values of the coupling $G_{V}$\footnote{From now on we will always assume the relation \eqref{agvrelation} among the couplings $a$ and $G_{V}$ so that choosing a value of $G_{V}$ means choosing the corresponding value of $a$.}.
In Fig.~\ref{Unitarityplots} we have plotted the allowed regions in the plane $\(m_{h},M_{V}\)$ for different values of the coupling $G_{V}$. From the figures we immediately see that for a value of $G_{V}$ around the minimal gauge model value $v/2$ (which corresponds to the choice $a=1/2$) we have a wide region in which the elastic unitarity is preserved up to the cutoff $\Lambda\approx 3$ TeV. In particular we see that for a wide range of values of $G_{V}$ a relatively light vector resonance allows a heavy composite Higgs boson with a mass in the TeV range. On the other hand, we see that the presence of a scalar, even if very light, allows for a relatively light vector resonance with a mass below $1$ TeV. Finally it is worth noting that the relation \eqref{agvrelation}, for which the cancellation of the contribution growing with $s$ in the amplitude \eqref{mainamp} occurs, always implies $a\leq 1$. As we have already noticed in Section~\ref{ElasticUnitarity2} (before Eq.~\eqref{cutoff1}) this relation is peculiar of models where the Higgs is a pseudo-GB of an extended symmetry (which are generally described by the effective Lagrangian \eqref{HSM}) \cite{Giudice:2007p85,Contino:2010vc}.

\subsection{ElectroWeak Precision Tests}\l{EWPT4}
In the previous section we have seen that the interplay between the scalar and the vector leads to a compensation of the contributions growing linearly with $s$ in the longitudinal elastic gauge boson scattering amplitude. In this section we investigate the effect of the vector-scalar interplay on the EWPT\footnote{We are assuming that the contributions to $S$ and $T$ are in this case the sum of the scalar and the vector contributions computed in the previous sections. We are however neglecting operators like $\mathcal{O}=g_{h\pi A}h \demua \pi^{a}A^{a}_{\mu}$ which can be present in this case, contributing to $S$ and $T$.}. 
\begin{figure}[h!]
\begin{minipage}[b]{4.9cm}
   \centering
   \includegraphics[width=4.8cm]{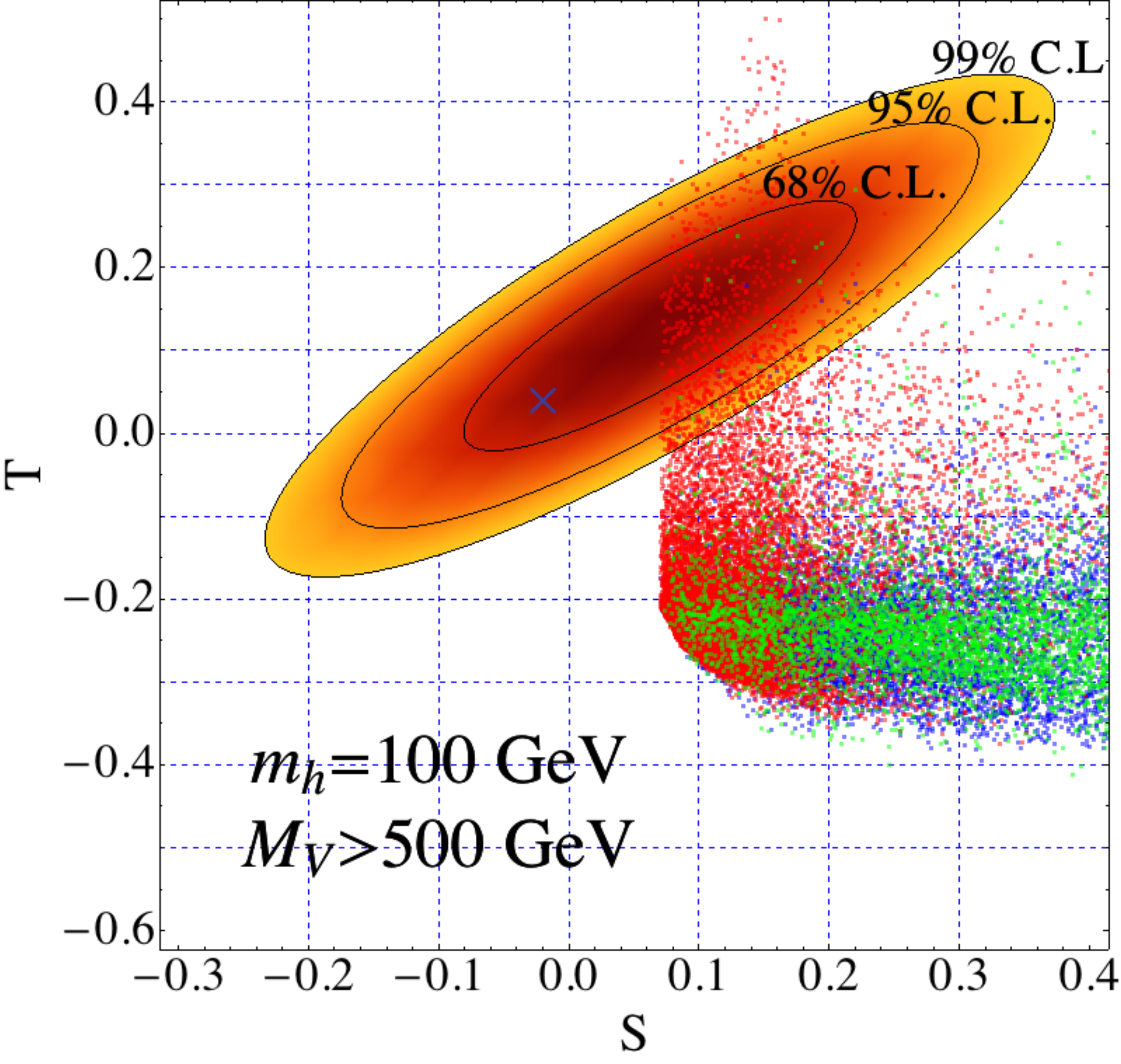}
 \end{minipage}
\ \hspace{3mm} \ 
\begin{minipage}[b]{4.9cm}
  \centering
   \includegraphics[width=4.8cm]{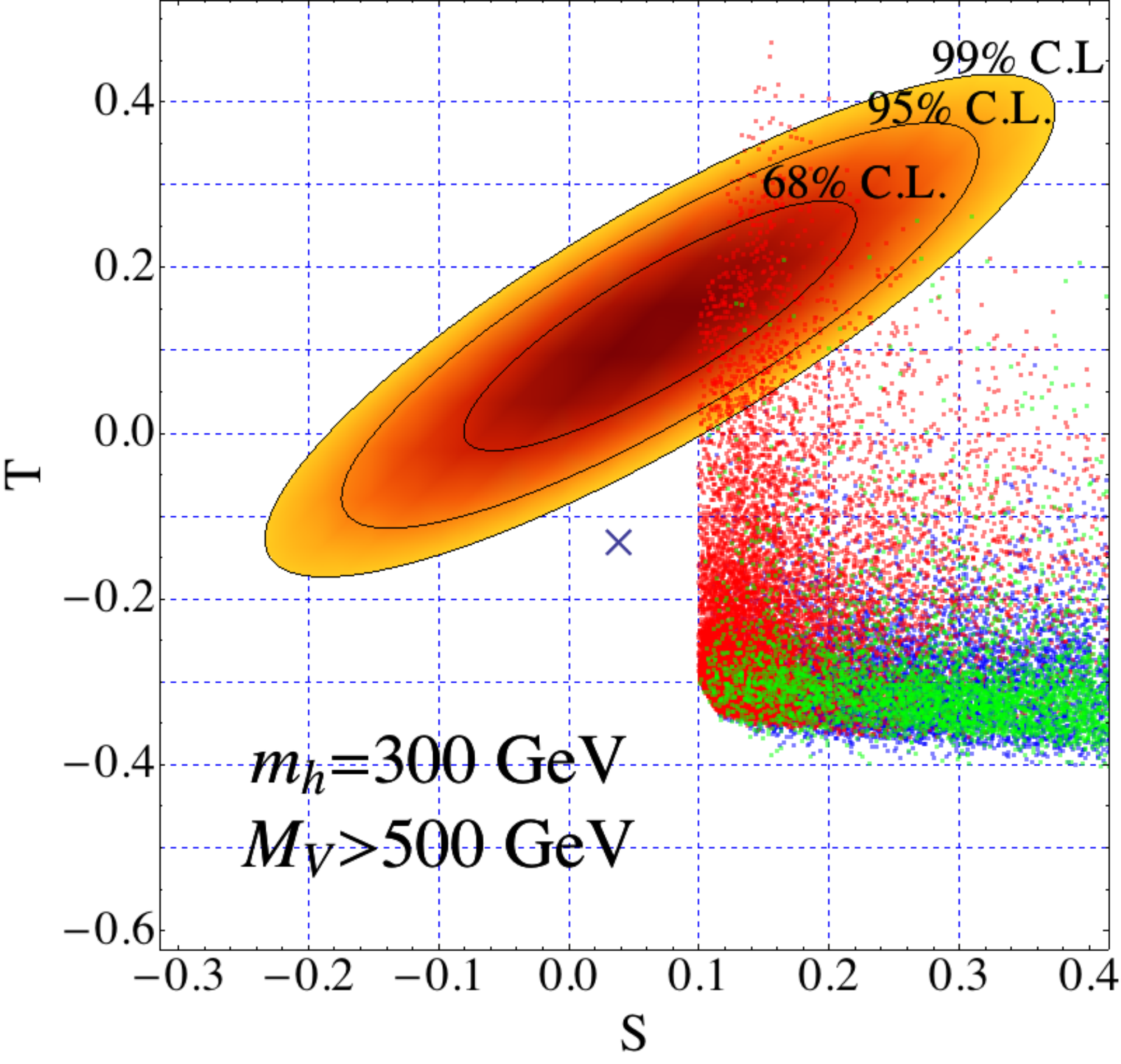}
 \end{minipage}
 \ \hspace{3mm} \ 
\begin{minipage}[b]{4.9cm}
  \centering
   \includegraphics[width=4.8cm]{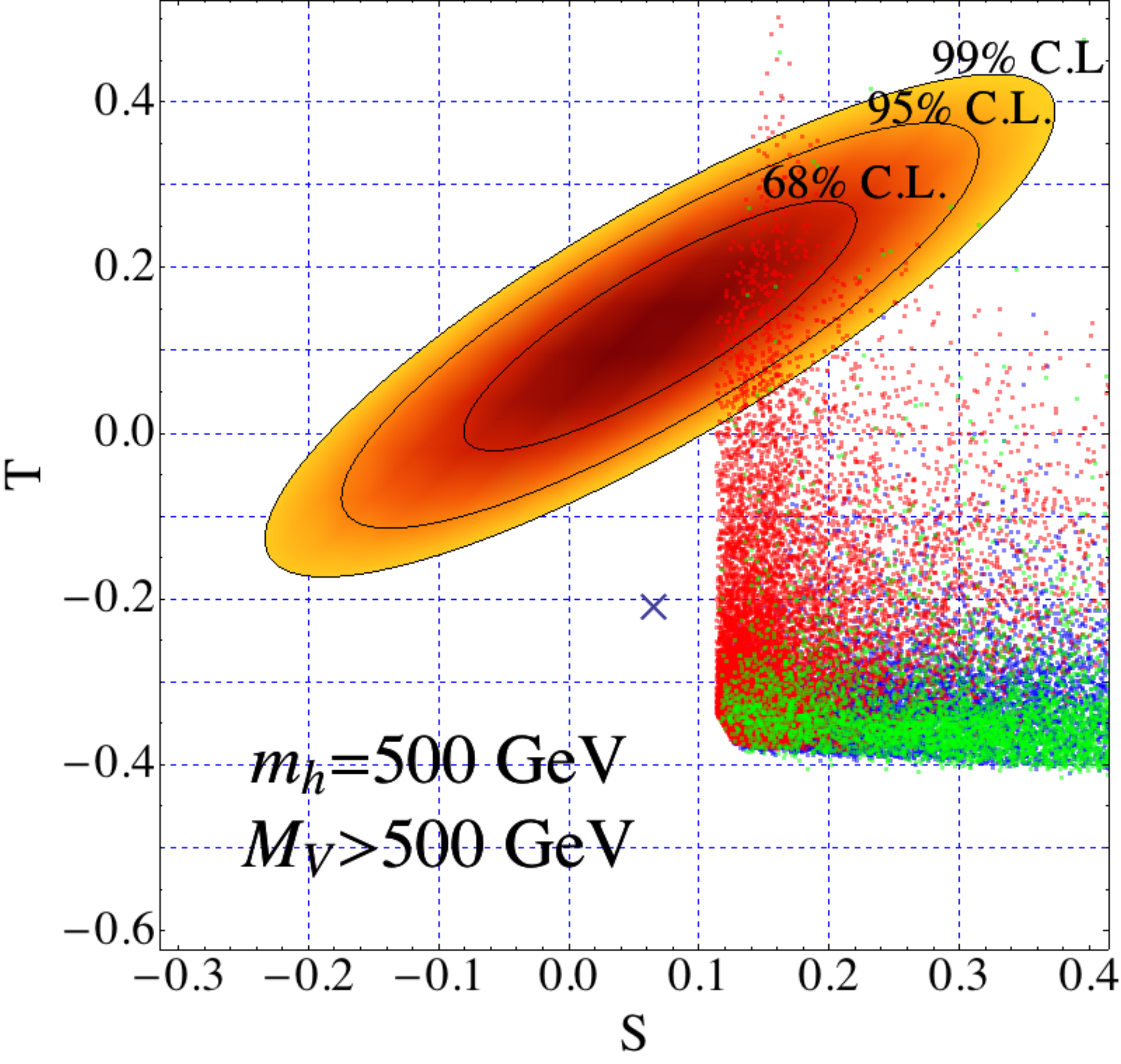}
 \end{minipage}
\caption{Experimental allowed regions and theoretical predictions for the $S$ and $T$ parameters in the SM with two composite vector bosons of opposite parity and a composite scalar for $m_{h}=100,\,300,\,500$ GeV, $M_{V}>500$ GeV and for three different regions of the parameter space: $F_{V}< G_{V}$ (blue), $G_{V}<F_{V}<2 G_{V}$ (red) and $2G_{V}<F_{V}<2.5 G_{V}$ (green). The small cross is the point $G_{V}=F_{V}=F_{A}=0$, i.e. SM contribution with a Higgs of mass $m_{h}$.}
\l{fig:EWPTHVHMscalar1}
\end{figure}
Since the parameter space now contains one more parameter, the mass of the scalar $m_{h}$, we can visualize the allowed regions in $S$ and $T$ by making two different plots: in Fig.~\ref{fig:EWPTHVHMscalar1} we have fixed the mass of the scalar $m_{h}$ to three different values and have explored the EWPO spanning over the rest of parameters, while in Fig.~\ref{fig:EWPTHVHMscalar2} we have done de same but fixing three different values of the mass of the vector $M_{V}$. The parameter space has been again constrained using the same conditions used for the heavy vectors (see Section \ref{EWPT3}) and the additional condition $G_{V}<v/\sqrt{3}$, which ensures that $a$ is real in Eq.~\eqref{agvrelation}.
\begin{figure}[h!]
\begin{minipage}[b]{4.9cm}
   \centering
   \includegraphics[width=4.8cm]{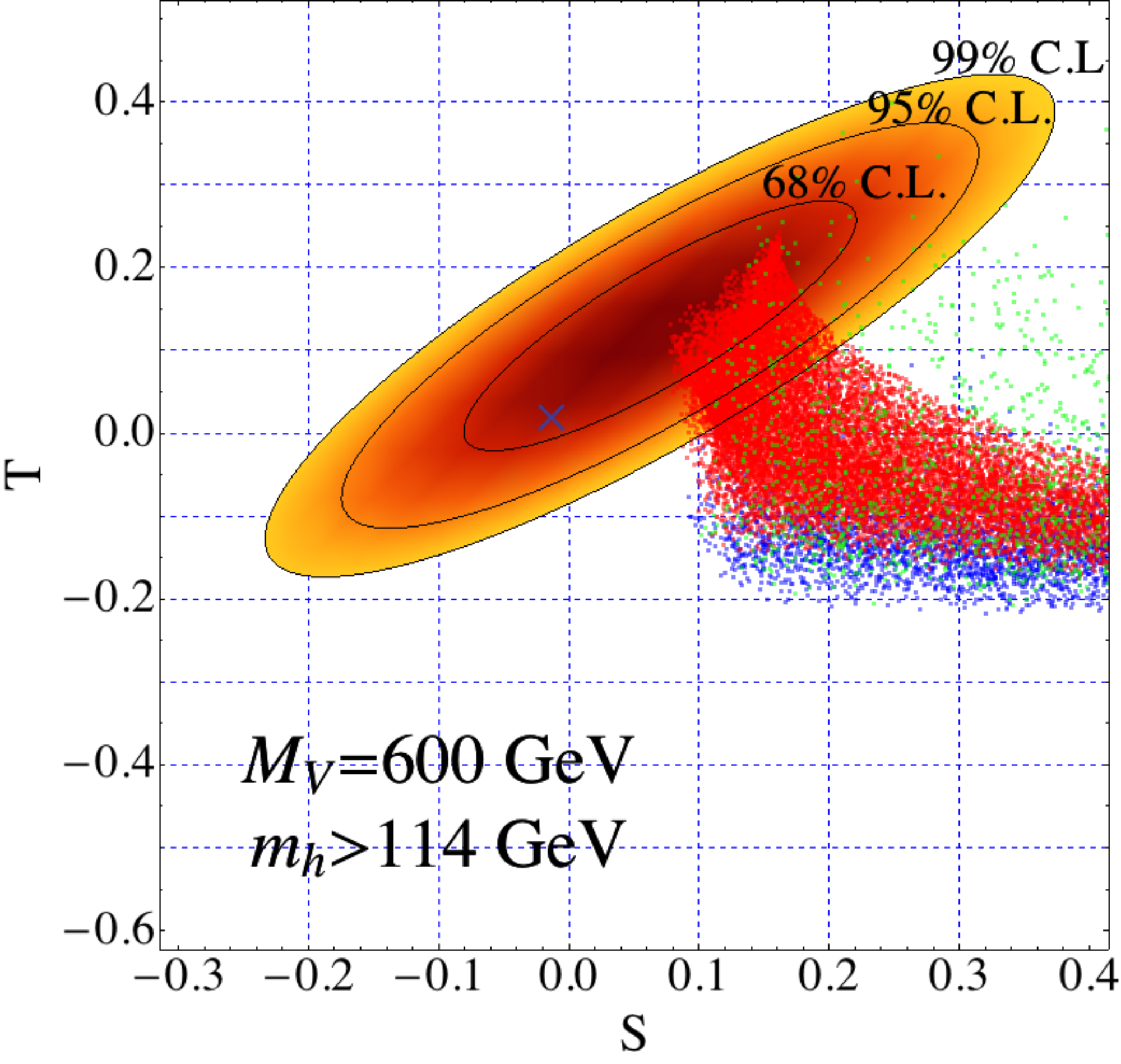}
 \end{minipage}
\ \hspace{3mm} \ 
\begin{minipage}[b]{4.9cm}
  \centering
   \includegraphics[width=4.8cm]{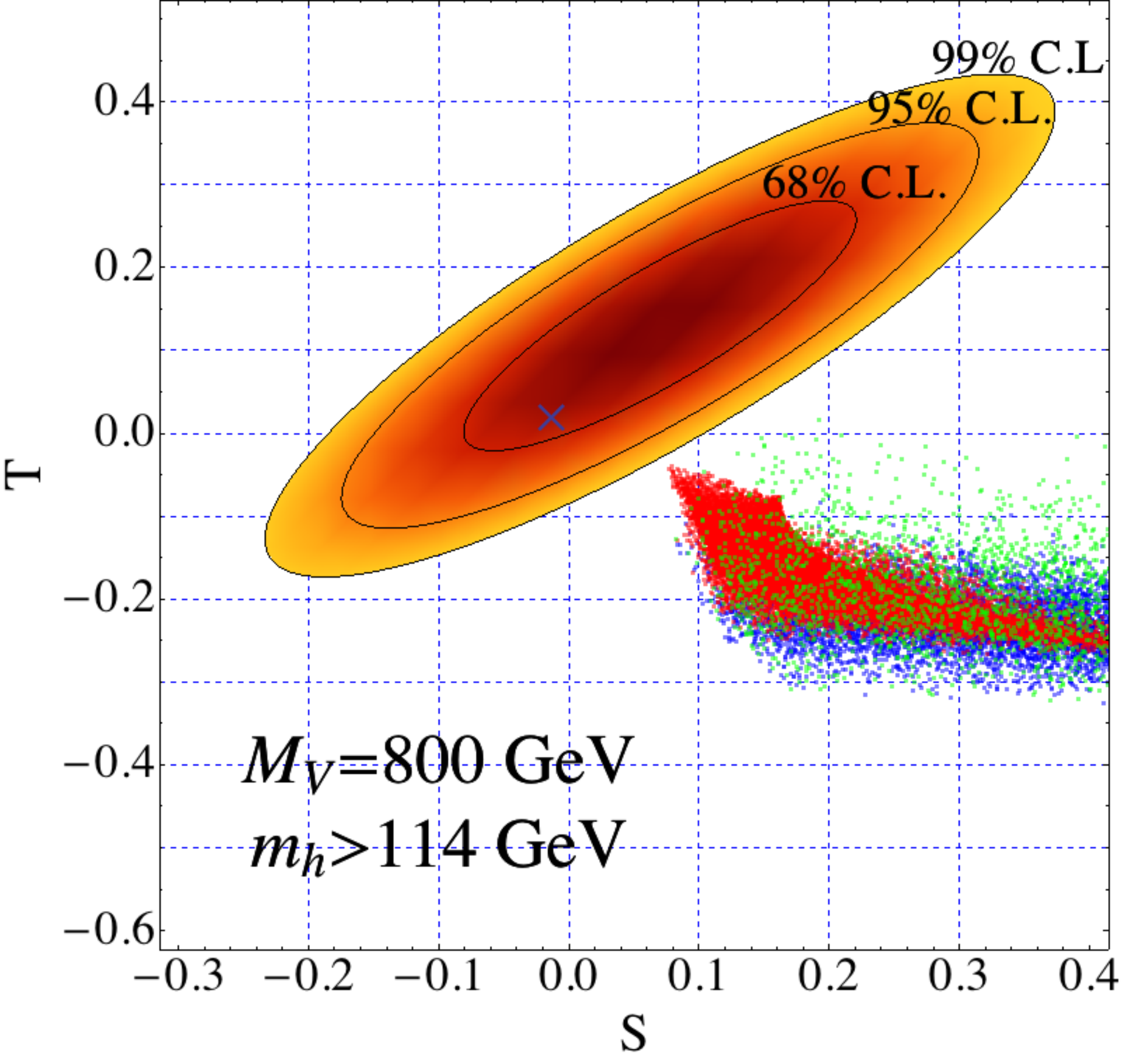}
 \end{minipage}
 \ \hspace{3mm} \ 
\begin{minipage}[b]{4.9cm}
  \centering
   \includegraphics[width=4.8cm]{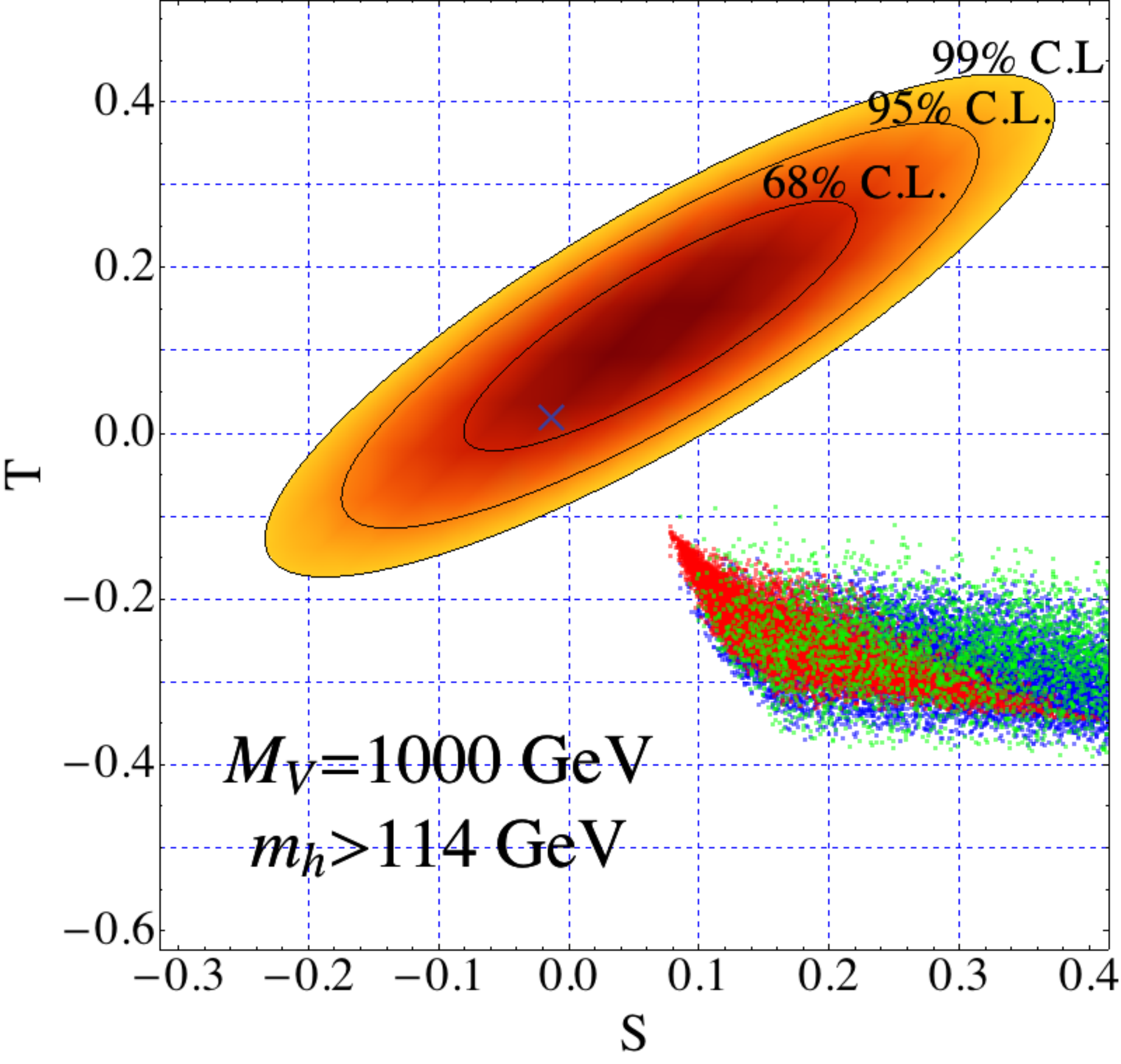}
 \end{minipage}
\caption{Experimental allowed regions and theoretical predictions for the $S$ and $T$ parameters in the SM with two composite vector bosons of opposite parity and a composite scalar for $M_{V}=600,\,800,\,1000$ GeV, $m_{h}>114$ GeV and for three different regions of the parameter space: $F_{V}< G_{V}$ (blue), $G_{V}<F_{V}<2 G_{V}$ (red) and $2G_{V}<F_{V}<2.5 G_{V}$ (green). The small cross inside the ellipses is the point $G_{V}=F_{V}=F_{A}=0$, $m_{h}=114$ GeV, i.e. the SM with a Higgs boson mass exactly equal to the LEP bound.}
\l{fig:EWPTHVHMscalar2}
\end{figure}
As in the case of previous section, from Figs.~\ref{fig:EWPTHVHMscalar1} and \ref{fig:EWPTHVHMscalar2} we see that the experimental ellipses are not the most populated regions in the $\(S,T\)$ plane. Moreover, as expected, the EWPT prefer a light scalar and a light vector. In particular a relatively light vector with $M_{V}\lesssim 500$ GeV can account for a scalar as heavy as $300$-$500$ GeV. In order to understand the fine tuning required to be in agreement with the EWPT we can span over the entire set of parameters $m_{h}$, $M_{V}$, $G_{V}$, $F_{V}$, $M_{A}$ and $F_{A}$ and plot the points in the $\(S,T\)$ plane using a coloring proportional to the fine-tuning $\Delta$ as defined by Eq.~\eqref{finetuning}.
\begin{figure}[htbp!]
\centering
\includegraphics[width=8cm]{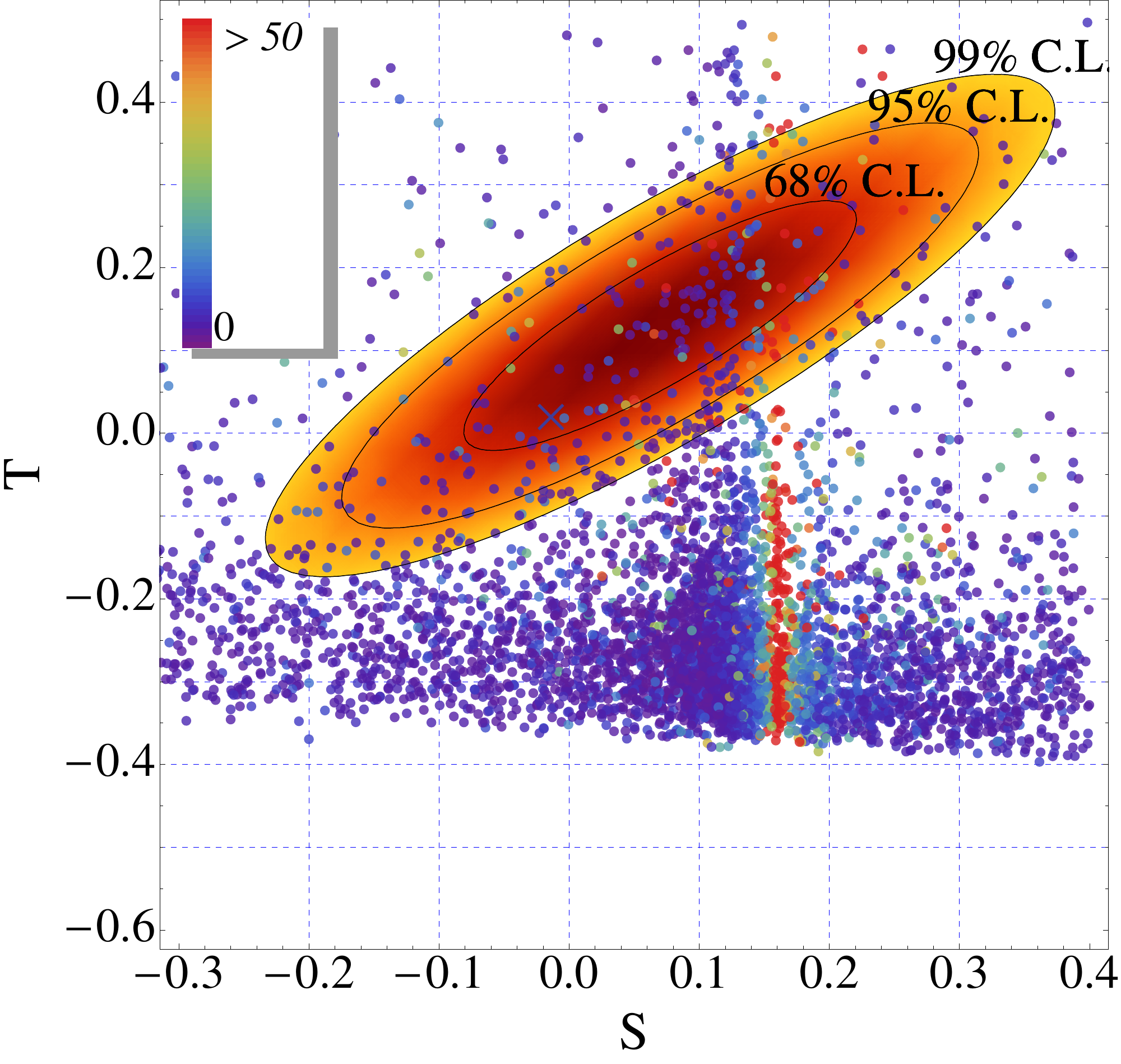}
\caption{Experimental allowed regions and theoretical predictions for the $S$ and $T$ parameters in the SM with two composite vector bosons of opposite parity and a composite scalar obtained spanning over all the parameter space. The coloring of the points represents the level of fine-tuning $\Delta$ (see the text for details). The parameter space is always constrained to satisfy the strongest unitarity bound and the scalar mass $m_{h}$ is constrained to be heavier than the LEP limit on the Higgs mass: $m_{h}>114$ GeV.}
\l{fig:EWPTHVHMscalarfinetuning}
\end{figure}
This is done in Fig.~\ref{fig:EWPTHVHMscalarfinetuning}. Again, even if the experimental ellipses are not the preferred region for the theoretical points, the EWPT can be satisfied in a region of the parameter space without a large fine-tuning, $\D\lesssim 10$-$20$.
\chapter{Composite vectors and scalars in a strongly interacting EWSB}\l{chap3}
\setlength{\epigraphwidth}{.4\textwidth}     
\vspace{-2mm}\epigraph{What we observe is not nature itself, but nature exposed to our mode of questioning.}%
              {\textsc{Werner Heisemberg}}
\vspace{3mm}     
\epigraph{God runs electromagnetics by wave theory on Monday, Wednesday, and Friday, and the Devil runs them by quantum theory on Tuesday, Thursday, and Saturday.}%
              {\textsc{Sir William Bragg}}
\vspace{7mm}
\lettrine{I}{n} this chapter we build a description of Strongly Interacting EWSB with a composite scalar and a composite vector only assuming that the new physics is parity and custodially invariant. We use the formalism of EWChL for the $SU\(2\)_{L}\times SU\(2\)_{R}/SU\(2\)_{L+R}$ SSB to construct a theory for the interactions of the SM particles (quarks, Gauge bosons and GBs) with a new vector  triplet and a new scalar singlet of $SU\(2\)_{L+R}$. We discuss the two possible formulations of the model for a vector boson, namely the Lorentz-vector and the antisymmetric-tensor formulations and we analyze the relations between composite and ``hidden'' gauge models.

The couplings in this effective Lagrangians will be of course strongly related to the mechanism that generates the new state. In fact, the measurement of the different cross sections that are sensitive to different combinations of the couplings, hopefully at the LHC, but eventually also at a future Linear Collider, could give information about this mechanism.

This chapter is organized as follows. In Section~\ref{sec3.1} we introduce the Lagrangian for the vectors using the Lorentz-vector formulation and in Section~\ref{sec3.2} we clarify the relations with antisymmetric-tensor formulation. Section~\ref{sec3.3} is devoted to the study of the relations between the effective model with composite vectors and the models with hidden gauge symmetry like those with $N$ sites and $N-1$ link fields. In Section~\ref{sec3.4} we introduce in the spectrum also a scalar singlet whose couplings, as we have already seen in the previous chapter, can interpolate between the SM with the Higgs boson and the completely Higgsless case. Finally in Section~\ref{sec3.5} we show that the complete scalar-vector system can arise from an $SU\(2\)_{L}\times SU\(2\)_{C}\times SU\(2\)_{R}$ gauge theory spontaneously broken by two Higgs doublets.

\section{The basic Lagrangian for heavy vectors in Higgsless models}\l{sec3.1}
The starting point is the usual lowest order EWChL for the $SU(2)_L\times SU(2)_R/SU(2)_{L+R}$ Goldstone fields (see Section~\ref{sec2.2.2})
\be\l{eq1}
\La_{\chi}=\f{v^{2}}{4} \<D_{\mu}\S\(D^{\mu}\S\)^{\dag}\> \,,
\ee
where 
\be\l{eq2}
\bry{l}
\S\(x\)=e^{i\f{\hat{\pi}\(x\)}{v}}\,,\qquad \hat{\pi}\(x\)=\pi^{a}\s^{a}=\(\bry{cc}\pi^{0} & \sqrt{2}\pi^{+} \\ \sqrt{2}\pi^{-} & -\pi^{0} \ery \)\,, \\\\ 
D_{\mu}\S=\demub \S-i\hat{B}_{\mu}\S+i\S \hat{W}_{\mu}\,,\qquad \dst \hat{W}_{\mu}=\f{g}{{2}}W_{\mu}^{a}\s^{a}\,,\qquad \hat{B}_{\mu}=\f{g^{\prime }}{{2}}B_{\mu}^{0} \s^{3}\,,
\ery
\ee
the $\s^{a}$ are the ordinary Pauli matrices and $\<\,\,\> $ denotes the trace over $SU\(2\)$. The transformation properties of the Goldstone fields under $SU\(2\)_{L}\times SU\(2\)_{R}$ are 
\be\l{eq5}
u \equiv \sqrt{\S} \to g_{R}u h^{\dag}=hug_{L}^{\dag}\,,
\ee
where $h=h\(u,g_{L},g_{R}\)$ is an element of $SU\(2\)_{L+R}$, as defined by this very equation \cite{Coleman:1969p1798,Callan:1969p1799}.

Especially in low-energy QCD studies, the heavy spin-1 states are most often described by antisymmetric tensors \cite{Ecker:1989p93,Ecker:1989p1559,Ecker:1989p67}. As it will be clear in the comparison of this model with a general gauge model of Section~\ref{sec3.3} it is convenient in this case to make use of the more conventional Lorentz vectors\footnote{The comparison with the antisymmetric-tensor formulation is discussed in the next section.}, belonging to the adjoint representation of $SU\(2\)_{L+R}$, 
\be
\hat{V}_{\mu}=\f{1}{\sqrt{2}}\s^{a}V_{\mu}^{a}\,,\qquad\qquad \hat{V}^{\mu}\to h \hat{V}^{\mu}h^{\dag}.
\ee
The $SU\(2\)_{L}\times SU\(2\)_{R}$ invariant kinetic Lagrangian for the heavy spin-1 fields is given by 
\be\l{eq7}
\La_{V-\text{kin}}=-\f{1}{4}\< \hat{\ovl{V}}^{\mu\nu}\hat{\ovl{V}}_{\mu\nu}\> +\f{M_{V}^{2}}{2}\<\hat{V}^{\mu}\hat{V}_{\mu}\> \,,
\ee
where $\hat{\ovl{V}}_{\mu\nu} =\nabla_{\mu} \hat{V}_{\nu} - \nabla_{\nu} \hat{V}_{\mu}$ is the $\hat{V}_{\mu}$ field strength written in terms of the covariant derivative 
\be\l{eq8}
\nabla_{\mu}{\hat{V}_\nu}=\demub \hat{V}_{\nu}+[\Gamma_{\mu},\hat{V}_{\nu}]\,,\quad\, \Gamma_{\mu}=\f{1}{2}\Big[u^{\dag}\(\demub-i\hat{B}_{\mu}\)u+u\(\demub-i\hat{W}_{\mu}\)u^{\dag}\Big]\,,\quad\, \Gamma_{\mu}^{\dag}=-\Gamma_{\mu}\,.
\ee
Note that this covariant derivative transforms homogeneously as $\hat{V}_\mu$ itself does. The other quantity that transforms covariantly is $u_{\mu}=u^{\dag}_{\mu}=iu^{\dag}D_{\mu}\S u^{\dag}$, so that indeed $u_{\mu}\to hu_{\mu}h^{\dag}$.

Assuming parity invariance of the new strong interaction, the effective Lagrangian for the interactions of the spin-1 field relevant for the LHC phenomenology we are interested in is
\be\l{int}
\La_{V-\text{int}}=\La_{\text{1V}-\text{Gauge}}+\La_{\text{1V}-\text{Goldstone}}+\La_{\text{2V}}+\La_{\text{3V}}\,,
\ee
where 
\bes
\begin{align}
&\La_{\text{1V}-\text{Goldstone}}=-\f{ig_{V}}{2\sqrt{2}}\< \hat{\ovl{V}}^{\mu\nu}[u_{\mu},u_{\nu}]\>\,,\l{L1VGoldstone} \\
\notag \\
&\La_{\text{1V}-\text{Gauge}}=-\f{f_{V}}{2\sqrt{2}}\< \hat{\ovl{V}}^{\mu\nu}f^{+}_{\mu\nu}\>\,,\l{L1VGauge} \\
\notag \\
&\La_{\text{2V}} =g_{1}\<\hat{V}_{\mu}\hat{V}^{\mu}u^{\nu}u_{\nu}\>+g_{2}\<\hat{V}_{\mu }u^{\nu}\hat{V}^{\mu}u_{\nu}\>+g_{3}\< \hat{V}_{\mu}\hat{V}_{\nu}[u^{\mu}, u^{\nu}]\>+g_{4}\<\hat{V}_{\mu}\hat{V}_{\nu}\{u^{\mu}, u^{\nu}\}\> \notag \\
& \qquad+g_{5}\<\hat{V}_{\mu}\(u^{\mu}\hat{V}_{\nu}u^{\nu}+u^{\nu} \hat{V}_{\nu}u^{\mu}\) \>+i g_{6}\<\hat{V}_{\mu}\hat{V}_{\nu }\(u\hat{W}^{\mu\nu}u^{\dag}+u^{\dag}\hat{B}^{\mu\nu}u\)\>\,,\l{L2V} \\
\notag \\
&\La_{\text{3V}}= \f{ig_K}{2\sqrt{2}} \< \hat{\ovl{V}}^{\mu\nu} \hat{V}^{\mu} \hat{V}^{\nu}\> \l{L3V}
\end{align}
\ees
and where we have defined the $SU\(2\)_{L+R}$ invariant field strength tensor
\be\l{fp}
f^{+}_{\mu\nu}=\(u\hat{W}_{\mu\nu}u^{\dag}+u^{\dag}\hat{B}_{\mu\nu}u\)\,.
\ee
Every parameter in Eq.~\eqref{int} is dimensionless.

Finally we have to take into account the contribution of the four independent operators containing only light fields, i.e. Goldstone and gauge fields, given by the Lagrangian\footnote{The need of these contact terms is due to the choice of the Lorentz-vector formulation for the new vector boson \cite{Ecker:1989p93} and it will be clarified in the following section.}
\be\l{Lcontact}
\La_{\text{contact}}=h_{1}\<f^{\mu\nu}_{+}f^{+}_{\mu\nu}\>+h_{2}\<\left[u^{\mu},u^{\nu}\right]\left[u_{\mu},u_{\nu}\right]\>+h_{3}\<\{u^{\mu},u^{\nu}\}\{u_{\mu},u_{\nu}\}\>+h_{4}\<f^{\mu\nu}_{+}\left[u_{\mu},u_{\nu}\right]\>\,. 
\ee     


We leave out from the total Lagrangian 
\be\l{Ltot}
\La_{V} = \La_{V-\text{kin}} + \La_{V-\text{int}}+\La_{\text{contact}}\,:
\ee
\begin{itemize}
\item Operators of dimension higher than 4, which we assume to be weighted by inverse powers of the cutoff $\Lambda\approx 3$ TeV, as suggested by Naive Dimensional Analysis (NDA) \cite{Georgi:1984er,Cohen:1997p1379}. As such, they would contribute to the LHC phenomenology of $V$ at c.o.m. energies sufficiently below $\Lambda$ by small terms relative to the ones generated by the Lagrangian we are considering.
\item Direct couplings between any SM fermion and the composite vectors. This is plausible if the SM fermions are elementary \cite{Chivukula:2004hw,Cacciapaglia:2004rb,Foadi:2004ps,Chivukula:2005bn,Casalbuoni:2005rs,2005PhRvD..72a5008C}. The
third generation doublet could be an exception. If this were the case, with a large enough coupling, this might lead to a dominant decay mode of the composite vectors into top and/or bottom quarks, rather than into $W, Z$ pairs.
\end{itemize}


\section{Vector versus tensor formulation}\l{sec3.2}
Especially in QCD, when discussing the low-energy pion dynamics, but also in applications to the EW interactions, it can be useful to describe spin-1 states by means of antisymmetric tensors rather than by Lorentz
vectors. At the level of linear spin-1 interaction terms only, i.e. $\La_{\text{1V}-\text{Goldstone}}$ and $\La_{\text{1V}-\text{Gauge}}$, it is easy to establish an exact correspondence of the vector formulation with the tensor one, as described by the Lagrangian 
\be
\La_T = \La_{T-\text{kin}} + \La_{\text{1T}-\text{Goldstone}}+ \La_{\text{1T}-\text{Gauge}}
\ee
in terms of the antisymmetric tensors $T^{\mu\nu}$, belonging to the adjoint representation of $SU(2)_{L+R}$, 
\be
\hat{T}_{\mu\nu}=\frac{1}{\sqrt{2}}\s^{a}T_{\mu\nu}^{a}\,,\qquad\qquad \hat{T}^{\mu\nu}\to h \hat{T}^{\mu\nu}h^{\dag}\,.
\ee
The kinetic Lagrangian for the heavy spin-1 fields is now given by 
\be  \l{eqkin}
\La_{T-\text{kin}}=-\frac{1}{2}\left\langle \nabla_{\mu}\hat{T}^{\mu\nu}\nabla^{\rho}\hat{T}_{\rho\nu}\right\rangle +\frac{M_{V}^{2}}{4}\left\langle \hat{T}^{\mu\nu}\hat{T}_{\mu\nu}\right\rangle \,,
\ee
with the covariant derivative $\nabla_{\mu}\hat{T}=\partial_{\mu} \hat{T}+[\Gamma_{\mu},\hat{T}]$. At the same time 
\bes  \l{eq9}
\begin{align}
& \La_{\text{1T}-\text{Goldstone}}= \frac{iG_{V}}{2\sqrt{2}}\left\langle \hat{T}^{\mu\nu}[u_{\mu},u_{\nu}]\right\rangle \\\notag\\
& \La_{\text{1T}-\text{Gauge}}=\frac{F_{V}}{2\sqrt{2}}\left\langle \hat{T}^{\mu\nu}f^{+}_{\mu\nu}\right\rangle,
\end{align}
\ees
where $G_{V}$ and $F_{V}$ are related to $g_V$ and $f_V$ by $G_V=g_V M_V$ and $F_V=f_V M_V$.

The correspondence of $\La_T $ with $\La_V$ stopped at the linear terms in $\hat{V}_\mu$ would be complete with the addition of the contact Lagrangian \eqref{Lcontact} only involving the GBs and the standard EW gauge bosons
\be
\La_{T}\leftrightarrow \La_{V}+\La_{\text{contact}}
\ee
with the parameters choice\footnote{As it will be shown in the next section these values are predicted in a minimal gauge model.}
\be\l{neww}
h_{1}=-\f{f_{V}^{2}}{8}\,,\qquad h_{2}=\f{g_{V}^{2}}{8}\,,\qquad h_{3}=0\,,\qquad h_{4}=-\f{if_{V}g_{V}}{4}\,.
\ee
A formal correspondence between the vector and the tensor formulations can also be established at the level of the multi spin-1 interaction terms \cite{Ecker:1989p93,Pallante:1992qe,Borasoy:1995ds,Harada:2003dz,Bijnens:1996p77,Cirigliano:2006p71,Kampf:2006p86,Kampf:2007p109}. However this would require the addition of an infinite number of terms. As discussed below the vector formulation proves more useful in discussing the LHC phenomenology of $V$ and the relation with gauge models.

\section{Composite versus gauge models}\l{sec3.3}
In this section we discuss the relation between the effective Lagrangian discussed in Section~\ref{sec3.1}, and the Lagrangian describing new gauge bosons of an hidden gauge symmetry \cite{Barbieri:2008p1580,Barbieri:2010p1579,Hernandez:2010p1578,Torre:2010p592}. In particular we consider a gauge theory based on $G=SU(2)_{L}\times SU(2)_{R}\times SU(2)^{N}$ broken to the diagonal subgroup $H=SU(2)_{L+R+\ldots}$ by a generic non-linear $\sigma$-model of the form 
\be\l{eq:one}
\La_{\chi}=\sum_{I,J}v_{IJ}^{2}\langle D_{\mu}\Sigma_{IJ}(D^{\mu}\Sigma_{IJ})^{\dagger}\rangle~,\qquad\Sigma_{IJ} \rightarrow g_{I}\Sigma _{IJ}g_{J}^{\dagger}~,  
\ee
where $g_{I,J}$ are elements of the various $SU(2)$ and $D_{\mu}$ is covariant derivative of $G$. Also in this case we assume that the gauge couplings of the various $SU(2)$ groups and $\La_{\chi}$ are parity invariant. This gauge model includes as special cases or approximates via deconstruction many of the models in the literature \cite{Casalbuoni:1985kq,Casalbuoni:1986vq,Csaki:2003p1576,Nomura:2003ju,Barbieri:2003jj,Foadi:2003p87,Georgi:2005hr}. The connection between a gauge model and a composite model for the spin-1 fields is best seen at the Lagrangian level by a suitable field redefinition, as we now show.
For the clarity of exposition we examine firstly the case $N=1$, i.e. a $3$-site model and then we generalize to the case of an $\(N+2\)$-site model for $N>1$.

\subsection{A single gauge vector}\l{sec3.3.1}
Consider now the simplest $N=1$ case, based on $SU(2)_L\times SU(2)_C \times SU(2)_R$, i.e. on the Lagrangian 
\be\l{gaugeL}
\La_{V-\text{gauge}}= \La_{\chi-\text{gauge}} -\frac{1}{2g_C^2}\left\langle \hat{v}_{\mu\nu} \hat{v}^{\mu\nu}\right\rangle -\frac{1}{2 g^2}\left\langle \hat{W}_{\mu\nu} \hat{W}^{\mu\nu}\right\rangle -\frac{1}{2 g^{\prime 2}}\left\langle \hat{B}_{\mu\nu} \hat{B}^{\mu\nu}\right\rangle \,,  
\ee
where 
\be
\hat{v}_\mu = \frac{g_C}{2} v_\mu^a \s^a
\ee
is the $SU(2)_C$-gauge vector, $g_{C}$ its gauge coupling and the chiral symmetry breaking Lagrangian is 
\be\l{symm-break}
\La_{\chi-\text{gauge}}=\frac{v^{2}}{2} \left\langle D_{\mu}\Sigma_{RC}\left(D^{\mu}\Sigma_{RC}\right)^{\dag}\right\rangle +\frac{v^{2}}{2} \left\langle D_{\mu}\Sigma_{CL}\left(D^{\mu}\Sigma_{CL}\right)^{\dag}\right\rangle\,.
\ee
We can denote collectively the three gauge vectors by 
\be
\hat{v}_\mu^I = \(\hat{W}_{\mu}, \hat{v}_{\mu}, \hat{B}_{\mu}\),~~ I = \(L, C, R\),
\ee
and write the covariant derivatives of the bi-fundamental scalars $\Sigma$ in the form
\be
D_\mu \Sigma_{IJ} = \partial_\mu \Sigma_{IJ} -i \hat{v}_\mu^I \Sigma_{IJ} +i \Sigma_{IJ} \hat{v}_\mu^J.
\ee
The $\Sigma_{IJ}$ can be put in the form $\Sigma_{IJ} = \sigma_I \sigma_J^\dag$, where $\sigma_I$ are the elements of $SU(2)_I/H$, transforming under the full $SU(2)_L\times SU(2)_C \times SU(2)_R$ as $\sigma_I \rightarrow g_I \sigma_I h^\dag$, i.e. the analogous of little $u$ defined in Eq.~\eqref{eq5}.
As the result of a gauge transformation
\be
\hat{v}_\mu^I \rightarrow \sigma_I^{\dag} \hat{v}_\mu^I \sigma_I  +i \sigma_I ^\dag \partial_\mu \sigma_I \equiv \Omega_\mu^I,,\qquad\qquad \Sigma_{IJ} \rightarrow \sigma_I^\dag \Sigma_{IJ} \sigma_J =1,
\ee
the chiral Lagrangian \eqref{gaugeL} reduces to
\be\l{new-symm-break}
\La_{\chi-\text{gauge}}=\frac{v^{2}}{2} \left\langle (\Omega_\mu^R -\Omega_\mu^C)^2 \right\rangle +\frac{v^{2}}{2} \left\langle (\Omega_\mu^L -\Omega_\mu^C)^2 \right\rangle,  
\ee
or, after the gauge fixing $\sigma_R =\sigma_L^\dag \equiv u$ and $\sigma_C=1$, to 
\be\l{new-new-symm-break}
\La_{\chi-\text{gauge}}={v^{2}} \left\langle \(\hat{v}_\mu -i\Gamma_\mu\)^2 \right\rangle +\frac{v^{2}}{4} \left\langle u_\mu^2 \right\rangle,
\ee
where 
\be\l{quantities}
\bry{l}
\dst u_\mu = \Omega_\mu^R - \Omega_\mu^L=iu^{\dag} D_{\mu} \S u^{\dag},\\\\
\dst \Gamma_\mu = -\frac{i}{2} \(\Omega_\mu^R + \Omega_\mu^L\)= \frac{1}{2} \Big[u^{\dag} \(\demub-i\hat{B}_\mu\) u+u \(\demub-i\hat{W}_\mu\) u^{\dag}\Big]\,,
\ery
\ee
are exactly the same vectors defined in Section~\ref{sec3.1}.

We can now express the Lagrangian \eqref{gaugeL} in terms of the new fields $\hat{V}_{\mu}$ by using the field redefinition 
\be\l{redef}
\hat{v}_\mu = \hat{V}_\mu + i \Gamma_\mu
\ee
and the identity \cite{Ecker:1989p93} 
\be\l{eckeridentity}
\hat{v}_{\mu\nu} = \hat{\ovl{V}}_{\mu\nu} -i [\hat{V}_\mu, \hat{V}_\nu] +\frac{i}{4} [u_\mu, u_\nu] + \frac{1}{2}f^{+}_{\mu\nu}.
\ee
With the further field rescaling $\hat{V}_{\mu}\to g_{C}/\sqrt{2} \hat{V}_{\mu}$, we can write the complete Lagrangian in the form
\be\l{v23}
\La_{V-\text{gauge}}=\La_{V=0}+\La_{V}\,,
\ee
where
\bes
\begin{align}
& \dst \bry{lll} \La_{V=0}&=&\dst \f{v^{2}}{4}\<\(D_{\mu}\S\)\(D^{\mu}\S\)^{\dag}\>-\f{1}{2g^{2}}\<\Hat{W}_{\mu\nu}\Hat{W}^{\mu\nu}\>-\f{1}{2g^{\prime\,2}}\<\Hat{B}_{\mu\nu}\Hat{B}^{\mu\nu}\>\ery\,,\l{lahless2}\\\nonumber\\
&\dst \bry{lll}
\dst \La_{V}
&=&\dst -\f{1}{4}\<\Hat{V}_{\mu\nu}\Hat{V}^{\mu\nu}\>+\f{M_{V}^{2}}{2}\<\Hat{V}_{\mu}\Hat{V}^{\mu}\>+\f{1}{16g_{V}^{2}}\<[\Hat{V}_{\mu},\Hat{V}_{\nu}][\Hat{V}^{\mu},\Hat{V}^{\nu}]\>+\f{i}{4\sqrt{2}g_{V}}\<\Hat{V}_{\mu\nu}[\Hat{V}^{\mu},\Hat{V}^{\nu}]\>\\\\
&&\dst -\f{ig_{V}}{2\sqrt{2}}\<\Hat{V}_{\mu\nu}[u^{\mu},u^{\nu}]\>-\f{g_{V}}{\sqrt{2}}\<\Hat{V}_{\mu\nu}f^{+\,\mu\nu}\>-\f{1}{8}\<[\Hat{V}_{\mu},\Hat{V}_{\nu}][u^{\mu},u^{\nu}]\>+\f{i}{4}\<[\Hat{V}_{\mu},\Hat{V}_{\nu}]f^{+\,\mu\nu}\>\\\\
&&\dst -\f{g_{V}^{2}}{2}\<f^{+}_{\mu\nu}f^{+\,\mu\nu}\>-\f{ig_{V}^{2}}{2}\<[u_{\mu},u_{\nu}]f^{+\,\mu\nu}\>+\f{g_{V}^{2}}{8}\<[u_{\mu},u_{\nu}][u^{\mu},u^{\nu}]\>\,,\l{lav2}
\ery
\end{align}
\ees
and where we have used $g_{C}=1/2g_{V}$. $\La_{V}$ in Eq.~\eqref{lav2} coincides as anticipated with $\La_{V}$ in \eqref{Ltot} for 
\be\l{gaugeparam}
\bry{l}
\dst g_C = \frac{1}{2 g_V}=\frac{g_{K}}{2}\,,\qquad f_{V} = 2 g_{V}\,,\qquad M_{V} = g_{K} \f{v}{2}\,\,\,(\text{or}\,\,G_{V} = \f{v}{2})\,,\\\\
\dst g_{1}=g_{2}=g_{4}=g_{5}=0\,,\qquad g_{3}=-\f{1}{4}\,,\qquad g_{6}=\f{1}{2}\,,\\\\
\dst h_{1}=-\f{f_{V}^{2}}{8}\,,\qquad h_{2}=\f{g_{V}^{2}}{8}\,,\qquad h_{3}=0\,,\qquad h_{4}=-\f{if_{V}g_{V}}{4}\,.
\ery
\ee
Note that the last line relations for $h_{i}$ come directly from the squared of the last two terms in Eq.~\eqref{eckeridentity}.

\subsection{More than a single gauge vector}\l{sec3.3.2}
To discuss the case of more than one vector, i.e. $N>1$, we can decompose the
vectors associated to $SU(2)^N$ with respect to parity as 
\be
\Omega_i^\mu = \hat{v}_i^\mu + \hat{a}_i^\mu,\quad\Omega_{P(i)}^\mu = \hat{v}_{i}^{\mu} - \hat{a}_{i}^{\mu},\quad i=1,\dots,N,
\ee
so that under $SU(2)_L\times SU(2)_R$ 
\be
\hat{v}_{i}^{\mu} \rightarrow h\hat{v}_{i}^{\mu}h^{\dagger}+ih\partial^{\mu}h^{\dagger},\quad \hat{a}_{i}^{\mu}\rightarrow h\hat{a}_{i}^{\mu}h^{\dagger}.
\l{gaugeinv}
\ee
In terms of these fields the gauge Lagrangian becomes 
\be
\La_{\mathrm{\text{gauge}}}=\La_{\text{gauge}-\text{SM}}-\sum_{i}\frac{1}{2g_{i}^{2}}\left[ \langle\left( \hat{v}_{i}^{\mu\nu}-i[\hat{a}_{i}^{\mu},\hat{a}_{i}^{\nu}]\right) ^{2}\rangle+\langle\left(D_{V}^{\mu}\hat{a}_{i}^{\nu}-D_{V}^{\nu}\hat{a}_{i}^{\mu}\right) ^{2}\rangle\right] ~,
\l{eq:Lgaugef}
\ee
where $\hat{v}_{i}^{\mu\nu}$ are the usual field strengths and 
\be
D_{V}^{\mu}\hat{a}_i^{\nu}=\partial^{\mu}\hat{a}_i^{\nu}-i[\hat{v}_i^{\mu},\hat{a}_i^{\nu}].
\ee
At the same time, as a generalization of Eq.~\eqref{new-new-symm-break} of
the $N=1$ case, the symmetry breaking Lagrangian will be the sum of two
separated quadratic forms in the parity-even and parity-odd fields of the
type 
\be
\La_{\chi-\text{gauge}}=\La_{m-V}(\hat{v}_i^\mu -i\Gamma^\mu)+ 
\La_{m-A}(u^\mu, \hat{a}_i^\mu) \,.  \l{multi-symm-break}
\ee
The dependence of $\La_{m-V}$ on the variables $\hat{v}_i^\mu -i\Gamma^\mu$
follows from Eq.~\eqref{gaugeinv}.

Concentrating on the parity-even fields only, we can generalize the redefinition \eqref{redef} to
\be
\hat{v}_i^\mu = \hat{V}_i^\mu + i \Gamma^{\mu}
\ee
and the identity \eqref{eckeridentity} to
\be\l{neweckeridentity}
\hat{v}_{\mu\nu}^{i} = \hat{\ovl{V}}_{\mu\nu}^{i} -i [\hat{V}_\mu^{i}, \hat{V}_\nu^{i}] +\frac{i}{4} [u_\mu, u_\nu] + \frac{1}{2}f^{+}_{\mu\nu}\,.
\ee
Now with the further rescaling $\hat{V}_i^\mu \rightarrow g_i/\sqrt{2} \hat{V}_i^\mu$, the
Lagrangian of the $SU(2)_{L}\times SU(2)_{R}\times SU(2)^{N}$ gauge model, restricted to the parity-even vectors, becomes a diagonal sum of $\La_{V_{i}}$,
each with 
\be\l{gaugeparam1}
\bry{l}
\dst g_i = \frac{1}{2 g_{V_{i}}}\,,\qquad f_{V_{i}} = 2 g_{V_{i}}\,,\qquad g_{K_{i}}=\f{1}{g_{V_{i}}} \,,\\\\
\dst g_{1}=g_{2}=g_{4}=g_{5}=0\,,\qquad g_{3}=-\f{1}{4}\,,\qquad g_{6}=\f{1}{2}\,,
\ery
\ee
but with the $V_i^\mu$ different from the mass eigenstates. Moreover, since the last two terms of the identity \eqref{neweckeridentity} do not depend on $i$, from the gauge Lagrangian \eqref{eq:Lgaugef} we find the contact Lagrangian
\be\l{contactnew}
\bry{lll}
\dst \La_{\text{contact}}
&=&\dst \Bigg[\frac{1}{4} \<[u_\mu, u_\nu][u^{\mu}, u^{\nu}]\>-\<f^{+}_{\mu\nu}f_{+}^{\mu\nu}\>-i \<[u_\mu, u_\nu]f_{+}^{\mu\nu}\>\Bigg]\sum_{i}\f{g_{V_{i}}^{2}}{2}\,,
\ery
\ee
that gives for the four parameters $h_{1},\ldots,h_{4}$:
\be\l{gaugeparam2}
\dst h_{1}=-\sum_{i}\f{f_{V_{i}}^{2}}{8}\,,\qquad h_{2}=\sum_{i}\f{g_{V_{i}}^{2}}{8}\,,\qquad h_{3}=0\,,\qquad h_{4}=-\sum_{i}\f{if_{V_{i}}g_{V_{i}}}{4}\,.
\ee
This is the generalization of the last line of Eq.~\eqref{gaugeparam} in the case of more than a single gauge vector. \\
Going to the mass-eigenstate basis, denoted by a tilde, with the rotation\footnote{For ease of reading in the discussion of the rotation to the mass-eigenstate basis we shall omit the Lorentz indices and the hat symbol over the $SU\(2\)$ matrix fields.}
\be\l{masseigen}
V^{i}=R^{ij}\tilde{V}_{j}\,,
\ee
maintains all the couplings quadratic in the $V_i^\mu$ unchanged. Moreover, we can define the new couplings $\tilde{f}_{V_{i}}$, $\tilde{g}_{V_{i}}$ and $\tilde{g}_{K_{i}}$ for the mass-eigenstate vectors as
\be\l{newcouplings}
\bry{l}
f_{V_{i}}V^{i}=\sum_{j}f_{V_{i}}R^{ij}\tilde{V}_{j}\qquad\Longrightarrow\qquad \tilde{f}_{V_{i}}=\sum_{j}R^{ij}f_{V_{j}}\,,\\\\
g_{V_{i}}V^{i}=\sum_{j}g_{V_{i}}R^{ij}\tilde{V}_{j}\qquad\Longrightarrow\qquad \tilde{g}_{V_{i}}=\sum_{j}R^{ij}g_{V_{j}}\,,\\\\
g_{K_{i}}V^{i}V^{i}V^{i}=\sum_{l,m,n}g_{K_{i}}R^{il}R^{im}R^{in}\tilde{V}_{l}\tilde{V}_{m}\tilde{V}_{n}\qquad\Longrightarrow\qquad \tilde{g}_{K_{i}}^{lmn}=\sum_{l,m,n}R^{il}R^{im}R^{in}g_{K_{i}}\,.
\ery
\ee
From the first two equations we see that the relation $\tilde{f}_{V_{i}}=2 \tilde{g}_{V_{i}}$ is preserved for the individual mass-eingenstate vectors, while the third relation implies that the trilinear couplings $g_{K_i}$ get spread among the mass eingenstates, so that the trilinear Lagrangian becomes
\be\l{L3Vgen}
\La_{\text{3V}}= \sum_{i}\frac{ig_{K_{i}}}{2\sqrt{2}} \left\langle \ovl{V}_{\mu\nu}^i V^{\mu}_i V^{\nu}_i\right\rangle= \sum_{i,l,m,n}\frac{i\tilde{g}_{K_{i}}^{lmn}}{2\sqrt{2}} \left\langle \ovl{\tilde{V}}_{\mu\nu}^l \tilde{V}^{\mu}_m \tilde{V}^{\nu}_n\right\rangle= \sum_{l,m,n}\frac{i\tilde{g}_{K}^{lmn}}{2\sqrt{2}} \left\langle \ovl{\tilde{V}}_{\mu\nu}^l \tilde{V}^{\mu}_m \tilde{V}^{\nu}_n\right\rangle\,,
\ee
where, in the last step we have defined $\sum_{i}\tilde{g}_{K_{i}}^{lmn}=\tilde{g}_{K}^{lmn}$.
Picking up the lightest vector only, form the relation 
\be\l{newgk}
\tilde{g}_{K_{i}}^{lmn}\tilde{g}_{V_{j}}=\sum_{k,l,m,n}R^{jk}R^{il}R^{im}R^{in}g_{V_{k}}g_{K_{i}}\,,
\ee
we find
\be\l{newgk1}
\tilde{g}_{K_{i}}^{111}\tilde{g}_{V_{1}}\neq 1\,.
\ee
However, by the orthogonality of the rotation matrix $R$ we have the following sum rule over the full set of vectors 
\be\l{sumrule}
\sum_{j}\tilde{g}_{K_{i}}^{jnn} \tilde{g}_{V_j}=\sum_{j,m,n}R^{ij}R^{in}R^{in}R^{jm}g_{K_{i}}g_{V_{m}}=\sum_{m}\d^{im}g_{K_{i}}g_{V_{m}}=g_{K_{i}}g_{V_{i}}=1
\ee
for any fixed $n$. Finally, since the coefficients $h_{1},h_{2},h_{4}$ of the contact Lagrangian given by Eq.~\eqref{gaugeparam2} are quadratic in $f_{V_{i}}$ and $g_{V_{i}}$, the orthogonality of the rotation matrix implies
\be\l{newcontactmass}
\dst h_{1}=-\sum_{i}\f{\tilde{f}_{V_{i}}^{2}}{8}\,,\qquad h_{2}=\sum_{i}\f{\tilde{g}_{V_{i}}^{2}}{8}\,,\qquad h_{3}=0\,,\qquad h_{4}=-\sum_{i}\f{i\tilde{f}_{V_{i}}\tilde{g}_{V_{i}}}{4}\,.
\ee

Note that this relations ensure that the asymptotic behavior of the amplitudes that we studied in Section~\ref{TCOAHLMWAHCV} and of those we will see in the following is not worse than in the case of a single gauge vector, but only at $s > M_{V_i}^2$ for any $i$.

\section{Adding a composite scalar}\l{sec3.4}
As mentioned at the beginning of this chapter, we also want to consider the possibility that a scalar particle exists below the cutoff $\Lambda\approx 4\pi v$. This light scalar could be a Strongly Interacting Light Higgs (SILH) boson in the sense of Ref.~\cite{Giudice:2007p85} or even a more complicated object arising from an unknown strong dynamics. Here we are only interested in constructing an effective Lagrangian that describes the interactions of the new scalar with the SM particles and with the heavy vector $V$. 

The most general renormalizable Lagrangian to describe the couplings of a new parity even scalar singlet of $SU\(2\)_{L+R}$ to the SM particles has been anticipated in Section~\ref{TROTHB} by Eq.~\eqref{HSM}. It is convenient to write it again here:
\be\l{lscalar2}
\La_{h}=\La_{h-\text{kin}}+\La_{h-\text{pot}}+\La_{h-\text{int}}\,,
\ee
where
\be\l{Hkin2}
\La_{h-\text{kin}}=\f{1}{2}\demub h \demua h\,,
\ee
\be\l{Hpot2}
\La_{h-\text{pot}}=-\f{m_{h}^{2}}{2}h^{2}-\f{d_{3}}{6}\(\f{3m_{h}^{2}}{v}\)h^{3}-\f{d_{4}}{24}\(\f{3m_{h}^{2}}{v}\)h^{4}\,,
\ee
and
\be\l{Hint2}
\La_{h-\text{int}}=\f{v^{2}}{4}\<D^{\mu}\S\(D_{\mu}\S\)^{\dag}\>\(2a\f{h}{v}+b\f{h^{2}}{v^{2}}\)-m_{i}\bar{\psi}_{L\,i}\S\(c\f{h}{v}\)\psi_{R\,i}\,.
\ee
We have also an interaction between the new scalar and two heavy vectors given by
\be\l{lvhinteraction}
\La_{h-V}=\f{k_{1}v}{8g_{V}^{2}}h\left<\hat{V}_{\mu}\hat{V}^{\mu}\right>+\f{k_{2}}{32g_{V}^{2}}h^{2}\left<\hat{V}_{\mu}\hat{V}^{\mu}\right>\,.
\ee
Here $a$, $b$, $c$, $d_{3}$, $d_{4}$, $k_{1}$, $k_{2}$ are dimensionless constants\footnote{The operators proportional to $d_{3}$, $d_{4}$ and $k_{2}$ have been written for completeness but are irrelevant for the following discussion.}.\\
Summarizing, the complete Lagrangian of our model, which we will consider for the following phenomenological studies, can be written in the compact form:
\be\l{lefftot}
\La_{\text{eff}}=\La_{\hsl-\text{SM}}+\La_{V}+\La_{h}+\La_{h-V}\,,
\ee
where the four contributions are given respectively by Eqs.~\eqref{HlessSM}, \eqref{Ltot}, \eqref{lscalar2}, \eqref{lvhinteraction}.\\
In the next section we show that the Lagrangian \eqref{lefftot}, for the special choice of parameters \eqref{gaugeparam} and for
\be\l{gaugeparameters}
a=\f{1}{2}\,,\quad b=\f{1}{4} \,,\quad k_{1}=1\,,\quad k_{2}=1\,,
\ee
is obtained from a gauge theory based on $SU\(2\)_{L}\times SU\(2\)_{C}\times U\(1\)_{Y}$ spontaneously broken to $U\(1\)_{Q}$ by two Higgs doublets (with the same VEV) in the limit $m_{H}\gg \Lambda$ for the mass of the parity odd scalar $H$\footnote{The mass of the parity odd scalar $H$ can be simply raised above the cutoff without any further hypothesis on the low energy physics.}.

\section{A gauge model with two Higgs doublets}\l{sec3.5}
Let us consider the following $SU(2)_{L}\times SU(2)_{C}\times U(1)_{Y}$ gauge invariant Lagrangian: 
\be\l{v1}
\La_{\text{tot}}=\La_{\chi}+\La_{\text{gauge}}+\La_{\text{pot}}\,,
\ee
where
\be\l{v2}
\La_{\text{gauge}}=-\f{1}{2g^{2}}\<\hat{W}_{\mu\nu}\hat{W}^{\mu\nu}\>-\f{1}{2g_{C}^{2}}\<\hat{v}_{\mu\nu}\hat{v}^{\mu\nu}\>-\f{1}{2g^{\prime\,2}}\<\hat{B}_{\mu\nu}\hat{B}^{\mu\nu}\>\,,
\ee
is the gauge Lagrangian,
\be\l{v3}
\La_{\text{SB}}=\frac{1}{2} \left\langle D_{\mu}\mathcal{H}_{YC}\left(D^{\mu}\mathcal{H}_{YC}\right)^{\dag}\right\rangle +\frac{1}{2} \left\langle D_{\mu}\mathcal{H}_{CL}\left(D^{\mu}\mathcal{H}_{CL}\right)^{\dag}\right\rangle\,,
\ee
is the symmetry breaking Lagrangian,
\be\l{v4}
\bry{lll}
\La_{\text{pot}}&=&\dst \f{\mu^{2}}{2}\<\mathcal{H}_{YC}\mathcal{H}_{YC}^{\dag}\>+\f{\mu^{2}}{2}\<\mathcal{H}_{CL}\mathcal{H}_{CL}^{\dag}\>-\f{\lambda}{4}\Big[\<\mathcal{H}_{YC}\mathcal{H}_{YC}^{\dag}\>\Big]^{2}\\\\
&&\dst -\f{\lambda}{4}\Big[\<\mathcal{H}_{CL}\mathcal{H}_{CL}^{\dag}\>\Big]^{2}-\kappa\<\mathcal{H}_{YC}\mathcal{H}_{CL}^{\dag}\mathcal{H}_{CL}\mathcal{H}_{YC}^{\dag}\>\,,
\ery
\ee
is the scalar potential and we have defined
\be\l{v5}
\bry{l}
\dst \mathcal{H}_{YC}=\(v+\f{h_{YC}}{\sqrt{2}}\)\S_{YC}=\(v+\f{h+H}{2}\)e^{\f{i\(\hat{\pi}_{W}+\hat{\pi}_{V}\)}{v}}\\\\
\dst \mathcal{H}_{CL}=\(v+\f{h_{CL}}{\sqrt{2}}\)\S_{CL}=\(v+\f{h-H}{2}\)e^{\f{i\(\hat{\pi}_{W}-\hat{\pi}_{V}\)}{v}}\,,
\ery
\ee
\be\l{v6}
\bry{l}
D_{\mu}\mathcal{H}_{YC}=\demub \mathcal{H}_{YC}-i\hat{B}_{\mu}\mathcal{H}_{YC}+i\mathcal{H}_{YC}\hat{v}_{\mu}\,,\\\\
D_{\mu}\mathcal{H}_{CL}=\demub \mathcal{H}_{CL}-i\hat{v}_{\mu}\mathcal{H}_{CL}+i\mathcal{H}_{CL}\hat{W}_{\mu}\,.
\ery
\ee
Now, considering the following gauge transformation
\be\l{v7}
\hat{v}_\mu^I \rightarrow \sigma_I^{\dag} \hat{v}_\mu^I \sigma_I  +i \sigma_I ^\dag \partial_\mu \sigma_I \equiv \Omega_\mu^I,\qquad\qquad \mathcal{H}_{IJ} \rightarrow \sigma_I^\dag \mathcal{H}_{IJ} \sigma_J =v+\f{h_{IJ}}{\sqrt{2}},
\ee
where $\s_{I}$ are the elements of $SU\(2\)_{I}/H$ and $\hat{v}_{\mu}^{I}=\(\hat{W}_{\mu},\hat{v}_{\mu},\hat{B}_{\mu}\)$, ($I=L,C,Y$) and the gauge fixing conditions $\s_{R}=\s_{L}^{\dag}=u=e^{\f{i\hat{\pi}}{v}}$ and $\s_{C}=1$ (i.e. the unitary gauge for the $v_{\mu}$ bosons) the SB Lagrangian can be written in the form
\be\l{v10}
\bry{lll}
\La_{\text{SB}}
&=&\dst \frac{1}{2} \demub h\demua h+\frac{1}{2}  \demub H\demua H\\\\ 
&&\dst +\frac{1}{4} \(\f{H^{2}+h^{2}}{2}+2vh+2v^{2}\)\left\langle 2\hat{v}_{\mu}\hat{v}^{\mu}+\O^{Y}_{\mu}\O^{Y\,\mu}+\O^{L}_{\mu}\O^{L\,\mu}-2\hat{v}_{\mu}\(\O^{Y\,\mu}+\O^{L\,\mu}\)\right\rangle\\\\
&&\dst +\frac{1}{4} \(hH+2vH\)\left\langle \O^{Y}_{\mu}\O^{Y\,\mu}-\O^{L}_{\mu}\O^{L\,\mu}-2\hat{v}_{\mu}\(\O^{Y\,\mu}-\O^{L\,\mu}\)\right\rangle\,.
\ery
\ee
while the gauge Lagrangian remains unchanged.
The minimum condition for the potential is
\be\l{nlin12}
v^{2}=\f{\mu^{2}}{2\(\lambda +\kappa\)}\,
\ee
and the scalar masses are
\be\l{nlin14}
\bry{l}
\dst m_{H}^{2}=4v^{2}\(\lambda-\kappa\)\,,\\\\
\dst m_{h}^{2}=4v^{2}\(\lambda+\kappa\)\,.
\ery
\ee 
Using these relations and our choice of the gauge fixing we can rewrite the potential Lagrangian \eqref{v4} in the form
\be\l{v9}
\bry{lll}
\La_{\text{pot}}
&=& \dst -\f{m_{h}^{2}}{8v^{2}}\(-8v^{4}+8v^{2}h^{2}+4vh^{3}+\f{H^{4}}{2}+\f{h^{4}}{2}\)-\f{m_{H}^{2}}{2} H^{2}-\f{m_{h}^{2}+2m_{H}^{2}}{16v^{2}}\(4v H^{2}h+H^{2}h^{2}\)\,.
\ery
\ee
The SB Lagrangian \eqref{v10} can be written as a function of the quantities in \eqref{quantities} as
\be\l{v12}
\bry{lll}
\La_{\text{SB}}
&=&\dst \frac{1}{2} \demub H\demua H+\frac{1}{2}  \demub h\demua h\\\\ 
&&\dst +\frac{1}{2} \(\f{h^{2}+H^{2}}{2}+2vh+2v^{2}\)\left\langle \(\Hat{v}_{\mu}-i\G_{\mu}\)^{2}+\f{1}{4}D_{\mu}\S\(D^{\mu}\S\)^{\dag}\right\rangle\\\\
&&\dst -\frac{1}{2} \(Hh+2vH\)\left\langle u_{\mu}\(\Hat{v}^{\mu}-i\G^{\mu}\)\right\rangle\,,
\ery
\ee
and with the further replacement $\Hat{V}_{\mu}=\Hat{v}_{\mu}-i\G_{\mu}$ we obtain
\be\l{v14}
\bry{lll}
\La_{\chi}
&=&\dst \frac{1}{2} \(\f{h^{2}+H^{2}}{2}+2vh+2v^{2}\)\( \<\Hat{V}_{\mu}\Hat{V}^{\mu}\>+\f{1}{4}\<\(D_{\mu}\S\)\(D^{\mu}\S\)^{\dag}\>\)\\\\
&&\dst -\frac{1}{2} \(Hh+2vH\)\left\langle u_{\mu}\Hat{V}^{\mu}\right\rangle+\frac{1}{2} \demub H\demua H+\frac{1}{2}  \demub h\demua h\,.
\ery
\ee
We can express the gauge Lagrangian \eqref{v2} in terms of the new field $\hat{V}_{\mu}$ by using the identity \eqref{eckeridentity}.
We obtain for the $\hat{v}_{\mu}$ kinetic term
\be\l{v17}
\bry{lll}
\La_{v-\text{gauge}}
&=&\dst -\f{1}{2g_{C}^{2}}\<\hat{\ovl{V}}_{\mu\nu}\hat{\ovl{V}}^{\mu\nu}-[\Hat{V}_{\mu},\Hat{V}_{\nu}][\Hat{V}^{\mu},\Hat{V}^{\nu}]-2i\Hat{V}_{\mu\nu}[\Hat{V}^{\mu},\Hat{V}^{\nu}]\right.\\\\
&&\dst +\f{i}{2}\Hat{V}_{\mu\nu}[u^{\mu},u^{\nu}]+\f{1}{2}[\Hat{V}_{\mu},\Hat{V}_{\nu}][u^{\mu},u^{\nu}]+\Hat{V}_{\mu\nu}f^{+\,\mu\nu}-i[\Hat{V}_{\mu},\Hat{V}_{\nu}]f^{+\,\mu\nu}\\\\
&&\dst \left.+\f{1}{4}f^{+}_{\mu\nu}f^{+\,\mu\nu}+\f{i}{4}[u_{\mu},u_{\nu}]f^{+\,\mu\nu}-\f{1}{16}[u_{\mu},u_{\nu}][u^{\mu},u^{\nu}]\>\,.
\ery
\ee
Finally, rescaling the field $\Hat{V}_{\mu}$ according to $\Hat{V}_{\mu}\to g_{C}/\sqrt{2} \hat{V}_{\mu}$, we can write the complete Lagrangian in the form
\be\l{v23}
\La_{\text{tot}}=\La_{V=0}+\La_{V}+\La_{h}+\La_{h-V}\,,
\ee
where the first two contributions are exactly those of Eqs.~\eqref{lahless2} and \eqref{lav2} and
\bes
\begin{align}
& \bry{lll}
\dst \La_{h}
&=&\dst \frac{1}{2} \demub H\demua H+\frac{1}{2}  \demub h\demua h+\frac{1}{16} \(h^{2}+H^{2}+4vh\)\<\(D_{\mu}\S\)\(D^{\mu}\S\)^{\dag}\>\\\\
&&\dst -\f{m_{h}^{2}}{16v^{2}}\(-8v^{4}+8v^{2}h^{2}+4vh^{3}+\f{H^{4}}{2}+\f{h^{4}}{2}\)-\f{m_{H}^{2}}{2} H^{2}-\f{m_{h}^{2}+2m_{H}^{2}}{16v^{2}}\(4v H^{2}h+H^{2}h^{2}\)\,,
\ery\\\nonumber\\
& \bry{lll}
\dst \La_{h-V}
&=&\dst \frac{1}{32g_{V}^{2}} \(h^{2}+H^{2}+4vh\)\<\Hat{V}_{\mu}\Hat{V}^{\mu}\>-\frac{1}{4\sqrt{2}g_{V}} \(Hh+2vH\)\left\langle u_{\mu}\Hat{V}^{\mu}\right\rangle\,,
\ery
\end{align}
\ees
and we have used $g_{C}=1/2g_{V}$ and $M_{V}=g_{C}v$. In the limit $m_{H}\gg \Lambda$ the parity odd scalar $H$ can be integrated out from the spectrum. In this limit the Lagrangian \eqref{v23} reduces to the Lagrangian \eqref{lefftot} with the values of the parameters \eqref{gaugeparam} and \eqref{gaugeparameters}.
\chapter{Heavy vectors pair production}\l{chap4}
\setlength{\epigraphwidth}{.4\textwidth}     
\vspace{-2mm}\epigraph{Real is what can be measured.}%
              {\textsc{Max Planck}}
\vspace{3mm}     
\epigraph{It is the theory that decides what we can observe.}%
              {\textsc{Albert Einstein}}
\vspace{7mm}
\lettrine{I}{n} this chapter we discuss the phenomenology of the model of strong EWSB that we have introduced in the previous chapter in the pair production channels.
If the new heavy vector is not too heavy, say below 1 TeV, the single production, either by Vector Boson Fusion (VBF) or by the Drell--Yan (DY) process, or its production in association with a standard gauge boson are very likely to be the first manifestations of $V$ at the LHC \cite{Birkedal:2005p1372,He:2008ey,Accomando:2009ei,Accomando:2008p1815,Belyaev:2009gl,Cata:2009ka}. The same consideration can be done for the scalar: if its couplings don't deviate too much from the coupling of the SM Higgs boson, its phenomenology in the single production is very similar to what we expect for the SM Higgs boson. However, to understand the underlying dynamics which generates the new states, further measurements and observations will certainly be required. The trilinear and quadrilinear couplings of the new particles are usually the most sensitive to the different explicit models and they should therefore be measured pretty well to gain insight on the structure of the new strong dynamics. This motivates the study of the pair production of $V$ and $h$ and the associated $hV$ production, which we are going to discuss in this chapter. In particular, since the scalar pair production was carefully studied in Ref.~\cite{Contino:2010vc}, we will focus our attention on the $V$ pair production \cite{Barbieri:2010p1579} and on the associated $hV$ production \cite{Hernandez:2010p1578,Torre:2010p592}. We are interested in estimating the values of the relevant couplings for which a discovery of these processes is possible at the LHC with its foreseen energy and luminosity, i.e. $\sqrt{s}=14$ TeV and $L=\int \La dt = 100$ fb$^{-1}$.

We will see that the heavy vector does not affect at all the scalar pair production and that the scalar, improving the asymptotic behavior of the longitudinal $WW\to VV$ amplitudes, can only decrease the rate in the vector pair production. For these reasons in discussing the vector pair production we assume that the scalar is integrated out from the spectrum, i.e. $a=0$. The associated $hV$ production is the only process which requires both the scalar and the vector in the spectrum and which is sensitive to the coupling $k_{1}$ in the Lagrangian \eqref{lvhinteraction}. From a phenomenological point of view, the pretty large number of different charge pair production channels, from VBF or from DY, is of potential interest. We shall present the cross sections for the $V$ pair production and for the $hV$ associated production and the expected rates of multi-lepton events from the decay of such heavy vectors at the LHC.

This chapter is organized as follows. In Section~\ref{sec4.1} we study the $W_{L} W_{L} \rightarrow V_{\lambda} V_{\lambda^{\prime}}$ helicity amplitudes and in Section~\ref{sec4.2} we discuss their high energy limit. In Section~\ref{sec4.3} we briefly summarize the expressions for the mass eigenstate $W_{L} W_{L} \rightarrow V_{\lambda} V_{\lambda^{\prime}}$ helicity amplitudes as functions of the weak isospin eigenstate amplitudes. Section~\ref{sec4.4} is devoted to the study of the VBF total cross sections in the Effective Vector Boson Approximation (EVBA), while in Section~\ref{sec4.5} we go beyond this approximation presenting the exact leading order total cross sections for the heavy vectors pair production as functions of $M_{V}$. In Section~\ref{sec4.6} we write down the squared amplitudes for the DY pair production and we present numerical results for the DY pair production total cross sections. Finally in Section~\ref{sec4.8} we show the predictions for the total number of same-sign di-lepton and tri-lepton events at the LHC with $100$ fb$^{-1}$ of integrated luminosity and we discuss the results.

\section{$W_{L} W_{L} \rightarrow V_{\lambda} V_{\lambda^{\prime}}$ helicity amplitudes}\l{sec4.1}

In this section we compute the scattering amplitudes for two longitudinal $W$ bosons into a pair of heavy vectors of any helicity $\lambda,\lambda^{\prime}= L, +, -$. To simplify the explicit formulae, we take full advantage of $SU(2)_{L+R}$ invariance by considering the $g^{\prime}=0$
limit, so that $Z \sim W^{3}$. Moreover at high energy, i.e. $s,M_{V}^{2} \gg M_{W}^{2}$, the Equivalence Theorem \eqref{eqteo1} implies \be\l{eqteo}
\mathcal{A}(W_{L}^{a} W_{L}^{b} \rightarrow V_{\lambda}^{c} V_{\lambda
^{\prime}}^{d}) \approx -\mathcal{A}(\pi^{a} \pi^{b} \rightarrow
V_{\lambda}^{c} V_{\lambda^{\prime}}^{d})\,.
\ee
There are in fact four such independent amplitudes: 
\begin{align}
& \mathcal{A}(W_{L}^{a} W_{L}^{b} \rightarrow V_{L}^{c} V_{L}^{d})\,, \\
& \mathcal{A}(W_{L}^{a} W_{L}^{b} \rightarrow V_{+}^{c} V_{-}^{d})\,, \\
& \mathcal{A}(W_{L}^{a} W_{L}^{b} \rightarrow V_{+}^{c} V_{+}^{d}) = 
\mathcal{A}(W_{L}^{a} W_{L}^{b} \rightarrow V_{-}^{c} V_{-}^{d})
\end{align}
and 
\be
\quad\mathcal{A}(W_{L}^{a} W_{L}^{b} \rightarrow V_{L}^{c} V_{+}^{d}) = -%
\mathcal{A}(W_{L}^{a} W_{L}^{b} \rightarrow V_{L}^{c} V_{-}^{d})\,.
\ee
By $SU(2)_{L+R}$ invariance the general form of these amplitudes is 
\be
\mathcal{A}(W_{L}^{a} W_{L}^{b} \rightarrow V_{\lambda}^{c} V_{\lambda
^{\prime}}^{d}) = \mathcal{A}_{\lambda\lambda^{\prime}}(s, t, u) \delta
^{ab}\delta^{cd} + \mathcal{B}_{\lambda\lambda^{\prime}}(s, t, u) \delta
^{ac}\delta^{bd} + \mathcal{C}_{\lambda\lambda^{\prime}}(s, t, u) \delta
^{ad}\delta^{bc}\,,  \l{invampl}
\ee
where, by Bose symmetry, it is simple to prove that 
\be\l{LL}
\mathcal{A}_{\lambda\lambda^{\prime}}(s, t, u)= \mathcal{A}_{\lambda
\lambda^{\prime}}(s, u, t)~ {\text{and}}~ \mathcal{C}_{\lambda\lambda^{%
\prime}}(s, t, u)= \mathcal{B}_{\lambda\lambda^{\prime}}(s, u, t)~{\text{for}%
}~ \lambda\lambda ^{\prime}= LL, + -, ++\,,
\ee
whereas 
\be\l{Lp}
\mathcal{A}_{L+}(s, t, u)= - \mathcal{A}_{L+}(s, u, t)~ {\text{and}}~ 
\mathcal{C}_{L+}(s, t, u)= -\mathcal{B}_{L+}(s, u, t)\,.
\ee
These amplitudes receive contributions from the Feynman diagrams of Fig.~\ref{fig4}, i.e.:

\bit
\item {\bf contact interactions}\\
Fig.~\ref{fig4}(a): contribution of the $\pi^2 V^2$ vertex contained in  $\La_{V-\text{kin}}$ of Eq.~\eqref{eq7};\\
Fig.~\ref{fig4}(b): contribution of the $\pi^2 V^2$ vertex contained in $\La_{2V}$ of Eq.~\eqref{L2V} and proportional to $g_i, i=1,\dots, 5$;

\item {\bf $t$ and $u$ channels $\pi$ exchange}\\
Fig.~\ref{fig4}(c)-(d): contributions proportional to $g_V^2$ coming from $\La_{1V-\text{Goldstone}}$ of Eq.~\eqref{L1VGoldstone};

\item {\bf $s$ channel $V$ exchange}\\
Fig.~\ref{fig4}(e): contribution proportional to $g_V g_K$ with $g_V$ contained in $\La_{1V-\text{Goldstone}}$ of Eq.~\eqref{L1VGoldstone} and $g_K$ contained in $\La_{3V}$ of Eq.~\eqref{L3V}.

\item {\bf $s$ channel $h$ exchange}\\
Fig.~\ref{fig4}(f): contribution proportional to $a k_{1}$ with $a$ contained in $\La_{h-\text{int}}$ of Eq.~\eqref{Hint} and $k_{1}$ contained in $\La_{h-V}$ of Eq.~\eqref{lvhinteraction}.
\eit

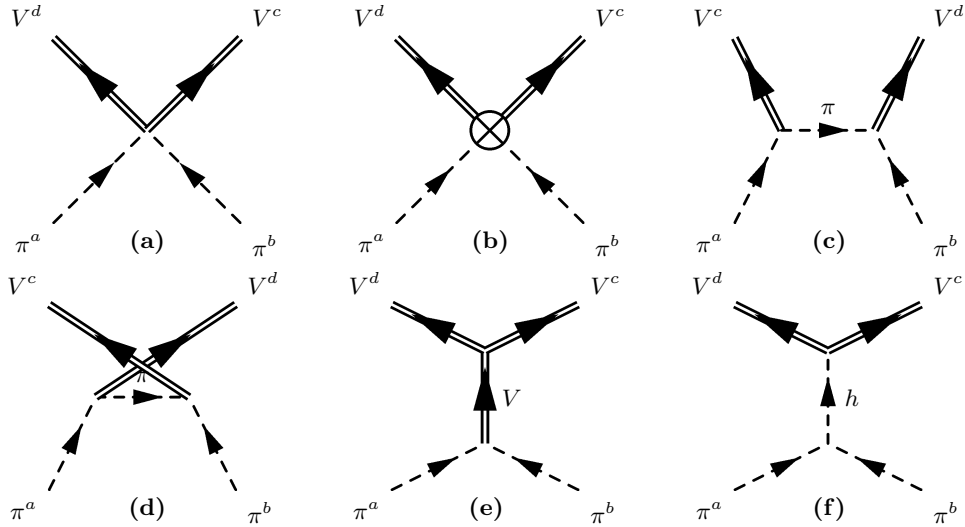
\begin{figure}[!htb]
\centering
\begin{minipage}[t]{3.5cm}
\begin{center}
\begin{fmffile}{ppVV}
  \fmfframe(1,7)(1,7){ 
          \begin{fmfgraph*}(70,70)
          \fmfstraight
          \fmflabel{\footnotesize{$V^{c}$}}{o2}
          \fmflabel{\footnotesize{$\pi^{a}$}}{i1}
          \fmflabel{\footnotesize{$V^{d}$}}{o1}
          \fmflabel{\footnotesize{$\pi^{b}$}}{i2}          
          \fmfbottom{i1,i2}
          \fmftop{o1,o2}
          \fmf{scalar}{i1,v1}
          \fmf{scalar}{i2,v1}
          \fmf{dbl_plain_arrow}{v1,o1}
          \fmf{dbl_plain_arrow}{v1,o2}
          \end{fmfgraph*}} \\\vspace{-4mm}  \footnotesize{{\bf (a)}}\l{ppVV}\vspace{4mm}
\end{fmffile}
\end{center}
\end{minipage}
\ \hspace{5mm} \
\begin{minipage}[t]{3.5cm}
\begin{center}
\begin{fmffile}{ppVVcounter}
  \fmfframe(1,7)(1,7){ 
          \begin{fmfgraph*}(70,70)
          \fmfstraight
          \fmflabel{\footnotesize{$V^{c}$}}{o2}
          \fmflabel{\footnotesize{$\pi^{a}$}}{i1}
          \fmflabel{\footnotesize{$V^{d}$}}{o1}
          \fmflabel{\footnotesize{$\pi^{b}$}}{i2}                 
          \fmfbottom{i1,i2}
          \fmftop{o1,o2}
          \fmf{phantom}{v1,v2}
          \fmf{scalar}{i1,v1}
          \fmf{scalar}{i2,v1}
          \fmf{dbl_plain_arrow}{v1,o1}
          \fmf{dbl_plain_arrow}{v1,o2}
          \fmfv{decor.shape=circle,decor.filled=empty}{v1}
          \fmfv{decor.shape=cross}{v2}
          \end{fmfgraph*}} \\\vspace{-4mm}  \footnotesize{{\bf (b)}}\l{ppVVcounter}\vspace{4mm}
\end{fmffile}
\end{center}
\end{minipage}
\ \hspace{5mm} \
\begin{minipage}[t]{3.5cm}
\begin{center}
        \begin{fmffile}{ppVVp1}
  \fmfframe(1,7)(1,7){ 
   \begin{fmfgraph*}(70,70)
          \fmfstraight
    \fmfleft{i1,i2}
    \fmfright{o1,o2}
    \fmflabel{\footnotesize{$\pi^{a}$}}{i1}
    \fmflabel{\footnotesize{$\pi^{b}$}}{o1}
    \fmflabel{\footnotesize{$V^{c}$}}{i2}
    \fmflabel{\footnotesize{$V^{d}$}}{o2}
    \fmf{scalar}{i1,v1}
    \fmf{dbl_plain_arrow}{v1,i2}
    \fmf{scalar}{o1,v2}
    \fmf{dbl_plain_arrow}{v2,o2}
    \fmf{scalar,label={\footnotesize{$\pi$}}}{v1,v2}
    \end{fmfgraph*}
        }  \\\vspace{-4mm}  \footnotesize{{\bf (c)}}  \l{ppVVp1}%
      \end{fmffile}
\end{center}
\end{minipage}
\ \hspace{5mm} \
\begin{minipage}[t]{3.5cm}
\begin{center}
        \begin{fmffile}{ppVVp2}
  \fmfframe(1,7)(1,7){ 
   \begin{fmfgraph*}(70,70)
          \fmfstraight
    \fmfleft{i1,i2}
    \fmfright{o1,o2}
    \fmflabel{\footnotesize{$\pi^{a}$}}{i1}
    \fmflabel{\footnotesize{$\pi^{b}$}}{o1}
    \fmflabel{\footnotesize{$V^{c}$}}{i2}
    \fmflabel{\footnotesize{$V^{d}$}}{o2}
    \fmf{scalar}{i1,v1}
    \fmf{phantom}{v1,i2}
    \fmf{scalar}{o1,v2}
    \fmf{phantom}{v2,o2}
    \fmf{scalar,label={\footnotesize{$\pi$}}}{v1,v2}
    \fmffreeze
    \fmf{dbl_plain_arrow}{v1,o2}
    \fmf{dbl_plain_arrow}{v2,i2}
    \end{fmfgraph*}
        }    \\\vspace{-4mm}  \footnotesize{{\bf (d)}}\l{ppVVp2}%
      \end{fmffile}
\end{center}
\end{minipage}
\ \hspace{5mm} \
\begin{minipage}[t]{3.5cm}
\begin{center}
        \begin{fmffile}{ppVVV1}
  \fmfframe(1,7)(1,7){ 
   \begin{fmfgraph*}(70,70)
          \fmfstraight
    \fmfbottom{i1,i2}
    \fmftop{o1,o2}
    \fmflabel{\footnotesize{$\pi^{a}$}}{i1}
    \fmflabel{\footnotesize{$V^{d}$}}{o1}
    \fmflabel{\footnotesize{$\pi^{b}$}}{i2}
    \fmflabel{\footnotesize{$V^{c}$}}{o2}
    \fmf{scalar}{i1,v1}
    \fmf{scalar}{i2,v1}
    \fmf{dbl_plain_arrow}{v2,o1}
    \fmf{dbl_plain_arrow}{v2,o2}
    \fmf{dbl_plain_arrow,label={\footnotesize{$V$}}}{v1,v2}
    \end{fmfgraph*}
        }  \\\vspace{-4mm}  \footnotesize{{\bf (e)}} \l{ppVVV1}%
      \end{fmffile}
\end{center}
\end{minipage}
\ \hspace{5mm} \
\begin{minipage}[t]{3.5cm}
\begin{center}
        \begin{fmffile}{ppVVfi}
  \fmfframe(1,7)(1,7){ 
   \begin{fmfgraph*}(70,70)
          \fmfstraight
    \fmfbottom{i1,i2}
    \fmftop{o1,o2}
    \fmflabel{\footnotesize{$\pi^{a}$}}{i1}
    \fmflabel{\footnotesize{$V^{d}$}}{o1}
    \fmflabel{\footnotesize{$\pi^{b}$}}{i2}
    \fmflabel{\footnotesize{$V^{c}$}}{o2}
    \fmf{scalar}{i1,v1}
    \fmf{scalar}{i2,v1}
    \fmf{dbl_plain_arrow}{v2,o1}
    \fmf{dbl_plain_arrow}{v2,o2}
    \fmf{scalar,label={\footnotesize{$h$}}}{v1,v2}
    \end{fmfgraph*}
        }  \\\vspace{-4mm}  \footnotesize{{\bf (f)}} \l{ppVVfi}%
      \end{fmffile}
\end{center}
\end{minipage}
\caption{The six Feynman diagrams contributing to the $\pi^{a} \pi^{b} \rightarrow
V_{\lambda}^{c} V_{\lambda^{\prime}}^{d}$ scattering: (a) and (b) are the contact interactions corresponding to $\La_{V-\text{kin}}$ and to $\La_{\text{2V}}$ respectively; (c) and (d) are the $t$ and $u$ channels of the contribution of the pion exchange; (e) and (f) are the $s$ channel contributions of the $V$ and $h$ exchange respectively.}\l{fig4}
\end{figure}

For ease of the reading, we keep first only the contributions with $\La_{2V}$, $\La_{3V}$ and $\La_{h-V}$ set to zero, so that\footnote{In all these functions the Mandelstam variables are in the order $(s, t, u)$ and are left understood.}:

\begin{itemize}
\item For $\lambda\lambda^{\prime}=LL$ 
\bes
\begin{align}
& \mathcal{A}_{LL}^{1V}=-\frac{g_{V}^{2}M_{V}^{2}s}{v^{4}\left( s-4M_{V}^{2}\right) }\left[ \frac{\left( t+M_{V}^{2}\right) ^{2}}{t}+\frac{\left(u+M_{V}^{2}\right) ^{2}}{u}\right]\,, \\
& \mathcal{B}_{LL}^{1V}=\frac{u-t}{2v^{2}}+\frac{g_{V}^{2}M_{V}^{2}s\left(u+M_{V}^{2}\right) ^{2}}{v^{4}u\left( s-4M_{V}^{2}\right) }\,.
\end{align}
\ees

\item For $\lambda\lambda^{\prime}=+-$ 
\bes
\begin{align}
& \mathcal{A}_{+-}^{1V}=\frac{2g_{V}^{2}M_{V}^{4}\left( t+u\right) \left(tu-M_{V}^{4}\right) }{v^{4}tu\left( s-4M_{V}^{2}\right) }\,, \\
& \mathcal{B}_{+-}^{1V}=\frac{2g_{V}^{2}M_{V}^{4}\left( M_{V}^{4}-tu\right)}{uv^{4}\left( s-4M_{V}^{2}\right) }\,.
\end{align}
\ees

\item For $\lambda\lambda^{\prime}=++$ 
\bes
\begin{align}
& \mathcal{A}_{++}^{1V}=\frac{2g_{V}^{2}M_{V}^{4}\left( t+u\right) \left(M_{V}^{4}-tu\right) }{v^{4}tu\left( s-4M_{V}^{2}\right) }\,, \\
& \mathcal{B}_{++}^{1V}=\frac{\left( t-u\right) }{2v^{2}}-\frac {2g_{V}^{2}M_{V}^{4}\left( M_{V}^{4}-tu\right) }{uv^{4}\left(s-4M_{V}^{2}\right) }\,.
\end{align}
\ees

\item For $\lambda\lambda^{\prime}=L+$ 
\bes
\begin{align}
& \mathcal{A}_{L+}^{1V}=\frac{\sqrt{2}g_{V}^{2}M_{V}^{5}\left( t-u\right)\sqrt{s\left( tu-M_{V}^{4}\right) }}{v^{4}tu\left( s-4M_{V}^{2}\right)}\,, \\
& \mathcal{B}_{L+}^{1V}=-\frac{\sqrt{s\left( tu-M_{V}^{4}\right) }\left\{v^{2}su+4M_{V}^{2}\left[ g_{V}^{2}M_{V}^{2}\left( M_{V}^{2}+u\right) -v^{2}u\right]\right\} }{2\sqrt{2}uv^{4}M_{V}\left( s-4M_{V}^{2}\right)}\,.
\end{align}
\ees
\end{itemize}

Switching on $\La_{2V}$, $\La_{3V}$ and $\La_{h-V}$ we find the extra contributions to the various amplitudes:

\begin{itemize}
\item For $\lambda \lambda ^{\prime }=LL$ 
\bes
\begin{align}
& \bry{lll}\dst \Delta \mathcal{A}_{LL}&=&\dst \left( g_{1}-g_{2}\right) \frac{s\left(s-2M_{V}^{2}\right) }{v^{2}M_{V}^{2}}+\left( g_{4}-g_{5}\right) \frac{s\left[2M_{V}^{2}\left( 3M_{V}^{2}-s\right) +t^{2}+u^{2}\right] }{v^{2}M_{V}^{2}\left( s-4M_{V}^{2}\right) }\\
&&\dst -\f{ak_{1}}{8g_{V}^{2}M_{V}^{2}}\(\f{s}{s-m_{h}^{2}}\)\(s-2M_{V}^{2}\)\,,\ery \\\notag\\
& \bry{lll}\dst \Delta \mathcal{B}_{LL}&=&\dst g_{2}\frac{s\left( s-2M_{V}^{2}\right) }{v^{2}M_{V}^{2}}+\frac{s\left( t-u\right) }{v^{2}M_{V}^{2}}\left( g_{3}+\frac{g_{K}g_{V}}{4}\frac{s+2M_{V}^{2}}{s-M_{V}^{2}}\right)\\
&&\dst +g_{5}\frac{s\left[2M_{V}^{2}\left( 3M_{V}^{2}-s\right) +t^{2}+u^{2}\right] }{v^{2}M_{V}^{2}\left( s-4M_{V}^{2}\right) }.\ery
\end{align}
\ees

\item For $\lambda \lambda ^{\prime }=+-$ 
\bes
\begin{align}
& \Delta \mathcal{A}_{+-}=4\left( g_{4}-g_{5}\right) \frac{\left(M_{V}^{4}-tu\right) }{v^{2}\left( s-4M_{V}^{2}\right) }\,, \\
& \Delta \mathcal{B}_{+-}=4g_{5}\frac{\left( M_{V}^{4}-tu\right) }{v^{2}\left( s-4M_{V}^{2}\right) }\,.
\end{align}
\ees

\item For $\lambda \lambda ^{\prime }=++$ 
\bes
\begin{align}
& \Delta \mathcal{A}_{++}=2\left( g_{1}-g_{2}\right) \frac{s}{v^{2}}+4\left(g_{4}-g_{5}\right) \frac{\left( tu-M_{V}^{4}\right) }{v^{2}\left(s-4M_{V}^{2}\right) }-\f{ak_{1}}{4g_{V}^{2}}\f{s}{s-m_{h}^{2}}\,, \\
& \Delta \mathcal{B}_{++}=2g_{2}\frac{s}{v^{2}}+\frac{4g_{5}\left(tu-M_{V}^{4}\right) }{v^{2}\left( s-4M_{V}^{2}\right) }-\frac{g_{K}g_{V}s(t-u)}{2v^{2}\left( s-M_{V}^{2}\right) }\,.
\end{align}
\ees

\item For $\lambda \lambda ^{\prime }=L+$ 
\bes
\begin{align}
& \Delta \mathcal{A}_{L+}=\left( g_{4}-g_{5}\right) \frac{\left( t-u\right)\sqrt{2s\left( tu-M_{V}^{4}\right) }}{v^{2}M_{V}\left( s-4M_{V}^{2}\right) }
\,, \\
& \Delta \mathcal{B}_{L+}=\frac{\sqrt{2s\left( tu-M_{V}^{4}\right) }}{v^{2}M_{V}}\left[ g_{5}\frac{t-u}{s-4M_{V}^{2}}+\left( g_{3}+\frac{g_{K}g_{V}}{2}\frac{s}{s-M_{V}^{2}}\right) \right] \,.
\end{align}
\ees
\end{itemize}

\section{Asymptotic behavior of the $W_{L} W_{L} \rightarrow V_{\lambda} V_{\lambda^{\prime}}$ amplitudes}\l{sec4.2}

For arbitrary values of the parameters all these amplitudes grow at least as 
$s/v^{2}$ and some as $s^2/\(v^2M_V^2\)$ or as $s^{3/2}/\(v^2 M_V\)$. We know from the previous chapter that in the case of a hidden gauge model, these amplitudes can grow at most as $s/v^{2}$. So we can check the conditions for the equivalence with the gauge model of Section~\ref{sec3.3.1} (or \ref{sec3.5} in presence of the scalar) by requiring the cancellation of the terms proportional to $s^2/\(v^2M_V^2\)$ or $s^{3/2}/\(v^2 M_V\)$ in the amplitudes of the previous section. As readily
seen from the amplitudes of previous section, there is a unique choice of the
various parameters that makes all these amplitudes growing at most like $%
s/v^2$, i.e. 
\be\l{special}
g_{V}M_{V}=\f{v}{2},~~g_V g_K= 1,~~g_3=-\frac{1}{4},~~ g_1 = g_2 =g_4 = g_5 = 0,~~ a=\f{1}{2},~~ k_{1}=1  
\ee
whereas $f_V$ and $g_6$ are irrelevant. This choice of parameter is exactly the same we found in Section~\ref{sec3.3.1} (\ref{sec3.5}) in the case of the minimal gauge model. With this choice of parameters the various helicity amplitudes simplify to\footnote{To keep track of the form of the amplitudes we have left explicit the dependence on $g_{V}$ and $M_{V}$ without using the first identity of Eq.~\eqref{special}. It can be seen very simply that using also this identity the growth with $s$ in the amplitude \eqref{ALL} is explicitly canceled.}
\begin{itemize}
\item For $\lambda\lambda^{\prime}=LL$%
\bes
\begin{align}
& \mathcal{A}_{LL}^{\text{gauge}}=-\frac{g_{V}^{2}M_{V}^{2}s}{v^{4}\left( s-4M_{V}^{2}\right) }\left[ \frac{\left( t+M_{V}^{2}\right) ^{2}}{t}+\frac{\left(u+M_{V}^{2}\right) ^{2}}{u}\right]-\f{1}{16g_{V}^{2}M_{V}^{2}}\(\f{s}{s-m_{h}^{2}}\)\(s-2M_{V}^{2}\)\,, \l{ALL}\\
& \mathcal{B}_{LL}^{\text{gauge}}=\frac{u-t}{2v^{2}}+\frac{g_{V}^{2}M_{V}^{2}s\left(u+M_{V}^{2}\right) ^{2}}{v^{4}u\left( s-4M_{V}^{2}\right) }-\frac{3s(u-t)}{4v^{2}\left( s-M_{V}^{2}\right) }\,.
\end{align}
\ees

\item For $\lambda\lambda^{\prime}=+-$
\bes
\begin{align}
& \mathcal{A}_{+-}^{\text{gauge}}=\frac{2g_{V}^{2}M_{V}^{4}\left( t+u\right)\left( tu-M_{V}^{4}\right) }{v^{4}tu\left( s-4M_{V}^{2}\right) }\,, \\
& \mathcal{B}_{+-}^{\text{gauge}}=\frac{2g_{V}^{2}M_{V}^{4}\left(M_{V}^{4}-tu\right) }{uv^{4}\left( s-4M_{V}^{2}\right) }\,.  \l{p6}
\end{align}
\ees

\item For $\lambda\lambda^{\prime}=++$ 
\bes
\begin{align}
& \mathcal{A}_{++}^{\text{gauge}}=\frac{2g_{V}^{2}M_{V}^{4}\left( t+u\right)\left( M_{V}^{4}-tu\right) }{v^{4}tu\left( s-4M_{V}^{2}\right) }-\f{1}{8g_{V}^{2}}\frac{s}{s-m_{h}^{2}}\,,
\l{p3} \\
& \mathcal{B}_{++}^{\text{gauge}}=-\frac{M_{V}^{2}\left( t-u\right) }{2v^{2}\left( s-M_{V}^{2}\right) }-\frac{2g_{V}^{2}M_{V}^{4}\left(M_{V}^{4}-tu\right) }{uv^{4}\left( s-4M_{V}^{2}\right) }\,.  \l{p4}
\end{align}
\ees

\item For $\lambda\lambda^{\prime}=L+$ 
\bes
\begin{align}
& \mathcal{A}_{L+}^{\text{gauge}}=\frac{\sqrt{2}g_{V}^{2}M_{V}^{5}\left(t-u\right) \sqrt{s\left( tu-M_{V}^{4}\right) }}{v^{4}tu\left(s-4M_{V}^{2}\right) }\,,  \l{p7} \\
& \mathcal{B}_{L+}^{\text{gauge}}=-\frac{\sqrt{2}g_{V}^{2}M_{V}^{3}\left(M_{V}^{2}+u\right) \sqrt{s\left( tu-M_{V}^{4}\right) }}{uv^{4}\left(s-4M_{V}^{2}\right) }+\frac{M_{V}\sqrt{s\left( tu-M_{V}^{4}\right) }}{\sqrt{2}v^{2}\left( s-M_{V}^{2}\right) }\,.  \l{p8}
\end{align}
\ees
\end{itemize}

These are exactly the $W_{L} W_{L} \rightarrow V_{\lambda}V_{\lambda^{\prime}}$ helicity amplitudes which arise in a minimal gauge model (the contribution of the scalar only arises if the minimal gauge model is spontaneously broken by two linear doublets as discussed in \ref{sec3.5}). On the other hand, in the generic framework considered here, some deviations from the relations \eqref{special} may occur. In such a case the asymptotic behavior of the various amplitudes will have to be improved, e.g., by the occurrence of heavier composite states, vectors and/or scalars, with appropriate couplings. Note in any event that, even sticking to the relations \eqref{special}, the amplitudes for longitudinally polarized vectors grow as $s/v^2$ for any value of $G_{V}$.

\section{Mass eigenstate $W_{L}W_{L}\to V_{\lambda}V_{\lambda'}$ helicity amplitudes}\l{sec4.3}
The scattering amplitudes for the processes $W_{L}^{a}W_{L}^{b}\to V_{\lambda}^{c}V_{\lambda'}^{d}$ have the general form of Eq.~\eqref{invampl}.
Using the superposition principle, and taking into account the definition of the mass eigenstates as functions of the weak isospin eigenstates for the EW gauge boson (see Eq.~\eqref{Eigen}) and for the heavy vector bosons
\be\l{Eigen2}
\bry{l}
V^{\pm}_{\mu}=\dst \f{1}{\sqrt{2}}\(V^{1}_{\mu}\mp iV^{2}_{\mu}\)\,,\\\\
V^{0}_{\mu}=V^{3}_{\mu}\,,
\ery
\ee
we can write the mass eigenstates scattering amplitudes as
\bes\l{i9}
\begin{align}
&\dst \Mamp\(W_{L}^{+}W_{L}^{-}\to V^{+}_{\lambda}V^{-}_{\lambda'}\)=\mathcal{A}_{\lambda\lambda'}\(s,t,u\)+\mathcal{B}_{\lambda\lambda'}\(s,t,u\)\,,\l{i9a}\\\nonumber\\
&\dst \Mamp\(W_{L}^{+}W_{L}^{-}\to V^{0}_{\lambda}V^{0}_{\lambda'}\)=\Mamp\(Z_{L}Z_{L}\to V^{+}_{\lambda}V^{-}_{\lambda'}\)=\mathcal{A}_{\lambda\lambda'}\(s,t,u\)\,,\l{i9b}\\\nonumber\\
&\dst \Mamp\(W_{L}^{\pm}Z_{L}\to V^{\pm}_{\lambda}V^{0}_{\lambda'}\)=\mathcal{B}_{\lambda\lambda'}\(s,t,u\)\,,\l{i9c}\\\nonumber\\
&\dst \Mamp\(Z_{L}Z_{L}\to V^{0}_{\lambda}V^{0}_{\lambda'}\)=\mathcal{A}_{\lambda\lambda'}\(s,t,u\)+\mathcal{B}_{\lambda\lambda'}\(s,t,u\)+\mathcal{C}_{\lambda\lambda'}\(s,t,u\)\,,\l{i9d}\\\nonumber\\
&\dst \Mamp\(W_{L}^{\pm}W_{L}^{\pm}\to V^{\pm}_{\lambda}V^{\pm}_{\lambda'}\)=\mathcal{B}\(s,t,u\)_{\lambda\lambda'}+\mathcal{C}\(s,t,u\)_{\lambda\lambda'}\l{i9e}\,.
\end{align}
\ees

\section{Vector boson fusion pair production cross sections: the Effective Vector Boson Approximation (EVBA)}\l{sec4.4}
In this section we want to consider the heavy vector pair production by VBF. This processes are represented diagrammatically in Fig.~\ref{fig:VBF}.
\begin{figure}[!htb]
\centering
\begin{fmffile}{ppWWqqVVqq1}
  \fmfframe(1,7)(1,7){ 
          \begin{fmfgraph*}(120,80)
          \fmfstraight
          \fmfleft{l1,l2} 
	  \fmfright{r1,r2,r3,r4,r5,r6,r7}
	  \fmflabel{\footnotesize{$p$}}{l1}
	  \fmflabel{\footnotesize{$p$}}{l2}
	  \fmflabel{\footnotesize{$q'$}}{r1}
	  \fmflabel{\footnotesize{$q$}}{r7}
	  \fmf{fermion,tension=.6}{l1,v1}
	  \fmf{plain}{v1,v2}
	  \fmf{fermion}{v2,r1} 
	  \fmf{fermion,tension=.6}{l2,v3}
	  \fmf{plain}{v3,v4}
	  \fmf{fermion}{v4,r7} 
	  \fmffreeze
	  \fmf{photon,label=\footnotesize{$G$},l.side=left}{v1,v5}
	  \fmf{phantom}{v5,v2}
	  \fmf{photon,label=\footnotesize{$G'$},l.side=right}{v3,v5}
	  \fmf{phantom}{v5,v4} 
	  \fmfblob{.15w}{v5}
	  \fmffreeze
 	  \fmf{dbl_plain,label=\footnotesize{$V'$},l.side=right}{v5,r3}
 	  \fmf{dbl_plain,label=\footnotesize{$V$},l.side=left}{v5,r5}
	  \fmffreeze 
	  \fmfi{plain}{vpath (__l1,__v1) shifted (thick*(0,-2))} 
	  \fmfi{plain}{vpath (__l1,__v1) shifted (thick*(0,-4))}
	  \fmfi{plain}{vpath (__v1,__v2) shifted (thick*(0,-2))} 
	  \fmfi{plain}{vpath (__v1,__v2) shifted (thick*(0,-4))}
	  \fmfi{plain}{vpath (__v2,__r1) shifted (thick*(0,-2))} 
	  \fmfi{plain}{vpath (__v2,__r1) shifted (thick*(0,-4))}
	  \fmfi{plain}{vpath (__l2,__v3) shifted (thick*(0,2))} 
	  \fmfi{plain}{vpath (__l2,__v3) shifted (thick*(0,4))}
	  \fmfi{plain}{vpath (__v3,__v4) shifted (thick*(0,2))} 
	  \fmfi{plain}{vpath (__v3,__v4) shifted (thick*(0,4))}
	  \fmfi{plain}{vpath (__v4,__r7) shifted (thick*(0,2))} 
	  \fmfi{plain}{vpath (__v4,__r7) shifted (thick*(0,4))}
          \end{fmfgraph*}} 
\end{fmffile}
\caption{Diagrammatic representation of the VBF process $pp\to GG'qq'\to VV'qq'$: $G$ and $G'$ are the two gauge bosons; the bubble represents the tree-level interaction between the two gauge bosons and the two $V$ bosons. In this figure we adopted a `left-right' convention to draw the Feynman diagram.}\l{fig:VBF}
\end{figure}
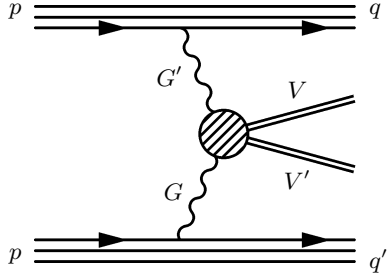
In order to compute the total cross sections for the VBF pair production of heavy vectors $V$ at the LHC, in the high energy limit $s\gg M_{W}^{2}$, we can use the EVBA. It consists in the assumption that the quarks contain a gauge boson ($W$ and $Z$) carrying a fraction of the quark momentum and that this gauge boson is radiated by the quark at a small angle, i.e. near on-shell. Thus the total cross section for $pp\to VVqq$ in the EVBA is generated by the subprocess $WW\to VV$ where the two initial $W$'s are on-shell. In order to use the EVBA we must know the momentum distributions of the $W$ bosons into the quarks constituting the proton. At the energy of the LHC we have to take into account that the proton is composed not only by its valence quarks, but also by the sea quarks, i.e. the $\bar{u}$, $\bar{d}$, $c$, $b$, $\bar{c}$ and $\bar{b}$. We can neglect the contributions of the $t$ and $\bar{t}$ to the proton momentum, since the top quark is very heavy and unstable.

In the previous sections we have seen that the amplitudes for longitudinal VBF have an asymptotic behavior that grows with the c.o.m.~energy. This is not the case for the longitudinal-transverse and transverse-transverse VBF that are therefore suppressed with respect to the longitudinal-longitudinal VBF. So we can compute the the total cross section in EVBA neglecting the contribution of the transverse polarizations\footnote{We evaluated the effect of the transverse polarizations that turned out to be negligible with respect to the longitudinal VBF (see the exact result in the next section).}. In this section we use the amplitudes of the previous section (in the limit $a=0$) and the EVBA to compute the numerical total cross sections summed over the heavy vectors polarizations as functions of the vector mass for the production of different type of $V$-pair final states: $V^{+}V^{-}$, $V^{+}V^{0}$, $V^{-}V^{0}$ and $V^{0}V^{0}$.

The computation of the Parton Distribution Functions (PDFs) of the $W$ and $Z$ bosons into the quarks is conceptually similar to the corresponding calculation for the PDFs of the electron into the photon in the equivalent photon approximation (Weizs\"acker-Williams approximation). In order to compute the distribution functions for the $W$ and $Z$ bosons let us recall what happens in the equivalent photon approximation.

Consider the following process: a final state $Y$ particle is produced in the process $e X\to e Y$ (where $e$ is either the electron or the positron) at small angles through the exchange of a photon described by the subprocess $\g X\to Y$. In the Weizs\"acker-Williams approximation the cross section of the complete process $e X\to e Y$ is obtained by the convolution of the subprocess cross section $\hat{\s}^{\g X\to Y}\(\hat{s}=zs\)$ with the probability of the initial electron to radiate a photon $f_{e/\g}\(z\)$:
\be\l{l1}
\s^{e X\to e Y}=\int_{0}^{1} dz f_{e /\g}\(z\)\hat{\s}^{\g X\to Y}\(\hat{s}=zs\)\,.
\ee
After the simple calculation of the matrix element for the splitting of an electron to an electron and a photon, Eq.~\eqref{l1} can be rewritten explicitly as
\be\l{l2}
\s^{e X\to e Y}=\int_{0}^{1} dz\int \f{dp_{\perp}^{2}}{p_{\perp}^{2}} \f{\a}{2\pi}\Bigg[\f{1+\(1-z\)^{2}}{z}\Bigg]\hat{\s}^{\g X\to Y}\(\hat{s}=zs\)\,,
\ee
where the integral over the transverse momentum $p_{\perp}^{2}$ runs from the electron mass $m_{e}^{2}$ that cuts off the soft singularity, to the c.m. energy of the complete process $s$, and the coefficient $\f{\a}{2\pi}\(\f{1+\(1-z\)^{2}}{z}\)$ comes from the matrix element for the electron splitting summed over final polarizations and averaged over initial polarizations. Performing the integration over $p_{\perp}^{2}$ in Eq.~\eqref{l2} we find
\be\l{l3}
f_{e/\g}\(z\)=\f{\a}{2\pi}\ln\f{s}{m_{e}^{2}}\Bigg[\f{1+\(1-z\)^{2}}{z}\Bigg]\,,
\ee
that is the desired PDF of the photon into the electron.

The computation of the PDFs for the transverse $W$ and $Z$ bosons (see e.g.~Refs.~\cite{Chanowitz:1984ne,Dawson:1984gx}) is very similar to the calculation for the photon. Using the notation $G_{\lambda}$ to indicate the gauge boson $G$ (either the $W$ or the $Z$) with polarization $\lambda$ and substituting the electron-photon coupling $e$ with the vector and axial couplings for the $\bar{q}qG$ interaction (the couplings, respectively $v_{q}^{G}$ and $a_{q}^{G}$, are summarized in Table~\ref{table_couplings}) we find\footnote{The photon case corresponds to $v_{q}^{\g}=e$ and $a_{q}^{\g}=0$.}
\be\l{l4}
\bry{lll}
f_{q/G_{\pm}}\(x\)&=&\dst \int_{M_{G}^{2}}^{\hat{s}}\f{dp_{\perp}^{2}}{p_{\perp}^{2}}\f{1}{16\pi^{2}}\sum_{q'}\Bigg[\f{\left|v_{q}^{G}\mp a_{q}^{G}\right|^{2}+\left|v_{q}^{G}\pm a_{q}^{G}\right|^{2}\(1-x\)^{2}}{x}\Bigg]\\\\
&=&\dst \f{1}{16\pi^{2}}\sum_{q'}\Bigg[\f{\left|v_{q}^{G}\mp a_{q}^{G}\right|^{2}+\left|v_{q}^{G}\pm a_{q}^{G}\right|^{2}\(1-x\)^{2}}{x}\Bigg]\ln\(\f{\hat{s}}{M_{G}^{2}}\)\,,
\ery
\ee
where $x$ is the fraction of the quark momentum carried by the gauge boson, $\hat{s}$ is the c.o.m.~energy squared of the incoming quarks, i.e. a fraction $z$ of the proton-proton system c.o.m.~energy and the sum over $q'$ corresponds to the sum over all the possible final state quarks in the splitting.
\begin{table}[h!]
\begin{center}
\begin{tabular}[c]{|c|c|c|}
		\hline
		Splitting & Vector coupling $v$ & Axial coupling $a$\\
		\hline
		$d^{g}\to W^{-}u^{f}$ & $v_{d^{g}}^{W^{-}}=\f{g}{2\sqrt{2}}V_{fg}=\f{e}{2\sin\theta_{W}\sqrt{2}}V_{fg}$ & $a_{d^{g}}^{W^{-}}=-\f{g}{2\sqrt{2}}V_{fg}=-\f{e}{2\sin\theta_{W}\sqrt{2}}V_{fg}$\\
		\hline
		$u^{f}\to W^{+}d^{g}$ & $v_{u^{f}}^{W^{+}}=\f{g}{2\sqrt{2}}V_{fg}^{*}=\f{e}{2\sin\theta_{W}\sqrt{2}}V_{fg}^{*}$ & $a_{u^{f}}^{W^{+}}=-\f{g}{2\sqrt{2}}V_{fg}^{*}=-\f{e}{2\sin\theta_{W}\sqrt{2}}V_{fg}^{*}$\\
		\hline
		$d^{f}\to Z d^{f}$ & $v_{d^{f}}^{Z}=\f{e}{4\cos\theta_{W}\sin\theta_{W}}\(-1+\f{4}{3}\sin^{2}\theta_{W}\)$ & $a_{d^{f}}^{Z}=\f{e}{4\cos\theta_{W}\sin\theta_{W}}$\\
		\hline
		$u^{f}\to Z u^{f}$ & $v_{u^{f}}^{Z}=\f{e}{4\cos\theta_{W}\sin\theta_{W}}\(1-\f{8}{3}\sin^{2}\theta_{W}\)$ & $a_{u^{f}}^{Z}=-\f{e}{4\cos\theta_{W}\sin\theta_{W}}$\\
		\hline
\end{tabular}\caption{Summary of the vector and axial couplings for the $\bar{q}qG$ quark-quark-gauge bosons interactions (for $g'\neq 0$). The indices $f$ and $g$ run over the quark families.}\l{table_couplings}
\end{center}
\end{table}
Note that now the integral is naturally cut off by the gauge boson mass $M_{G}$. Moreover a difference in a factor of $1/2$ from the photon case arises since we are not summing over the gauge bosons polarizations. 

The PDFs of longitudinal gauge bosons can be computed in the same way with the only difference that the amplitude for the quark-gauge boson splitting involves the longitudinal polarizations vectors. The final result is
\be\l{l5}
f_{q/G_{L}}\(x\)= \f{1}{4\pi^{2}}\f{1-x}{x}\sum_{q'}\(\left|v_{q}^{G}\right|^{2}+ \left|a_{q}^{G}\right|^{2}\)\,.
\ee
The absence of the logarithmic enhancement in the longitudinal gauge bosons distribution functions induces a suppression in the longitudinal channels at high energy $s\gg M_{G}^{2}$. On the other hand the growth with the c.o.m.~energy squared of the longitudinal VBF amplitudes computed in the previous sections compensates this suppression, making the longitudinal-longitudinal channel the most important at high energy.

Since we are interested in computing cross sections for the longitudinal-longitudinal VBF, we can neglect the transverse PDFs. The longitudinal PDFs can be summarized as follows
\bes
\begin{align}
& \dst f_{q/W^{\pm}_{L}}\(x\)
=\dst \f{g^{2}}{16\pi^{2}}\f{1-x}{x}=\f{\a}{4\pi\sin^{2}\theta_{W}}\f{1-x}{x}\,,\l{l6c}\\
& \dst f_{u^{f}/Z_{L}}\(x\)
=\f{\a}{16\pi\cos^{2}\theta_{W}\sin^{2}\theta_{W}}\f{1-x}{x}\(2-\f{16}{3}\sin^{2}\theta_{W}+\f{64}{9}\sin^{4}\theta_{W}\)\,,\l{l9c}\\
& \dst f_{d^{f}/Z_{L}}\(x\)
= \f{\a}{16\pi\cos^{2}\theta_{W}\sin^{2}\theta_{W}}\f{1-x}{x}\(2-\f{8}{3}\sin^{2}\theta_{W}+\f{16}{9}\sin^{4}\theta_{W}\)\,.\l{l9e}
\end{align}
\ees
Using these distributions we can write the effective luminosity for the emission of two longitudinal gauge bosons $G_{L}$ and $G'_{L}$ from a pair of quarks $q_{i}$ and $q_{j}$ in the form
\be\l{l15}
\bry{lll}
\dst \f{dL}{d\xi}\bigg|_{q_{i}q_{j}/G_{L}G'_{L}}\(\xi=\f{z}{y}\)&=&\dst \int_{\xi}^{1}\f{dx}{x}f_{q_{i}/G_{L}}\(x\)f_{q_{j}/G'_{L}}\(\f{\xi}{x}\)\,,
\ery
\ee
where $z$ is the fraction of the proton momentum carried by the $G$ boson, $y$ is the fraction of the proton momentum carried by the quark and $\xi=z/y$ is the fraction of the quark momentum carried by the $G$ boson. Therefore, the effective luminosity for the emission of two longitudinal gauge bosons $G_{L}$ and $G'_{L}$ from a proton-proton system is given by
\be\l{l22}
\bry{lll}
\dst \f{dL}{dz}\bigg|_{pp/G_{L}G'_{L}}&=&\dst \sum_{i,j}\int_{z}^{1}\f{dy}{y}\int_{y}^{1}\f{dx}{x}\Bigg[f_{q_{i}}\(x,\mu^{2}\)f_{q_{j}}\(\f{y}{x},\mu^{2}\)+f_{\bar{q}_{i}}\(x,\mu^{2}\)f_{\bar{q}_{j}}\(\f{y}{x},\mu^{2}\)\\\\
&&\dst +f_{\bar{q}_{i}}\(x,\mu^{2}\)f_{q_{j}}\(\f{y}{x},\mu^{2}\)+f_{q_{i}}\(x,\mu^{2}\)f_{\bar{q}_{j}}\(\f{y}{x},\mu^{2}\)\Bigg]\f{dL}{d\xi}\bigg|_{q_{i}\bar{q}_{j}/G_{L}G'_{L}}\(\xi=\f{z}{y}\)\,,
\ery
\ee
where the sum runs over all the valence and sea quark flavors and $f_{q_{i}}\(x,\mu^{2}\)$ is the PDF of the quark $q_{i}$ in the proton with momentum fraction $x$ defined at the factorization scale $\mu^{2}$. In our case, the factorization scale is conveniently chosen to be of order of $2M_{V}$. At this point it is straightforward to write the total cross section for the process $pp\to G_{L}G_{L}'qq\to VVqq$. We have
\be\l{l23}
\s^{pp\to G_{L}G'_{L}qq\to VVqq}\(s\)=\int_{\f{4M_{V}^{2}}{s}}^{1}dz\f{dL}{dz}\bigg|_{pp/G_{L}G'_{L}}\s^{G_{L}G'_{L}\to VV}\(zs\)\,,
\ee
where $\s^{G_{L}G'_{L}\to VV}\(zs\)$ is the partonic cross section for the $G_{L}G'_{L}\to VV$ process and is given by the phase space integration of the sum over the $V$ polarizations of the squared amplitude:
\be\l{l24}
\bry{lll}
\dst \s^{G_{L}G'_{L}\to VV}\(z s\)&=&\dst \int_{\hat{t}_{\text{min}}}^{\hat{t}_{\text{max}}}d\hat{t}\,\,\f{d\s^{G_{L}G'_{L}\to VV}}{d\hat{t}}=\int_{\hat{t}_{\text{min}}}^{\hat{t}_{\text{max}}}d\hat{t}\,\,\f{\sum_{\lambda,\lambda'}\left|\Mamp^{G_{L}G'_{L}\to V_{\lambda}V_{\lambda'}}\right|^{2}}{64\pi zs \left|\pvec_{1}\right|^{2}}\\\\
&=&\dst \f{1}{64\pi z s\(\f{z s}{4}-M_{G}^{2}\)}\int_{\hat{t}_{\text{min}}}^{\hat{t}_{\text{max}}}d\hat{t}\sum_{\lambda,\lambda'}\left|\Mamp^{G_{L}G'_{L}\to V_{\lambda}V_{\lambda'}}\right|^{2}\,.
\ery
\ee
The minimum and maximum values of $\hat{t}$ are fixed by the kinematics to
\be\l{l25}
\hat{t}_{\text{min}}=m_{G}^{2}+m_{V}^{2}-2zE_{p_{1}}E_{k_{1}}-2z\left|\pvec_{1}\right|\left|\kvec_{1}\right|\qquad\qquad\qquad \hat{t}_{\text{max}}=m_{G}^{2}+m_{V}^{2}-2zE_{p_{1}}E_{k_{1}}+2z\left|\pvec_{1}\right|\left|\kvec_{1}\right|\,.
\ee
In particular, in the case of two $W_{L}$, where $\hat{t}$ is given by
\be\l{l25bis}
\hat{t}=M_{V}^{2}+M_{W}^{2}-\f{zs}{2}+\f{\sqrt{zs-4M_{V}^{2}}\sqrt{zs-4M_{W}^{2}}\cos\theta_{CM}}{2}\,,
\ee
we find
\be\l{l26}
\hat{t}_{\text{min}}=-\(\sqrt{\f{z s}{4}-M_{W}^{2}}+\sqrt{\f{z s}{4}-M_{V}^{2}}\)^{2}\,,\qquad\qquad\qquad \hat{t}_{\text{max}}=-\(\sqrt{\f{z s}{4}-M_{W}^{2}}-\sqrt{\f{z s}{4}-M_{V}^{2}}\)^{2}\,,
\ee
where the minimum and the maximum correspond respectively to $\theta_{CM}=\pi$ and $\theta_{CM}=0$. These latter relations can be written, in the high energy limit $s\gg M_{W}^{2}$, as
\be\l{l27}
\hat{t}_{\text{min}}=-\(\sqrt{\f{z s}{4}}+\sqrt{\f{z s}{4}-M_{V}^{2}}\)^{2}\,,\qquad\qquad\qquad \hat{t}_{\text{max}}=-\(\sqrt{\f{z s}{4}}-\sqrt{\f{z s}{4}-M_{V}^{2}}\)^{2}\,.
\ee
Finally, substituting Eqs.~\eqref{l24} and \eqref{l22} into Eq.~\eqref{l23} and taking into account Eq.~\eqref{l15} we can write the total cross section for the process $pp\to G_{L}G'_{L}qq\to VVqq$ as
\small\be\l{l28}
\bry{lll}
\dst \s^{pp\to G_{L}G'_{L}qq\to VVqq}\(s\)&=&\dst \int_{\f{4M_{V}^{2}}{s}}^{1}dz\Bigg\{\sum_{i,j}\int_{z}^{1}\f{dy}{y}\int_{y}^{1}\f{dx}{x}\bigg[f_{q_{i}}\(x,\mu^{2}\)f_{q_{j}}\(\f{y}{x},\mu^{2}\)+f_{\bar{q}_{i}}\(x,\mu^{2}\)f_{\bar{q}_{j}}\(\f{y}{x},\mu^{2}\)\\\\
&&\dst \hspace{-1.2cm}+f_{\bar{q}_{i}}\(x,\mu^{2}\)f_{q_{j}}\(\f{y}{x},\mu^{2}\)+f_{q_{i}}\(x,\mu^{2}\)f_{\bar{q}_{j}}\(\f{y}{x},\mu^{2}\)\bigg]\bigg[\int_{\f{z}{y}}^{1}\f{dx'}{t}f_{q_{i}/G_{L}}\(x'\)f_{q_{j}/G'_{L}}\(\f{\f{z}{y}}{x'}\)\bigg]\\\\
&&\dst \hspace{-1.2cm}\times\bigg[\f{1}{64\pi z s\(\f{z s}{4}-M_{G}^{2}\)}\int_{\hat{t}_{\text{min}}}^{\hat{t}_{\text{max}}}d\hat{t}\sum_{\lambda,\lambda'}\left|\Mamp^{G_{L}G'_{L}\to V_{\lambda}V_{\lambda'}}\right|^{2}\bigg]\Bigg\}\,.
\ery
\ee\normalsize

\subsection{Numerical results}\l{sec4.4.1}
Using the master formula \eqref{l28} and the amplitudes of Section~\ref{sec4.3} we have computed the numerical total cross sections for the following processes:
\begin{align}
& pp \rightarrow W_{L}^+W_{L}^-qq \rightarrow V^+V^-qq~(\rightarrow W^+ Z~W^-Z qq), \\
& pp \rightarrow Z_{L}Z_{L}qq \rightarrow V^+V^-qq ~(\rightarrow W^+ Z~W^-Z qq), \\
& pp \rightarrow W_{L}^+W_{L}^-qq \rightarrow V^0V^0qq ~(\rightarrow
W^+W^-W^+W^- qq), \\
& pp \rightarrow W_{L}^\pm W_{L}^\pm qq \rightarrow V^\pm V^\pm qq ~(\rightarrow W^\pm Z~W^\pm Z qq), \\
& pp \rightarrow W_{L}^\pm Z_{L}  qq \rightarrow V^\pm V^0 qq~(\rightarrow W^\pm Z~W^+ W^- qq),
\end{align}
where the intermediate states indicate that we are using the EVBA. In the last step of these equations we have also indicated the final state due to the largely dominant decay modes of the heavy vectors into $WW$ or $WZ$ (see Section~\ref{sec4.8}). The cross sections are summed over all the polarizations of the heavy spin-1 fields.

\begin{figure}[htbp]
\begin{minipage}[b]{7.5cm}
   \centering
   \includegraphics[height=5.1cm]{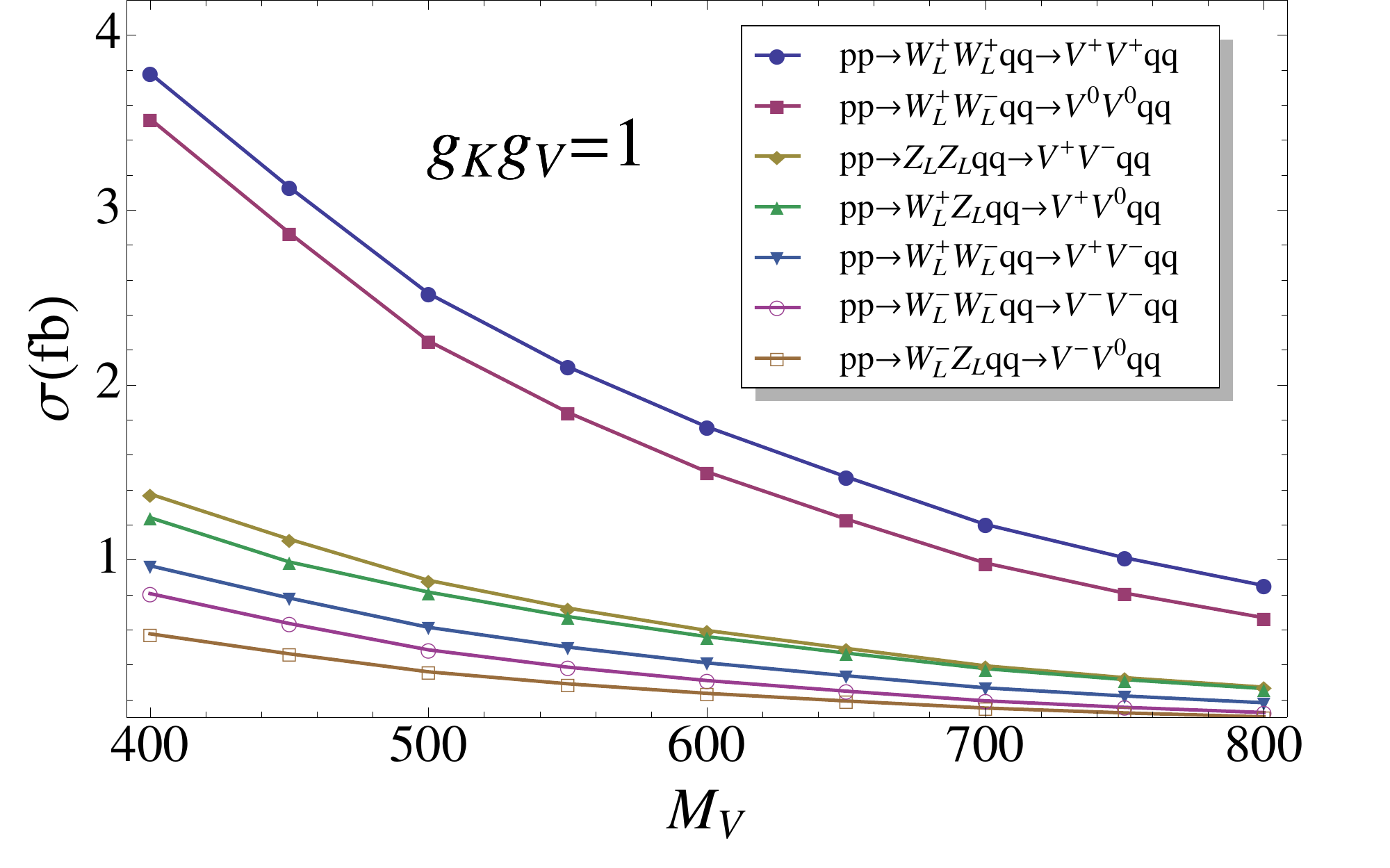}
 \end{minipage}
\ \hspace{1mm} \ 
\begin{minipage}[b]{7.5cm}
  \centering
   \includegraphics[height=5.1cm]{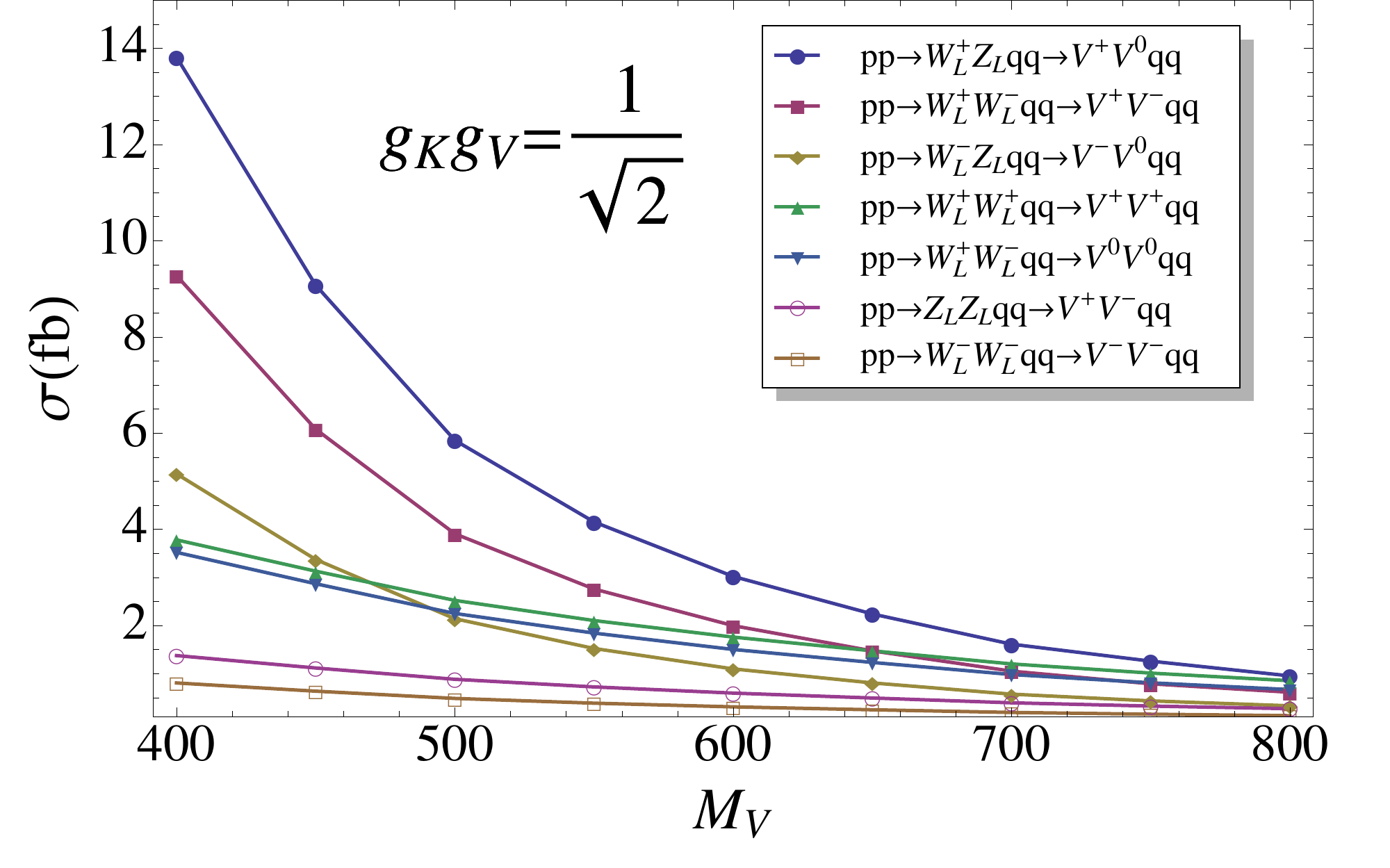}
 \end{minipage}
\caption{Total cross sections for the pair productions of heavy vectors by VBF in the EVBA in a minimal gauge model (left panel) and in a composite model (right panel) as functions of the heavy vector mass. See the text for the choice of parameters.}
\l{Fig_VBF_EVBA}
\end{figure}

These cross sections depend in general on a number of parameters. The left panel of Fig.~\ref{Fig_VBF_EVBA} shows the total cross sections for the different charge channels with all the parameters fixed as in the minimal gauge model, i.e. as in Eq.~\eqref{special}, $G_V =g_V M_V = 200$ GeV and $a=0$. As discussed in Section~\ref{TCOAHLMWAHCV} a value of $G_V$ between $150$ and $200$ is in agreement with the unitarity bound for a vector mass below $1.5$ TeV. $M_V$ is taken to range from 400 to 800 GeV. A value of $M_V$ above 800 GeV would lead to a threshold for the VBF subprocess dangerously close to the cutoff scale of the effective Lagrangian \eqref{Ltot}. We have checked that the typical c.o.m.~energy of $WW\to VV$ is on average well below 2.5 TeV, even for the highest $M_V$ that we consider.

As discussed in Sections~\ref{TCOAHLMWAHCV} and \ref{sec4.2}, the parameters of the minimal gauge model damp the high energy behavior of the different amplitudes. Not surprisingly, therefore, any deviation from them leads to significantly larger cross sections, as it may be the case already in a gauge model with more than one vector. As an example, this is shown in the right panel of Fig.~\ref{Fig_VBF_EVBA}, where all the parameters are kept as in the minimal gauge model, except for $g_K g_V = 1/\sqrt{2}$ rather than 1, having in mind a compensation of the growing amplitudes by the occurrence of (a) significantly heavier vector(s) according to the sum rule \eqref{sumrule}. Although there are only three amplitudes that depends on $g_{K}$ (see Section~\ref{sec4.1}), it is manifest from the comparison of the two plots in Fig.~\ref{Fig_VBF_EVBA} that a small deviation from the minimal gauge model results in a sharp increase of the total cross sections.
Furthermore, both in the VBF case and in the DY case, to be discussed below, it must be stressed that the deviations from the minimal gauge model are quite dependent on the choice of the parameters, with cross sections that can be even higher than those in Fig.~\ref{Fig_VBF_EVBA}. In turn, these cross sections have to be considered as indicative, given the limitations of the effective Lagrangian approach. The results obtained with the EVBA will be compared in the next section with the results obtained using the exact tree-level $2\to 4$ matrix element.

\section{Beyond the EVBA with the CalcHEP matrix element generator}\l{sec4.5}
In this section we discuss the exact total cross sections for the VBF heavy vectors pair production obtained using the CalcHEP matrix element generator \cite{Belyaev:Df-Kj9yc,Pukhov:2010p147}. It is a free package, based on the same code of CompHEP \cite{Pukhov:1999p1261}, which allows the user to define its own model of new physics and to compute partonic total and differential cross sections and widths with a maximum of $8$ ($2\to 6$ or $1\to 7$) external particles\footnote{This limitation is due to the necessary amount of memory.}. Furthermore, CalcHEP can generate partonic events that can be passed to a Monte Carlo parton shower like Pythia \cite{Sjostrand:2006p92} or Herwig \cite{Corcella:478828}, in a proper Les Houches format \cite{Skands:2003cj,Alwall:2007p58,Allanach:2008qq}. Finally, CalcHEP allows the user to completely parallelize the calculation, eventually using computer clusters. 

We implemented the model described in Chapter~\ref{chap3} (plus the strong and the flavor sectors of the SM) into CalcHEP using the Mathematica package FeynRules \cite{Christensen:2009tf,Duhr:tm}. It allows the user to automatically generate new model files for different matrix element generators simply defining fields and parameters and writing down the Lagrangian in a proper Mathematica format. CalcHEP allows the user to work both in the unitary and in the 't Hooft-Feynman gauge ($\xi=1$) and the model was implemented in such a way to offer both possibilities. The implementation in both the gauges gives the possibility to perform non-trivial cross checks. However, from the point of view of the calculations, CalcHEP is very much faster in the Feynman gauge than in the unitary gauge. This fact is due to the different complexity of the squared matrix element related to the different form of the sum over gauge boson polarizations in the two gauges. Therefore the unitary gauge is usually used in CalcHEP only to cross-check the model by computing some different processes in the two gauges and comparing the results. We checked the implemented model as just described and then we computed the exact total pair production cross sections corresponding to the diagram of Fig.~\ref{fig:VBF}.

\subsection{Numerical results}\l{sec4.5.1}

We computed the LHC production cross sections at $\sqrt{s}= 14$ TeV by VBF of two heavy vectors in the different charge configurations:
\begin{align}
& pp \rightarrow \(W^+W^-, ZZ, \gamma\gamma, \gamma Z\)qq \rightarrow V^+V^-
qq ~(\rightarrow W^+ Z~W^-Zqq), \\
& pp \rightarrow \(W^+W^-, ZZ \) qq \rightarrow V^0V^0qq ~(\rightarrow
W^+W^-W^+W^- qq), \\
& pp \rightarrow W^\pm W^\pm qq \rightarrow V^\pm V^\pm qq ~(\rightarrow
W^\pm Z~W^\pm Z qq), \\
& pp \rightarrow \(W^\pm Z, W^\pm \gamma\) qq \rightarrow V^\pm V^0 qq
~(\rightarrow W^\pm Z~W^+ W^- qq).
\end{align}
where as before, we have explicitly indicated the four gauge bosons final states. As in the EVBA the cross sections are summed over the polarizations of the heavy spin-1 fields. However, in the exact calculation of the cross sections we have reintroduced the hypercharge coupling $g^\prime \neq 0$ considering both the photon and the $Z$ and we have made standard acceptance cuts for the forward quark jets\footnote{In presence of photon fusion contributions the cuts are indispensable to avoid infrared divergences when the photon is emitted near on-shell.}, 
\be
p_T > 30~\text{GeV},~~|\eta|< 5.
\ee

\begin{figure}[htbp]
\begin{minipage}[b]{7.5cm}
   \centering
   \includegraphics[height=5.1cm]{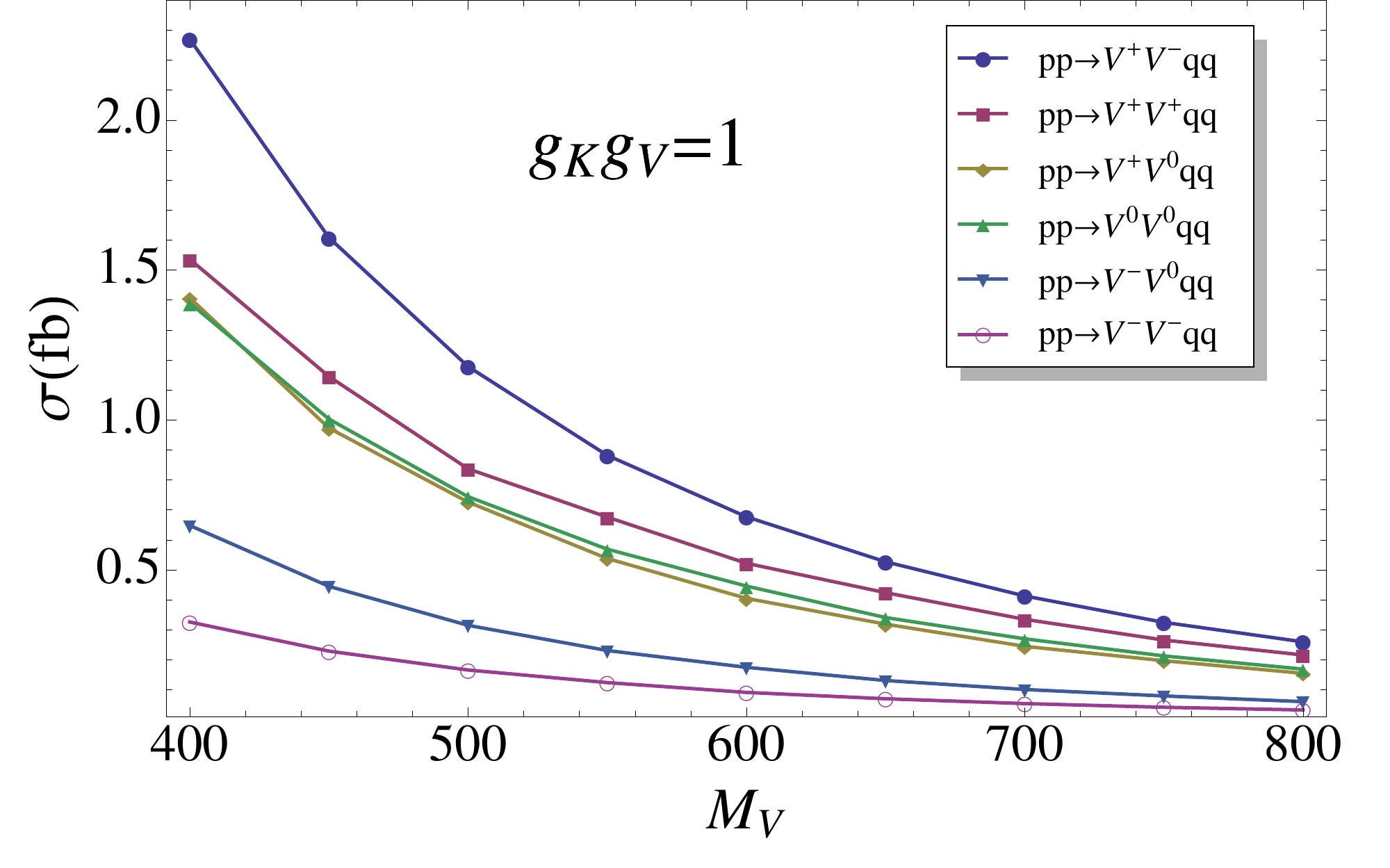}
 \end{minipage}
\ \hspace{1mm} \ 
\begin{minipage}[b]{7.5cm}
  \centering
   \includegraphics[height=5.1cm]{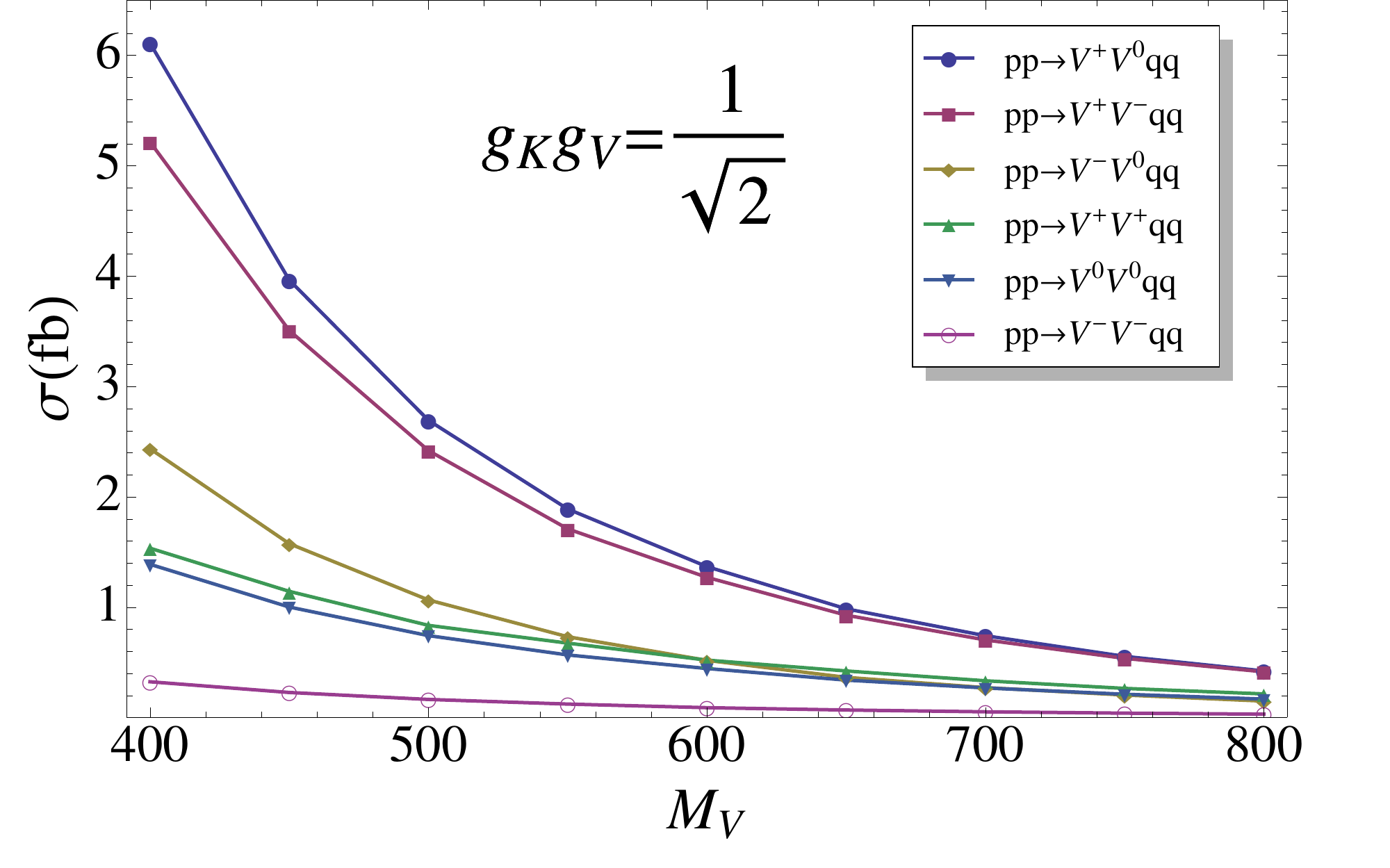}
 \end{minipage}
\caption{Total cross sections for the pair productions of heavy vectors by VBF in a minimal gauge model (left panel) and in a composite model (right panel) as functions of the heavy vector mass. See the text for the choice of parameters and acceptance cuts.}
\l{Fig_VBF}
\end{figure}

Fig.~\ref{Fig_VBF} shows the results obtained with the same choice of the parameters of Fig.~\ref{Fig_VBF_EVBA}. While being a factor of $\sim 1.5\div 2$ systematically lower, the exact results reproduce the $M_V$ dependence and the relative size of the different charge channels obtained in the previous section using the EVBA.

\section{Drell--Yan pair production}\l{sec4.6}

At the parton level there are four DY production amplitudes, related to each other by $SU(2)_{L+R}$ invariance (in the $g^\prime=0$ limit, as usual): 
\be
|\mathcal{A}(u\bar{d}\rightarrow V^+ V^0)| = |\mathcal{A}(d\bar{u}%
\rightarrow V^- V^0)|= \sqrt{2} |\mathcal{A}(u\bar{u}\rightarrow V^+ V^-)|= 
\sqrt{2}|\mathcal{A}(d\bar{d}\rightarrow V^+ V^-)|.
\ee
They receive contributions from: \newpage
\bit
\item {\bf $W (Z)$-exchange}\\
Fig.~\ref{fig6}(a): contribution proportional to $gg_{6}$ with these couplings contained in $\La^V_{\text{kin}}$ and $\La_{2V} $ of Eq.~\eqref{eq7} and Eq.~\eqref{L2V} respectively;
\item {\bf $W (Z)-V$ mixing}\\
Fig.~\ref{fig6}(b): contribution proportional to $f_V g_K$
with these couplings contained in $\La_{1V}$ and $\La_{3V}$ of Eqs.~\eqref{L1VGauge} and \eqref{L3V} respectively.
\eit
\begin{figure}[!b]
\centering
\begin{minipage}[t]{7cm}
\begin{center}
        \begin{fmffile}{DY1}
  \fmfframe(1,7)(1,7){ 
   \begin{fmfgraph*}(70,70)
          \fmfstraight
    \fmflabel{\footnotesize{$q$}}{i1}
    \fmflabel{\footnotesize{$V^{b}$}}{o1}
    \fmflabel{\footnotesize{$q$}}{i2}
    \fmflabel{\footnotesize{$V^{a}$}}{o2}
    \fmfbottom{i1,i2}
    \fmftop{o1,o2}
    \fmf{fermion}{i1,v1}
    \fmf{fermion}{v1,i2}
    \fmf{dbl_plain_arrow}{v2,o1}
    \fmf{dbl_plain_arrow}{v2,o2}
    \fmf{photon,label={\footnotesize{$W$}}}{v1,v2}
    \fmf{phantom_arrow}{v1,v2}
    \end{fmfgraph*}
        }  \\\vspace{-4mm}  \footnotesize{{\bf (a)}} \l{DY1}%
      \end{fmffile}
\end{center}
\end{minipage}
\ \hspace{5mm} \
\begin{minipage}[t]{7cm}
\begin{center}
        \begin{fmffile}{DY2}
  \fmfframe(1,7)(1,7){ 
   \begin{fmfgraph*}(70,70)
          \fmfstraight
    \fmflabel{\footnotesize{$q$}}{i1}
    \fmflabel{\footnotesize{$V^{b}$}}{o1}
    \fmflabel{\footnotesize{$q$}}{i2}
    \fmflabel{\footnotesize{$V^{a}$}}{o2}
    \fmfbottom{i1,i2}
    \fmftop{o1,o2}
    \fmf{fermion}{i1,v1}
    \fmf{fermion}{v1,i2}
    \fmf{dbl_plain_arrow}{v2,o1}
    \fmf{dbl_plain_arrow}{v2,o2}
    \fmf{dbl_plain_arrow,label={\footnotesize{$V$}}}{v3,v2}
    \fmf{photon,tension=.1,label={\footnotesize{$W$}}}{v3,v1}
    \fmf{phantom_arrow}{v1,v3}
    \end{fmfgraph*}
        } \\\vspace{-4mm}  \footnotesize{{\bf (b)}} \l{DY2}%
      \end{fmffile}
\end{center}
\end{minipage}
\caption{The two Feynman diagrams contributing to the $qq \rightarrow V^{a} V^{b}$ DY pair production: (a) is the contribution of $W$ exchange and (b) is the contribution of $WV$ mixing generated by the Lagrangian \eqref{L1VGauge}.}\l{fig6}
\end{figure}
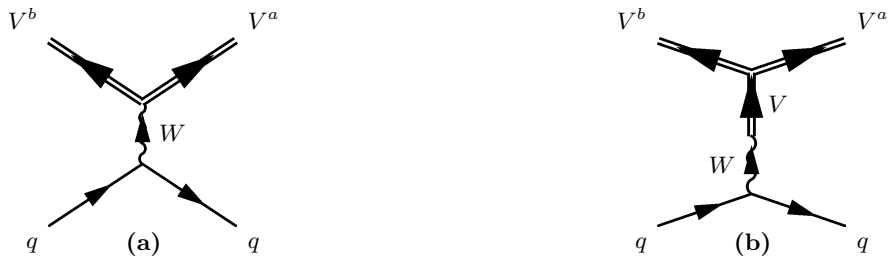
Their modulus squared, summed over the polarizations of the final-state vectors and averaged over color and polarization of the initial fermions, can be written as 
\be
\<|\mathcal{A}(u\bar{d}\rightarrow V^+ V^0)|^2\> = \frac{g^4}{1536 M_V^6 s^2 (s-M_V^2)^2} F(s, t-u, M_V^2),  \l{DYsquared}
\ee
with $F$ organized in different powers of $s$: 
\be
F(s, t-u, M_V^2) = F^{(6)}(s, t-u, M_V^2) + F^{(5)}(s, t-u, M_V^2) + F^{(\leq 4)}(s, t-u, M_V^2)\,,
\ee
where 
\bes
\begin{align}
& F^{(6)}= (g_K f_V - 4 g_6)^2 M_V^2 s^4[s^2-(t-u)^2], \\
& F^{\left( 5\right) } = 4M_{V}^{4}s^{3}\left\{ \left(g_{K}f_{V}-4g_{6}\right) {}^{2}\left[ 2s^{2}+(t-u)^{2}\right] \right.  \notag
\\
&\quad +\left. \left( g_{K}f_{V}-4g_{6}\right) {}\left[ 2\left(7g_{6}-3\right) s^{2}+2\left( g_{6}-1\right) (t-u)^{2}\right] +2\left(1-2g_{6}\right) ^{2}\left[ s^{2}+(t-u)^{2}\right] \right\},  \l{z7a} \\
& F^{(\leq 4)} = 4M_{V}^{6}\left\{ -3s^{2}f_{V}^{2}g_{K}^{2} \left[3s^{2}+(t-u)^{2}+4M_{V}^{2}s\right] -4M_{V}^{4}\left[ \left( 8g_{6}\left(g_{6}+2\right) -25\right) s^{2}+3(t-u)^{2}\right] \right.  \notag \\
&\quad+\left. 2f_{V}g_{K}s\left[ s\left\{ \left( 26g_{6}+9\right)s^{2}+\left( 2g_{6}+7\right) (t-u)^{2}\right\} -6M_{V}^{2}\left[ \left(4g_{6}-3\right) s^{2}+(t-u)^{2}\right] -24sM_{V}^{4}\right] \right.  \notag
\\
&\quad+\left. 2M_{V}^{2}s\left[ \left( 28g_{6}^{2}+9\left( 8g_{6}-3\right)\right) s^{2}+\left( 4g_{6}^{2}+13\right) (t-u)^{2}\right] \right.  \notag \\
&\quad-\left. 4s^{2}\left[ 3g_{6}\left( g_{6}+8\right) s^{2}+\left(5g_{6}^{2}+4\right) (t-u)^{2}\right] -48M_{V}^{6}s\right\}.  \l{z7b}
\end{align}
\ees
$F^{(5)}$ is written in such a way as to make evident what controls its high-energy behavior after the dominant $F^{(6)}$ is set to zero by taking $g_K f_V = 4 g_6$. In general, these amplitudes squared grow at high energy as $(s/M_V^2)^2$, which is turned to a constant behavior for 
\be
g_K f_V= 2, ~~ g_6 = \frac{1}{2}.
\ee
As usual, these values of the parameters are predicted in a minimal gauge model (see Eq.~\eqref{gaugeparam}).
In this special case the function $F$ in Eq.~\eqref{DYsquared} acquires the
form 
\be
F^{\text{gauge}} =4M_{V}^{6}\left\{ s^{2}\left[ s^{2}-\left( t-u\right) ^{2}%
\right] +4M_{V}^{2}s\left[ 2s^{2}+(t-u)^{2}\right] -12M_{V}^{4}\left[
3s^{2}+(t-u)^{2}\right] -48M_{V}^{6}s\right\}\,.  \l{z7}
\ee

\subsection{Numerical results}\l{sec4.7}

The DY process is an additional source of $V$ pair production at the LHC.
From the elementary parton-level squared amplitudes $q\bar q\to V^+V^-$ and $q_i\bar q_j\to V^\pm V^0$ of the previous section, the physical cross sections for the different charge channels 
\begin{align}
& pp \rightarrow V^+V^- , \\
& pp \rightarrow V^\pm V^0
\end{align}
are readily computed. In general, the cross sections depend in this case on 3 parameters other than $M_V$: $f_V, g_K$ and $g_6$.

\begin{figure}[tbp]
\begin{minipage}[b]{7.5cm}
   \centering
   \includegraphics[height=5.1cm]{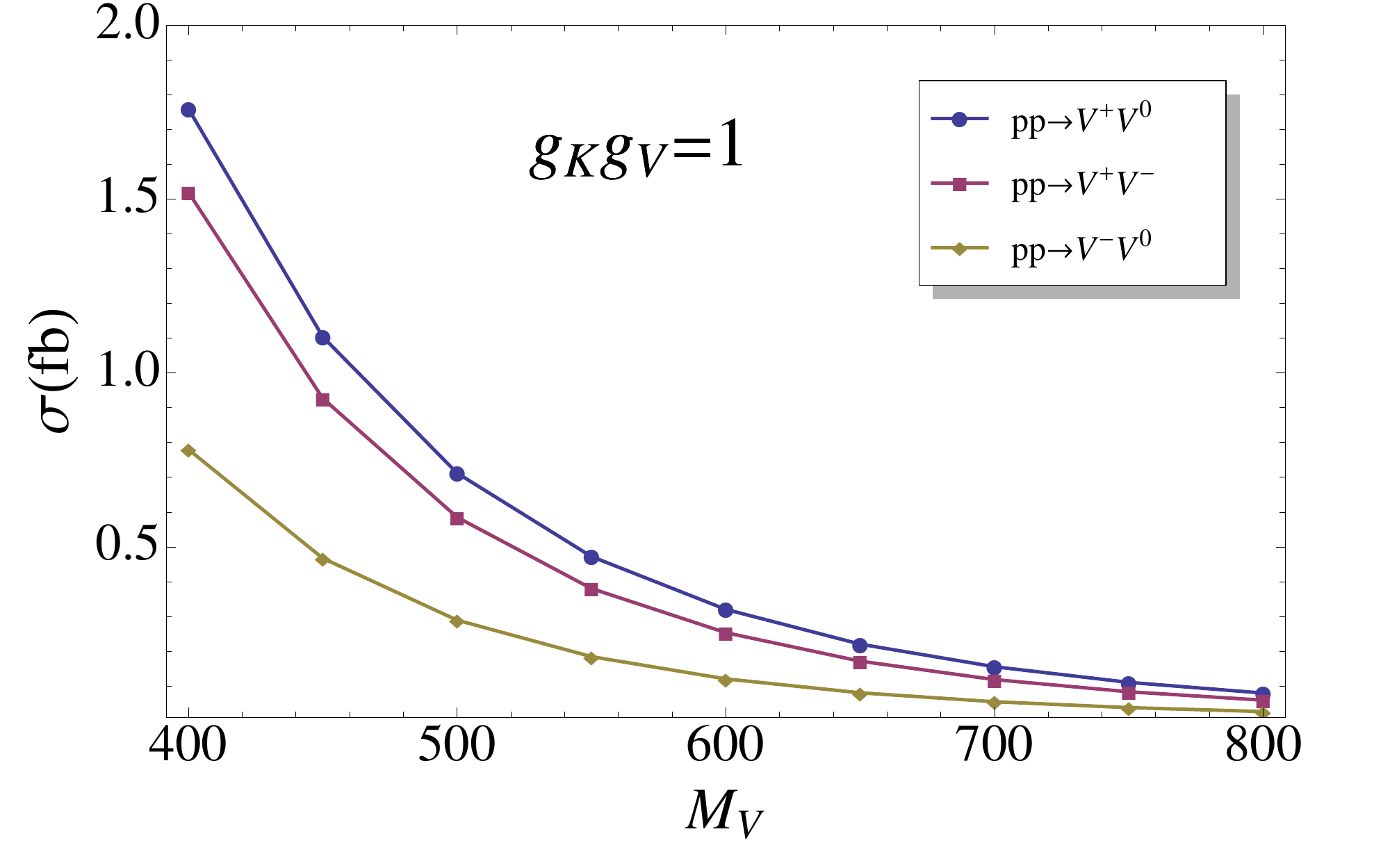}
 \end{minipage}
\ \hspace{1mm} \ 
\begin{minipage}[b]{7.5cm}
  \centering
   \includegraphics[height=5.1cm]{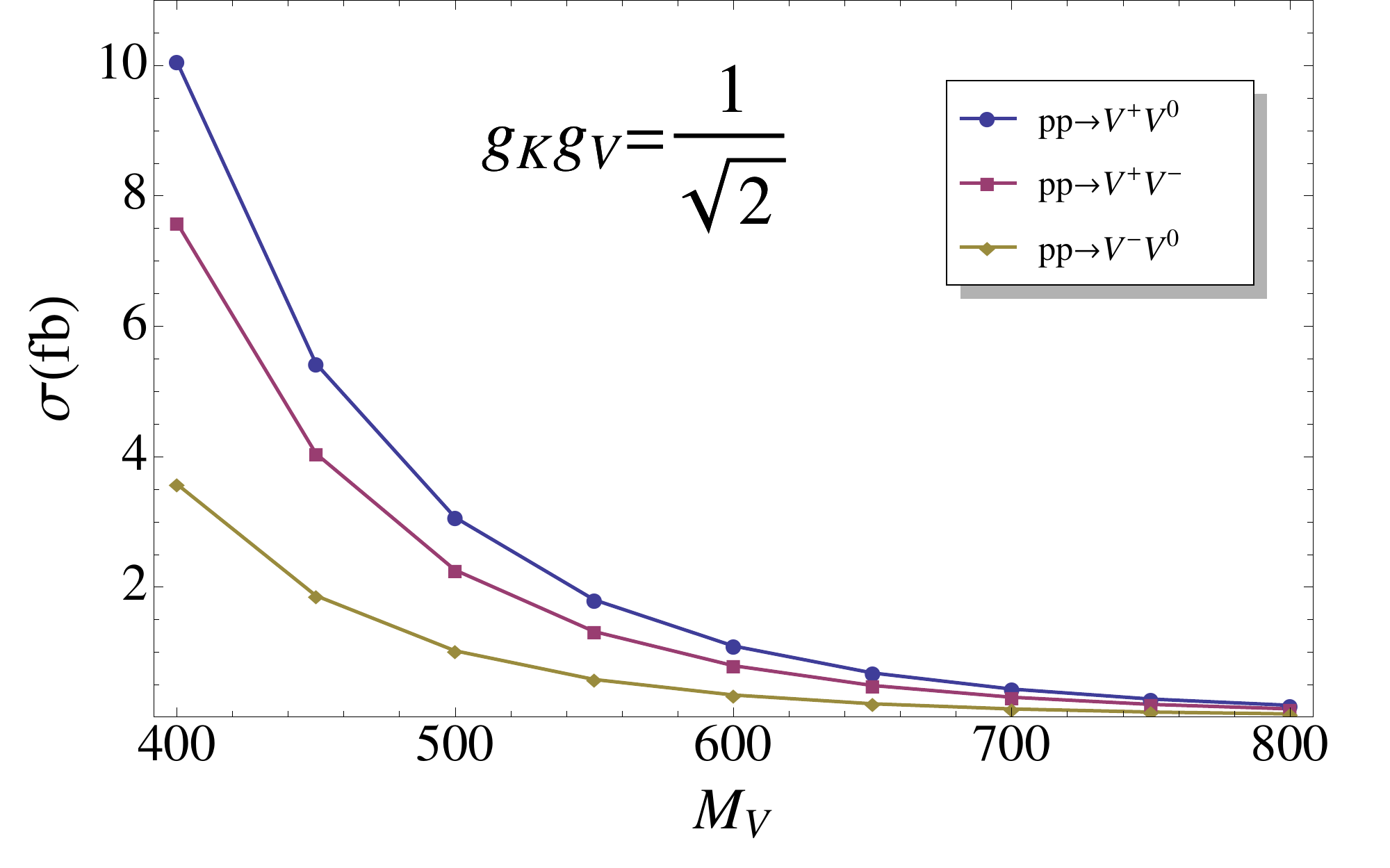}
 \end{minipage}
\caption{Total cross sections for the pair productions of heavy vectors via DY $q\bar{q}$ annihilation in a minimal gauge model (left panel) and a composite model (right panel) as functions of the heavy vector mass. See the text for the choice of parameters.}
\l{Fig_DY}
\end{figure}

As for the VBF, we show in Fig.~\ref{Fig_DY} the three cross sections for the values taken by the parameters in the minimal gauge model (left panel), $f_V g_K =2, g_6=1/2$ and $F_V = f_V M_V = 400$ GeV (corresponding to $f_V = 2 g_V$ and $G_V = g_V M_V = 200$ GeV as in Fig.~\ref{Fig_VBF}) and in a non minimal model (right panel), $f_V g_K =\sqrt{2}, g_6=1/2$ and still $F_V = f_V M_V = 400$ GeV.

\section{Same-sign di-lepton and tri-lepton events}\l{sec4.8}

In the limit $g'\approx 0$ the width of the heavy vectors into two fermions is related to the width into two gauge bosons by the relation \cite{Barbieri:2008p1580}
\be
\f{\Gamma\(V^{0}\to \bar{\psi}\psi\)}{\Gamma\(V^{0}\to W_{L}^{+}W_{L}^{-}\)}\approx \f{4M_{W}^{4}}{M_{V}^{4}}\,,
\ee
so that the Branching Ratios (BRs) of the new vectors are given by
\be\l{BR1}
\text{BR}\(V^{+}\to W_{L}^{+}Z_{L}\)\approx \text{BR}\(V^{0}\to W_{L}^{+}W_{L}^{-}\)\approx 1\,.
\ee
Therefore, after the decay of the composite vectors, each $VV$-production channel, either from VBF or from DY, leads to final states containing 2 $W$'s and 2 $Z$'s, from $V^+ V^-$ and $V^\pm V^\pm$, 3 $W $'s and 1 $Z$, from $V^+ V^0$, or 4 $W$'s from $V^0 V^0$\footnote{Or in fact multi-top events (see Section~\ref{sec3.1}).}. In fact, all final states, except for $V^+ V^-$, contain at least a pair of equal sign $W$'s, i.e., after $W \rightarrow e \nu, \mu \nu$, a pair of same-sign leptons. In many cases there are at least 3 $W$'s, i.e. also 3 leptons.

\begin{table}[ht!]
\begin{center}
{\small 
\begin{tabular}{|c|c|c|}
\hline
& di-leptons & tri-leptons \\ \hline
VBF (MGM) & 16 & 3 \\ \hline
DY (MGM) & 5 & 1 \\ \hline
VBF (comp) & 28 & 6 \\ \hline
DY (comp) & 18 & 4 \\ \hline
\end{tabular}
}
\end{center}
\caption{Total number of events with at least two same-sign leptons or three leptons ($e$ or $\mu$ from $W$ decays) from VBF or DY at the LHC for $\sqrt{s}=14$ TeV and $L=\int \La dt = 100$~fb$^{-1}$ in the minimal gauge model (MGM) or in a composite model (comp) with the parameters as in Figs.~\ref{Fig_VBF}-\ref{Fig_DY} and $M_V= 500$ GeV.}
\l{tabcut}
\end{table}

\begin{table}[ht!]
\begin{center}
{\small 
\begin{tabular}{|c|c|c|}
\hline
& di-leptons($\%$) & tri-leptons($\%$) \\ \hline
$V^0 V^0 $ & 8.9 & 3.2 \\ \hline
$V^\pm V^\pm $ & 4.5 & - \\ \hline
$V^\pm V^0$ & 4.5 & 1.0 \\ \hline
\end{tabular}
}
\end{center}
\caption{Cumulative BRs for at least two same-sign di-leptons or tri-leptons ($e$ or $\mu$) in the $W$ decays from two vectors in the given charge configuration.}
\l{BRs}
\end{table}

At the LHC with an integrated luminosity of 100 fb$^{-1}$ and $\sqrt{s}$ = 14 TeV, putting together all the different charge configurations, one obtains from $W \rightarrow e \nu, \mu \nu$ decays the number of same-sign di-lepton and tri-lepton events given in Table~\ref{tabcut} for $M_V$ = 500 GeV. The other parameters are fixed as in the Minimal Gauge Model (and labelled `MGM') or as in the right panels of Figs.~\ref{Fig_VBF} and \ref {Fig_DY} for VBF and for DY in the previous two sections (and labelled `comp'). These numbers of events are based on the cross sections in Figs.~\ref{Fig_VBF}-\ref{Fig_DY} and on the BRs for the various charge channels listed in Table~\ref{BRs}. The numbers of events for different values of $M_V$ are also easily obtained. As already noticed, depending on the parameters, the number of events in the non minimal case could also be significantly higher. No attempt is made, at this stage, to compare the signal with the background from SM sources. To see if a signal can be observed a careful analysis will be required, with a high cut on the scalar sum, $H_t$, of all the transverse momenta and of the missing energy in each event probably playing a crucial role. The use of the leptonic decays of the $Z$ might also be important.

\chapter{Associated scalar-vector production at the LHC}\l{chap5}
\setlength{\epigraphwidth}{.4\textwidth}     
\vspace{-2mm}\epigraph{It is tragic, but now, we have the string theorists, thousands of
them, that also dream of explaining all the features of nature. They
just celebrated the 20th anniversary of superstring theory. So when
one person spends 30 years, itÕs a waste, but when thousands waste 20
years in modern day, they celebrate with champagne. I find that
curious.}%
              {\textsc{Sheldon Lee Glashow}}
\vspace{3mm}     
\epigraph{Common sense is the collection of prejudices acquired by age eighteen.}%
              {\textsc{Albert Einstein}}
\vspace{7mm}
\lettrine{T}{he} associated $hV$ production is the most peculiar pair production in the model of EWSB described in Chapter~\ref{chap3}. It is the only process sensitive to the coupling $k_{1}$ of the $hV^{2}$ operator in the Lagrangian \eqref{lvhinteraction}\footnote{As we have already noticed in Chapter~\ref{chap4}, also the $V$ pair production is sensitive to this coupling. However, the interference with the other channels is always negative, making it very difficult to observe the process and to access the corresponding coupling.}. The associated $hV$ production can be generated both by VBF and by DY $q\bar{q}$ annihilation\footnote{The gluon-gluon fusion associated production was also considered in Ref.~\cite{Hernandez:2010qp}.}. In this chapter we study the total cross sections for the associated production by VBF and by DY annihilation.

The chapter is organized al follows. In Section \ref{sec5.1} we study the asymptotic amplitude for the $W_{L}W_{L}\to V_{L}h$ process, while in Section \ref{sec5.2} we give numerical results for the VBF and the DY $hV$ associated production total cross sections. Finally in Section \ref{sec5.3} we discuss the expected number of same-sign di-lepton and tri-lepton events.

\section{$W_{L} W_{L} \rightarrow V_{L}h $ asymptotic amplitude}\l{sec5.1}
Using the Equivalence Theorem the amplitude for the process $W_{L} W_{L} \rightarrow V_{L}h $ can be written as
\be\l{WWVhamp1}
\Mamp\(W^{a}_{L}W^{b}_{L}\to V^{c}_{L}h\) \approx -\Mamp\(\pi^{a}\pi^{b}\to V^{c}_{L}h\)
\ee
As in the previous chapter, to simplify the explicit formulae we take the limit $g'=0$ so that the $SU\(2\)_{L+R}$ invariance is preserved by the scattering amplitudes. This implies
\be
\Mamp\(W_{L}^{a}W_{L}^{b}\to V^{c}_{L}h \)=\Mamp\(s,t,u\)\epsilon^{abc}\,.
\ee
These amplitudes receive contributions from the Feynman diagrams of Fig.~\ref{fig:ppVhdiag}, i.e.:
\bit
\item {\bf $s$-channel $V$ exchange}\\
Fig.~\ref{fig:ppVhdiag}(a): contribution proportional to $g_{V}k_{1}$ with these couplings contained in $\La_{1V-\text{Goldstone}}$ and $\La_{h-V}$ of Eq.~\eqref{L1VGoldstone} and Eq.~\eqref{lvhinteraction} respectively;
\item {\bf $t$ and $u$ channels $\pi$ exchange}\\
Fig.~\ref{fig:ppVhdiag}(b)-(c): contributions proportional to $g_{V} a$ with these couplings contained in $\La_{1V-\text{Goldstone}}$ and $\La_{h-\text{int}}$ of Eq.~\eqref{L1VGoldstone} and Eq.~\eqref{Hint} respectively.
\eit
\begin{figure}[!htb]
\centering
\begin{minipage}[t]{3.5cm}
\begin{center}
        \begin{fmffile}{ppVfi1}
  \fmfframe(1,7)(1,7){ 
   \begin{fmfgraph*}(70,70)
          \fmfstraight
    \fmfbottom{i1,i2}
    \fmftop{o1,o2}
          \fmflabel{\footnotesize{$h$}}{o2}
          \fmflabel{\footnotesize{$\pi^{a}$}}{i1}
          \fmflabel{\footnotesize{$V^{c}$}}{o1}
          \fmflabel{\footnotesize{$\pi^{b}$}}{i2}   
    \fmf{scalar}{i1,v1}
    \fmf{scalar}{i2,v1}
    \fmf{dbl_plain_arrow}{v2,o1}
    \fmf{dbl_dashes_arrow}{v2,o2}
    \fmf{dbl_plain_arrow,label={\footnotesize{$V$}}}{v1,v2}
    \end{fmfgraph*}
        }   \\\vspace{-4mm}  \footnotesize{{\bf (a)}}\l{ppVfi4}%
      \end{fmffile}
\end{center}
\end{minipage}
\ \hspace{5mm} \
\begin{minipage}[t]{3.5cm}
\begin{center}
        \begin{fmffile}{ppVfi2}
  \fmfframe(1,7)(1,7){ 
   \begin{fmfgraph*}(70,70)
          \fmfstraight
    \fmfleft{i1,i2}
    \fmfright{o1,o2}
    \fmflabel{\footnotesize{$\pi^{a}$}}{i1}
    \fmflabel{\footnotesize{$\pi^{b}$}}{o1}
    \fmflabel{\footnotesize{$h$}}{o2}
    \fmflabel{\footnotesize{$V^{c}$}}{i2}
    \fmf{scalar}{i1,v1}
    \fmf{dbl_plain_arrow}{v1,i2}
    \fmf{scalar}{o1,v2}
    \fmf{dbl_dashes_arrow}{v2,o2}
    \fmf{scalar,l.side=right,label={\footnotesize{$\pi$}}}{v1,v2}
    \end{fmfgraph*}
        }   \\\vspace{-4mm}  \footnotesize{{\bf (b)}} \l{ppfifi2}%
      \end{fmffile}
\end{center}
\end{minipage}
\ \hspace{5mm} \
\begin{minipage}[t]{3.5cm}
\begin{center}
        \begin{fmffile}{ppVfi3}
  \fmfframe(1,7)(1,7){ 
   \begin{fmfgraph*}(70,70)
          \fmfstraight
    \fmfleft{i1,i2}
    \fmfright{o1,o2}
    \fmflabel{\footnotesize{$\pi^{a}$}}{i1}
    \fmflabel{\footnotesize{$\pi^{b}$}}{o1}
    \fmflabel{\footnotesize{$V^{c}$}}{o2}
    \fmflabel{\footnotesize{$h$}}{i2}
    \fmf{scalar}{i1,v1}
    \fmf{phantom}{v1,i2}
    \fmf{scalar}{o1,v2}
    \fmf{phantom}{v2,o2}
    \fmf{scalar,l.side=right,label={\footnotesize{$\pi$}}}{v1,v2}
    \fmffreeze
    \fmf{dbl_plain_arrow}{v1,o2}
    \fmf{dbl_dashes_arrow}{v2,i2}
    \end{fmfgraph*}
        }  \\\vspace{-4mm}  \footnotesize{{\bf (c)}} \l{ppfifi3}\vspace{4mm}%
      \end{fmffile}
\end{center}
\end{minipage}
\caption{The three Feynman diagrams contributing to the $W_{L} W_{L} \rightarrow V_{L}\phi$ scattering: (a) represents the $s$-channel $V$ exchange contribution; (b) and (c) are respectively the $t$ and $u$ channels of the $\pi$ exchange contribution.}\l{fig:ppVhdiag}
\end{figure}
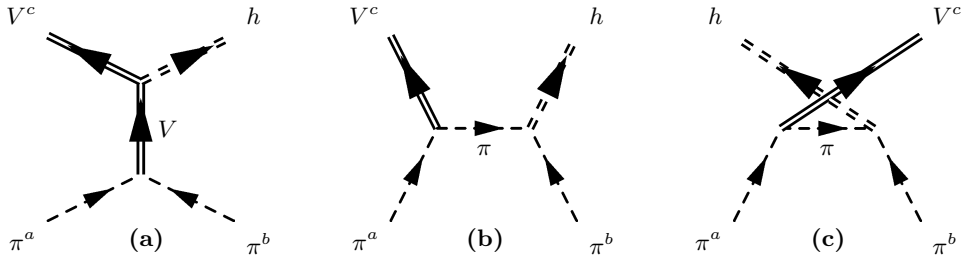
The amplitude is given by
\be\l{pipiVhL}
\bry{lll}
\dst \Mamp\(s,t,u\)
&=&\dst \f{i\(u-t\)}{2v\sqrt{\(M_{V}^{2}+m_{h}^{2}-s\)^{2}-4m_{h}^{2}M_{V}^{2}}}\Bigg[\f{k_{1}}{4g_{V}M_{V}}\f{s}{s-M_{V}^{2}}\(m_{h}^{2}-M_{V}^{2}-s\)\\
&&\dst +\f{2ag_{V}M_{V}}{v^{2}tu}\Big[m_{h}^{2}M_{V}^{2}\(m_{h}^{2}-M_{V}^{2}+s\)+tu\(M_{V}^{2}-m_{h}^{2}+s\)\Big]\Bigg]\,.
\ery
\ee
In the very high energy limit in which $s\gg M_{V}^{2}\gg m_{h}^{2}$ it can be rewritten in the form
\be\l{asymptamplitudes5}
\bry{lll} 
\dst \Mamp\(s,t,u\) &\approx&\displaystyle \f{ig_{V}M_{V}\(t-u\)}{v}\Bigg[\f{a}{v^{2}}-\f{k_{1}}{8g_{V}^{2}M_{V}^{2}}\Bigg]\vspace{2mm}\\
&&\displaystyle+\f{ig_{V}M_{V}\(t-u\)}{vs}\Bigg[\f{a}{v^{2}}\(M_{V}^{2}-m_{h}^{2}\)+\f{k_{1}}{8g_{V}^{2}M_{V}^{2}}\(m_{h}^{2}-2M_{V}^{2}\)\Bigg]\,.
\end{array}
\ee
For generic values of the parameters this amplitude grows, as usual, with the c.o.m.~energy squared $s$. On the other hand, with the parameters as in Eqs.~\eqref{gaugeparam} and \eqref{gaugeparameters} (or equivalently as in Eq.~\eqref{special}) the amplitude reduces to
\be\l{asymptamplitudesgauge5}
\Mamp\(s,t,u\) \approx \f{iM_{V}^{2}\(u-t\)}{4v^{2}s}+O\(\f{m_{h}^{2}}{v^{2}}\)\,.
\ee
This amplitudes is exactly the same one would obtain in the $SU\(2\)_{L}\times SU\(2\)_{C}\times U\(1\)_{Y}$ gauge model spontaneously broken by two Higgs doublets that we have discussed in Section~\ref{sec3.5} and, as expected, has a constant asymptotic behavior. 
In the next section we discuss the total cross sections for the associated $hV$ production by VBF and by DY.

\section{$hV$ associated production total cross sections}\l{sec5.2}
There are three possible final states for the associated production, corresponding to the three charge states of the $V$: $hV^{-}$, $hV^{0}$ and $hV^{+}$. According to the constraint \eqref{agvrelation}, which ensures that the strongest unitarity constraint is satisfied by the scalar-vector system up to $\Lambda\approx 3$ TeV in a wide region of the parameter space, we can compute the total cross sections for some reference values of the independent parameters, that we choose to be $G_{V}$ and $d$. Some values of the total cross sections at the LHC for $\sqrt{s}=14$ TeV for different values of the parameters and for a scalar mass $m_{h}=180$ GeV are listed in Tables~\ref{table1}, \ref{table2} and \ref{table3} for the production of $hV^{-}$, $hV^{0}$ and $hV^{+}$ respectively. 
\begin{table}[htb!]
\begin{minipage}[b]{7.5cm}
\centering
\resizebox{6cm}{!} {
\begin{tabular}[c]{|c|c|c|c|c|}
		\hline
		$G_{V}$ &  $a$ &  $k_{1}$ & VBF (fb) & DY (fb) \\
		\hline
		$\sqrt{5} v/4$  &$1/4$ & $0$ & $0.05$& $0$ \\
		\hline
		$\sqrt{5} v/4$  &$1/4$ & $1$ & $0.09$ & $3.31$ \\
		\hline
		$\sqrt{5} v/4$  &$1/4$ & $2$ & $0.62$& $13.24$ \\
		\hline
		$v/2$  & $1/2$ & $0$ & $0.15$ & $0$ \\
		\hline
		$v/2$  & $1/2$ & $1$ & $0.05$ & $4.14$ \\
		\hline
		$v/2$  & $1/2$ & $2$ & $0.56$ & $16.56$ \\
		\hline
		$v/\sqrt{6}$  & $1/\sqrt{2}$ & $0$ & $0.20$  & $0$ \\
		\hline
		$v/\sqrt{6}$  & $1/\sqrt{2}$ & $1$ & $0.08$  & $6.20$ \\
		\hline
		$v/\sqrt{6}$  & $1/\sqrt{2}$ & $2$ & $0.89$  & $24.80$ \\
		\hline
\end{tabular}}
 \end{minipage}
\ \hspace{2mm} \hspace{3mm} \ 
\begin{minipage}[b]{7.5cm}
\centering
\resizebox{6cm}{!} {
\begin{tabular}[c]{|c|c|c|c|c|}
		\hline
		$G_{V}$ &  $a$ &  $k_{1}$ & VBF (fb) & DY (fb) \\
		\hline
		$\sqrt{5} v/4$  &$1/4$ & $0$ & $0.02$ & $0$ \\
		\hline
		$\sqrt{5} v/4$  &$1/4$ & $1$ & $0.08$ & $1.23$ \\
		\hline
		$\sqrt{5} v/4$  &$1/4$ & $2$ & $0.49$ & $4.92$ \\
		\hline
		$v/2$  & $1/2$ & $0$ & $0.07$ & $0$ \\
		\hline
		$v/2$  & $1/2$ & $1$ & $0.06$ & $1.54$ \\
		\hline
		$v/2$  & $1/2$ & $2$ & $0.48$ & $6.16$ \\
		\hline
		$v/\sqrt{6}$  & $1/\sqrt{2}$ & $0$ & $0.09$  & $0$ \\
		\hline
		$v/\sqrt{6}$  & $1/\sqrt{2}$ & $1$ & $0.09$  & $2.30$ \\
		\hline
		$v/\sqrt{6}$  & $1/\sqrt{2}$ & $2$ & $0.75$  & $9.20$ \\
		\hline
\end{tabular}}
 \end{minipage}
\caption{Total cross sections for the associated production of $hV^{-}$ final state by VBF and DY at the LHC for $\sqrt{s}=14$ TeV as functions of the different parameters for $M_{V}=700$ GeV (left panel) and $M_{V}=1$ TeV (right panel). The parameter $a$ is fixed by the value of $G_{V}$ (and vice versa) according to Eq.~\eqref{agvrelation}.}\l{table1}
\end{table}
We have chosen $m_{h}=180$ GeV to maximize both the total cross sections and the BR for $h\to W^{+}W^{-}$. In this case signals of the associated productions can appear in the multi-lepton channels. In particular, if the final state contains at least a pair of equal sign $W$'s there can be signals in the same-sign di-lepton and tri-lepton final states from W decays that are much simpler to be separated from the background than those ones corresponding to the hadronic final states.
\begin{table}[htb!]
\begin{minipage}[b]{7.5cm}
\centering
\resizebox{6cm}{!} {
\begin{tabular}[c]{|c|c|c|c|c|}
		\hline
		$G_{V}$  &$a$ &  $k_{1}$ & VBF(fb) & DY(fb) \\
		\hline
		$\sqrt{5} v/4$ & $1/4$ & $0$ & $0.08$& $0$ \\
		\hline
		$\sqrt{5} v/4$ & $1/4$ & $1$ & $0.14$ & $6.14$ \\
		\hline
		$\sqrt{5} v/4$ & $1/4$ & $2$ & $0.99$& $24.56$ \\
		\hline
		$v/2$  & $1/2$ & $0$ & $0.24$  & $0$ \\
		\hline
		$v/2$ & $1/2$ & $1$ & $0.08$ & $7.67$ \\
		\hline
		$v/2$ & $1/2$ & $2$ & $0.90$  & $30.68$ \\
		\hline
		$v/\sqrt{6}$ & $1/\sqrt{2}$ & $0$ & $0.32$  & $0$ \\
		\hline
		$v/\sqrt{6}$ & $1/\sqrt{2}$ & $1$ & $0.13$ & $11.51$ \\
		\hline
		$v/\sqrt{6}$ & $1/\sqrt{2}$ & $2$ & $1.42$ & $46.04$ \\
		\hline
\end{tabular}}
 \end{minipage}
\ \hspace{2mm} \hspace{3mm} \ 
\begin{minipage}[b]{7.5cm}
\centering
\resizebox{6cm}{!} {
\begin{tabular}[c]{|c|c|c|c|c|}
		\hline
		$G_{V}$  &$a$ &  $k_{1}$ & VBF(fb) & DY(fb) \\
		\hline
		$\sqrt{5} v/4$ & $1/4$ & $0$ & $0.04$ & $0$ \\
		\hline
		$\sqrt{5} v/4$ & $1/4$ & $1$ & $0.13$ & $2.43$ \\
		\hline
		$\sqrt{5} v/4$ & $1/4$ & $2$ & $0.79$ & $9.74$ \\
		\hline
		$v/2$  & $1/2$ & $0$ & $0.11$  & $0$ \\
		\hline
		$v/2$ & $1/2$ & $1$ & $0.09$ & $3.04$ \\
		\hline
		$v/2$ & $1/2$ & $2$ & $0.78$ & $12.16$ \\
		\hline
		$v/\sqrt{6}$ & $1/\sqrt{2}$ & $0$ & $0.15$  & $0$ \\
		\hline
		$v/\sqrt{6}$ & $1/\sqrt{2}$ & $1$ & $0.15$ & $4.57$ \\
		\hline
		$v/\sqrt{6}$ & $1/\sqrt{2}$ & $2$ & $1.22$ & $18.28$ \\
		\hline
\end{tabular}}
 \end{minipage}
\caption{Total cross sections for the associated production of $hV^{0}$ final state by VBF and DY at the LHC for $\sqrt{s}=14$ TeV as functions of the different parameters for $M_{V}=700$ GeV (left panel) and $M_{V}=1$ TeV (right panel). The parameter $a$ is fixed by the value of $G_{V}$ (and vice versa) according to Eq.~\eqref{agvrelation}.}\l{table2}
\end{table}
Obviously different values of $m_{h}$ are possible: in that case the detection of a signal can be disfavored by the large BR for $h\to b\bar{b}$ for $m_{h}<2M_{W}$, by the large BR for $h\to ZZ$ for $m_{h}>2M_{Z}$ and by the small cross sections for $m_{h}\apprge 250$ GeV (see Fig.~\ref{fig:hVDY}).
\begin{table}[htb!]
\begin{minipage}[b]{7.5cm}
\centering
\resizebox{6cm}{!} {
\begin{tabular}[c]{|c|c|c|c|c|}
		\hline
		$G_{V}$  & $a$ &  $k_{1}$ & VBF(fb) & DY(fb) \\
		\hline
		$\sqrt{5} v/4$ & $1/4$ & $0$ & $0.10$ & $0$ \\
		\hline
		$\sqrt{5} v/4$ & $1/4$ & $1$ & $0.18$  & $7.30$ \\
		\hline
		$\sqrt{5} v/4$ & $1/4$ & $2$ & $1.28$ & $29.20$ \\
		\hline
		$v/2$ & $1/2$ & $0$ & $0.33$  & $0$ \\
		\hline
		$v/2$ & $1/2$ & $1$ & $0.10$  & $9.12$ \\
		\hline
		$v/2$ & $1/2$ & $2$ & $1.15$  & $36.48$ \\
		\hline
		$v/\sqrt{6}$ & $1/\sqrt{2}$ & $0$ & $0.43$  & $0$ \\
		\hline
		$v/\sqrt{6}$ & $1/\sqrt{2}$ & $1$ & $0.17$  & $13.68$ \\
		\hline
		$v/\sqrt{6}$ & $1/\sqrt{2}$ & $2$ & $1.82$  & $54.72$ \\
		\hline
\end{tabular}}
 \end{minipage}
\ \hspace{2mm} \hspace{3mm} \ 
\begin{minipage}[b]{7.5cm}
\centering
\resizebox{6cm}{!} {
\begin{tabular}[c]{|c|c|c|c|c|}
		\hline
		$G_{V}$  & $a$ &  $k_{1}$ & VBF(fb) & DY(fb) \\
		\hline
		$\sqrt{5} v/4$ & $1/4$ & $0$ & $0.05$ & $0$ \\
		\hline
		$\sqrt{5} v/4$ & $1/4$ & $1$ & $0.18$  & $3.03$ \\
		\hline
		$\sqrt{5} v/4$ & $1/4$ & $2$ & $1.10$ & $12.12$ \\
		\hline
		$v/2$ & $1/2$ & $0$ & $0.16$  & $0$ \\
		\hline
		$v/2$ & $1/2$ & $1$ & $0.12$  & $3.79$ \\
		\hline
		$v/2$ & $1/2$ & $2$ & $1.07$  & $15.16$ \\
		\hline
		$v/\sqrt{6}$ & $1/\sqrt{2}$ & $0$ & $0.22$  & $0$ \\
		\hline
		$v/\sqrt{6}$ & $1/\sqrt{2}$ & $1$ & $0.20$  & $5.69$ \\
		\hline
		$v/\sqrt{6}$ & $1/\sqrt{2}$ & $2$ & $1.66$  & $22.76$ \\
		\hline
\end{tabular}}
 \end{minipage}
\caption{Total cross sections for the associated production of $hV^{+}$ final state by VBF and DY at the LHC for $\sqrt{s}=14$ TeV as functions of the different parameters for $M_{V}=700$ GeV (left panel) and $M_{V}=1$ TeV (right panel). The parameter $a$ is fixed by the value of $G_{V}$ (and vice versa) according to Eq.~\eqref{agvrelation}.}\l{table3}
\end{table}

The total cross sections have been computed using the CalcHEP matrix element generator with the model implemented using the FeynRules Mathematica package. For the calculation of the VBF total cross sections the acceptance cuts $p_{T\,j}>30$ GeV and $|\eta|<5$ for the forward quark jets have been imposed. From the tables we immediately see that the DY total cross sections are much larger than the corresponding VBF ones.
This is due in part to the structure of the phase space, that for the DY is a $2\to2$ and for the VBF is a $2\to4$ and in part to the structure of the squared amplitudes that for the DY is proportional to
\be
|\Mamp\(q\bar{q}\to hV\)|^{2}\propto g_{V}^{2}\f{k_{1}^{2}}{g_{V}^{4}}=\f{k_{1}^{2}}{g_{V}^{2}}\,,
\ee
while the VBF has a more complicated structure that has a strong dependance on $k_{1}-a$. 

One important result that emerges from the tables is that if the VBF total cross sections are too small to expect a signal at the LHC, the DY ones can give rise to a signal for a large region of the parameter space. In the next section we give the expected rates of multi-lepton events coming from the total cross sections listed in Tables~\ref{table1}-\ref{table3}.

\begin{figure}[tbp]
\begin{minipage}[b]{7.5cm}
   \centering
   \includegraphics[height=5.1cm]{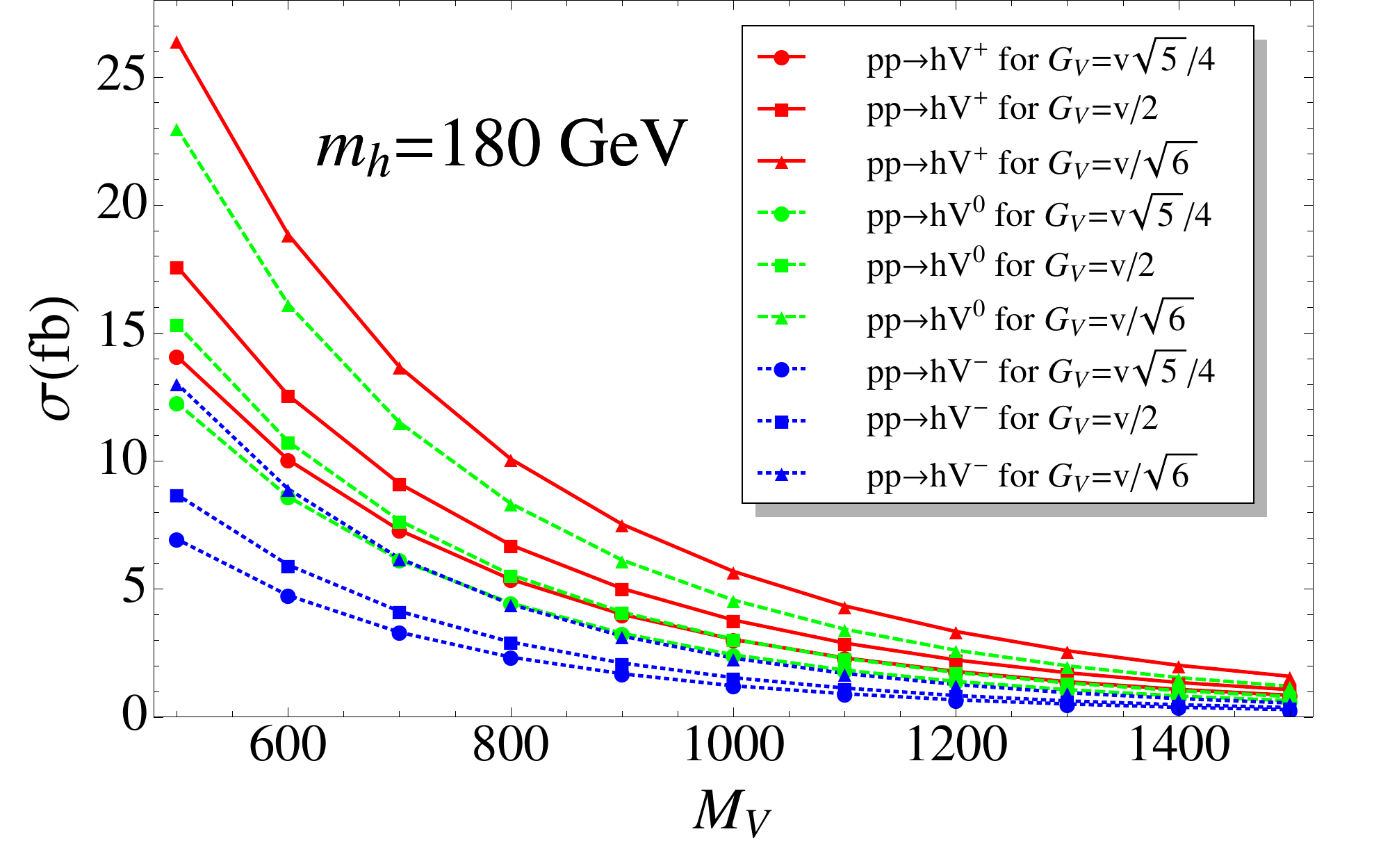}
 \end{minipage}
\ \hspace{1mm} \ 
\begin{minipage}[b]{7.5cm}
  \centering
   \includegraphics[height=5.1cm]{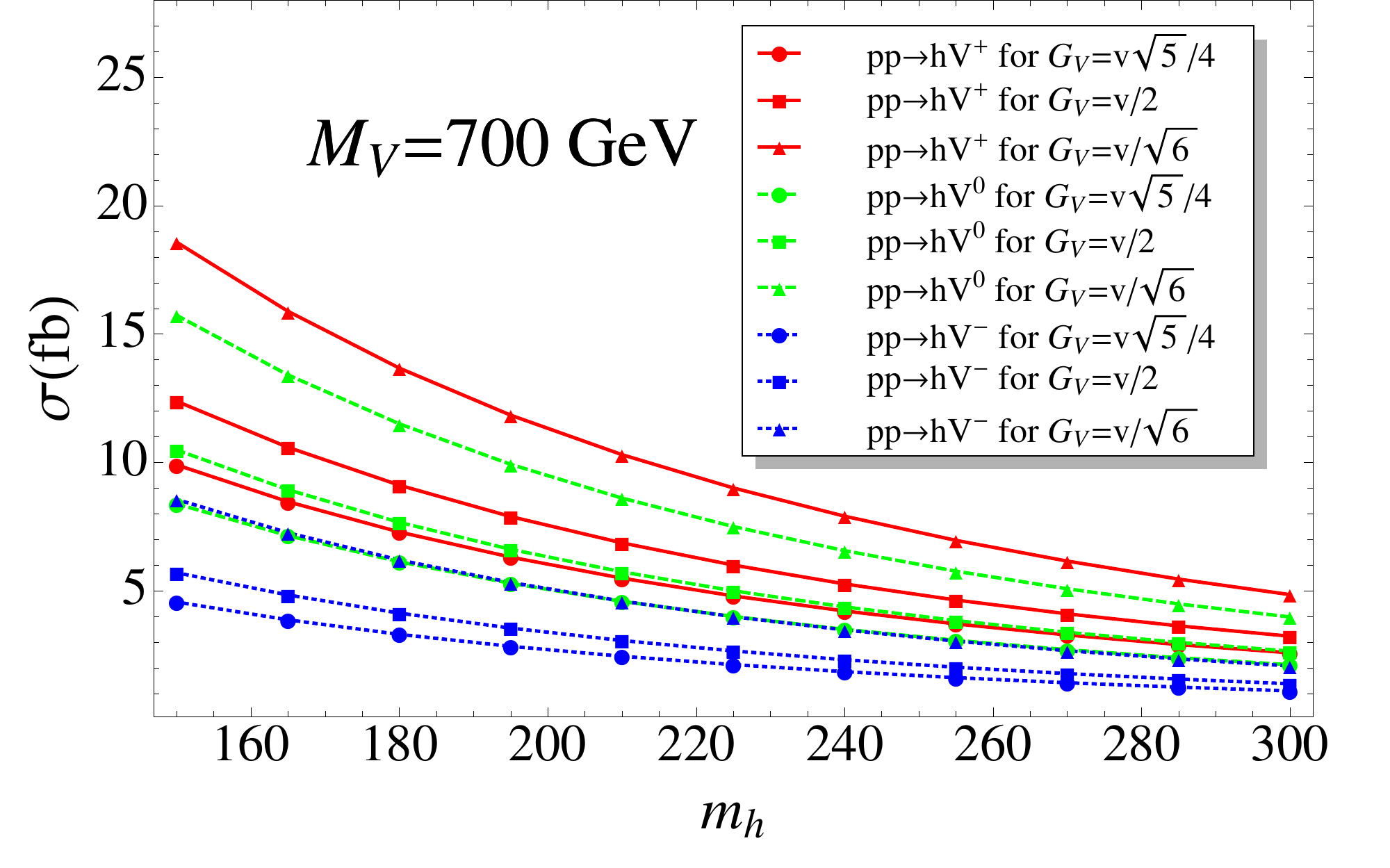}
 \end{minipage}
\caption{Total cross sections for the $hV$ associated productions via DY $q\bar{q}$ annihilation as functions of the heavy vector mass for $m_{h}=180$ GeV (left panel) and of the scalar mass for $M_{V}=700$ GeV (right panel) at the LHC for $\sqrt{s}=14$ TeV, for different values of $G_{V}$ (corresponding to different values of $a$ according to Eq.~\eqref{agvrelation}) and for $k_{1}=1$. Since the DY total cross sections are proportional to $k_{1}^{2}$ the results can be simply generalized to different values of $k_{1}$.}
\l{fig:hVDY}
\end{figure}

In the left panel of Fig.~\ref{fig:hVDY} we show the total cross sections for the DY associated production at the LHC for $\sqrt{s}=14$ TeV as functions of the heavy vector mass for different values of the parameter $G_{V}$ (and therefore of $a$ according to Eq.~\eqref{agvrelation}). We see that even for $k_{1}=1$, which corresponds to minimal gauge model value (see Section~\ref{sec3.5}) the total cross sections are of order of $10$ fb for a vector mass between $500$ GeV and $800$ GeV. Furthermore, since  the DY total cross sections grow with $k_{1}^{2}$, deviations from $k_{1}=1$ could result in a strong increase of  the values given in Fig.~\ref{fig:hVDY}.

Finally, to give an idea of the dependence of the total cross sections on the scalar mass $m_{h}$ we have plotted in the right panel of Fig.~\ref{fig:hVDY} the total cross sections for the $hV$ associated production as functions of the scalar mass for $150\,\text{GeV}<m_{h}<300\, \text{GeV}$. From the figure we see that the total cross sections have almost halved going form $m_h=180$ GeV to $m_h=270$ GeV. Taking also into account the relevant BR of $h$ we can conclude that $2M_{W}<m_{h}<2M_{Z}$ is the most favorable situation to find a signal of the associated $hV$ production, while it can be much more difficult to access a signal for $m_{h}<2M_{W} $ or $m_{h}>2M_{Z}$.

\section{Same-sign di-lepton and tri-lepton events}\l{sec5.3}
The number of multi-lepton events is strongly dependent on the decay modes of the light scalar and the heavy vector. We have already seen in Section~\ref{sec4.8} that the heavy vectors decay almost only into di-boson final states (see Eq.~\eqref{BR1}). For what concerns the scalar $h$, we can neglect $\Gamma\(h\to \bar{\psi}\psi\)$ with respect to $\Gamma\(h\to W^{+}W^{-}\)$ and $\Gamma\(h\to ZZ\)$ in the range of interest for the mass of the scalar, $2M_{W}\lesssim m_{h}\lesssim 2M_{Z}$.  Here we assume as a reference value  $BR\(h\to W^{+}W^{-}\)\approx 1$.
\begin{table}[htb!]
\centering
\small\begin{tabular}[c]{|c|c|c|c|}
		\hline
		Decay Mode & di-leptons ($\%$)& tri-leptons ($\%$)\\
		\hline  
		$V^{0}h\to W^{+}W^{-}W^{+}W^{-}$ & $8.9$ & $3.2$\\
		\hline  
		$V^{\pm}h\to W^{\pm}ZW^{+}W^{-}$ & $4.5$ & $1.0$\\
		\hline  
\end{tabular}\caption{Decay modes and cumulative BRs for the different charge configurations of the $hV$ system assuming $BR\(h\to W^{+}W^{-}\)\approx 1$. For the same-sign di-lepton and tri-lepton BRs we consider only the $e$ and $\mu$ leptons coming from the $W$ decays.}\l{BR}
\end{table}

These values of the BRs for the vector and the scalar lead to the cumulative BRs given in Table~\ref{BR}. For these values of the BRs and for a reference integrated luminosity $L=\int\mathcal{L} dt=100~\text{fb}^{-1}$, we obtain the total number of same-sign di-lepton and tri-lepton events given in Tables~\ref{events}.

\begin{table}[htb!]
\small\begin{minipage}[b]{8.2cm}
\centering
\begin{tabular}[c]{|c|c|c|c|}
		\hline
		$G_{V}$ &  $a$ & di-leptons & tri-leptons \\
		\hline
		$\sqrt{5}v/4$  &$1/4$ & $102.4$ & $30.3$ \\
		\hline
		$v/2$  & $1/2$ & $128.0$ & $37.8$ \\
		\hline
		$v/\sqrt{6}$  & $1/\sqrt{2}$ & $192.0$  & $56.7$ \\
		\hline
\end{tabular}
 \end{minipage}
\ \hspace{2mm} \hspace{3mm} \ 
\begin{minipage}[b]{8.2cm}
\centering
\begin{tabular}[c]{|c|c|c|c|c|}
		\hline
		$G_{V}$ &  $a$ & di-leptons & tri-leptons \\
		\hline
		$\sqrt{5}v/4$  &$1/4$ & $41.0$ & $12.0$ \\
		\hline
		$v/2$  & $1/2$ & $51.0$ & $15.1$ \\
		\hline
		$v/\sqrt{6}$  & $1/\sqrt{2}$ & $76.6$  & $22.6$ \\
		\hline
\end{tabular}
 \end{minipage}
\caption{Total number of same-sign di-lepton and tri-lepton events ($e$ or $\mu$ from $W$ decays) for the DY $hV$ associated production at the LHC for $\sqrt{s}=14$ TeV and $L=\int\mathcal{L}dt=100$ fb$^{-1}$ for $M_{V}=700$ GeV (left panel) and $M_{V}=1$ TeV (right panel) for different values of the parameter $G_{V}$ (or $a$ according to Eq.~\eqref{agvrelation}) and for $k_{1}=1$. Since the DY total cross sections are proportional to $k_{1}^{2}$ the results can simply be generalized to different values of $k_{1}$.}\l{events}
\end{table}

\chapter{Summary and conclusions of Part~\ref{PARTI}}\l{chap6}
\setlength{\epigraphwidth}{.4\textwidth}     
\vspace{-2mm}\epigraph{Hope is nature's veil for hiding truth's nakedness.}%
              {\textsc{Alfred Nobel}}
\vspace{3mm}     
\vspace{0mm}\epigraph{In the Middle Ages people believed that the earth was flat, for which they at least had the evidence of their senses: we believe it to be round, not because as many as one percent of us could give the physical reasons for so quaint a belief, but because modern science has convinced us that nothing that is obvious is true, and that everything that is magical, improbable, extraordinary, gigantic, microscopic, heartless, or outrageous is scientific.}%
              {\textsc{George Bernars Shaw}}
\vspace{7mm}
\lettrine{I}{f} the EWSB is triggered by a strong dynamics, relatively light degrees of freedom may occur which play a special role in preserving perturbative unitarity in longitudinal $WW$ scattering. 
To describe the phenomenology of an unspecified strong dynamics, responsible for the EWSB, we
have adhered to the general program based on:

\begin{itemize}
\item 1. Keep $SU(2)\times U(1)$ gauge invariance but leave out the Higgs boson, while insisting on $SU(2)_L\times SU(2)_R \rightarrow SU(2)_{L+R}$ as the relevant (approximate) symmetry;

\item 2. Introduce new composite particles of mass less than $\Lambda \approx 4\pi v$ consistently with 1 and study the related phenomenology.
\end{itemize}

More specifically, we have considered the case in which perturbative unitarity is partially kept under control by a scalar signet $h$ and/or by a vector triplet $V^{a}$ under the custodial $SU\(2\)_{L+R}$. In this hypothesis we have studied the perturbative unitarity bounds and the constraints coming from EWPT in the presence of the new states, one at a time or simultaneously .

The interactions of these states among themselves and with the standard gauge bosons and GBs can be approximately described by an effective Lagrangian invariant under $SU\(2\)_{L}\times SU\(2\)_{R}/SU\(2\)_{L+R}$. This effective Lagrangian has several free parameters and gives rise in general to scattering amplitudes with a bad asymptotic behavior. This does not come as a surprise, given the consolidated knowledge about massive vectors in field theory. Suitable properties/relations among the various parameters must at least approximately exist to keep the asymptotic properties under control. We have found these relations and used them to partially constrain the parameter space. We have also shown how these constraints relate to the properties of a gauge vector and a fundamental scalar from a $SU(2)_L\times SU(2)_R\times SU(2)^N$ gauge theory spontaneously broken to the diagonal $SU(2)$ subgroup by a generic non linear $\sigma$-model or a two Higgs doublets model. As such, the approach followed here can be used to analyze in a unified way several different models proposed in the literature. It should also serve as a useful and unbiased mean to analyze the LHC data, if these vectors exist in nature.\\
In general, the extent to which the various parameters deviate from the minimal gauge model relations is a relevant open issue that can in principle be addressed experimentally by studying and comparing the single and pair production processes. With $M_V$ and $m_{h}$ below one TeV, large deviations are both unlikely and a threat to the very use of the effective Lagrangian approach described here.
They are unlikely if an underlying theory (a `UV completion') exists with a meaningful asymptotic behavior of the physical amplitudes. They constitute a threat to our effective Lagrangian approach since the cutoff would be reduced to an unacceptably low level.  As far as we can tell, however, moderate deviations can exist, still leading to potentially significant signatures for $M_V$ and $m_{h}$ below one TeV. In the particular QCD case, which need not be copied by the putative strong dynamics of EWSB, the $\rho$ has a mass of about 2/3 of the cutoff and couplings which deviate from the gauge model at the $20\div 30\%$ level \cite{Ecker:1989p1559}.
From a phenomenological point of view we have studied the heavy vector pair production and the associated production of a scalar and a heavy vector by VBF and DY annihilation. At the LHC at $\sqrt{s}=14$ TeV we have found that for the heavy vector pair production the VBF and DY processes give comparable cross sections of the order of some femto barns, while for the associated production the DY annihilation is the main production mechanism giving rise to cross sections of some tens of femto barns. These values of the total cross sections can also be strongly increased if the couplings depart from their minimal gauge model value. The expected number of same-sign di-lepton and tri-lepton events are in general around $10$ for the heavy vector pair production and of order of $10-100$ for  the associated production for an integrated luminosity of $100$ fb$^{-1}$. 
It remains to be seen to what extent these signatures can be made to emerge at the LHC from the background.
\part{New physics at the early LHC}\l{PARTII}
\chapter{Composite particles at the earliest LHC}\l{chap7}
\setlength{\epigraphwidth}{.4\textwidth}     
\vspace{-2mm}\epigraph{Theory is when we know everything but nothing works. Praxis is when everything works but we do not know why. We always end up by combining theory with praxis: nothing works and we do not know why.}%
              {\textsc{Albert Einstein}}
\vspace{3mm}     
\vspace{0mm}\epigraph{All truths are easy to understand once they are discovered; the point is to discover them.}%
              {\textsc{Galileo Galilei}}
\vspace{7mm}

\section{Probing new physics at the earliest LHC}\l{sec7.1}

\lettrine{T}{he} beginning of the LHC operation at half the foreseen c.o.m energy and with an unavoidably low initial luminosity brings the focus on possible signals of new physics that might show up at all in the early stages of operation. Model building prejudices normally play an important role in determining the search strategies. While this is understandable, here we  set aside such prejudices as much as possible. We base our considerations on simple phenomenological Lagrangians, fulfilling reasonable consistency conditions. 
We are generally guided by the possible existence of composite states produced by a putative strong dynamics responsible for the EWSB, but we aim mostly at a neat and simple definition of the interactions responsible for the signals under discussion.
We think that this should also represent an appropriate guide  in the presentation of the experimental results, to maximize their usefulness for any subsequent consideration.

Single production of a relatively narrow resonance is the most obvious candidate for a copious source of new physics signals in the early stages of the LHC operation. As we have already noticed in the first part of this work (and as we shall show explicitly in the next chapter), very much studied is the case of a vector resonance produced in the $q \bar{q}$-channel, either neutral or charged, with its leptonic decay modes \cite{Langacker:2008yv,Salvioni:2009mt,Salvioni:2010p2769,Grojean:2011wi}. In fact, the multiplicity of  partonic channels in  a $p p$ collider makes several different cases possible. As a matter of fact, in the competition with the Tevatron with its integrated luminosity already available, the $q \bar{q}$-channel is definitely less favorable relative to the $qq$-, $qg$- and $gg$-channels in view of the corresponding parton luminosities. We therefore concentrate our attention to these channels. If one considers in each of them the lowest possible spin and QCD representation, for matter of simplicity, the cases of interest are \cite{Barbieri:2011p2759}:
\begin{itemize}
\item {\large \bf $g g$-channel}: a spin-less totally neutral scalar $S$;
\item {\large \bf $q g$-channel}: a $J=1/2$ color-triplet "heavy quark", either "$U$" or "$D$";
\item {\large \bf $q q$-channel}: a spin-less color-triplet or color-sextet $\phi$, with various possible charges.
\end{itemize}
We assume that one single new particle is available at a time, which therefore can only decay into SM particles. We concentrate our attention to the first two cases, since the scalar triplet, or sextet \cite{Bauer:2009p1295}, can only decay into a pair of jets, consistently with known constraints, whereas we find relatively more promising the final states containing at least one photon. The resonances in the $q q$ channel also suffer of problems with flavor physics (see below).

\section{Neutral scalar singlet $S$}

The reference Lagrangian that we adopt is the following,
\be
\La_{S}=c_{3}\frac{g_{S}^{2}}{\Lambda}G_{\mu\nu}^{a}G^{\mu\nu\,a}S+c_{2}\frac{g^{2}}{\Lambda}W_{\mu\nu}^{i}W^{\mu\nu\,i}S+c_{1}\frac{g^{\prime 2}}{\Lambda}B_{\mu\nu}B^{\mu\nu}S+\sum_{f}c_{f}\frac{m_{f}}{\Lambda}\bar{f}fS\,,
\l{LS}
\ee
where $\Lambda$ is an energy scale, the $c_i,~i=1,2,3,f$, are dimensionless coefficients and $f$ is any SM fermion, of mass $m_f$. $S$ is a scalar of mass $M_S$, totally neutral under the SM gauge group. We leave out from the Lagrangian \eqref{LS}:
\begin{itemize}
\item A coupling of $S$ to the Higgs doublet of the form $m_S S H^{\dag} H$. This coupling could also be present, as it would actually be induced by radiative corrections. In a wide range of $M_S$ and $\Lambda$ it is however consistent to assume that such coupling is irrelevant.
\item A ``Higgs-like'' coupling of $S$ to the SM gauge bosons of the form $m_{W} S W_{\mu}^{a}W^{\mu\,a}$. As we have seen in Section \ref{sec3.4} such coupling is present in general for a composite Higgs boson (see Eq.~\eqref{lvhinteraction}). If present, this coupling is likely to dominate the width of the singlet, making its phenomenology similar to that of a composite Higgs boson and making the decays into di-jet and di-photon final states subdominants. However, if this coupling is very suppressed, i.e. if the new singlet is not directly related to the sector which breaks the EW symmetry and/or other states contribute to keep perturbative unitarity of longitudinal $WW$ scattering under control up to the cutoff $\Lambda$, the phenomenology of the singlet can become much less ``Higgs-like'', making the di-jet and the di-photon final states the most favorable at the LHC.
\end{itemize}

The two body widths of $S$ are given by
\bes\l{eq2}
\begin{align}
&\dst \Gamma\(S\to gg\)=\f{2c_{3}^{2}g_{S}^{4}M_{S}^{3}}{\pi\Lambda^{2}}\,,\\
&\dst \Gamma\(S\to W^{+}W^{-}\)=\f{c_{2}^{2}g^{4}\sqrt{M_{S}^{2}-4M_{W}^{2}}\(M_{S}^{4}-4M_{S}^{2}M_{W}^{2}+6M_{W}^{4}\)}{2\pi M_{S}^{2}\Lambda^{2}}\,,\\
&\dst \Gamma\(S\to ZZ\)=\f{\(c_{2}g^{2}\cos^{2}\theta_{W}+c_{1}g^{\prime 2}\sin^{2}\theta_{W}\)^{2}\sqrt{M_{S}^{2}-4M_{Z}^{2}}\(M_{S}^{4}-4M_{S}^{2}M_{Z}^{2}+6M_{Z}^{4}\)}{4\pi M_{S}^{2}\Lambda^{2}}\,,\\
&\dst \Gamma\(S\to Z\gamma\)
=\f{\sin^{2}\theta_{W}\cos^{2}\theta_{W}\(c_{1}g^{\prime 2}-c_{2}g^{2}\)^{2}\(M_{S}^{2}-M_{Z}^{3}\)^{2}}{2\pi M_{S}^{3}\Lambda^{2}}\,,\\
&\dst \Gamma\(S\to \gamma\gamma\)=\f{\(c_{1}+c_{2}\)^{2}e^{4}M_{S}^{3}}{4\pi\Lambda^{2}}\,,\\
&\dst \Gamma\(S\to t\bar{t}\)=\f{3c_{t}^{2}m_{t}^2\(M_{S}^{2}-4m_{t}^{2}\)\sqrt{M_{S}^{2}-4m_{t}^{2}}}{8\pi M_{S}^{2}\Lambda^{2}}\,,
\end{align}
\ees
with the corresponding BRs shown in Fig.~\ref{BR_scalar} as functions of $M_S$ for all $c_i=1$, which we take as reference case. 
\begin{figure}[htbp]
\begin{center}
\includegraphics[height=6.5cm]{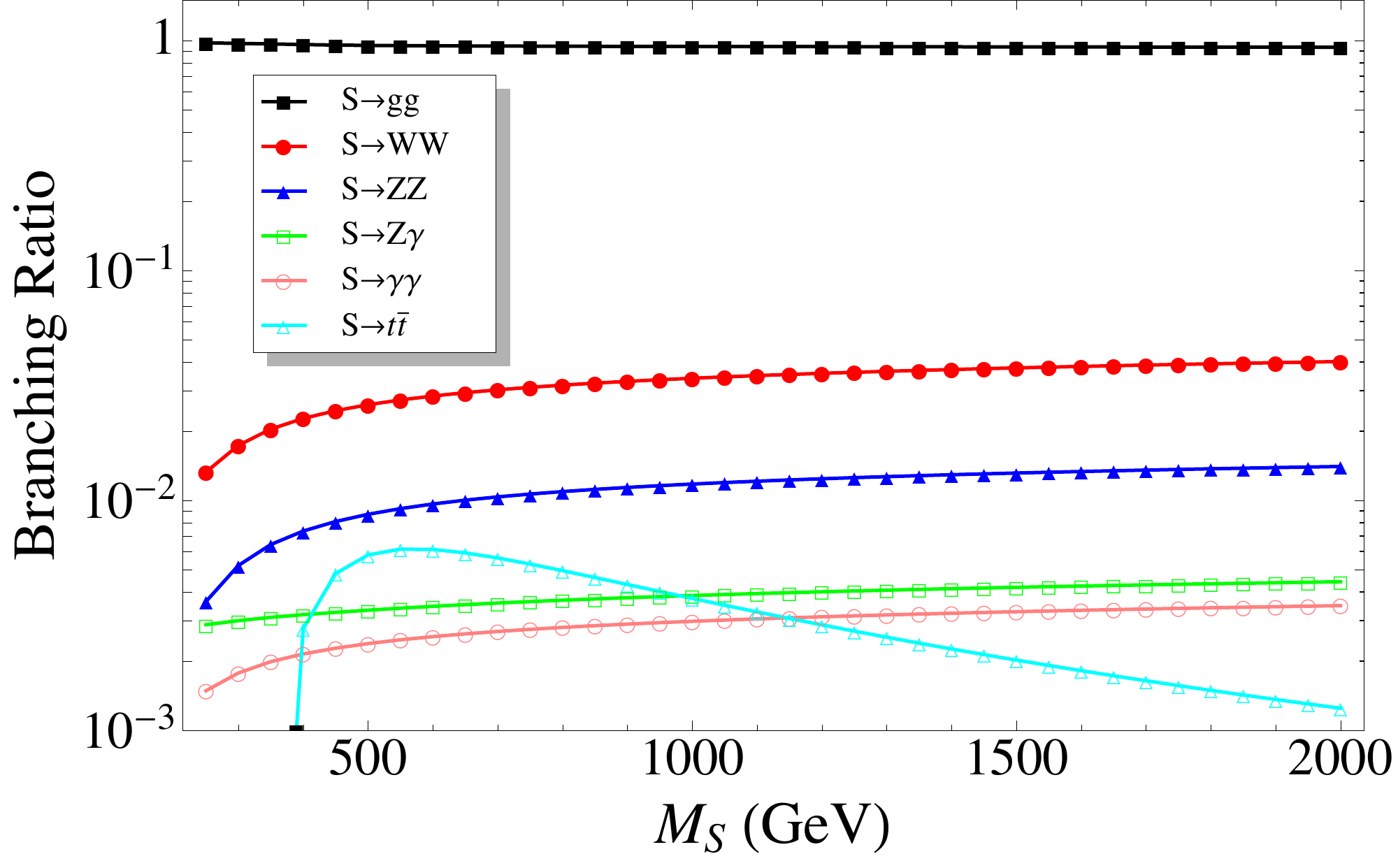}
\vspace{-2mm}\caption{BRs of $S$ as functions of $M_{S}$ for $c_i=1$.}
\l{BR_scalar}
\end{center}
\end{figure}
The total width of $S$ scales as $1/\Lambda^2$. $\Lambda$ is an arbitrary mass parameter, which we take for illustrative purposes at 3 TeV, reminiscent of a  new strong interaction at $\Lambda \approx 4\pi v$, possibly responsible for EWSB. If $S$ were a composite particle in Eq.~\eqref{LS} by the strong dynamics, NDA \cite{Georgi:1984er,Cohen:1997p1379} would suggest $c_i\approx 1/4\pi$. Larger values could however arise from large N and/or from more drastic assumptions about the nature of the gauge bosons.
 For $\Lambda = 3$ TeV and $c_i = 1$, $\Gamma_S$ goes from about 10 GeV at $M_S= 500$ GeV to about 250 GeV for $M_S=1.5$ TeV.

We have made a preliminary study of the sensitivity to the search 
for $S$ in the di-$jet$, $\gamma\gamma$ and $\gamma Z$ channels. The results for the first two cases are illustrated in Fig.~\ref{Scalar_1} for $M_S=0.5$ and $1$ TeV respectively. While it appears difficult to see an emerging signal in the di-$jet$ final state, taking into account the large SM background and the systematic uncertainties, a discovery  looks possible in the $\gamma\gamma$ channel with a modest luminosity of about $10$ to $100$ pb$^{-1}$, as illustrated in Fig.~\ref{Scalar_2}. In the case of $\gamma Z$ channel one would have to pay for an extra factor of about $20\div40$ in the needed luminosity.
\begin{figure}[ptbh]
\begin{minipage}[b]{7.5cm}
\centering
\includegraphics[height=5.1cm]{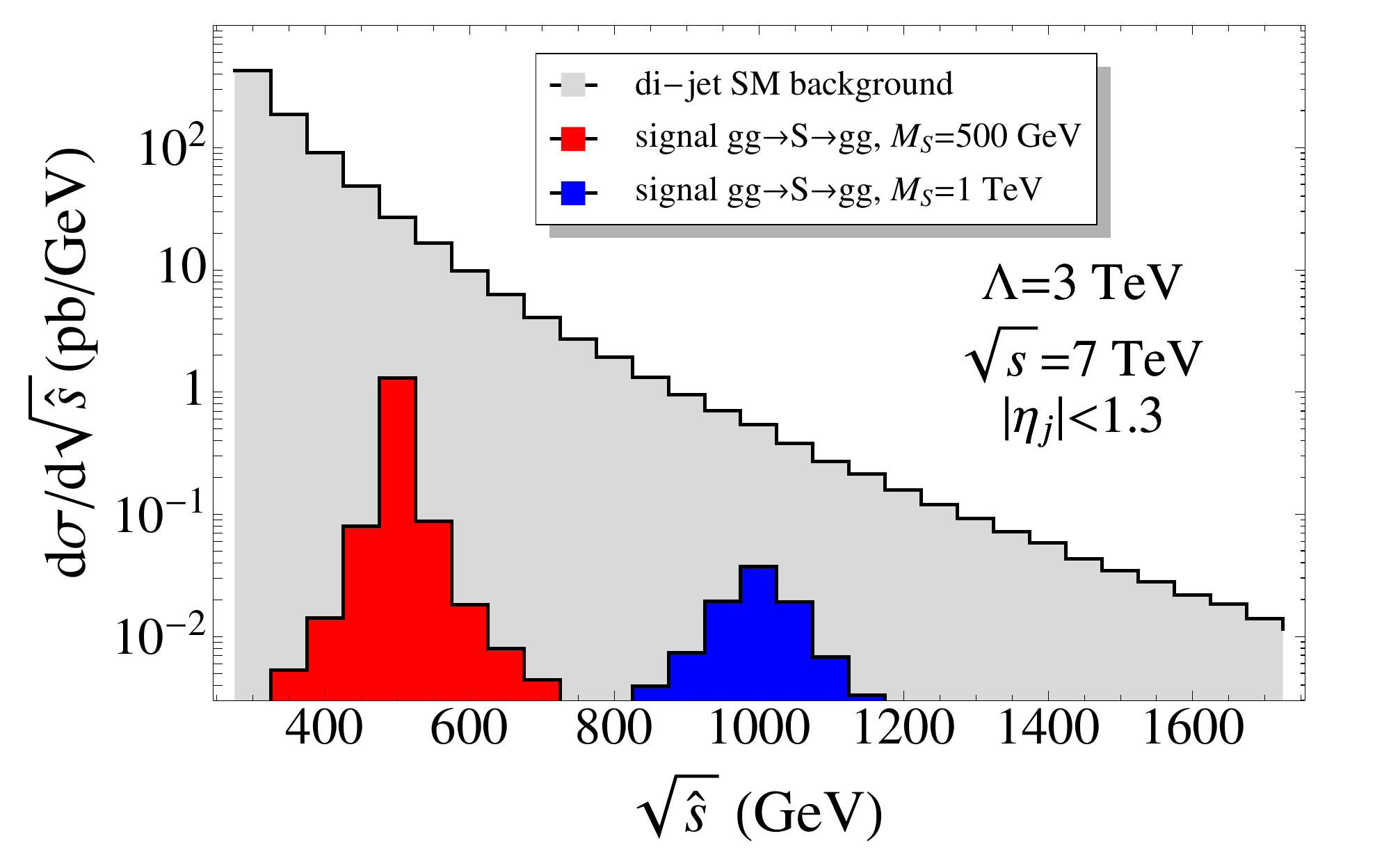}
\end{minipage}
\ \hspace{1mm} \ 
\begin{minipage}[b]{7.5cm}
\centering
\includegraphics[height=5.1cm]{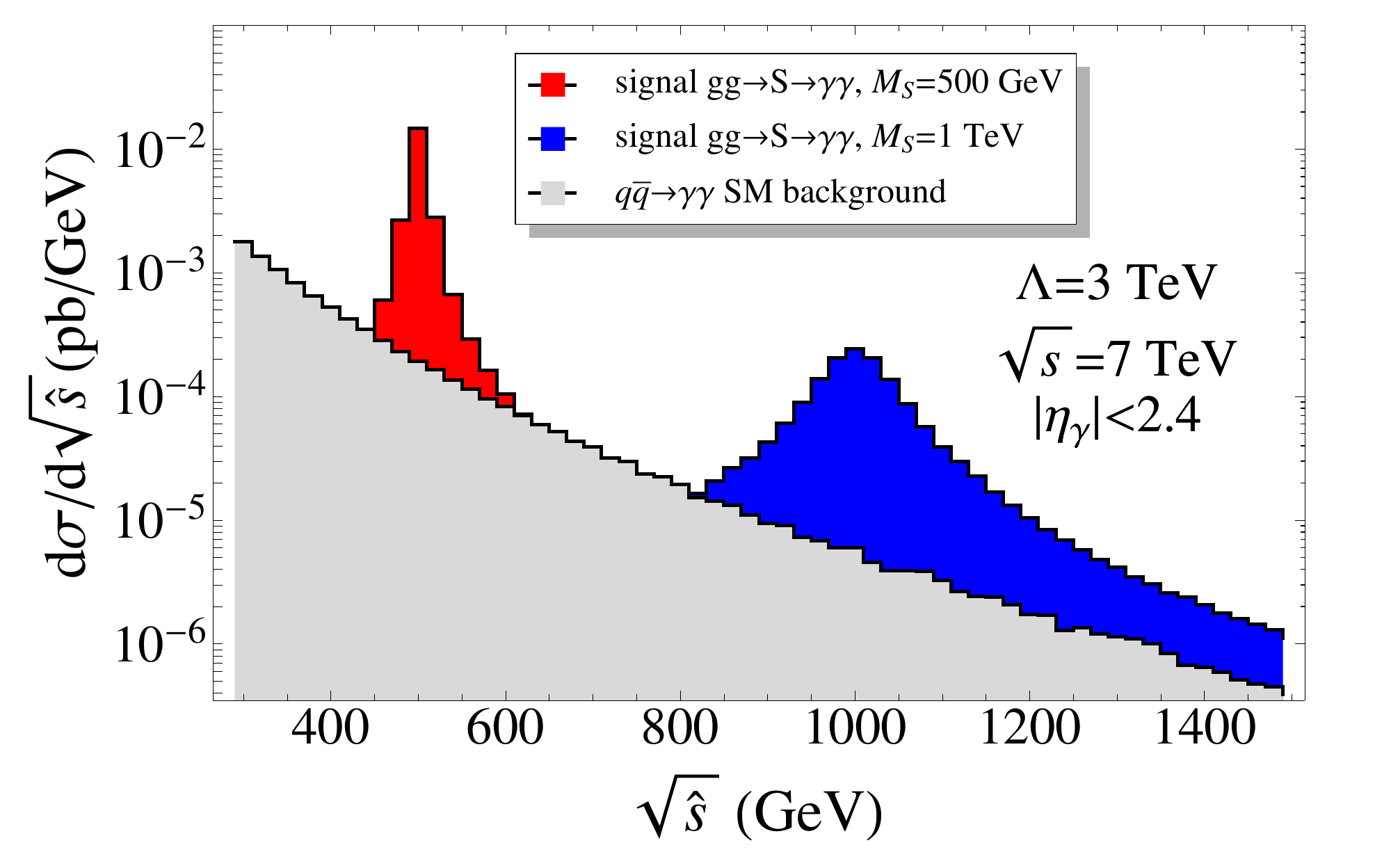}
\end{minipage}
\caption{Signals of a $500$ GeV (blue) and $1$ TeV (red) scalar singlet $S$ in the di-$jet$ (left panel) and $\gamma\gamma$ (right panel) invariant mass distributions vs the SM background (gray) at the early LHC ($\sqrt{s}=7$ TeV) for $ \Lambda = 3$ TeV and $c_i=1$.}%
\l{Scalar_1}%
\end{figure}
\begin{figure}[ptb]
\begin{minipage}[b]{7.5cm}
\centering
\includegraphics[height=5.1cm]{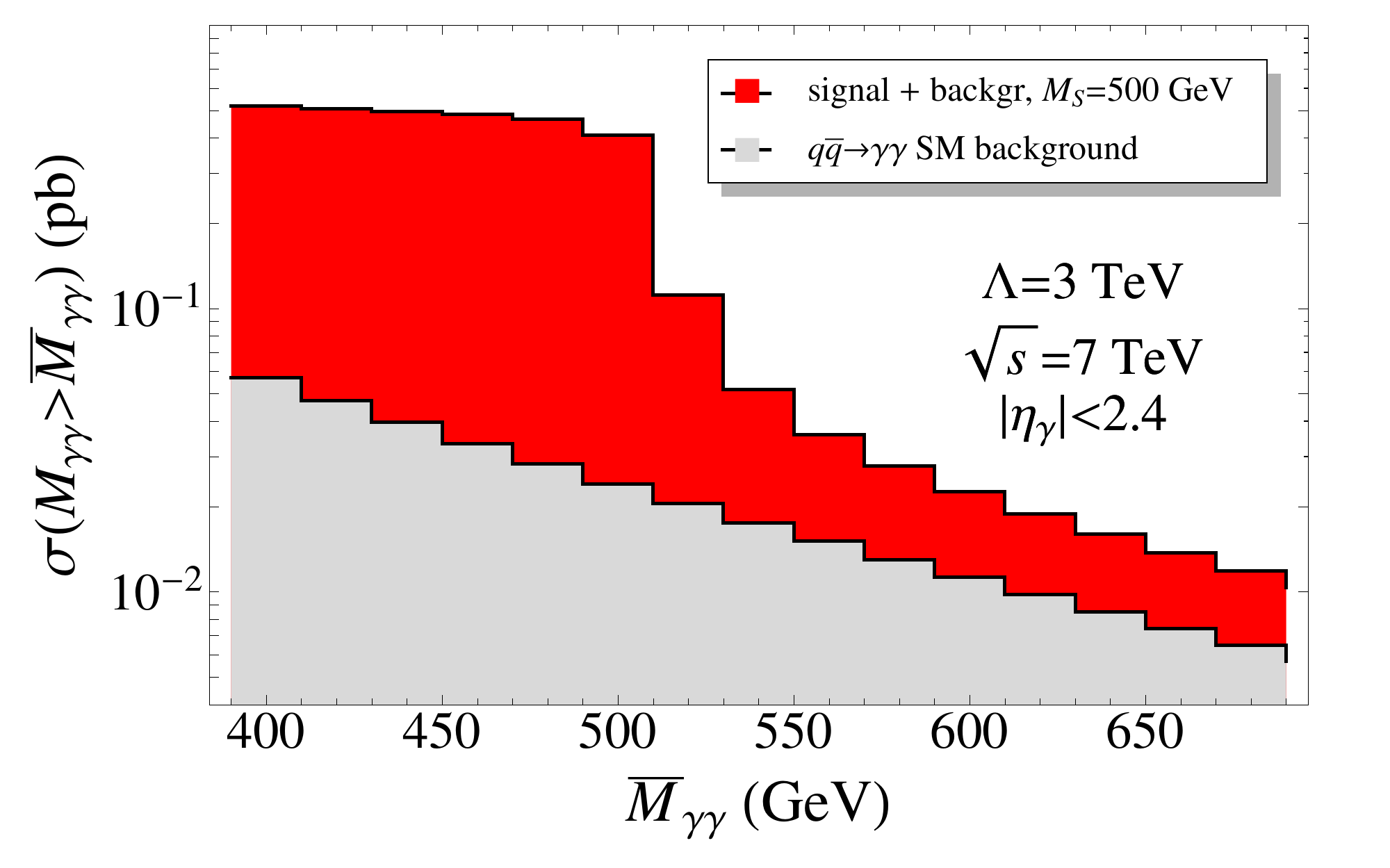}
\end{minipage}
\ \hspace{1mm} \ 
\begin{minipage}[b]{7.5cm}
\centering
\includegraphics[height=5.1cm]{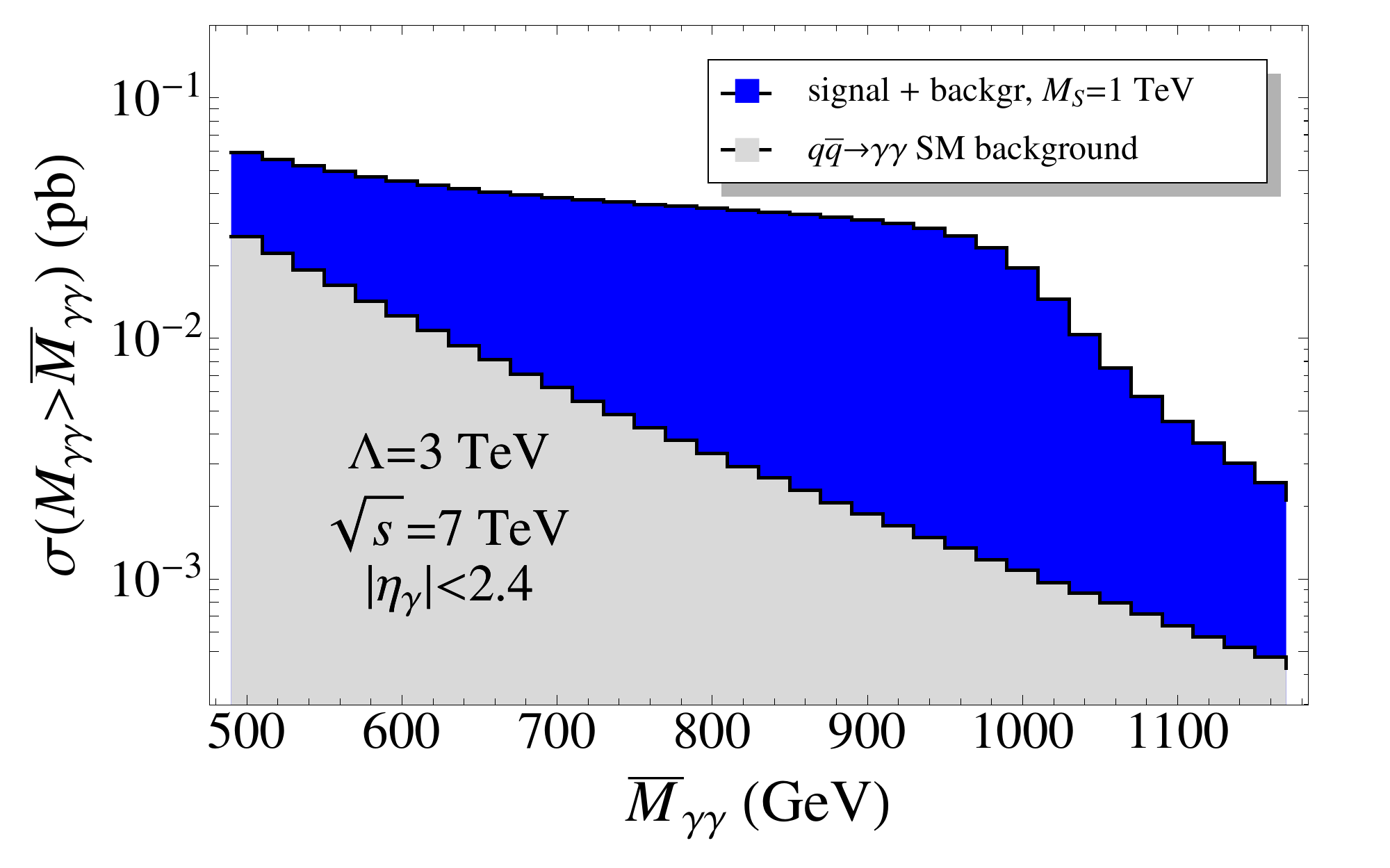}
\end{minipage}
\caption{Number of events for $L=\int{\La dt} = 1$ pb$^{-1}$ at the early LHC ($\sqrt{s}=7$ TeV) with a $\gamma \gamma$ invariant mass greater than $\ovl{M}_{\gamma \gamma}$ for a $500$ GeV (left panel) and a $1$ TeV (right panel) scalar singlet $S$ and for $\Lambda = 3$ TeV and $c_i=1$.}
\l{Scalar_2}%
\end{figure}

All this is trivially extendible to arbitrary values of $\Lambda$ and $c_i$. In particular, assuming the dominance of the $S\to gg$ channel and the validity of the Narrow Width Approximation\footnote{The NWA allows one to decompose the total cross section as the product of the production cross section times the decay BR, i.e. $\s\(pp\to X\to 2x\)\approx \s\(pp\to X\)\times BR\(X\to 2x\)$.} (NWA), an absence of signal for a given integrated luminosity is easily translated in a lower bound on $\Lambda/c_{3}$ for the di-$jet$ channel and $\Lambda/\(c_{1}+c_{2}\)$ for the $\gamma\gamma$ channel.\\
\indent We have used the recent search for a narrow resonance in the di-$jet$ channel performed by the CMS experiment\footnote{We could not directly use the searches of the ATLAS Collaboration of Refs.~\cite{ATLASCollaboration:2010p1746,ATLASCollaboration:2010wv,ATLASCollaboration:2011vna} since they were performed assuming the $qg$ final state that gives rise to a different quantity of QCD radiation. In the last reference they have put model independent upper limits on the total number of events for a gaussian assumed signal. However, since these limits are strongly dependent on the width of the resonance for a width around the $10\%$ of the mass, we cannot use them for large values of $c_{i}$ ($c_{i}\sim 1$) in the case of our scalar singlet $h$.}\l{notaCMSATLAS} in Ref.~\cite{CMSCollaboration:2010p2315} to set an upper bound on the coupling $c_{3}$ for fixed $\Lambda=3$ TeV as a function of $M_{S}$ \cite{Torre:2011xw}. The upper bound on the total cross section compared with the prediction for our reference values of the parameters and the corresponding bound on the coupling $c_{3}$ are depicted in Fig.~\ref{Scalar_3}.
\begin{figure}[h]
\begin{minipage}[b]{7.5cm}
\centering
\includegraphics[height=5.1cm]{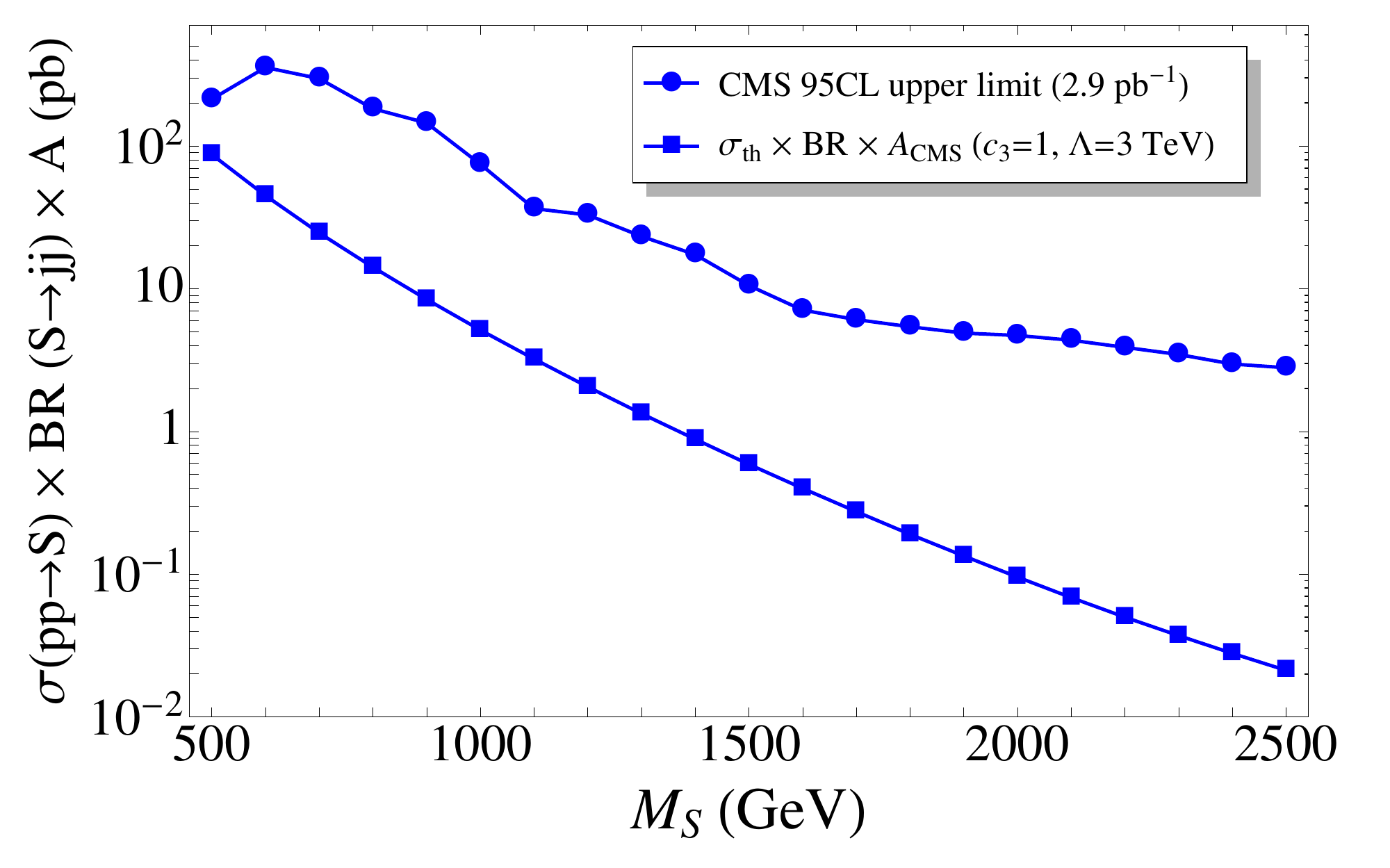}
\end{minipage}
\ \hspace{1mm} \ \begin{minipage}[b]{7.5cm}
\centering
\includegraphics[height=5.1cm]{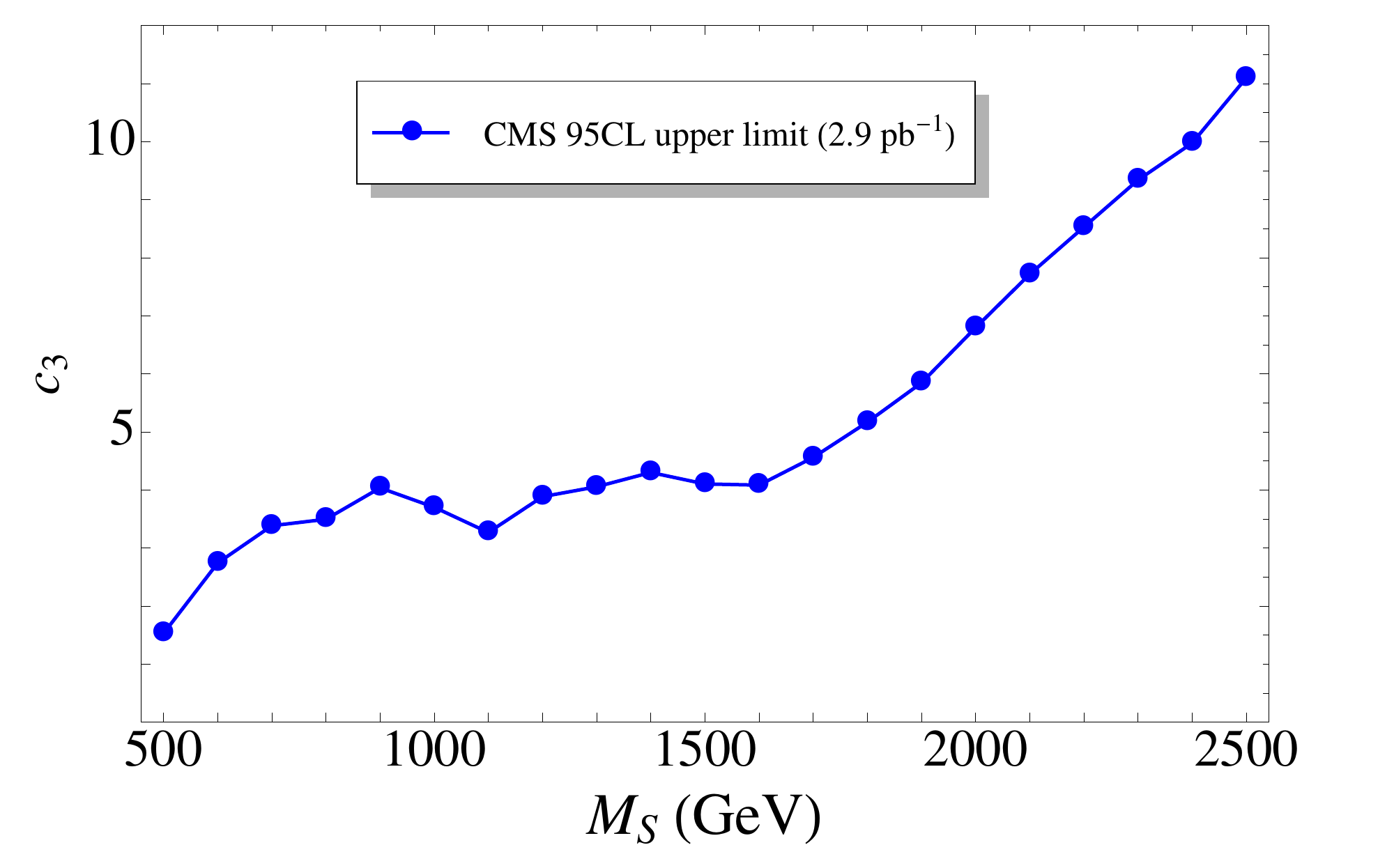}
\end{minipage}
\caption{Left panel: experimental upper limit at $95\%$ CL on the total cross section for a $gg$ resonance measured by the CMS experiment and total cross section predicted with the Lagrangian \eqref{LS} for the scalar $S$, as functions of the resonance mass; Right panel: upper bound at $95\%$ CL on the coupling $c_{3}$ as a function of the resonance mass for $\Lambda=3$ TeV, assuming $BR\(S\to gg\)\approx 1$.}
\l{Scalar_3}%
\end{figure}
In Fig.~\ref{Scalar_4} we also show a preliminary estimation of the sensitivity of the early LHC to the couplings combination $c_{1}+c_{2}$ relevant for the $\gamma\gamma$ final state. From this figure we see that an integrated luminosity larger than $10$ fb$^{-1}$ is required to approach the NDA limit $1/4\pi$ for the couplings $c_{1}$ and $c_{2}$. Note that the upper limits in Figs.~\ref{Scalar_3} and \ref{Scalar_4} were obtained assuming the NWA. However, when the couplings become large ($c_{i}\gtrsim 1$), the resonance can become significantly broad, making the NWA no longer reliable. Moreover, threshold effects can spoil the NWA in the large resonance mass region \cite{Grojean:2011wi,Berdine:2007p2702,Kauer:2007p2703}. For these reasons the limit in the high mass region can differ significantly from those in Figs.~\ref{Scalar_3} and \ref{Scalar_4} and an analysis for broad resonances should be done to improve them.
\begin{figure}[htbp]
\begin{center}
\includegraphics[height=5.1cm]{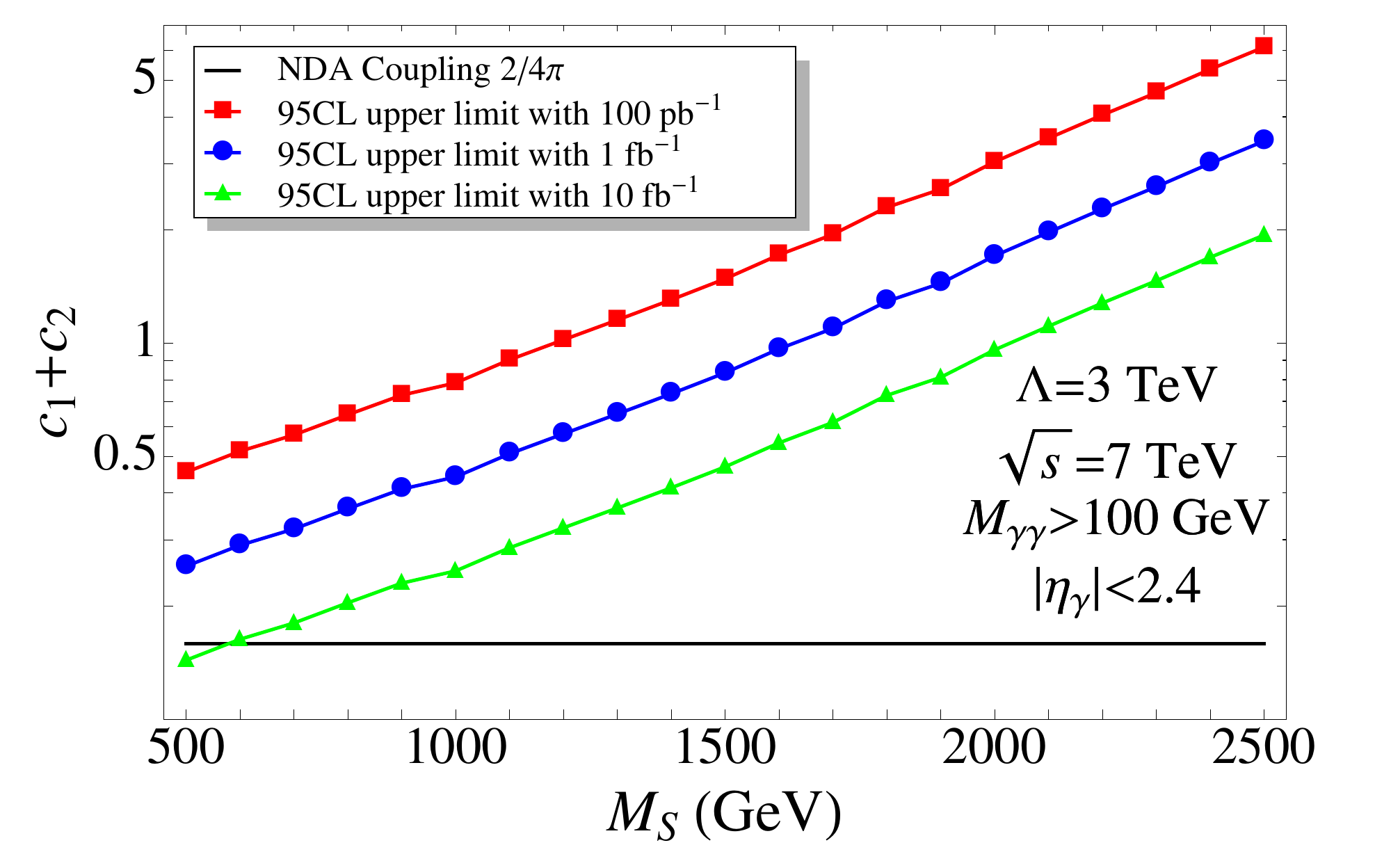}
\vspace{-2mm}\caption{Upper limits on $c_{1}+c_{2}$ as functions of the scalar mass $M_{S}$ for $\Lambda=3$ TeV assuming $BR\(S\to \g\g\)\ll BR\(S\to gg\)$.}
\l{Scalar_4}
\end{center}
\end{figure}

\section{Heavy quark $U$}

In analogy with \ref{LS}, to describe the interactions of a heavy $U$-type quark  of mass $M_U$ with the SM particles we consider the interaction Lagrangian
\be\l{LU}
\La_{U}=c_{G}\frac{g_{S}}{\Lambda}\bar{U}_{L}\sigma^{\mu\nu} T^{a} u_{R} G_{\mu\nu}^{a}+c_{B}\frac{g^{\prime}}{\Lambda}\bar{U}_{L}\sigma^{\mu\nu} u_{R} B_{\mu\nu}+\text{h.c.}\,,
\ee
where $\sigma^{\mu\nu}=i/2 [\gamma^{\mu},\gamma^{\nu}]$ and $T^{a}$ are the generators of the  fundamental representation of SU(3) ($T^{a} = \lambda^a/2$). 
$U$ transforms therefore as a $(3,1)_{2/3}$ of the SM gauge group. 
As in the case of the scalar $S$, $U$ could be a composite state by a strong dynamics responsible for EWSB, in which case NDA suggests $\Lambda \approx 4\pi v \approx 3$ TeV and $c_G \approx c_B \approx 1/4\pi$.

Note that the Lagrangian \eqref{LU} does not break the chirality of the standard $u$-quark, which is crucial to preserve its lightness. A problem with flavor arises, however, since we pretend that $U$ couples to the physical standard $u$-quark but has no (significant) coupling to the $c$-quark, not to cause unobserved $\Delta C=2$ flavor changing effects. In other words the coupling in Eq.~\eqref{LU} would have to be particularly aligned with the standard Yukawa matrix for the up-type quarks. While this is not excluded, a neater way to get around this problem would be to introduce three $U$-fields, one per generation, and assume that the Lagrangian \eqref{LU} respects a global $SU(3)_R$ acting on the standard three up-type quarks. In either case the following considerations apply\footnote{Analogous considerations hold for the scalar states in the $q q$ channel. For example in the case of the $u u$ channel one would have to introduce a $SU(3)_R$-sextet of highly degenerate $\phi$'s.}.

\begin{figure}[ptbh]
\begin{minipage}[b]{7.5cm}
\centering
\includegraphics[height=5.1cm]{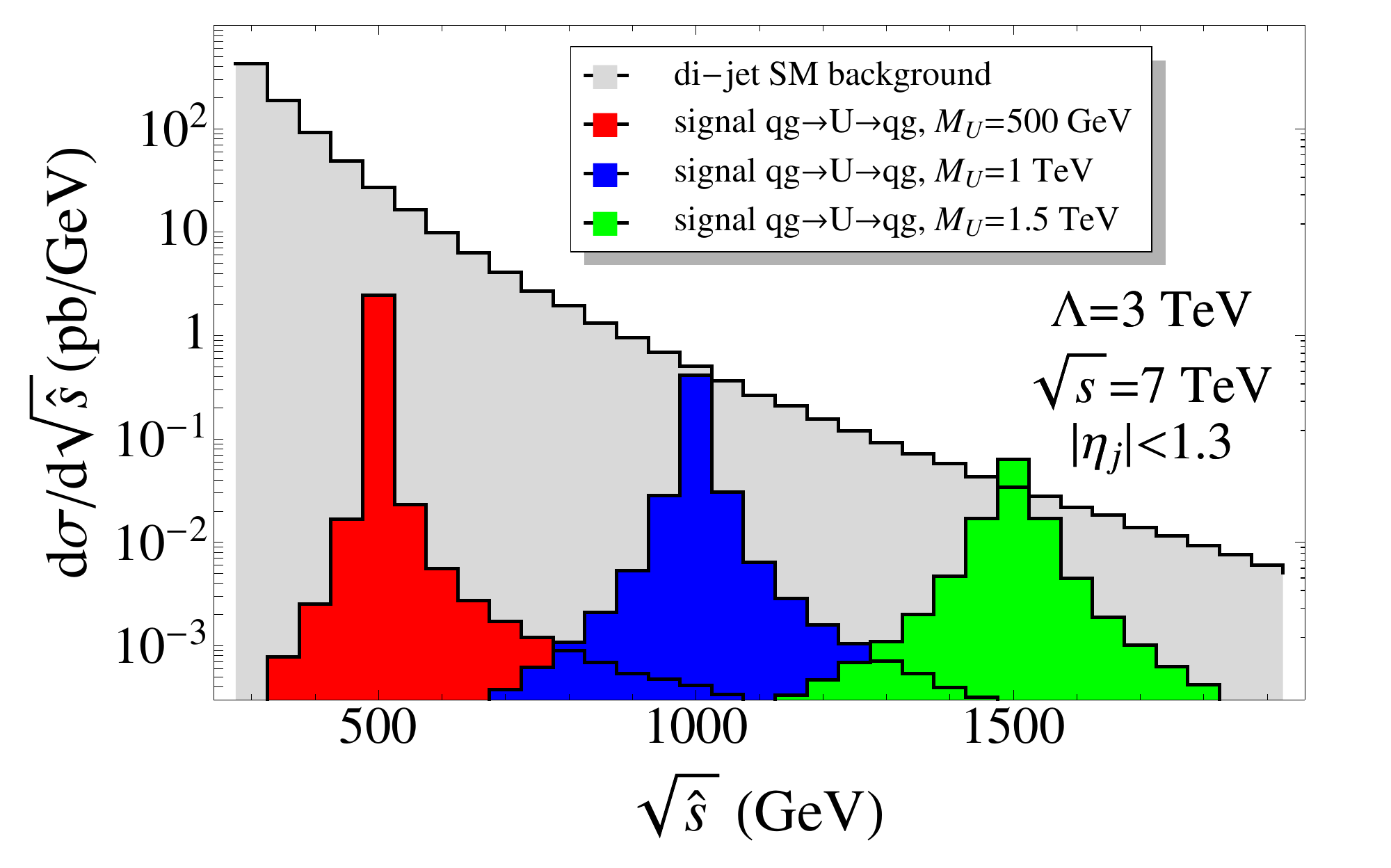}
\end{minipage}
\ \hspace{1mm} \ \begin{minipage}[b]{7.5cm}
\centering
\includegraphics[height=5.1cm]{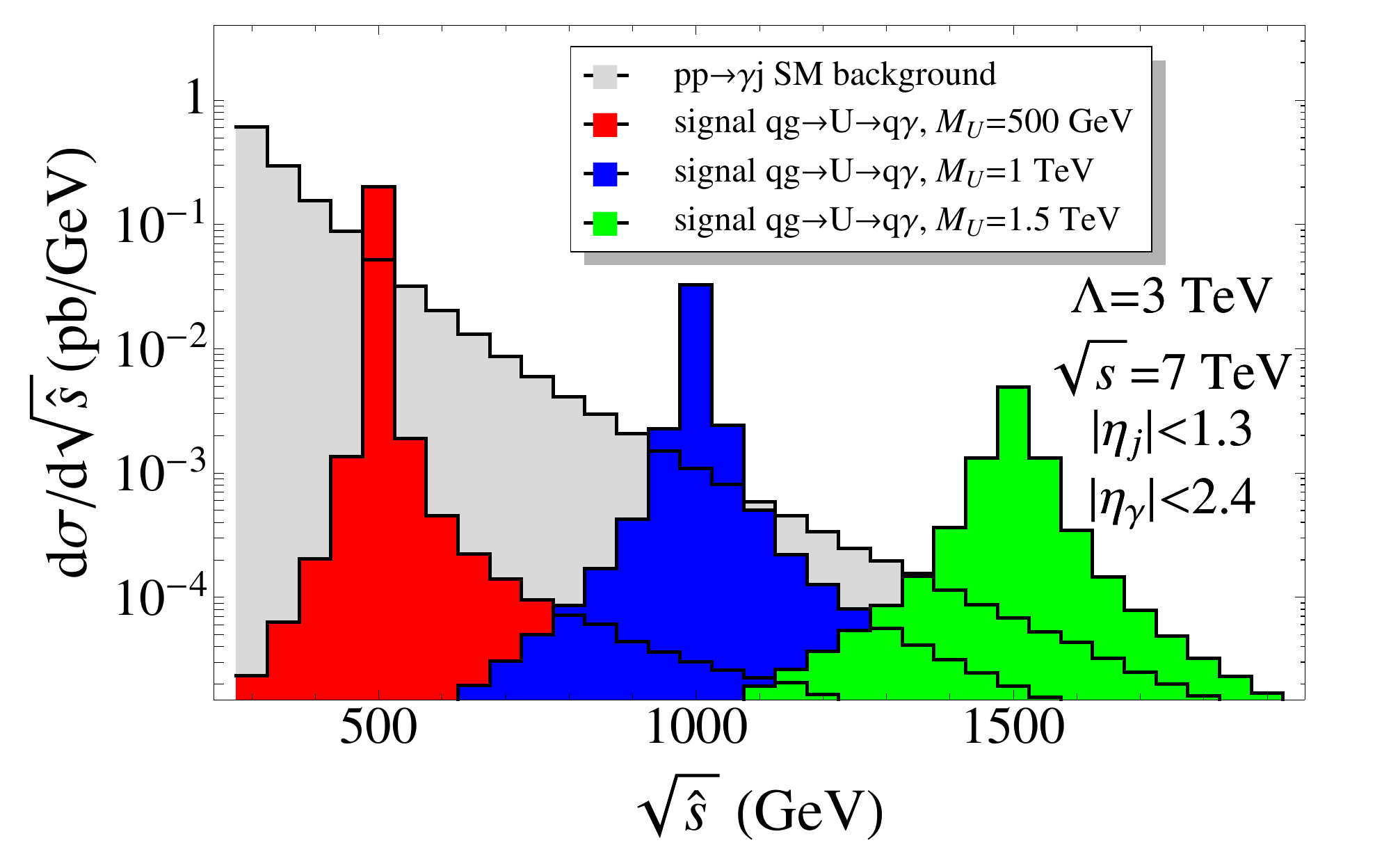}
\end{minipage}
\caption{Signals of a $500$ GeV (blue), $1$ TeV (red) and $1.5$ TeV (green) heavy quark in the di-$jet$ (left panel) and $\gamma+jet$ (right panel) invariant mass distribution vs the SM background (gray) at the early LHC ($\sqrt{s}=7$ TeV) for $\Lambda = 3$ TeV and $c_i=1$.}%
\l{Fermion_1}%
\end{figure}

To study the sensitivity to $U$, as in the case of the singlet $S$ we consider the reference values $\Lambda=3$ TeV and $c_G = c_B =1$. The two body widths of $U$ are given by
\bes\l{eq5new}
\begin{align}
&\dst \Gamma\(U\to ug\)
=\f{4\alpha_{S}M_{U}^{3}}{3\Lambda^{2}}\,,\\
&\dst \Gamma\(U\to uZ\)
=\f{g^{\prime\,2}\sin^{2}\theta_{W}\sqrt{M_{U}^{2}-4M_{Z}^{2}}\(2M_{U}^{4}-M_{Z}^{2}M_{U}^{2}-M_{Z}^{4}\)}{8\pi M_{U}^{2}\Lambda^{2}}\approx \f{\alpha M_{U}^{3}\tan^{2}\theta_{W}}{\Lambda^{2}}\,,\\
&\dst \Gamma\(U\to u\gamma\)
=\f{\alpha M_{U}^{3}}{\Lambda^{2}}\,,
\end{align}
\ees
so that the total width,
\be
\Gamma_U \approx \frac{M_{U}^{3}\alpha_{S}}{\Lambda^{2}} (\frac{4}{3}
+\frac{\alpha}{\alpha_{S}\cos^{2}\theta_{W}})\,,
\ee
ranges from about 2 GeV to about 30 GeV for $M_U=0.5$ to $M_U=1.5$ TeV. Almost irrespective of the $U$-mass, 
the dominant $u g$ decay mode has a BR of about 92$\%$, whereas the $u \gamma$ and $u Z$ modes have BRs of about 6$\%$ and 2$\%$ respectively.

\begin{figure}[ptbh]
\begin{minipage}[b]{7.5cm}
\centering
\includegraphics[height=5.1cm]{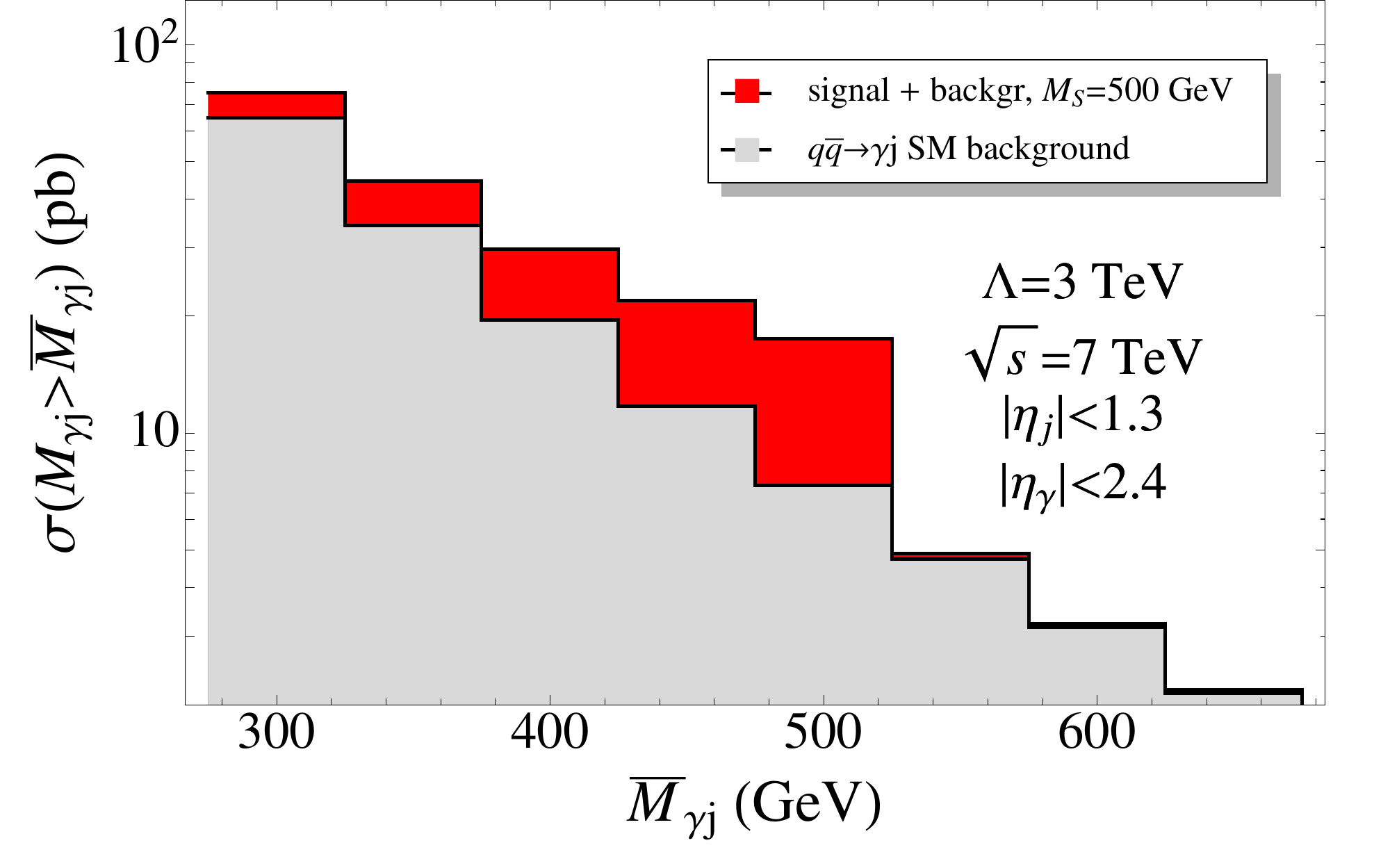}
\end{minipage}
\ \hspace{1mm} \ \begin{minipage}[b]{7.5cm}
\centering
\includegraphics[height=5.1cm]{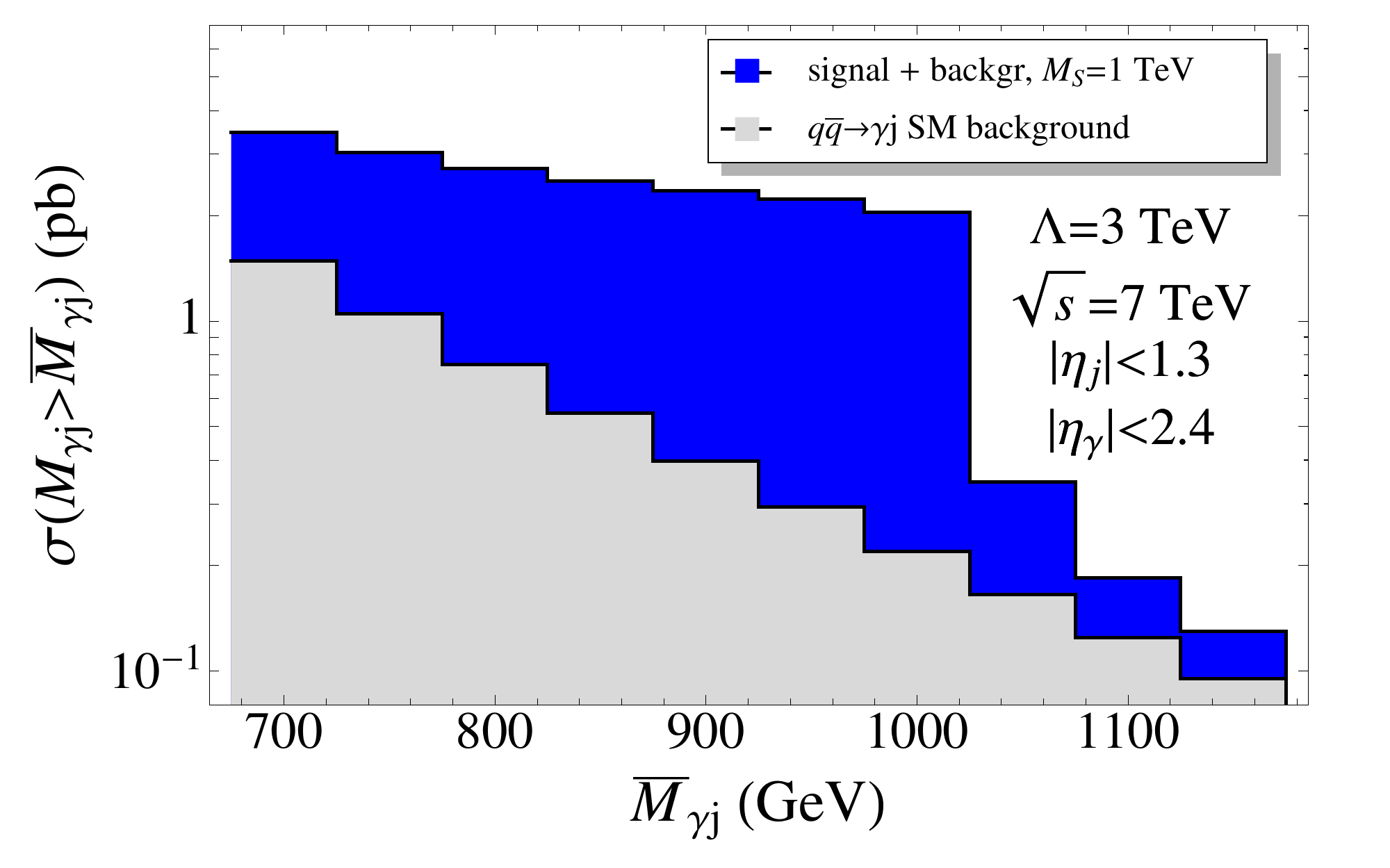}
\end{minipage}
\\
\begin{minipage}[b]{16.5cm}
\centering
\includegraphics[height=5.1cm]{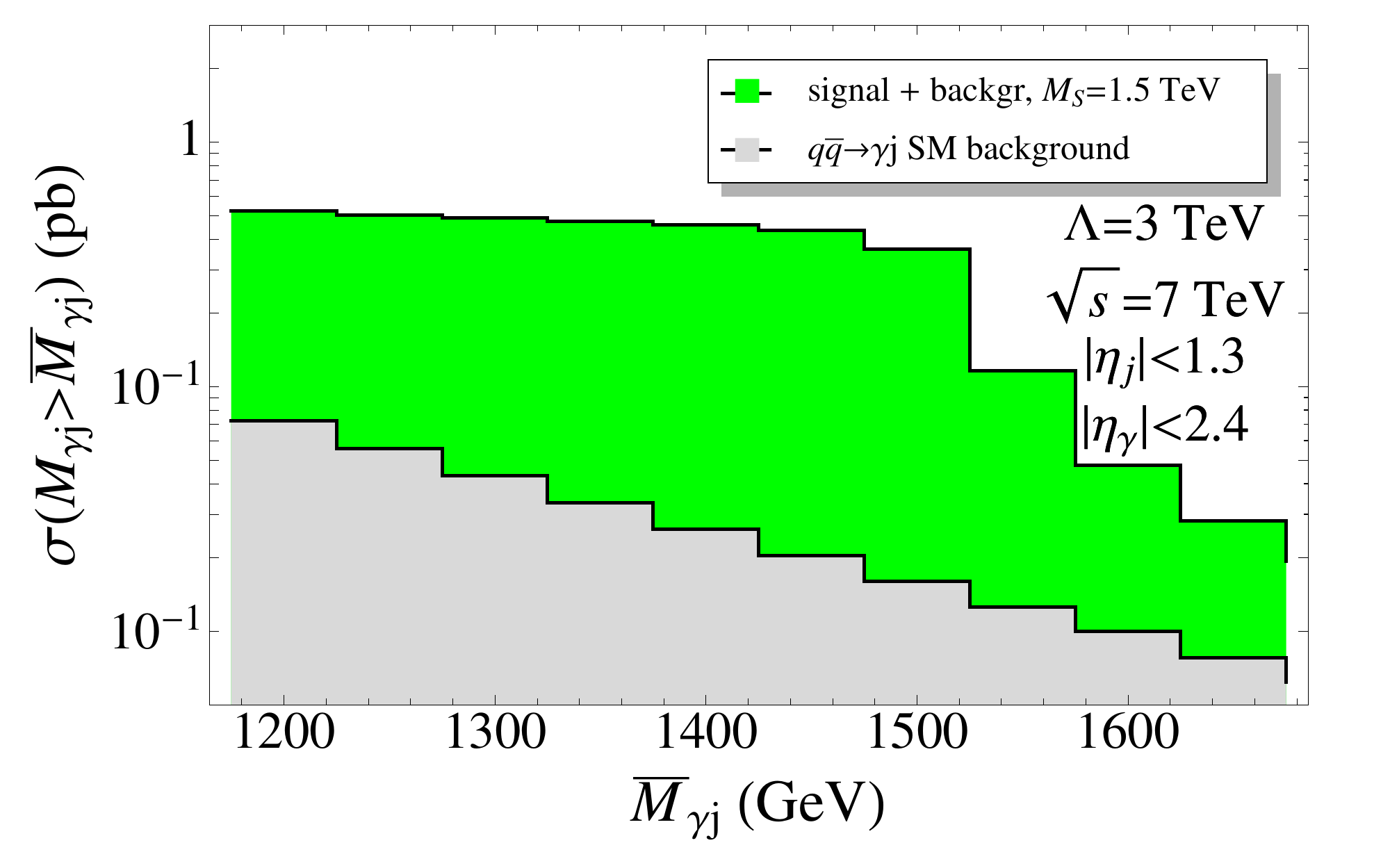}
\end{minipage}
\caption{Number of events for $L=\int{\La dt} = 1$ pb$^{-1}$ at the early LHC ($\sqrt{s}=7$ TeV) with a $\gamma + jet$ invariant mass greater than $\ovl{M}_{\gamma j}$ for a $500$ GeV (top left panel), a $1$ TeV (top right panel) and a $1.5$ TeV (bottom panel) heavy quark and for $\Lambda = 3$ TeV and $c_i=1$.}
\l{Fermion_2}%
\end{figure}

As in the case of $S$, we have made a preliminary study of the sensitivity to the search 
for the heavy $U$ in the di-$jet$, $\gamma+jet$ and $Z+jet$ channels. The results for the first two cases are illustrated in Fig.~\ref{Fermion_1} for $M_U=0.5,~1$ and 1.5 TeV respectively. 
While it appears difficult to see an emerging signal in the di-$jet$ final state, taking into account the large SM background and the systematic uncertainties, a discovery  looks possible in the $\gamma +jet$ channel with a modest luminosity of about $5$ or $10$ pb$^{-1}$, as illustrated in Fig.~\ref{Fermion_2}.
The $Z+jet$ channel  would require luminosities in the hundreds of inverse picobarns.
Assuming the dominance of the $U\to qg$ channel and the validity of the NWA, an absence of signal for a given integrated luminosity is easily translated in a lower bound on $\Lambda/c_{G}$ for the di-$jet$ channel and $\Lambda/c_{B}$ for the $\gamma+jet$ channel.

In this case we have used both the searches of the CMS \cite{CMSCollaboration:2010p2315} and the ATLAS \cite{ATLASCollaboration:2010wv,ATLASCollaboration:2011vna} experiments in the di-$jet$ channel to set an upper bound on the coupling $c_{G}$ (for fixed $\Lambda=3$ TeV) as a function of $M_{U}$ \cite{Torre:2011xw}. The upper bound on the total cross section compared with the prediction for our reference values of the parameters and the bounds on the corresponding coupling $c_{G}$ are depicted in Fig.~\ref{Fermion_3}.
\begin{figure}[ptbh]
\begin{minipage}[b]{7.5cm}
\centering
\includegraphics[height=5.1cm]{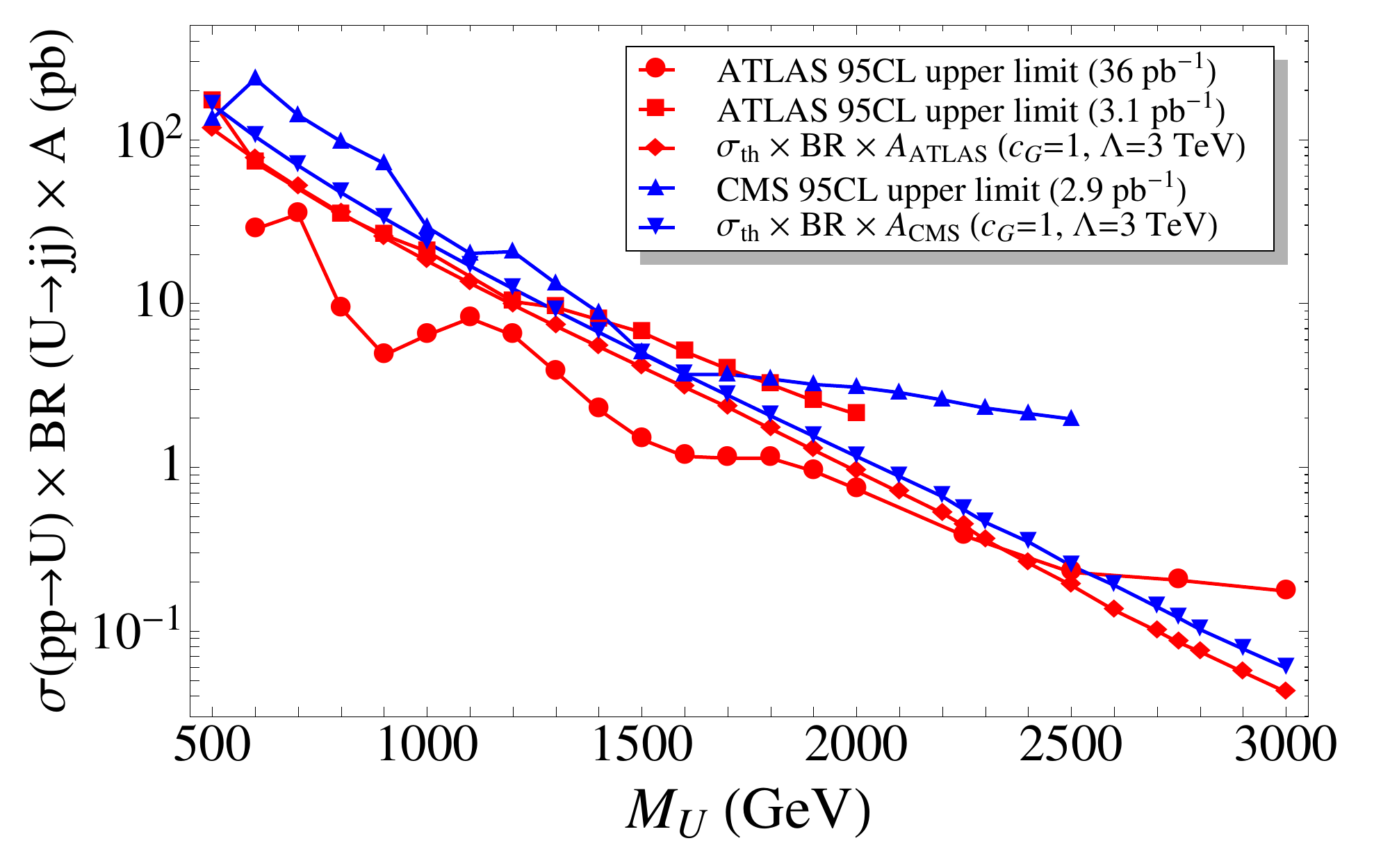}
\end{minipage}
\ \hspace{1mm} \ \begin{minipage}[b]{7.5cm}
\centering
\includegraphics[height=5.1cm]{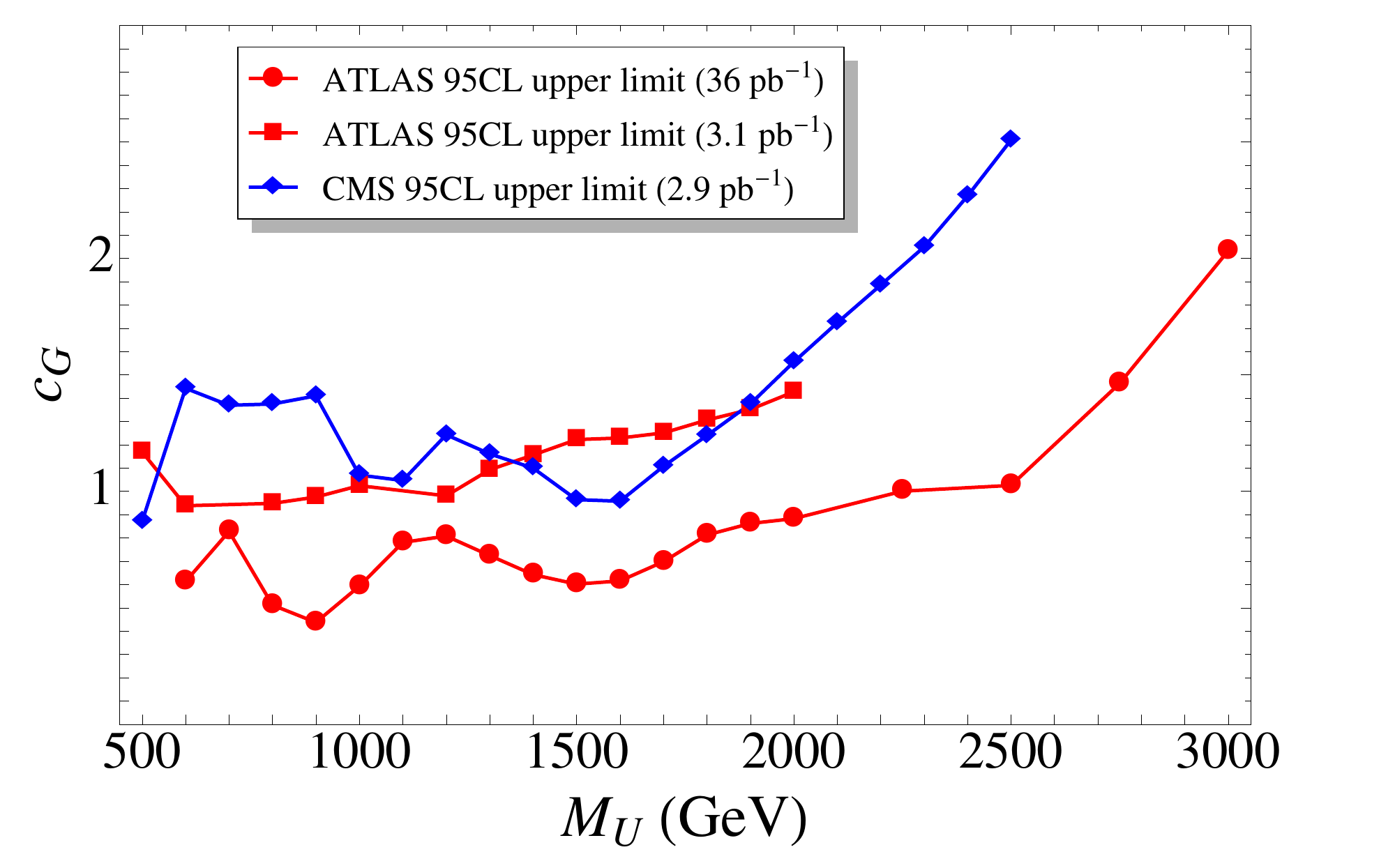}
\end{minipage}
\caption{Left panel: experimental upper limits at $95\%$ CL on the total cross section for a $qg$ resonance measured by the CMS (blue) and ATLAS (red) experiments and cross sections predicted with the Lagrangian \eqref{LU} for the heavy quark $U$, as functions of the resonance mass; Right panel: upper bound at $95\%$ CL on the coupling $c_{G}$ as function of the resonance mass for $\Lambda=3$ TeV, assuming $BR\(S\to qg\)\approx 1$.}
\l{Fermion_3}%
\end{figure}
In Fig.~\ref{Fermion_4} we also show a preliminary estimation of the sensitivity of the early LHC to the coupling $c_{B}$ relevant for the $\gamma+jet$ final state.
\begin{figure}[htbp]
\begin{center}
\includegraphics[height=5.1cm]{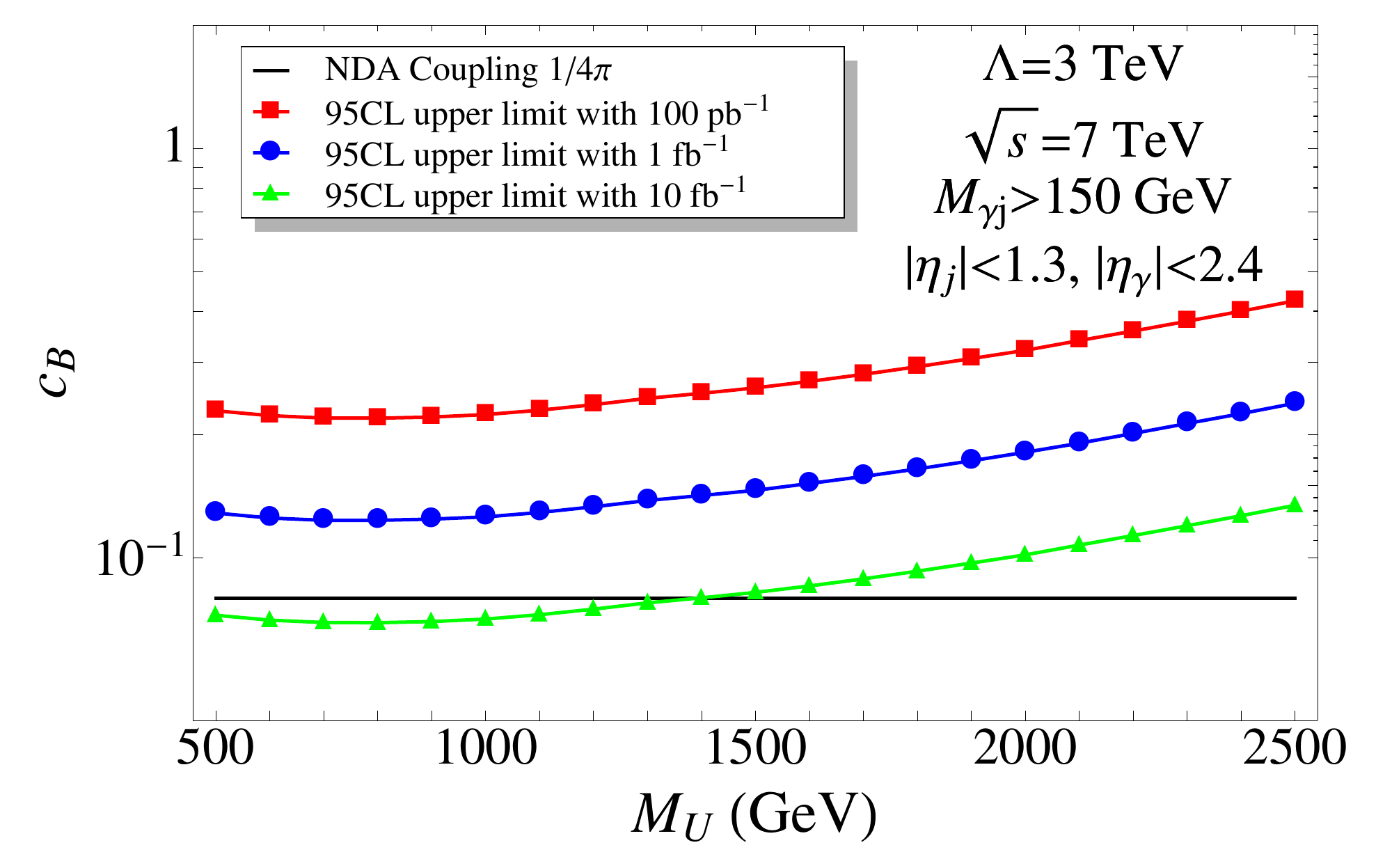}
\vspace{-2mm}\caption{Upper limits on $c_{B}$ as functions of the heavy quark mass $M_{U}$ for $\Lambda=3$ TeV assuming $BR\(S\to q\g\)\ll BR\(S\to qg\)$.}
\l{Fermion_4}
\end{center}
\end{figure}
From this figure we can see that the NDA coupling $1/4\pi$ can be excluded, in a wide region of $M_{U}$, with an integrated luminosity of about $1$ to $10$ fb$^{-1}$. Moreover, Fig.~\ref{Fermion_3} shows the expected behavior of the scaling with the luminosity of the limits on the cross section and on the coupling for the ATLAS experiment: from pure statistical considerations, for a gaussian distribution, we expect the limit on the cross section to scale roughly as $L^{1/2}$ and the limit on the coupling as $L^{1/4}$. Taking into account the systematic uncertainties this is compatible with the two ATLAS analyses.

\section{Conclusions}

The lack so far of a thorough experimental exploration of the energy range at or well above the Fermi scale, the most fundamental scale in particle physics as presently known,  calls for 
an open attitude in the expectation for possible signals of new physics. In turn this has a twofold general implication. On one side one must be ready for surprises \cite{CDFCollaboration:2011vq}. On the other side it is essential that the experimental results be analyzed 
and presented in the neatest possible way with a minimum of biases.

In this chapter we have attempted to apply this attitude to the discussion of possible signals of new physics in the early stages of the LHC operation, concentrating our attention to the production of a relatively narrow resonance that can compete with the well studied case of a neutral vector. While we have generally in mind a possible composite state produced by a putative strong dynamics responsible for EWSB, the important thing is that the interaction that produces one of these states be clearly and simply defined. This is the case for the phenomenological Lagrangians 
\eqref{LS} and \eqref{LU} with a single interaction term responsible for the production of the singlet scalar $S$ or of the triplet fermion $U$ respectively. 

With a suitable guess for the decay BRs we have found that final states in a pair of photons for $S$ or in a photon plus jet for $U$ may lead to detectable signals already with a few tens of inverse picobarns of integrated luminosity at $\sqrt{s}=7$ TeV. For this to be the case we may have used an optimistic value for the strength of the relevant couplings. While such values cannot in any case be excluded a priori and, to the best of our knowledge, have not been excluded at the Tevatron so far, the progression of the integrated luminosity achievable at the LHC can quickly explore fully relevant regions for  these couplings. Meanwhile, with a defined and simple single coupling for the production of $S$ or $U$,  the experimental results for a narrow resonance in the $\gamma \gamma$ or in the $\gamma+ jet$ channels can be usefully given as plots of $d\sigma/dM\cdot BR$ versus the invariant mass of the final state channel\footnote{Possibly unfolded by acceptance effects.}. We think that the interpretation of a possible positive signal in terms of more elaborate and more defined theoretical models would be eased by this way of presenting the data.
\chapter{A weakly constrained $W'$ at the early LHC}\l{chap8}
\setlength{\epigraphwidth}{.4\textwidth}     
\vspace{-2mm}\epigraph{Experience without theory is blind, but theory without experience is mere intellectual play.}%
              {\textsc{Immanuel Kant}}
\vspace{3mm}     
\vspace{0mm}\epigraph{Nothing can come of nothing.}%
              {\textsc{William Shakespeare}}
\vspace{7mm}
\lettrine{I}{n} the previous chapter we have discussed possible signals of new physics in the $gg$- and $qg$-channels. As we have noticed, in competition with the Tevatron, these channels are more favorable at the LHC with respect to the $q\bar{q}$-channel. In this chapter we are going to show that a vector state in the $q\bar{q}$-channel can appear as a signal at the early LHC only if it is very weakly constrained from present bounds, i.e. $M<1$ TeV and $g_{i}\sim 1$. In particular, we apply an effective approach to study the early LHC phenomenology of a $W'$ transforming in the representation 
\be \l{111}
(\mathbf{1},\mathbf{1})_{1}
\ee
of the SM group, where the notation $(SU(3)_{c},SU(2)_{L})_{Y}$ has been adopted. A similar approach has been employed in Ref.~\cite{Aguila:2010p1781}, where however the focus was on computing constraints from EW data. In Ref.~\cite{Aguila:2010p1781}, bounds from EWPT were discussed for all the irreducible representations of the SM gauge group which can have linear and up to dimension four interactions with the SM fields. There it was shown that the only such representations containing a color-singlet $W'$ (for a study of colored resonances at the early LHC, see Ref.~\cite{Han:2010p2696}) coupled to the SM fermions, in addition to that in Eq.~\eqref{111}, are $(\mathbf{1},\mathbf{3})_{0}$ and $(\mathbf{1},\mathbf{2})_{-3/2}$. The $(\mathbf{1},\mathbf{2})_{-3/2}$ multiplet does not have any dimension four interaction with quarks or gluons, and as a consequence its production at the early LHC would be very suppressed. The choice to discuss the representation $(\mathbf{1},\mathbf{1})_{1}$ is motivated by the fact that in this case we can add to the SM only a charged resonance, without any associated neutral state. This is in contrast with the other representation commonly obtained in specific models, namely the $SU(2)_{L}$ triplet $(\mathbf{1},\mathbf{3})_{0}$. We have encountered this latter representation in the first part of this thesis, where a custodially symmetric extension of the SM have been considered. The single production at the LHC, either by VBF or by DY process, or the production in association with a standard gauge boson for states belonging to a triplet of the custodial $SU\(2\)$ have been extensively studied in the literature \cite{Birkedal:2005p1372,He:2008ey,Accomando:2009ei,Accomando:2008p1815,Belyaev:2009gl,Cata:2009ka}. 
On the other hand, a $W'$ transforming as $(\mathbf{1},\mathbf{1})_{1}$, is usually less studied \cite{Grojean:2011wi}. However, since its couplings to leptons only arise through $W$-$W'$ mixing it is strongly suppressed\footnote{Since we only consider the SM field content, we do not include right-handed neutrinos; or, equivalently for our purposes, we assume them to be heavier than the $W'$, so that the decay $W'\rightarrow \ell_{R}\nu^{\ell}_{R}$ is forbidden.} and the new state is only constrained by hadronic processes (except for the oblique $T$ parameter). Moreover, if particular forms for the right-handed quark mixing matrix are chosen as to evade constraints from $\Delta F=2$ transitions, the coupling of the $W'$ to quarks is only constrained by Tevatron direct searches \cite{Langacker:1989p2578}, and therefore it can be sizable, without violating any existing constraint, even for a $W'$ mass below one TeV, making a discovery of the resonance at the early LHC possible. Furthermore, as discussed in Refs.~\cite{Hsieh:2010p2579,Schmaltz:2010p2610}, in Left-Right (LR) models, which give a $(\mathbf{1},\mathbf{1})_{1}$ charged state after LR symmetry breaking, the splitting between the masses of the $W'$ and $Z'$ (with the latter being a singlet under the SM group) can be large, without violating EWPT constraints, if one takes $g_{X}\gg g_{R}$, where $g_{X}$ and $g_{R}$ are the couplings of the abelian factor and of $SU(2)_{R}$, respectively. Also assigning the Higgs responsible for $SU(2)_{R}\times U(1)_{X}\rightarrow U(1)_{Y}$ breaking to a higher dimensional representation (for example, introducing a $SU(2)_{R}$ triplet Higgs) can help in increasing the mass splitting between the $W'$ and $Z'$. If such splitting is large enough, constraints from the $Z'$ can be made negligible, and one can study the phenomenology of the $W'$ using an effective theory for a $(\mathbf{1},\mathbf{1})_{1}$ state. Another example of a construction where the $W'$ we consider arises is the Littlest Higgs with custodial symmetry \cite{Chang:2003bd} (incidentally, we remark that several Little Higgs models contain in the spectrum a spin-1 $SU(2)_{L}$ triplet). While these provide specific examples of $W'$ that are described by the effective theory we consider, the interest of our approach goes much further, as it encompasses any composite state, whose properties could depart significantly from those of the gauge boson of a minimal non-abelian extension of the SM group.

We discuss the prospects of the early LHC to discover the $W'$ in the dijet channel, which, together with the $tb$ final state \cite{Gopalakrishna:2010p2723}, is the main avenue to look for the lepto-phobic $W'$ we are considering. A particularly striking difference between gauge models and the effective theory we consider is the presence in the latter case of a sizable $W'W\gamma$ interaction, which is very suppressed if the $W'$ is a fundamental gauge boson. As a consequence, observation of the $W'\rightarrow W\gamma$ decay at the LHC would be a hint of the compositeness of the resonance. In this light, we discuss the LHC prospects for discovery of the $W'\rightarrow W\gamma$ decay. We also present the prospects for observing the $W'\rightarrow WZ$ decay at the early LHC, and compare the reach in this channel to that in the $W\gamma$ final state. For previous relevant work on the phenomenology of a $W'$ at the LHC, see Refs.~\cite{Schmaltz:2010p2610,Frank:2010p2250,Rizzo:2007bk,Nemevvsek:2011wd}. In Ref.~\cite{Schmaltz:2010p2610}, the early LHC reach on two simple $W'$ models was discussed. The $W'$ we are discussing here differs from the discussion of a right-handed $W'$ in Ref.~\cite{Schmaltz:2010p2610} in two ways: firstly, as already detailed above we adopt an effective approach, without relying on any specific model; secondly, we make the `pessimistic' assumption that the decay of $W'$ into right-handed neutrinos, which was studied in Ref.~\cite{Schmaltz:2010p2610} (see also Ref.~\cite{Nemevvsek:2011wd}), be kinematically closed, and discuss the reach in the dijet and diboson final states. 

This chapter is organized as follows. After introducing the effective Lagrangian in Section~\ref{model-indep approach}, we discuss the early LHC reach on the $W'$ in Section~\ref{secLHC}: in Section~\ref{LHC} we present results for the dijet final state and in Section~\ref{LHCWgamma} we study the $W'\rightarrow W\gamma$ channel and we discuss how it could be used to obtain information on the theoretical nature of the resonance; the complementary search for $W'\rightarrow WZ$ is discussed in Section~\ref{WZsearch}. Finally, a summary of the results is given in Section~\ref{summary}. 

\section{The effective Lagrangian} \l{model-indep approach}
We consider, in addition to the SM field content, a complex spin-1 state transforming as a singlet under color and weak isospin, and with hypercharge equal to unity, according to Eq.~\eqref{111}. The extra vector is therefore electrically charged, with unit charge. We do not make any assumption on the theoretical origin of the extra state, and in particular we do not assume it to be a gauge boson associated with an extended gauge symmetry. Taking a \mbox{model-independent} approach, we write down all the operators up to dimension four describing the interactions between the new vector and the SM fields which are allowed by the $SU(3)_{c}\times SU(2)_{L}\times U(1)_{Y}$ gauge symmetry. Higher Dimensional Operators (HDO) would be suppressed with respect to renormalizable ones by the cutoff of the theory; we neglect them in our analysis. We expect HDO to give corrections roughly of order $M^{2}_{W'}/\Lambda^{2}$ to our results: in Section~\ref{LHCWgamma} we show that the cutoff always satisfies $\Lambda\gtrsim 5M_{W'}$, so we can conservatively estimate our results to hold up to 10 percent corrections due to HDO. Within this framework, we write down the Lagrangian     
\be  \l{lagrangianwprime}
\La_{\text{eff}}=\La_{\text{SM}}+\La_{V}+\La_{V-SM}\,,
\ee
where $\La_{\text{SM}}$ is the SM Lagrangian, and\footnote{To be general, we should also include the operators $V_{\mu}^{+}V^{+\mu}V_{\nu}^{-}V^{-\nu}$ and $V_{\mu}^{+}V^{-\mu}V_{\nu}^{+}V^{-\nu}$; however, these operators only contribute to quartic interactions of vectors and can thus be neglected for the scope of this study. On the other hand, a cubic self-interaction of $V_{\mu}$ is forbidden by gauge invariance.}
\begin{align}
\La_{V}\,=\,& \nonumber D_{\mu}V_{\nu}^{-}D^{\nu}V^{+\mu}-D_{\mu}V_{\nu}^{-}D^{\mu}V^{+\nu}+\tilde{M}^{2}V^{+\mu}V_{\mu}^{-}\\ 
&+\frac{g_{4}^{2}}{2}|H|^{2}V^{+\mu}V_{\mu}^{-}-ig_{B}B^{\mu\nu}V^{+}_{\mu}V^{-}_{\nu}\,,\\ \l{V-SM}
\La_{V-\text{SM}}\,=\,& V^{+\mu}\left(ig_{H}H^{\dagger}(D_{\mu}\tilde{H})+ \frac{g_{q}}{\sqrt{2}}(V_{R})_{ij}\overline{u_{R}^{i}}\gamma_{\mu}d_{R}^{j}\right) +\text{h.c.}\,, 
\end{align}
where we have denoted the extra state with $V^{\pm}_{\mu}$ and we have defined $\tilde{H}\equiv i\sigma_{2}H^{\ast}$. We remark that we have not introduced right-handed neutrinos, in order to avoid making any further assumptions about the underlying model. The coupling of $V_{\mu}$ to left-handed fermionic currents is forbidden by gauge invariance. The covariant derivative is referred to the SM gauge group: for a generic field $\mathcal{X}$, neglecting color we have
\be
D_{\mu}\mathcal{X}=\partial_{\mu}\mathcal{X}-igT^{a}\hat{W}^{a}_{\mu}\mathcal{X}-ig'YB_{\mu}\mathcal{X}\,,
\ee
where $T^{a}$ are the generators of the $SU(2)_{L}$ representation where $\mathcal{X}$ lives, and we have denoted the $SU(2)_{L}$ gauge bosons with a hat, to make explicit that they are gauge (and not mass) eigenstates. In fact, upon EWSB the coupling $g_{H}$ generates a mass mixing between $\hat{W}_{\mu}^{\pm}$ and $V_{\mu}^{\pm}$. This mixing is rotated away by introducing mass eigenstates 
\be
\begin{pmatrix} W^{+}_{\mu} \\
W^{\prime\,+}_{\mu} \end{pmatrix} = \begin{pmatrix}
\cos\hat\theta\,\,\, & \sin\hat\theta \\
-\sin\hat\theta\,\,\, & \cos\hat\theta 
\end{pmatrix} \begin{pmatrix}
\hat{W}_{\mu}^{+} \\
V_{\mu}^{+}
\end{pmatrix}\,.
\ee 
The expression of the mixing angle is
\be
\tan(2\hat{\theta})=\frac{2\Delta^{2}}{m^{2}_{\hat{W}}-M^{2}}\,,
\ee
where 
\be
m_{\hat{W}}^{2}=\frac{g^{2}v^{2}}{4},\quad \Delta^{2}=\frac{g_{H}gv^{2}}{2\sqrt{2}},\quad M^{2}=\tilde{M}^{2}+\frac{g_{4}^{2}v^{2}}{4}\,.
\ee
We assume that Eq.~\eqref{lagrangianwprime} is written in the mass eigenstate basis for fermions. We have written the heavy vector mass explicitly: the details of the mass generation mechanism will not affect our phenomenological study, as long as additional degrees of freedom possibly associated with such mechanism are heavy enough. We assume that the standard redefinition of the phases of the quark fields has already been done in $\La_{\text{SM}}$, thus leaving only one CP-violating phase in the Cabibbo--Kobayashi--Maskawa (CKM) mixing matrix $V_{\text{CKM}}$. The \mbox{right-handed} mixing matrix $V_{R}$ does not need to be unitary in the framework we adopt here: it is in general a complex $3\times 3$ matrix. This is a relevant difference with respect to LR models, where $V_{R}$ must be unitary, as a consequence of the gauging of $SU(2)_{R}$. We normalize $g_{q}$ in such a way that $|\det (V_{R})|=1$ (a generalization of this condition can be applied if $V_{R}$ has determinant zero). Moreover, for all the phenomenological studies we are going to discuss, we assume the least constrained form of $V_{R}$ \cite{Langacker:1989p2578}
\be\l{VR}
V_{R}=\mathbb{I}\,.
\ee
We also note that, for suitable choices of $V_{R}$, our effective approach encompasses the class of $W'$ with flavor-violating couplings that has been recently called for as an explanation of the anomaly observed by CDF in the top pair forward-backward asymmetry \cite{CDFCollaboration:2010tt,Jung:2010fr,Cheung:2009dq,Shelton:2011wf,Barger:1980fx}.

In the mass eigenstate basis both for spin-$1/2$ and spin-1 fields, the charged current interactions for quarks read:  
\be
\La^{q}_{\text{cc}}\,=\,W^{+}_{\mu}\,\overline{u}^{i}\left(\gamma^{\mu}v_{ij}+\gamma^{\mu}\gamma_{5}a_{ij}\right)d^{j}+
W^{\prime\,+}_{\mu}\,\overline{u}^{i}\left(\gamma^{\mu}v^{\prime}_{ij}+\gamma^{\mu}\gamma_{5}a^{\prime}_{ij}\right)d^{j}+\text{h.c.}\,,
\ee
where $u^{i}, d^{j}$ are Dirac fermions, and the couplings have the expressions
\begin{align} \nonumber
v_{ij}=& \frac{1}{2\sqrt{2}}\left(g_{q}\sin\hat\theta(V_{R})_{ij}+g\cos\hat\theta(V_{\text{CKM}})_{ij}\right)\,, \\
a_{ij}=& \frac{1}{2\sqrt{2}}\left(g_{q}\sin\hat\theta(V_{R})_{ij}-g\cos\hat\theta(V_{\text{CKM}})_{ij}\right)\,, \nonumber\\
v^{\prime}_{ij}=& \frac{1}{2\sqrt{2}}\left(g_{q}\cos\hat\theta(V_{R})_{ij}-g\sin\hat\theta(V_{\text{CKM}})_{ij}\right)\,, \nonumber\\
a^{\prime}_{ij}=& \frac{1}{2\sqrt{2}}\left(g_{q}\cos\hat\theta(V_{R})_{ij}+g\sin\hat\theta(V_{\text{CKM}})_{ij}\right)\,. \nonumber
\end{align}
On the other hand, the charged current interactions for leptons have the form
\be 
\La^{\ell}_{\text{cc}}\,=\,W^{+}_{\mu}\cos\hat\theta\frac{g}{\sqrt{2}}\overline{\nu}^{i}_{L}\gamma^{\mu}e^{i}_{L}-W^{\prime\,+}_{\mu}\sin\hat\theta \frac{g}{\sqrt{2}}\overline{\nu}^{i}_{L}\gamma^{\mu}e^{i}_{L}\,.
\ee
The trilinear couplings involving the $W'$, the $W$ and the Higgs and the $W'$ and two SM gauge bosons read
\bes
\begin{align} \nonumber
\La_{W'Wh}&=\Big[-\frac{1}{2}g^{2}vh\sin\hat\theta\cos\hat\theta+\frac{g_{H}g}{\sqrt{2}}vh(\cos^{2}\hat\theta -\sin^{2}\hat\theta)+\frac{g_{4}^{2}}{2}hv\sin\hat\theta\cos\hat\theta\, \Big] \\
&\hspace{4mm}\times (W^{+\,\mu}W_{\mu}^{\prime\,-}+W^{-\,\mu}W_{\mu}^{\prime\,+})\,, \\
\l{W'Wgamma-111}
\La_{W'W\gamma}&=-i\,e(c_{B}+1)\sin\hat\theta \cos\hat\theta F_{\mu\nu}(W^{+\,\mu}W^{\prime\,-\,\nu}+W^{\prime\,+\,\mu}W^{-\,\nu})\,, \\
\La_{W'WZ}&=\, i\sin\hat\theta  \cos\hat\theta\Big[(g\cos\theta_{W}+g'\sin\theta_{W})(W^{-\,\mu}W^{\prime\,+}_{\nu\mu}+W^{\prime\, -\, \mu}W^{+}_{\nu\mu}- W^{\prime\,+\, \mu}W^{-}_{\nu\mu} \nonumber\\ \l{W'WZvertex}
&\hspace{4mm} -W^{+\,\mu} W^{\prime\,-}_{\nu\mu})Z^{\nu} -(g\cos\theta_{W}-g'\sin\theta_{W}c_{B})\left(W^{+\,\mu} W^{\prime\,-\,\nu}+ W^{\prime\,+\,\mu} W^{-\,\nu}\right) Z_{\mu\nu}\Big]\,,  
\end{align}
\ees
where $\theta_{W}$ is the weak mixing angle.

In summary, in addition to the $W'$ mass, 4 couplings appear in our phenomenological Lagrangian: $g_{q}$, $g_{H}$ (or equivalently the mixing angle $\hat\theta$), $g_{B}$ and $g_{4}$. We find it useful to normalize $g_{B}$ to the SM hypercharge coupling, so we will refer to $c_{B}\equiv g_{B}/g'$ in what follows. Our phenomenological Lagrangian describes the low energy limit of a LR model\footnote{Here we are assuming the $Z'$ to be sufficiently heavier than the $W'$, and we are neglecting effects coming from a different scalar spectrum.} for the following values of the parameters (see Ref.~\cite{Grojean:2011wi}):
\be
\begin{array}{llll} \l{LRmodelcorrespondence}
g=g_{L}\,, \qquad &\displaystyle g^{\prime}=\frac{g_{X}g_{R}}{\sqrt{g_{X}^{2}+g_{R}^{2}}}\,, \qquad &g_{q}=g_{R}\,, \qquad &\displaystyle g_{H}=-2\sqrt{2}g_{R}\frac{kk'}{v^{2}}\,,\\\\
c_{B}=-1\,,\qquad &\displaystyle g_{4}^{2}=2g_{R}^{2}\frac{k^{2}+k^{\prime\,2}}{v^{2}}\,, \qquad &\displaystyle \tilde{M}^{2}=\frac{g_{R}^{2}v_{R}^{2}}{4}\,,\qquad  &v^{2}=v_{L}^{2}+2(k^{2}+k^{\prime\,2})\,.  
\end{array}
\ee
where $g_{L}$, $g_{R}$, $g_{X}$ are the gauge couplings of the $SU\(2\)_{L}\times SU\(2\)_{R}\times U\(1\)_{X}$ gauge groups, $k$ and $k'$ are the VEVs of the $SU\(2\)_{L}\times SU\(2\)_{R}$ bi-doublet which breaks the EW symmetry and $v_{L}$ and $v_{R}$ are the VEVs respectively of an additional $SU\(2\)_{L}$ scalar doublet and of an $SU\(2\)_{R}$ scalar doublet which realizes the $SU\(2\)_{R}\times U\(1\)_{X}\to U\(1\)_{Y}$ symmetry breaking.

Using the Lagrangian \eqref{lagrangianwprime} we can compute the two body widths of the $W'$. Defining
\begin{equation*}
p=\frac{1}{2M_{W'}}\sqrt{M_{W'}^{4}+M_{1}^{4}+M_{2}^{4}-2M_{W'}^{2}M_{1}^{2}-2M_{W'}^{2}M_{2}^{2}-2M_{1}^{2}M_{2}^{2}}\,,
\end{equation*}
with $M_{1,2}$ the masses of the final state particles, we find:
\bes
\begin{align} 
& \bry{lll}
\dst \Gamma(W^{\prime\,+}\rightarrow u^{i}\overline{d}^{j})&=&\dst \frac{p}{2\pi M_{W'}^{2}}\Big[|(v^{\prime})_{ij}|^{2}(3\sqrt{m_{d}^{2}+p^{2}}\sqrt{m_{u}^{2}+p^{2}}+3m_{d}m_{u}+p^{2}) \\
&&\dst +|(a^{\prime})_{ij}|^{2}(3\sqrt{m_{d}^{2}+p^{2}}\sqrt{m_{u}^{2}+p^{2}}-3m_{d}m_{u}+p^{2})\Big]\,,\ery\\\notag\\
& \bry{lll}
\dst \Gamma(W^{\prime\,+}\to \nu^{i}\overline{e}^{i})&=&\dst \frac{M_{W'}}{48\pi}g^{2}\sin^{2}\hat{\theta}\,.\ery\\\notag\\
& \bry{lll}
\dst \Gamma(W'\rightarrow W\gamma)&=&\dst \frac{e^{2}}{96\pi}(c_{B}+1)^{2}\sin^{2}\hat\theta\cos^{2}\hat\theta \left(1-\frac{M^{2}_{W}}{M^{2}_{W'}}\right)^{3}\left(1+\frac{M^{2}_{W'}}{M^{2}_{W}}\right)M_{W'}\,,\l{widthWgamma}\ery\\\notag\\
&\bry{lll}
\dst \Gamma(W^{\prime}\rightarrow WZ)&=&\dst \frac{p}{8\pi M_{W'}^{2}}\frac{g^2\cos^{2}\theta_{W}}{3}\sin^{2}\hat\theta\cos^{2}\hat\theta\times \mathcal{T}(M_{W'}^{2},M^{2}_{Z},M^{2}_{W}; c_{B})\,,\ery\\\notag\\
& \bry{lll}
\dst \Gamma(W^{\prime}\rightarrow Wh)&=&\dst \frac{p}{8\pi M_{W'}^{2}}\frac{v^{2}}{3}K^{2}\left(3+\frac{p^{2}}{M_{W}^{2}}\right)\,,\ery
\end{align}
\ees
where we have defined
\be
K=\frac{g_{4}^{2}-g^{2}}{2}\sin\hat{\theta}\cos\hat\theta+\frac{g_{H}g}{\sqrt{2}}(\cos^{2}\hat\theta-\sin^{2}\hat\theta)\,,
\ee
\small\begin{align*}
\mathcal{T}&(M_{W'}^{2},M^{2}_{Z},M^{2}_{W}; c_{B})= \frac{1}{M_W^2 M_Z^2}p^2 \Bigg\{\tan^2\theta_{W} \Bigg[\tan^2\theta_{W} \Big[c_B^2 M_Z^2 (4 E_W E_Z+3 M_Z^2+4 p^2)\\ &+6 M_{W'} (E_W+E_Z)(-c_B M_Z^2+M_W^2)+M_W^2 (2 (-c_B (-2 c_B+3)+2) M_Z^2+4 E_W E_Z+4 p^2)\\ &+M_{W'}^2 (2 E_W E_Z+3 M_W^2+3 M_Z^2+2 p^2)+3 M_W^4\Big]+2 \Big[3 M_{W'} (E_W+E_Z) ((-c_B+1) M_Z^2\\ &+2 M_W^2)+M_W^2 (7 (-c_B+1) M_Z^2 +4 E_W E_Z+4 p^2)-c_B M_Z^2 (4 E_W E_Z+3 M_Z^2+4 p^2)\\ &+M_{W'}^2 (2 E_W E_Z+3 M_W^2+3 M_Z^2+2 p^2) +3 M_W^4\Big]\Bigg]+M_{W'}^2 \Big[2 (E_W E_Z+p^2)+3 M_W^2\\ &+3 M_Z^2\Big]+6 M_{W'} (E_W+E_Z) (M_W^2+M_Z^2) +2 M_W^2 \Big[2 (E_W E_Z+p^2)+7 M_Z^2\Big]\\ &+M_Z^2 \Big[4 (E_W E_Z+p^2)+3 M_Z^2\Big]+3 M_W^4\Bigg\}\,,
\end{align*}\normalsize
and $E_{W,Z}\equiv\sqrt{M^{2}_{W,Z}+p^{2}}$. Note that in the limit $M_{W'}\gg M_{W,\,Z}$, the latter expression simplifies to
\begin{equation*}
\mathcal{T}(M^{2}_{W'}\gg M_{W,\,Z}^{2})\approx \frac{M^{6}_{W'}}{4M^{2}_{Z}M^{2}_{W}}(1+\tan^{2}\theta_{W})^{2}\,,
\end{equation*}
which is independent of $c_{B}$.

The dependence of the ratio $\Gamma_{W'}/M_{W'}$ on the coupling ratio $g_{q}/g$ is plotted in the left panel of Fig.~\ref{fig:totalwidth}, while the BRs as functions of $M_{W'}$ are shown in the right panel of the same figure, for representative values of the parameters.
\begin{figure}[t]
\includegraphics[scale=0.34]{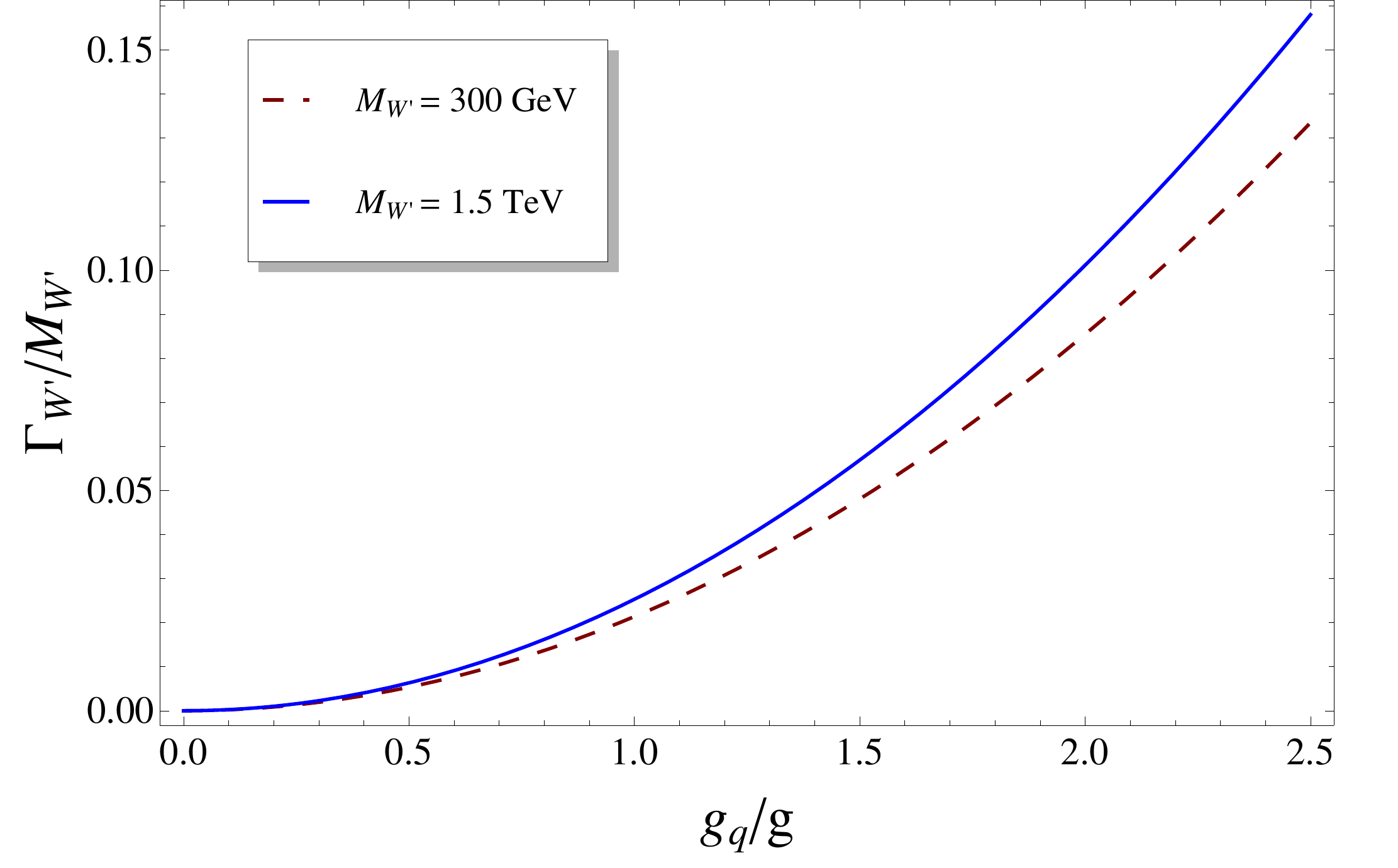}
\hspace{1mm}\includegraphics[scale=0.34]{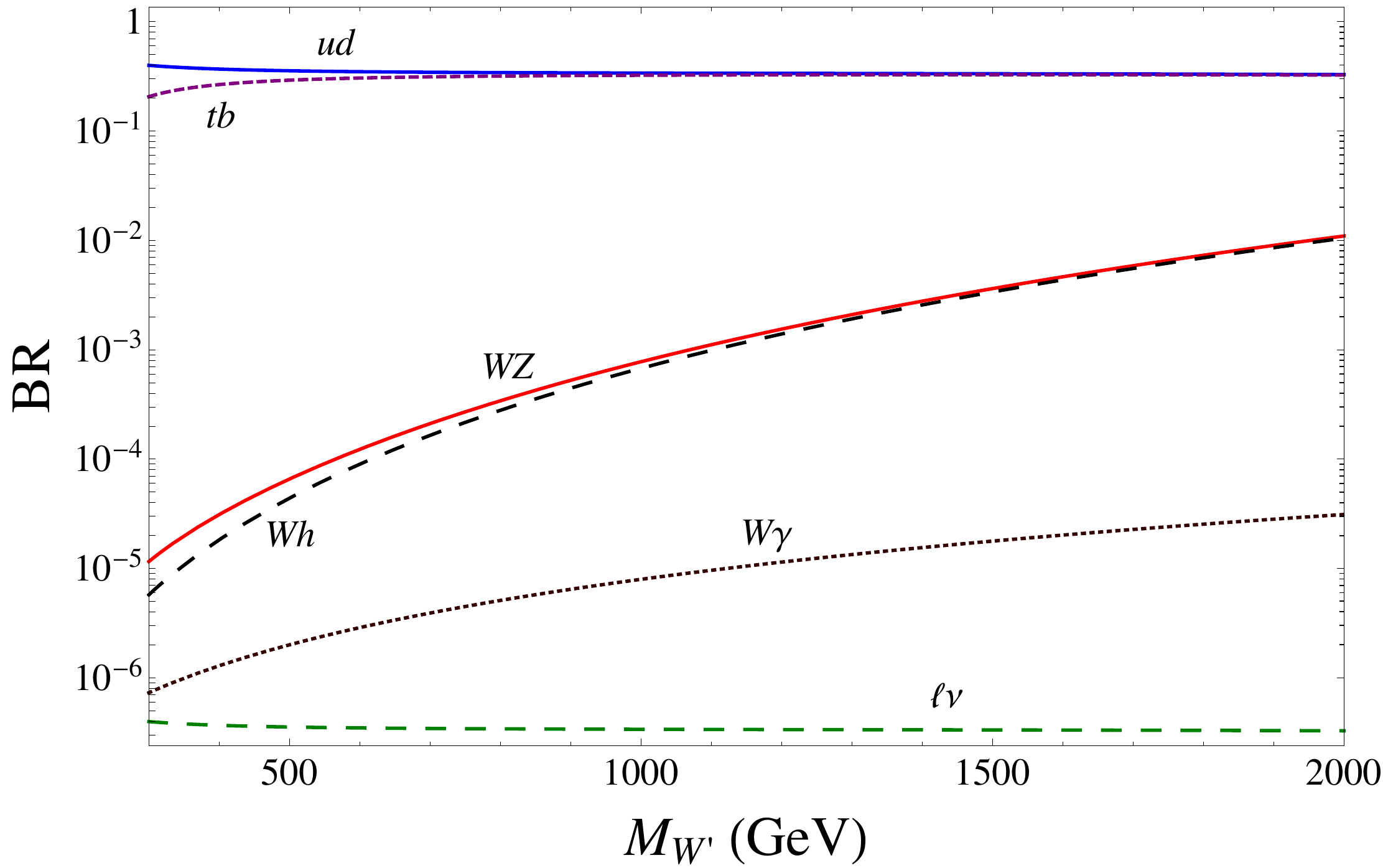}
\caption{\textsl{Left panel.} $W'$ width over mass ratio as a function of $g_{q}/g$  for negligible mixing, $\hat\theta \approx 0$, for $M_{W'}=300\,\mathrm{GeV}$ (dashed, red) and 1.5 TeV (blue). \textsl{Right panel.} BRs of the $W'$ as a function of its mass, for the following choice of the remaining parameters: $g_{q}=g$, $\hat\theta=10^{-3}$, $c_{B}=-3$, $g_{4}=g$. From top to bottom: $ud$, $tb$, $WZ$, $Wh$, $W\gamma$, $\ell\nu$ (the latter includes all the three lepton families).}
\l{fig:totalwidth}
\end{figure}

The present bounds on the $W'$ we are considering have been discussed in Ref.~\cite{Grojean:2011wi}. We will always assume, in the discussion of the phenomenology, that such bounds are satisfied\footnote{We will exceptionally relax the bounds on $\hat{\theta}$ coming from EWPT in discussing the $W'\to W \gamma$ channel.}.

\section{LHC phenomenology} \l{secLHC}

In this section we discuss the reach of the early LHC on the composite $W'$ introduced in the previous section. We analyze first the prospects for discovery of the resonance as an excess of events in the dijet invariant mass spectrum, and subsequently move on to discuss decays into two gauge bosons. We study first the $W'\rightarrow W\gamma$ decay, which is of special interest since it is strongly suppressed in gauge models. As a consequence, its observation would be a hint of the compositeness of the $W'$. Finally, we discuss the $W'\rightarrow WZ$ channel.   

\subsection{Dijet searches} \l{LHC}

The search for resonances in the dijet mass spectrum is one of the first new physics analyses performed by the CMS \cite{CMSCollaboration:2010p2315} and ATLAS \cite{ATLASCollaboration:2010p1746,ATLASCollaboration:2010wv,ATLASCollaboration:2011vna} experiments at the LHC, with an integrated luminosity of 2.9 and 3.1 (36) pb${}^{-1}$ respectively at 7 TeV. Due to the very small data sample analyzed so far, such searches are not competitive yet with those performed at the Tevatron: from Fig.~\ref{fig:LHCcouplingVSmass} we see that only in a very narrow interval around $M_{W'}\sim 500$ GeV does the CMS search place a meaningful (even if weaker than the Tevatron one) upper limit on the $W'$ coupling to quarks. For larger masses, the CMS upper bound on the $W'$ cross section is saturated for values of the coupling $g_{q}>2g$, which implies that the width of the resonance is larger than the dijet mass resolution, and as a consequence the experimental analysis would need to be modified to account for a broad resonance. We have only used the CMS results for the same reasons discussed in the previous chapter (see e.g. the Footnote~\ref{notaCMSATLAS}).

Future LHC analyses, however, will soon overtake the Tevatron results, so it is interesting to discuss the reach of dijet searches on the $W'$ we are considering. We assume the CMS kinematic cuts, namely on the pseudorapidity $|\eta|<2.5$ of each jet, and on the pseudorapidity difference $|\Delta\eta|<1.3$ \cite{CMSCollaboration:2010p2315}. For values of $M_{W'}$ between \mbox{300 GeV} and 2.6 TeV, in intervals of 100 GeV, we compute as a function of the coupling $g_{q}$ the integral of the signal differential invariant mass distribution $d\sigma_{S}/dM_{jj}$ over the region $M_{jj}>M_{W'}(1-\epsilon/2)$, and compare the result with the integral of the background distribution over the same range, to obtain $5\sigma$ discovery and 95$\%\, \mathrm{CL}$ exclusion contours in the $(M_{W'},g_{q}/g)$ plane. Here $\epsilon$ is the dijet mass resolution, which following Ref.~\cite{CMSCollaboration:2010p2315} we assume to vary from 8$\%$ at $M_{W'}=500$ GeV to 5$\%$ at $2.5\,\mathrm{TeV}$. The results are shown in Fig.~\ref{fig:LHCcouplingVSmass} for three different integrated luminosities, namely \mbox{$L=\int\La=0.1,1,5$ fb${}^{-1}$}, and for two LHC c.o.m. energies, namely 7 and 8 TeV\footnote{It has recently been decided that the LHC will run at 7 TeV in 2011. However, a higher energy for 2012 cannot be excluded at the time of writing \cite{Myers:v84LmGZE}.}. We find that \mbox{100 pb${}^{-1}$} are not sufficient for a discovery, even at 8 TeV (except perhaps for a very small region around $M_{W'}=1\,\mathrm{TeV}$). On the other hand, if we focus on the exclusion contours, we see that the LHC can do better than the Tevatron already with 100 pb${}^{-1}$ for \mbox{$M_{W'}\gtrsim 700\,\text{GeV}$}, and for essentially all $W'$ masses if the luminosity is increased to $1\,\text{fb}^{-1}$.
We also report in Fig.~\ref{fig:LHCluminosityVSmass}, as a function of $M_{W'}$, the integrated luminosity needed for discovery or exclusion of a $W'$ with coupling to quarks equal to that of the SM $W$($g_{q}=g$), both for the 7 and 8 TeV LHC.  
 
We choose to compare the integrals over $M_{jj}>M_{W'}(1-\epsilon/2)$ of the signal and background differential dijet mass distributions rather than their integrals in a finite interval centered on the $W'$ mass, because the former method is less sensitive to smearing effects generated by hadronization and jet reconstruction, which we cannot take into account in our parton-level analysis. In this way, we expect our estimate of the reach of the early LHC to be closer to the actual experimental results than it would be if we compared signal and background in an interval centered around the $W'$ mass.  
\begin{figure}[t]
\includegraphics[scale=0.34]{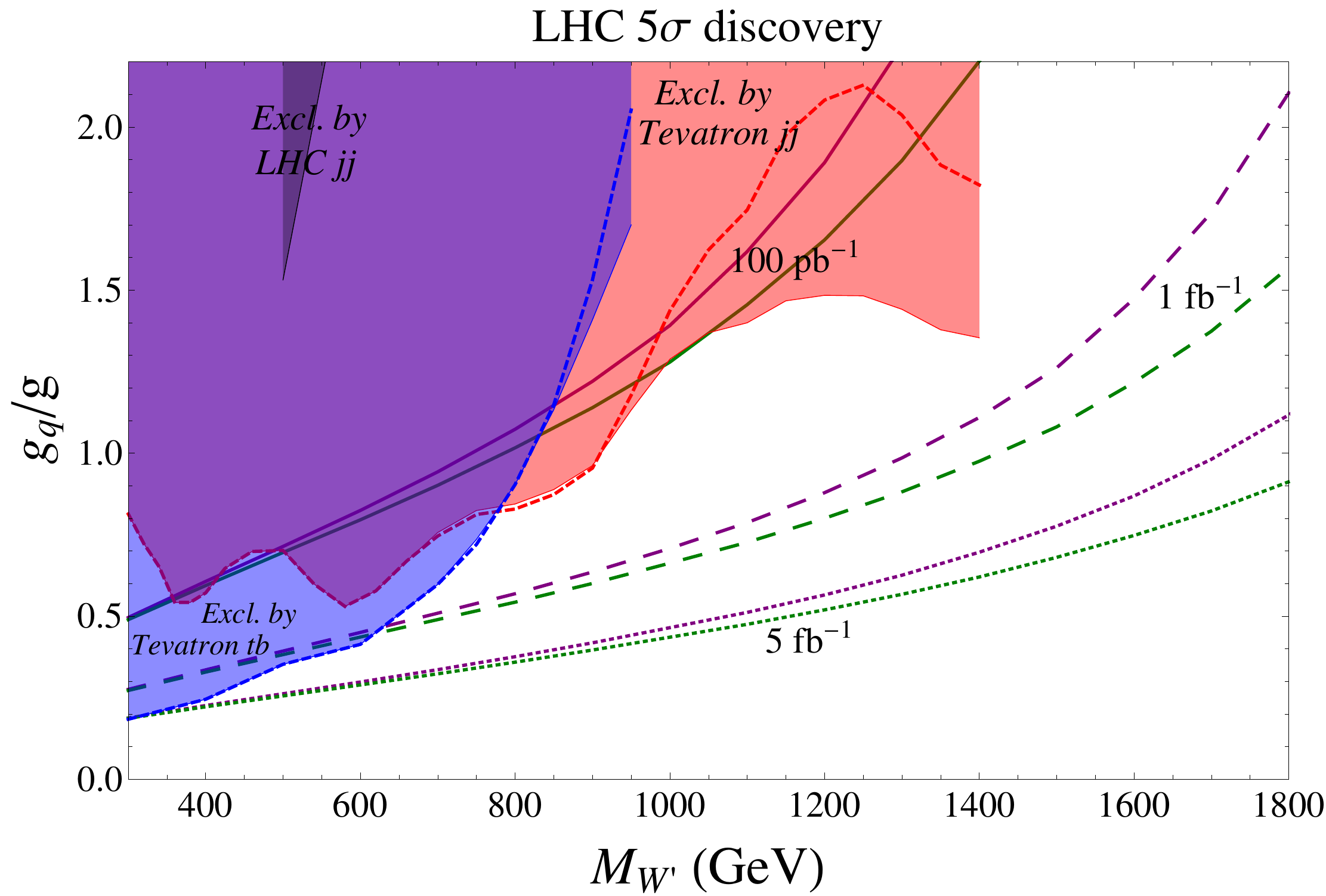}
\hspace{2mm}\includegraphics[scale=0.33]{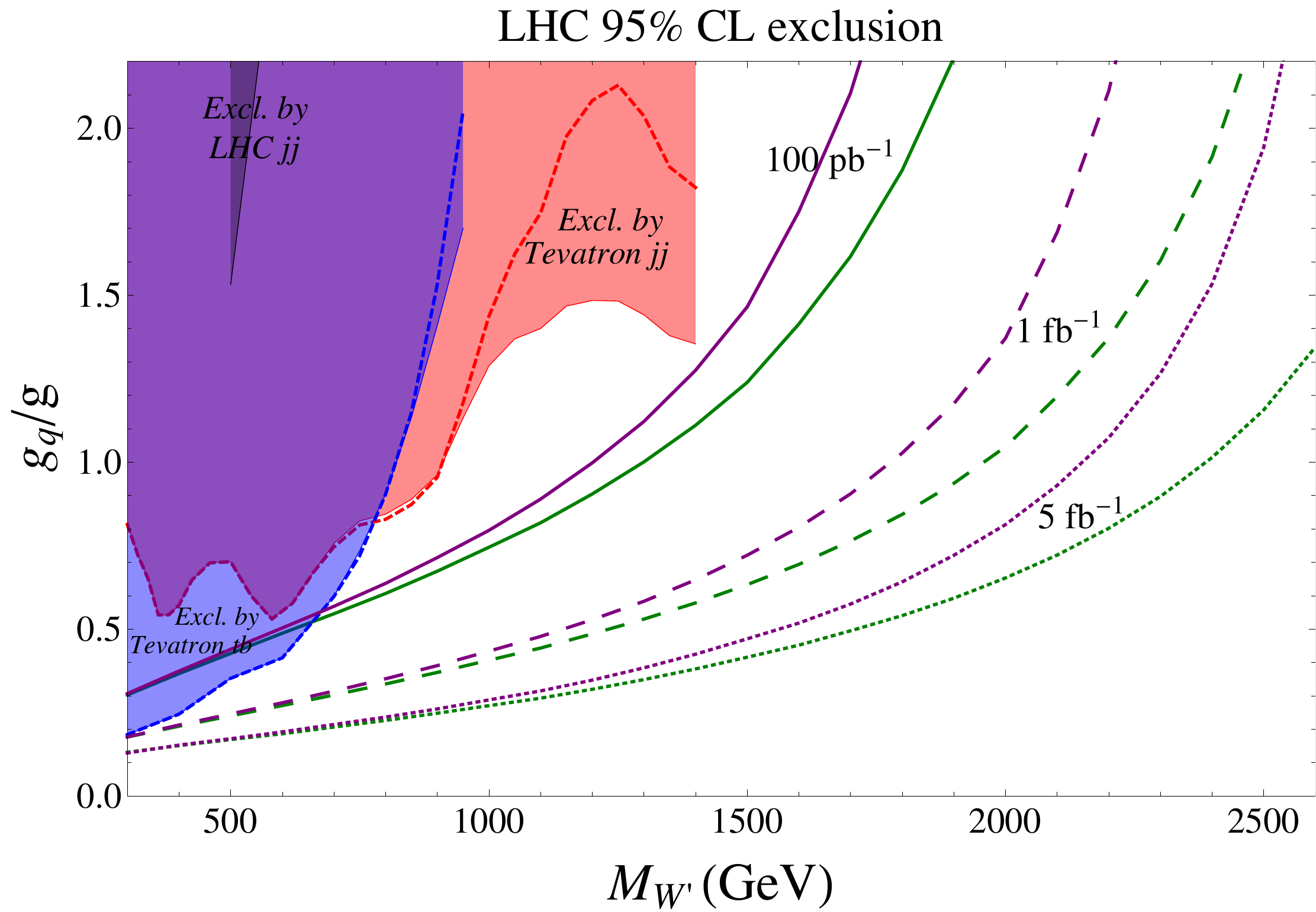}
\caption{Contours in the $(M_{W'},g_{q}/g)$ plane for $5\sigma$ discovery (left) and $95\%$ CL exclusion (right) at the 7 TeV and 8 TeV LHC, for an integrated luminosity of $L=\int \mathcal{L}dt =0.1,\,1$ and $5$ fb${}^{-1}$, corresponding to the continuous, dashed and dotted lines, respectively (for each different dashing, the upper, purple line is for $7$ TeV and the lower, green line is for $8$ TeV).  Also shown are the Tevatron dijet (red) and $tb$ (blue) exclusions, together with the CMS exclusion with $2.9$ pb${}^{-1}$ (grey).}
\l{fig:LHCcouplingVSmass}
\end{figure}
\begin{figure}[t]
\includegraphics[scale=0.335]{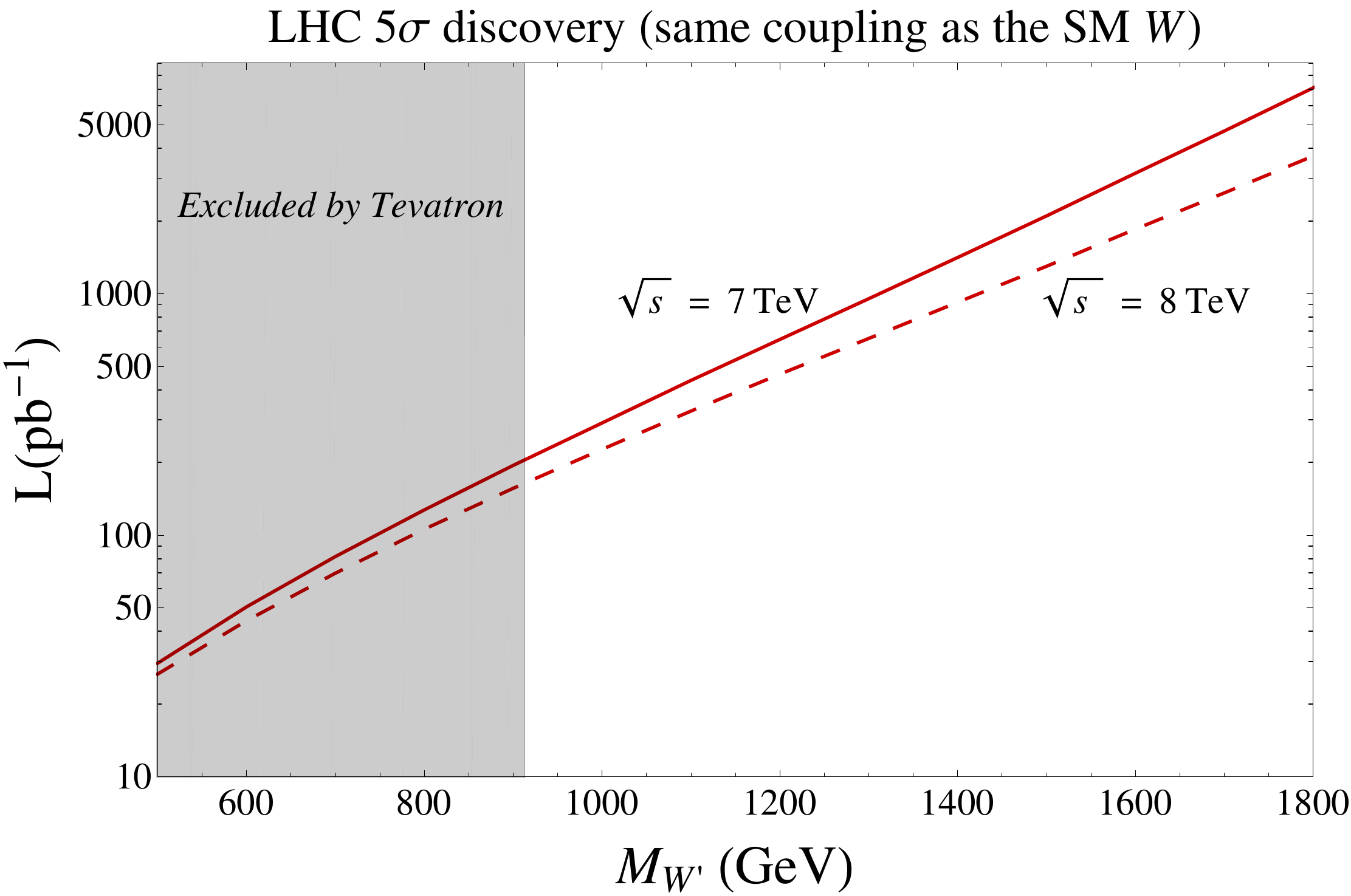}
\hspace{3mm}\includegraphics[scale=0.33]{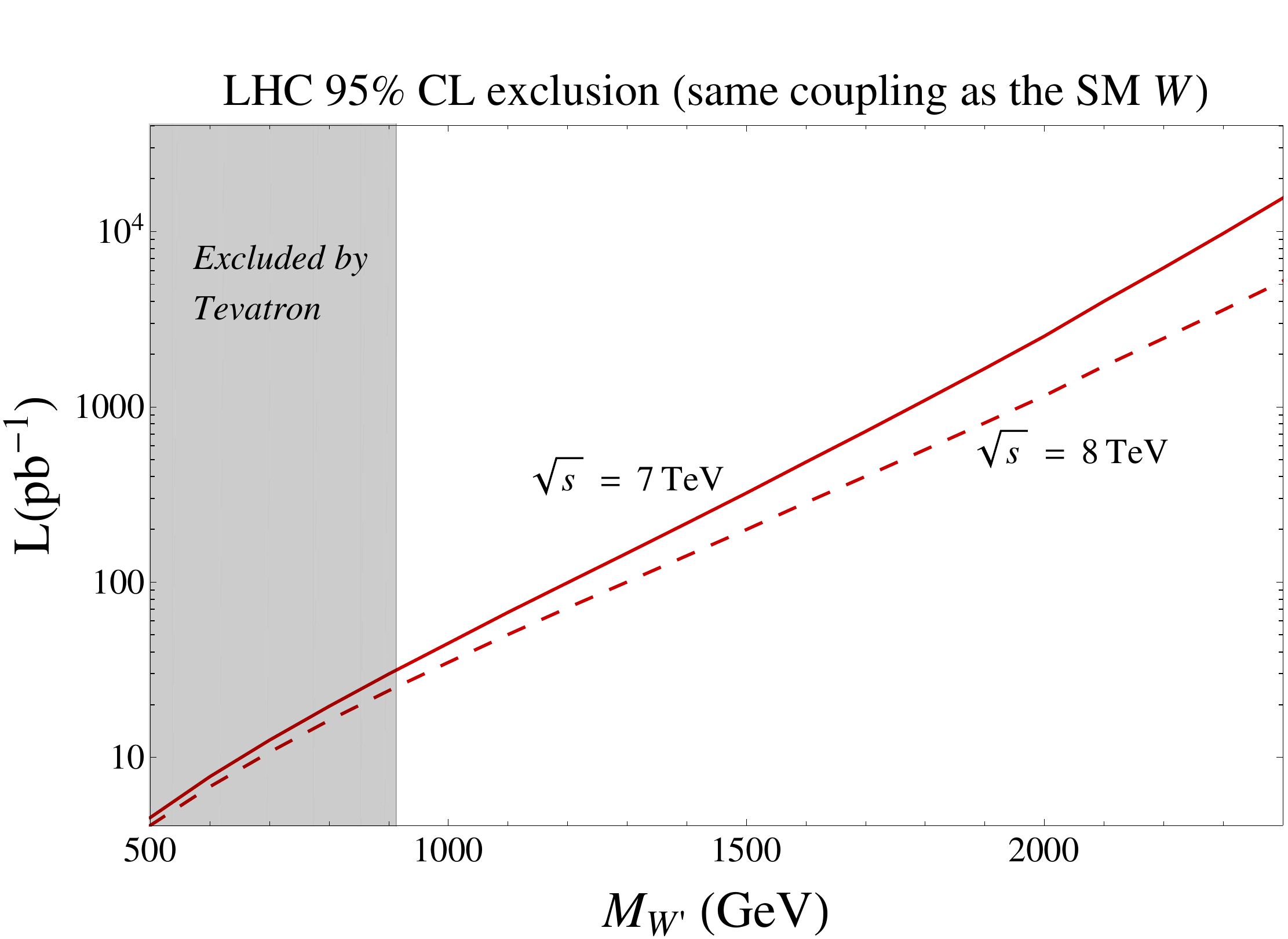}
\caption{Integrated luminosity needed for $5\sigma$ discovery (left) and $95\%$ CL exclusion (right) as a function of the $W'$ mass, for the 7 TeV (continuous) and 8 TeV (dashed) LHC. The region shaded in grey, corresponding to $M_{W'}<913$ GeV, is excluded at $95\%$ CL by Tevatron searches \cite{CDFCollaboration:2009hy}.}
\l{fig:LHCluminosityVSmass}
\end{figure}
In addition to those in the dijet final state, also the LHC searches in the $tb$ channel will be of course relevant to the $W'$ we are studying. We do not discuss them here, and refer the reader to the recent, extensive analysis of Ref.~\cite{Gopalakrishna:2010p2723}.

\subsection{Search for $W'\rightarrow W\gamma$} \l{LHCWgamma}

We now move on to consider decay channels of the $W'$ which have partial widths proportional to the $W$-$W'$ mixing angle $\hat{\theta}$. These include $WZ,\,Wh$ and $W\gamma$ final states. We will focus first on the last channel, which is of special interest since it is very suppressed in the gauge models containing a $(\mathbf{1},\mathbf{1})_{1}$ $W'$, such as for instance LR models. Therefore, observation of $W'\to W\gamma$ would point to a composite nature of the $W'$. The partial width for decay into $W\gamma$ reads
\be \l{widthWgamma}
\Gamma(W'\rightarrow W\gamma)=\frac{e^{2}}{96\pi}(c_{B}+1)^{2}\sin^{2}\hat\theta\cos^{2}\hat\theta \left(1-\frac{M^{2}_{W}}{M^{2}_{W'}}\right)^{3}\left(1+\frac{M^{2}_{W'}}{M^{2}_{W}}\right)M_{W'}\,.
\ee
Since the width for decay into this channel is controlled by $\hat\theta$ and $c_{B}$, it is interesting to estimate which values of these parameters will be accessible to the LHC in its first run. To assess the discovery potential, we choose two benchmark values for the $W'$ mass, namely 800 and 1200 GeV, and we assume two representative values of the integrated luminosity, namely 1 and 5 fb${}^{-1}$, at a c.o.m.~energy $\sqrt{s}=7$ TeV. We set the coupling to quarks to $g_{q}=0.84\,(1.48)g$ for $M_{W'}=800\,(1200)$ GeV, that is, to the largest value allowed by Tevatron $jj$ and $tb$ searches (see Fig.~\ref{fig:LHCcouplingVSmass}). Notice that the upper limit on $g_{q}$ from Tevatron searches in quark final states was computed for $\hat{\theta}=0$; when the mixing is introduced, the bound on the coupling weakens, due to the smaller BR of the resonance into quarks. 

A direct constraint on the mixing angle $\hat\theta$ comes from the non-observation of resonances decaying into $WZ$ in a search performed by the D0 collaboration \cite{DCollaboration:2010td}: we take such constraint into account in our analysis for the $W\gamma$ final state. On the other hand, the CDF Collaboration has performed a search in the $\ell\gamma\MET$ ($\ell=e,\mu$) final state \cite{CDFCollaboration:2007dj}, without observing any discrepancies with the SM prediction. Also the constraints coming from this channel were taken into account; however, they turn out to be less stringent than those obtained from the $WZ$ channel, because of the smaller dataset analyzed.

We select decays of the $W$ into an electron and a neutrino, and apply the following cuts on the $e\gamma\MET$ final state: $p_{T}^{\gamma}>250\,(400)$ GeV, $p_{T}^{e}>50$ GeV, $\MET >50\,\,\textrm{GeV}$, $|\eta_{e,\gamma}|<2.5$, and $|M(W\gamma)-M_{W'}|<0.05\,(0.10)M_{W'}\,$, for $M_{W'}=800\,(1200)$ GeV. We note that, even though the neutrino longitudinal momentum $p_{z}^{\,\nu}$ is not measured experimentally, it can be reconstructed by imposing that the lepton and neutrino come from an on-shell $W$: a quadratic equation for $p_{z}^{\,\nu}$ is thus obtained. It follows that a criterion must be chosen to unfold this ambiguity. The assessment of the effects of such choice on the cuts on $\MET$ and on the total invariant mass $M(W\gamma)$ goes beyond the scope of this work, and we leave it to the experimental collaborations\footnote{In this regard, we also note that, at the detector level, fluctuations in the measured $\MET$ can lead to events where no solution for $p_{z}^{\,\nu}$ can be found even though the lepton and neutrino come from the decay of a $W$ (see, e.g., the section on top quark mass measurements in Ref.~\cite{ATLASCollaboration:2008ut}).}. We neglect the interference between $W$ and $W'$, which is due to the $O(\hat\theta)$ coupling of $W'$ to left-handed quark currents. The main background process is the SM $W\gamma$ production, which we include in our analysis, while we leave out the $W+j$ production with the jet misidentified as a photon. We have checked that applying the rejection factor for misidentification into a $\gamma$ of very high-$p_{T}$ jets, which is of the order of $5\times 10^{3}$ if photon identification and isolation cuts are applied (see, e.g., Ref.~\cite{ATLASCollaboration:2008ut}), the $W+j$ background contribution is roughly one order of magnitude smaller than the irreducible $W\gamma$ process. This estimate suffers from the fact that we are not including NLO corrections to $W+j$, and from the fact that requiring photon identification and isolation has an efficiency of $\sim80\%$ on `real' photons \cite{ATLASCollaboration:2008ut}, which would slightly reduce the number of signal events detected. Other possibly relevant instrumental backgrounds that we do not include in our exploratory study are $ee\MET$ with $e$ misidentified as a photon, and QCD jets faking $e+\MET$. We leave the proper treatment of such detector-dependent backgrounds to the experimental analyses; we just note that doubling the statistics by including also the $W\rightarrow \mu\nu$ channel would help in balancing the sensitivity loss, in case the sum of instrumental backgrounds -- such as those mentioned above -- happened to be of the same order of magnitude of the irreducible $W\gamma$ background (for example, in the D0 $\ell\gamma\MET$ search, the total background was estimated to be roughly twice as large as the irreducible $W\gamma$, see Ref.~\cite{CDFCollaboration:2007dj}).

In Fig.~\ref{fig:LHCWgammaDISTRIB}, we show the distributions of the reconstructed invariant mass of the $W'$ and of the $p_{T}$ of the photon, compared to the SM $W\gamma$ background. We stress that experimentally, reconstruction of the longitudinal component of the neutrino momentum by imposing the on-shell condition for the $W$ will have an impact on the resolution of the $W'$ invariant mass. From the invariant mass distribution, it is also evident that for the values of the parameters chosen, the $W'$ of mass 1.2 TeV has a quite large width, which motivated the use of a broader cut around the peak, as discussed above. While the number of events predicted at the early LHC is clearly small, these distributions can be used as a guideline also for searches at higher integrated luminosity, after rescaling cross section to higher LHC c.o.m.~energy. 
\begin{figure}[t]
\centering
\begin{minipage}{20cm}
\includegraphics[scale=0.36]{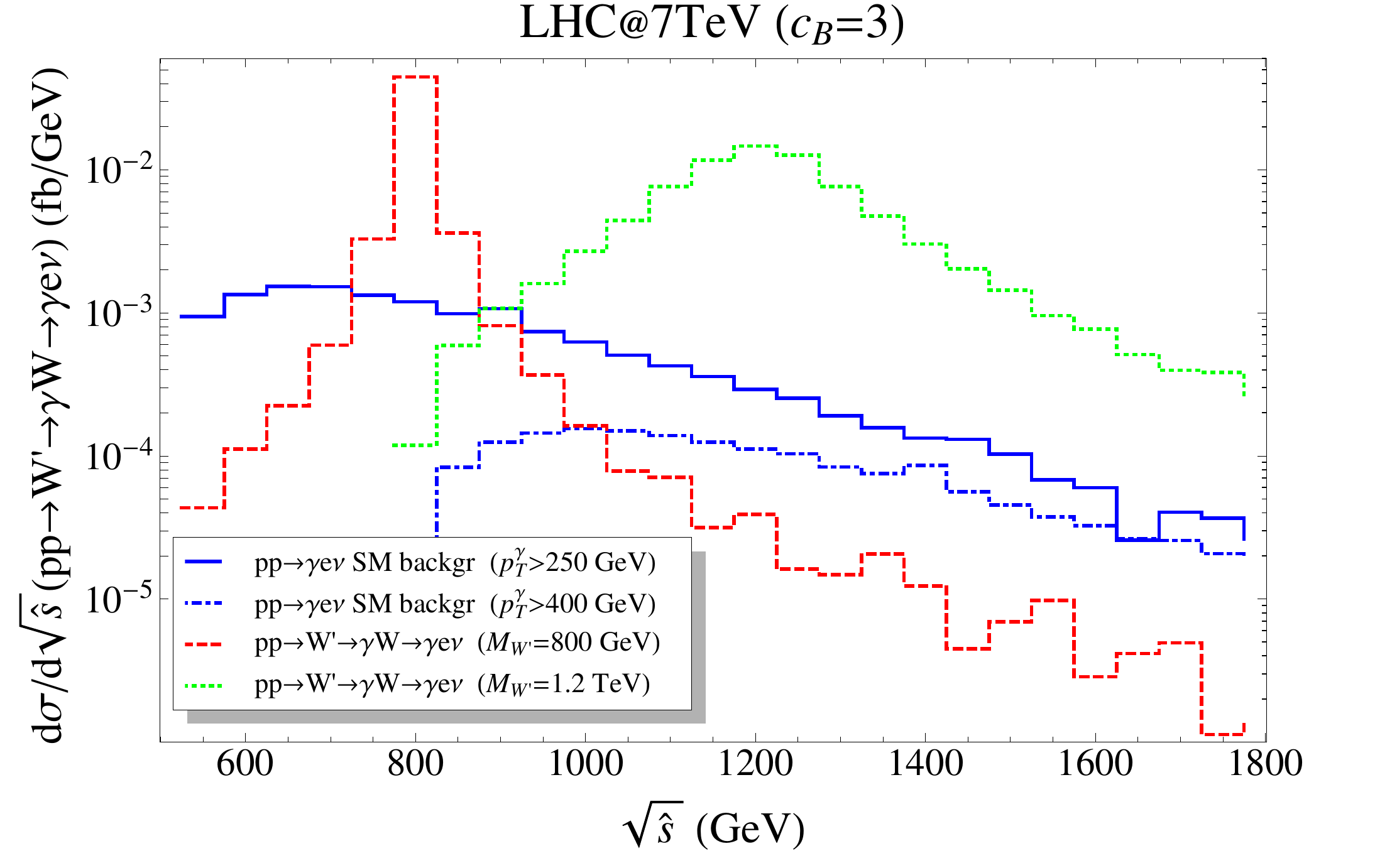}\hspace{-3mm}
\includegraphics[scale=0.36]{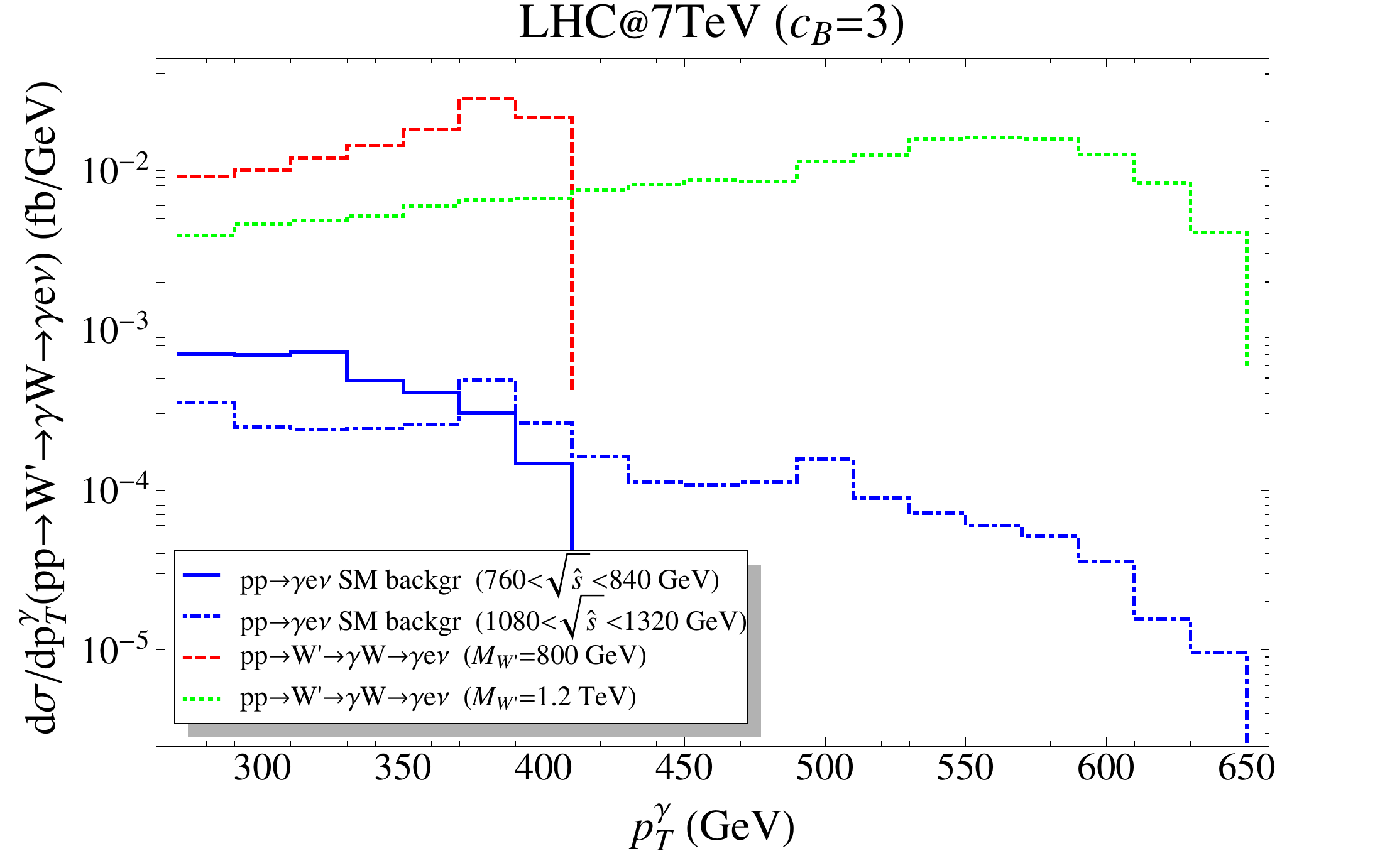}
\end{minipage}
\caption{Invariant mass (left) and photon $p_{T}$ (right) distributions for the \mbox{$W'\rightarrow W\gamma\rightarrow e\nu\gamma$} signal and for the irreducible background. The values of the couplings are as follows: \mbox{$g_{q}=0.84g$} and $\hat\theta = 10^{-2}$ for $M_{W'}=800\,\mathrm{GeV}$, and $g_{q}=1.48g$ and $\hat\theta =4\times 10^{-2}$ for $M_{W'}=1.2\,\mathrm{TeV}$.}
\l{fig:LHCWgammaDISTRIB}
\end{figure}

Our main results are shown in Fig.~\ref{fig:LHCWgamma}. As can be read off the left side of the figure, for $M_{W'}=800$ GeV, assuming $c_{B}=5$ (which corresponds to $g_{B}=5g'\sim 1.8$), the interval $5\times 10^{-3}<\hat\theta<1.25\times 10^{-2}$ is accessible for a discovery with 5 fb${}^{-1}$. Such values of $\hat\theta$, while being excluded by EWPT if we assume the $W'$ is the only new physics contributing to precision data, are however allowed by $u\rightarrow d$ and $u\rightarrow s$ transitions if the CP phases are not small \cite{Grojean:2011wi}. It is conceivable that a positive contribution to the $T$ parameter coming from additional new physics (such as, for example, a heavy neutral spin-1 state) relaxes the bound from EWPT, allowing for such relatively large values of $\hat\theta$. On the other hand, from the right side of Fig.~\ref{fig:LHCWgamma} we see that setting the mixing angle to the value $\hat\theta = 10^{-2}$, discovery of a $W'$ with mass \mbox{$800$ GeV} is possible with 5 fb$^{-1}$ for $c_{B}\gtrsim 2$, which corresponds to a moderate value of the coupling $g_{B}\sim 0.7$. The prospects for a heavier $M_{W'}=1200$ GeV are similar, except that in this case there is no relevant bound from Tevatron searches. 
\begin{figure}[h]
\centering
\begin{minipage}{20cm}
\includegraphics[scale=0.37]{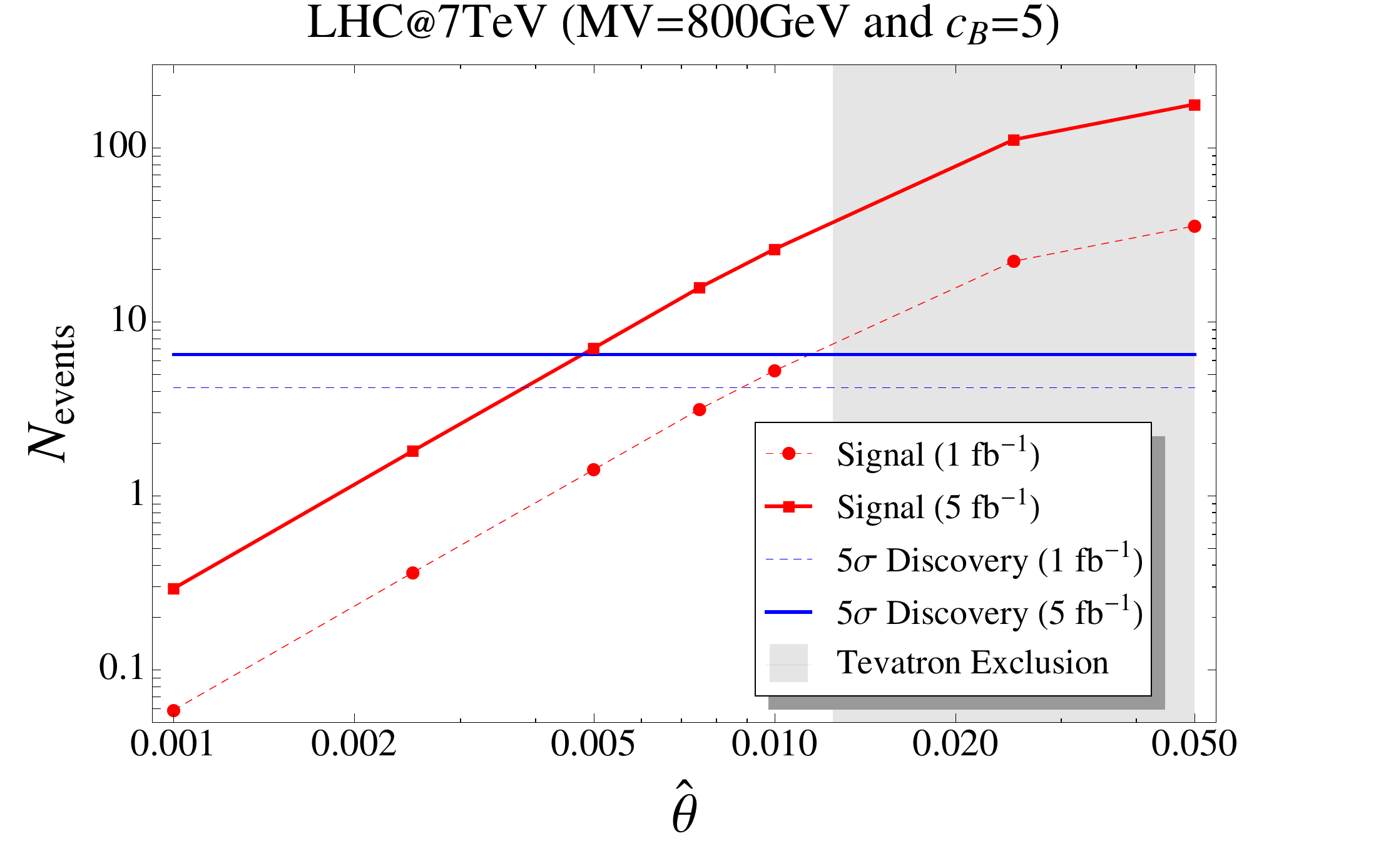}\hspace{-5mm}
\includegraphics[scale=0.37]{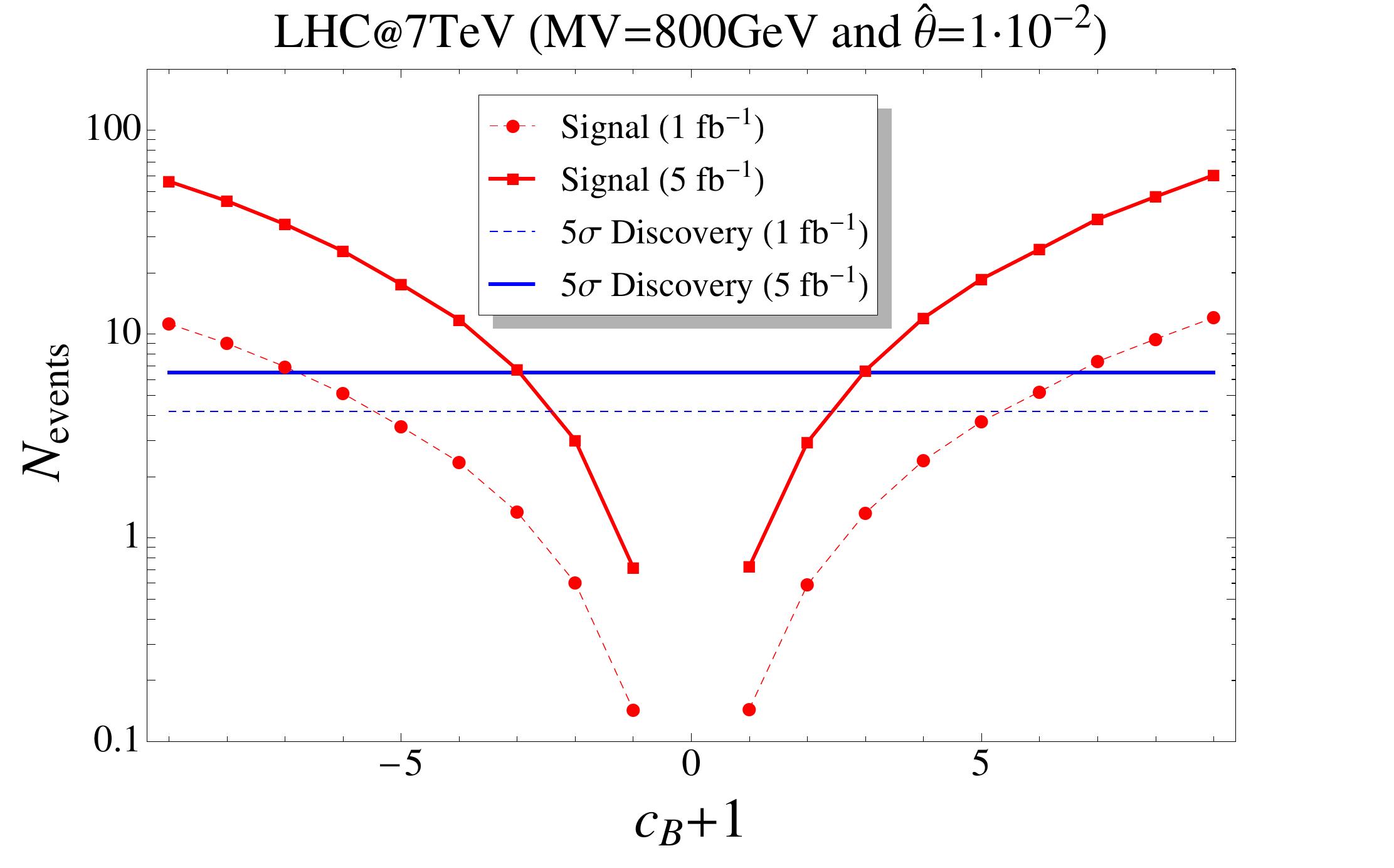} \\
\end{minipage}
\begin{minipage}{20cm}
\includegraphics[scale=0.37]{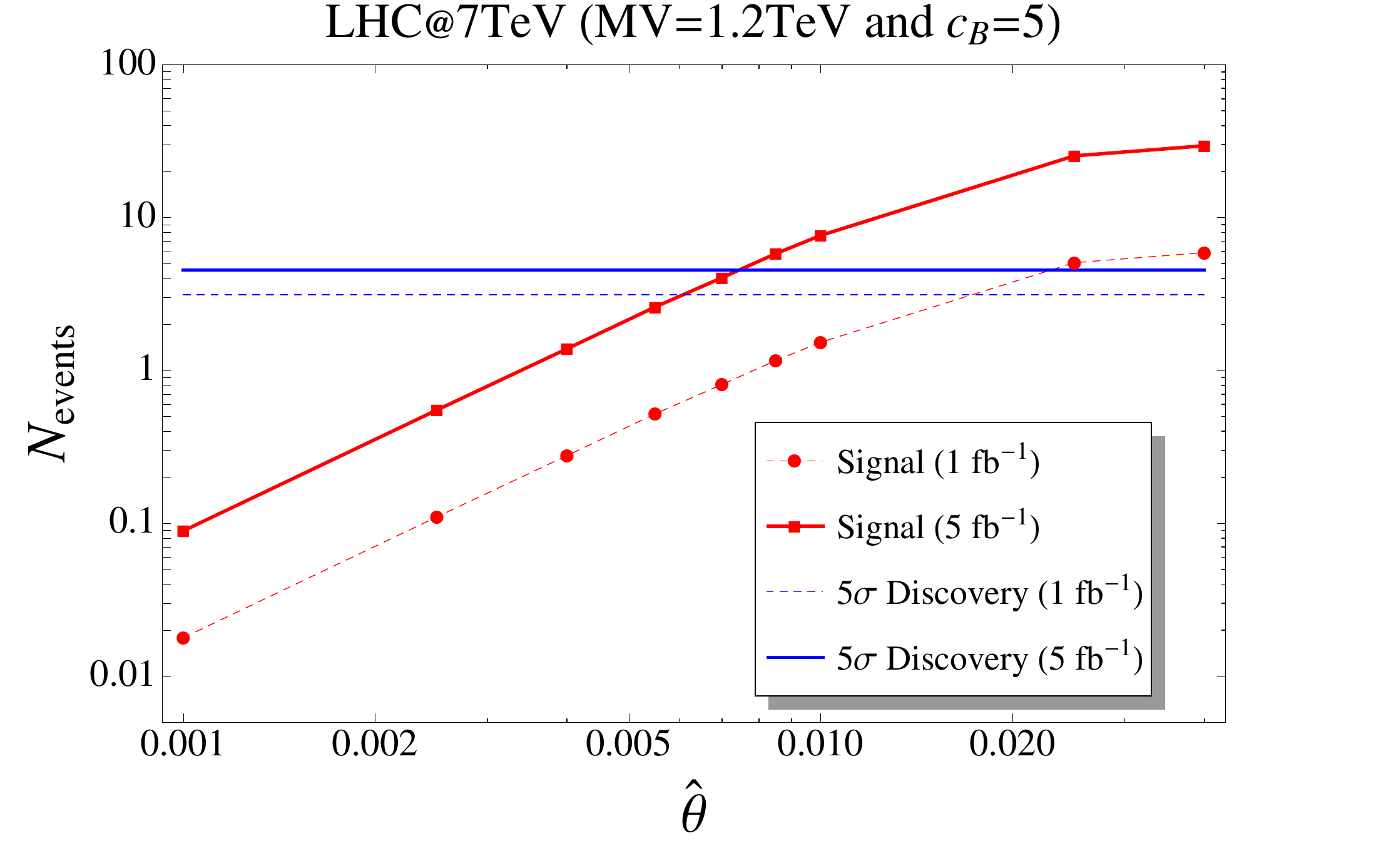} \hspace{-7mm}
\includegraphics[scale=0.37]{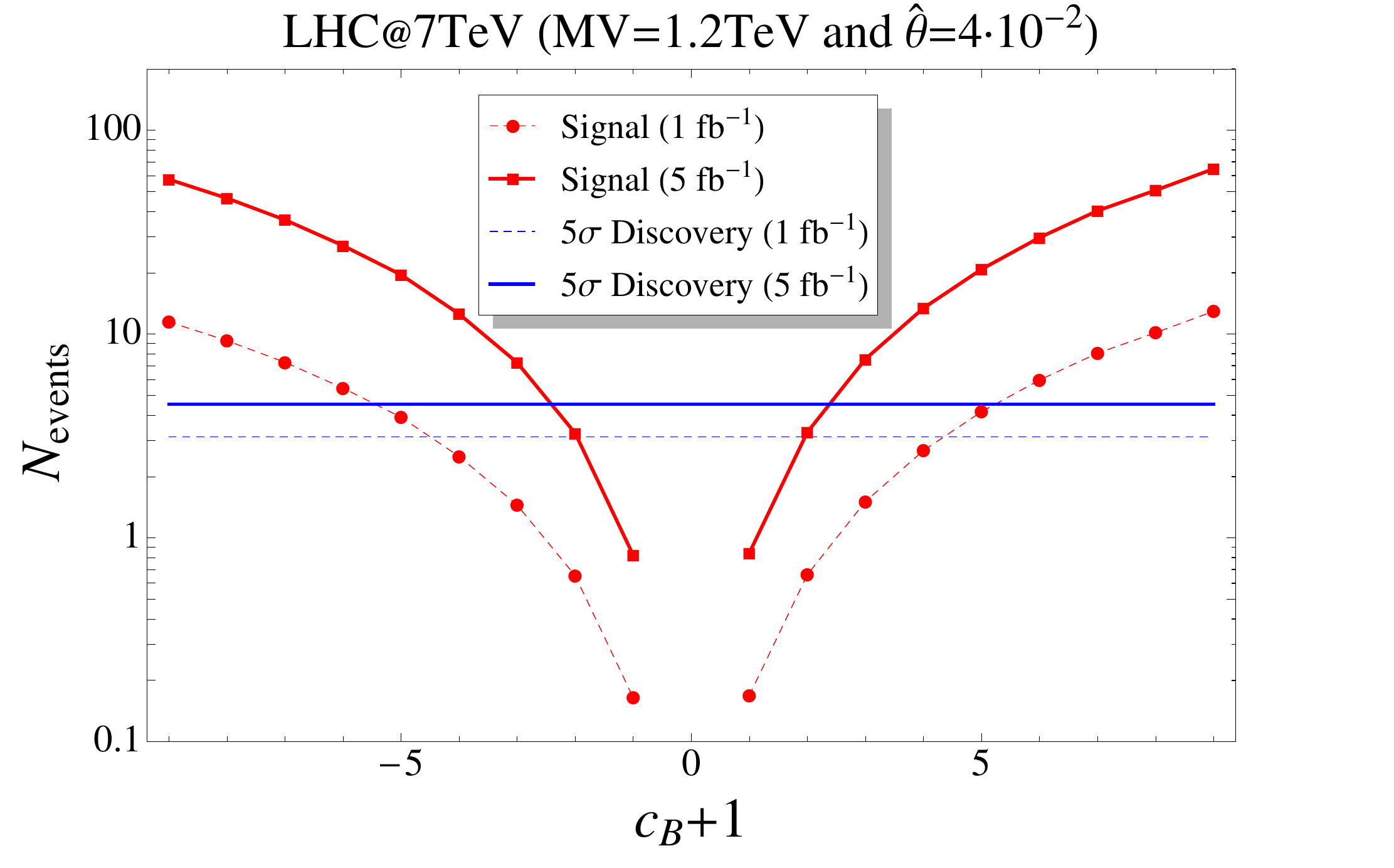}
\end{minipage}
\caption{`$5\sigma$' discovery prospects of the 7 TeV LHC for the $W'\rightarrow W\gamma\rightarrow e\nu \gamma$ process, for $M_{W'}=800$ GeV, $g_{q}=0.84g$ (top row) and $M_{W'}=1200$ GeV, $g_{q}=1.48g$ (bottom). The region shaded in grey is excluded at $95\%$ CL by Tevatron searches for resonances decaying into $WZ$.}
\l{fig:LHCWgamma}
\end{figure}

For illustrative purposes, we also give in Table~\ref{14TeV800GeV} an estimate of the sensitivity on $c_{B}$ for the 14 TeV LHC with 300 fb${}^{-1}$ luminosity. Background events are due to the irreducible SM $W\gamma$ process only. Cuts on the final state kinematics are the same as for the early LHC case discussed above.
\begin{table}[t]
\centering
\begin{tabular} {| c | c | c | c || c | c | c | c |}
\hline
$c_{B}+1$ & $N_{\mathrm{s}}$ & $N_{\mathrm{bckgr}}$ & $N_{\sigma}$ & $c_{B}+1$ & $N_{\mathrm{s}}$ & $N_{\mathrm{bckgr}}$ & $N_{\sigma}$  \\
\hline
$0.6$ & $57$  & $102$ & $5.7$ & $0.4$ & $34$  & $45$ & $5.0$ \\   
\hline
$0.5$ & $40$  & $102$ & $4.0$ & $0.3$ & $23$  & $45$ & $3.4$ \\   
\hline
$0.4$ & $26$  & $102$ & $2.6$ & $0.2$ & $9$  & $45$ & $1.5$ \\   
\hline
\end{tabular}
\caption{Sensitivity on $c_{B}$ at the 14 TeV LHC with 300 fb${}^{-1}$, for $M_{W'}=800$ GeV, $g_{q}=0.84g$ and $\hat\theta = 10^{-2}$ (left), and for $M_{W'}=1.2$ TeV, $g_{q}=1.48g$ and $\hat\theta = 4\times 10^{-2}$ (right).}
\l{14TeV800GeV}
\end{table}

Clearly, it is very interesting to understand what are the predictions for the strength of the $W'W\gamma$ coupling in extensions of the SM. In Ref.~\cite{Ferrara:1992ep} it was shown that the gyromagnetic ratio of any elementary particle of mass $M$ (of any spin) coupled to the photon has to take the value $g=2$, which can be equivalently written as $c_{B}=-1$ in our effective language, in order for perturbative unitarity to be preserved up to energies $E\gg M$. As a consequence, in any gauge extension of the SM, where the $W'$ is the fundamental gauge boson of some extra symmetry, $g=2$ has to be expected, since perturbative unitarity is preserved up to much larger scales. Indeed, in the `minimal' gauge model containing an iso-singlet $W'$, namely a LR model, we find that $c_{B}=-1$ at the renormalizable level. Including dimension-6 operators, we expect $c_{B}=-1+O(v_{R}^{2}/\Lambda^{2})$, where $\Lambda$ is the cutoff of the LR model. Therefore, $c_{B}\approx -1$ will still hold, and observation of $W'\to W\gamma$ is likely to be out of the reach of the LHC. 

On the other hand, if the $W'$ is a composite state of some new strong interaction, then the requirement of preservation of perturbative unitarity is relaxed, and significant departures from $c_{B}=-1$ can be envisaged. The only condition that needs to be satisfied even in the composite case is that the scale of violation of perturbative unitarity be sufficiently larger than the $W'$ mass. To verify that this is indeed the case, and since $c_{B}$ only appears in the $BVV$ vertex (see Eq.~\eqref{V-SM}), where $B$ is the hypercharge gauge boson and $V$ is the extra vector, we compute the amplitude for $BB\to VV$ scattering. The two independent amplitudes that grow the most with energy are $B_{+}B_{\pm}\to V_{L}V_{L}$, where $B_{\pm}$ are the two transverse polarizations of the $B$, and $V_{L}$ is the longitudinally polarized $V$. The leading term of these amplitudes in the high-energy limit reads
\be \l{UVamplitudes}
A_{++\to LL}\approx \frac{(1-c_{B}^{2})g^{\prime\,2}s}{2M^{2}}\,,\qquad A_{+-\to LL}\approx \frac{(1+c_{B})^{2}g^{\prime\,2}s}{4M^{2}}\,. 
\ee 
Notice that for $c_{B}\to -1$, the dangerous high-energy behavior is removed, as it was anticipated above. Requiring the amplitudes in Eq.~\eqref{UVamplitudes} not to exceed $16\pi^{2}$, we find the cutoff $\Lambda$ at which perturbative unitarity is lost\footnote{Other definitions of the perturbative unitarity bound are possible, and have been used in the literature. A different choice would simply change the numerical factors appearing in the definition of the cutoff.}, as a function of $c_{B}$: taking the maximum value we used in the phenomenological analysis, namely $c_{B}=10$, we find $\Lambda \approx 5 M$; for smaller values of $c_{B}$, the cutoff is obviously larger. This result guarantees that we can safely study the phenomenology at scale $M$ with relatively large values of $c_{B}$, without encountering any perturbative unitarity violation issues. 

We conclude that, since the size of the $W'W\gamma$ coupling is expected to be very small if the $W'$ is a fundamental gauge boson, observation of $W'\to W\gamma$ at the LHC would be a hint of the composite nature of the $W'$.    

\subsection{Search for $W'\rightarrow WZ$} \l{WZsearch}

We also discuss the $W'\rightarrow WZ$ decay, which is complementary to $W'\rightarrow W\gamma$ because, being the rate for resonant $WZ$ production almost independent of the parameter $c_{B}$, its measurement would allow one to estimate the size of the mixing angle $\hat\theta$. Since we consider the early LHC reach, where integrated luminosity will be $\lesssim 5$ fb$^{-1}$, the most promising final state is $WZ\rightarrow \ell \MET jj$, which has a larger rate with respect to the purely leptonic channel; on the contrary, selecting leptonic decays of the $Z$ together with a hadronic $W$ has been shown to be less promising \cite{Alves:2009gh}. 
Therefore, we implement simple cuts on the $e\nu jj$ final state (we only consider $W$ decays into an electron, in analogy to what we did for the $W'\rightarrow W\gamma$ process) to enhance the ratio of signal over background, namely: $p_{T}^{e,j}>50$ GeV, $\MET > 50$ GeV, $|\eta_{e,j}|<2.5$, and in addition we require the invariant mass of the dijet system to reconstruct a $Z$, $|M(jj)-M_{Z}|<20\,\mathrm{GeV}$. Finally, we select events which have an invariant mass compatible with $M_{W'}$ as follows: $|M(e\nu jj)-M_{W'}|<0.10M_{W'}\,$. The background we consider is the SM $pp\rightarrow e\nu jj$, which includes a large contribution from $W+jj$. The $t\overline{t}$ background can be efficiently reduced to roughly one order of magnitude less than the QCD background by applying a central jet veto \cite{Alves:2009gh}, and we do not consider it here. The invariant mass distributions of signal and background for this channel are shown in Fig.~\ref{fig:WZdistrib}.
Our results are shown in Fig.~\ref{fig:WZ} for the same choices of the $W'$ mass and couplings that we already discussed when studying $W'\rightarrow W\gamma$, so that a direct comparison between the two searches can be made. We can see that with 5 fb${}^{-1}$, a mixing angle larger than $\hat\theta\approx 3\div 4\times 10^{-3}$ is accessible for discovery; this result is to a good approximation independent of the size of $c_{B}$. We also notice that the number of signal events can be sizable, which is the main reason why this channel is more favorable than the purely leptonic one for limited LHC luminosity. 

Finally, we do not discuss $W'$ decays into $Wh$, because the choice of the most relevant final states is strongly dependent on the Higgs boson mass. We refer the interested reader to Refs.~\cite{Azuelos:2005kw,Bao:1334933} and to the references cited therein. 

\begin{figure}[t]
\centering
\includegraphics[scale=0.37]{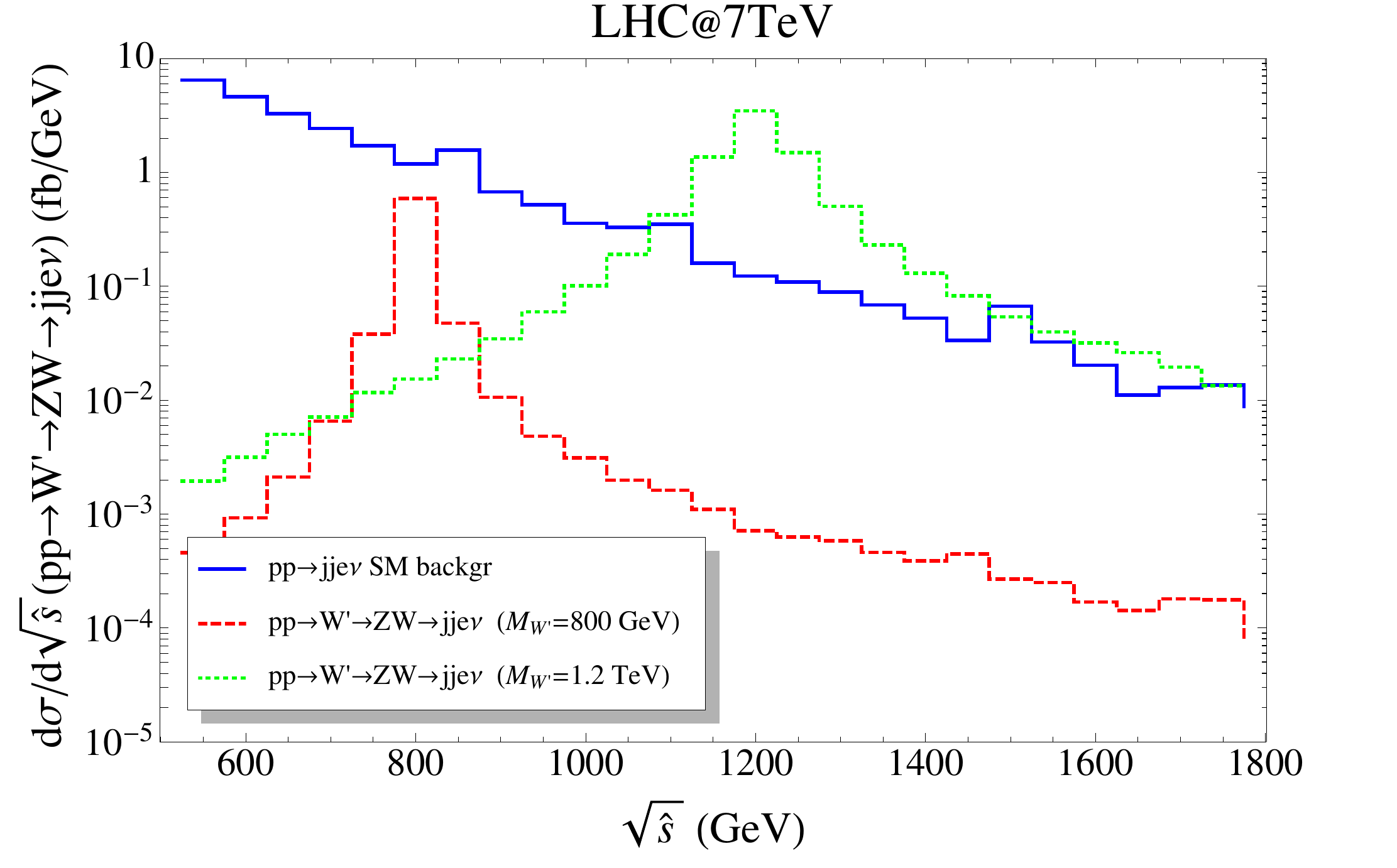}
\caption{Invariant mass distribution of the $e\nu jj$ system for the $W'$ signal and for the $pp\rightarrow e\nu jj$ background. The values of the coupling $g_{q}$ are the same as in Fig.~\ref{fig:LHCWgammaDISTRIB}.}
\l{fig:WZdistrib}
\end{figure}
\begin{figure}[t]
\centering
\begin{minipage}{20cm}
\includegraphics[scale=0.37]{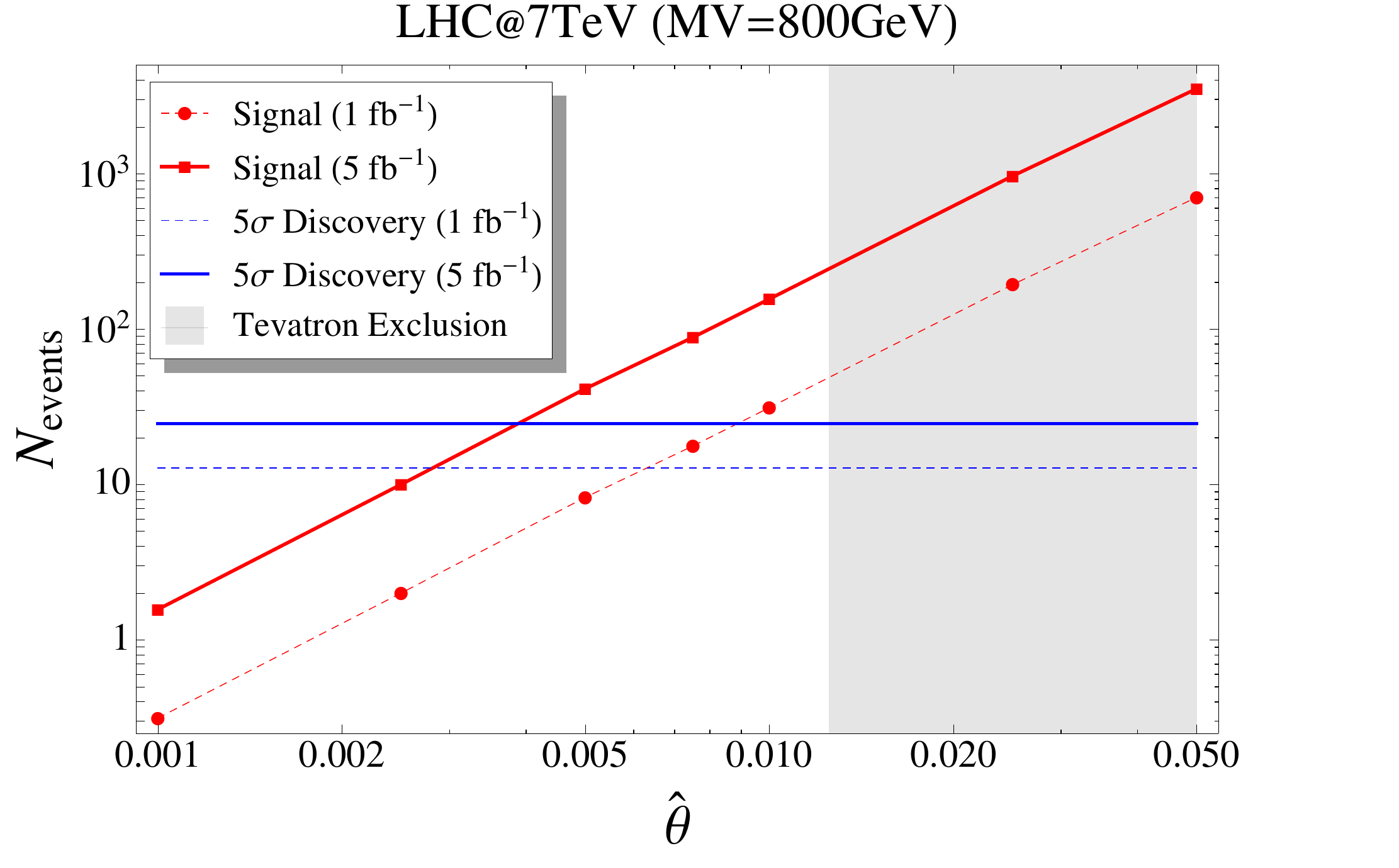}\hspace{-5mm}
\includegraphics[scale=0.37]{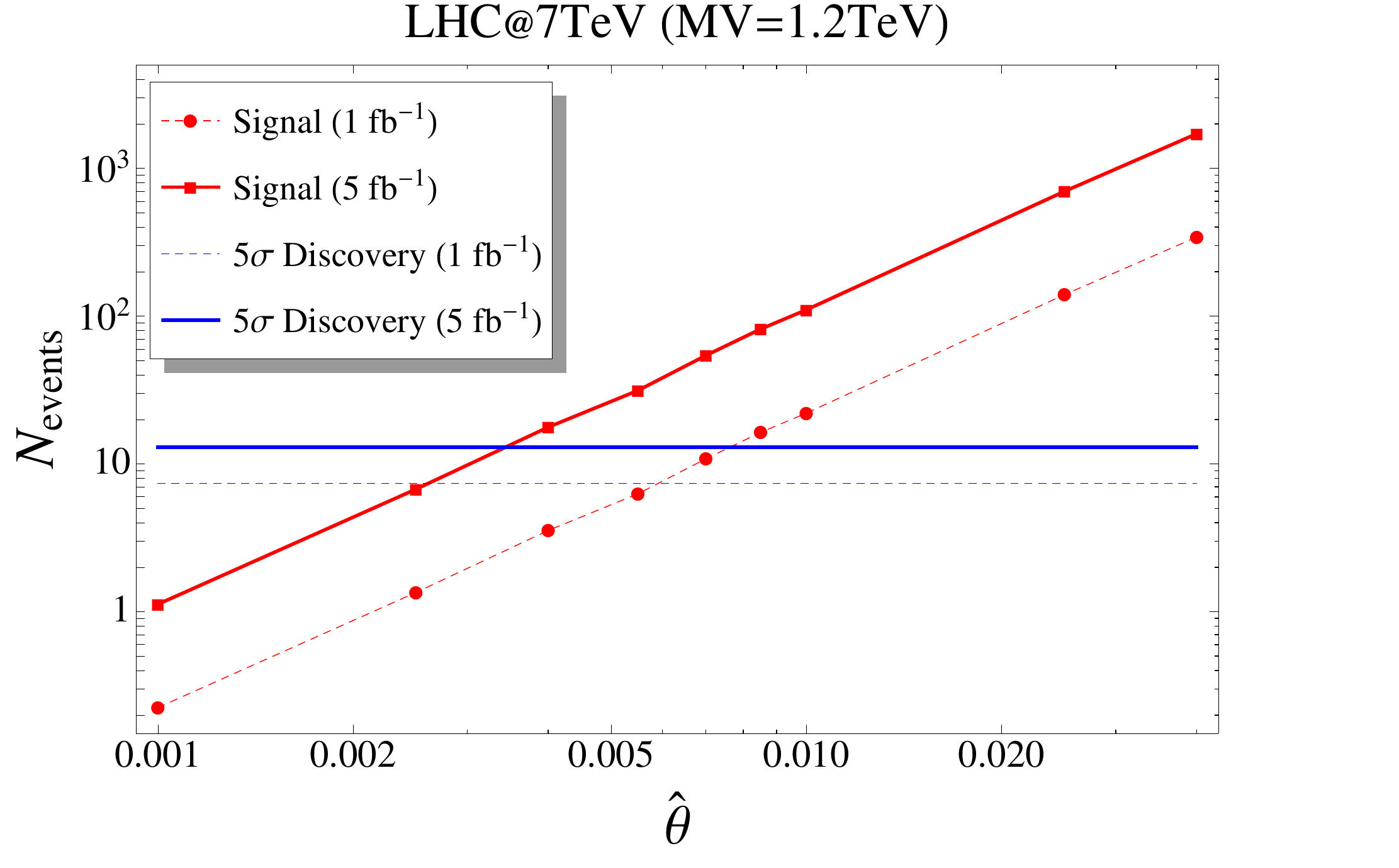}
\end{minipage}
\caption{`$5\sigma$' discovery prospects on the mixing angle $\hat\theta$ via the $W'\rightarrow WZ\rightarrow e\nu jj$ process at the 7 TeV LHC, for $M_{W'}=800\,\mathrm{GeV}$, $g_{q}=0.84g$ (left) and $M_{W'}=1200\,\mathrm{GeV}$, $g_{q}=1.48g$ (right). The region shaded in grey is excluded at $95\%$ CL by Tevatron searches for resonances decaying into $WZ$. The results are almost independent of $c_{B}$.}
\l{fig:WZ}
\end{figure}

\section{Summary and conclusions} \l{summary}

In this chapter we have applied an effective approach to study the phenomenology of a heavy $W'$ transforming as a singlet under weak isospin. Such $W'$ with mass even below a TeV and sizable coupling to quarks is allowed by present data \cite{Grojean:2011wi}. We have estimated the early LHC reach in the dijet channel on such a resonance. 

Subsequently we have discussed the possibility that the $W$-$W'$ mixing angle be large enough to allow observation of the decays $W'\rightarrow W\gamma$ and $W'\to WZ$ at the early LHC. We have shown that the $W'\to W\gamma$ channel is of significant relevance to gain insight on the nature of the $W'$ after a discovery in the dijet (or $tb$) final state. In particular it would be a hint of the composite nature of the resonance. Finally we have studied the experimentally accessible values of the parameters $(c_{B},\hat\theta)$ at the early LHC.
\part{Composite vectors in Dark Matter models}\l{PARTIII}
\chapter{Composite vectors in a composite ElectroWeak neutral Dark Matter model}\l{chap9}
\setlength{\epigraphwidth}{.4\textwidth}     
\vspace{-2mm}\epigraph{An expert is someone who knows some of the worst mistakes that can be made in his subject and how to avoid them.
}%
              {\textsc{Werner Heisemberg}}
\vspace{3mm}     
\vspace{0mm}\epigraph{Science is the belief in the ignorance of the experts.}%
              {\textsc{Richard Feynman}}
\vspace{7mm}
\lettrine{T}{he} possibility that the Dark Matter (DM) be related, directly or indirectly, to the physics of the EWSB deserves the highest consideration. Indeed this has been and is being extensively discussed both in weak-coupling and in strong-coupling scenarios of EWSB. We are interested here to the strong-coupling case, without specific reference to any detailed model.

We consider the case in which the sector responsible for EWSB respects a global (quasi-) conserved charge $X$ which enforces the (quasi-)stability of the lightest particle, $\Phi$, with non vanishing $X$. $\Phi$ is a candidate DM particle. Its mass, $m_{\Phi}$, is in the TeV range, characteristic of the strong forces that may give rise to EWSB. This particle is made of constituents that feel the strong force and carry non-vanishing EW quantum numbers, but is itself EW neutral. This is needed in order to suppress the tree-level coupling of $\Phi$ to the Z-boson, which would be in conflict with direct DM searches.

At face value, the cosmological relic abundance of the $\Phi$ particles is too low to explain the observed DM energy density, $\Omega_{\text{DM}}$, normalized as usual to the critical cosmological density. We have in mind the effect of two
body processes, $\Phi\bar{\Phi}\leftrightarrow Q\bar{Q}$, where $Q$ is any unstable particle lighter than $\Phi$, also feeling the new strong force. For example, longitudinal W and Z bosons may play the role of $Q$. The associated
thermally averaged cross section is far bigger than the needed $\left\langle \sigma v\right\rangle \approx1$pb, since
\be
\left\langle \sigma v\right\rangle \approx\frac{\lambda^{4}}{4\pi m_{\Phi}^{2}}f\left(  \frac{m_{\Phi}^{2}}{\Lambda^{2}}\right)  \approx10^{6}\,\text{pb}\left(  \frac{\lambda}{4\pi}\right)  ^{4}\left(  \frac{\text{TeV}}{m_{\Phi}}\right)  ^{2}f\left(  \frac{m_{\Phi}^{2}}{\Lambda^{2}}\right)  \,,
\ee
where ${\lambda}\approx{4\pi}$ is the NDA estimate of the strong coupling ${\lambda}$, and the model-dependent function $f$ of the ratio between the $\Phi$ mass and the scale $\Lambda$ characteristic of
the new strong interaction is of order unity for $m_{\Phi}\approx\Lambda $.\footnote{In Ref.~\cite{Mardon:2009kd}, strongly interacting DM belonging to a non-EWSB hidden sector was considered, with a thermal freezout as the source of DM abundance. However, much higher DM\ masses up to 100 TeV were considered and smaller than NDA couplings were assumed.}

We are thus led to consider the case where the relic abundance of $\Phi$ particles originates from a $X-\bar{X}$ asymmetry, analogous to the standard $B-\bar{B}$ asymmetry responsible for the dominance of matter over anti-matter in the present universe. An interesting aspect of this hypothesis is that one can try to relate $\Omega_{X}=\Omega_{\text{DM}}$ to the standard $\Omega_{B}$ by assuming that $X$, like $B$ or $L$, are all broken by mixed EW anomalies. In this case, non-perturbative EW sphaleron interactions at a critical temperature $T^{\ast}\approx100\div200$ GeV may redistribute any original asymmetry, leading in particular today to \cite{Kaplan:1992dn,Barr:1990ca,Nardi:2008cf} 
\be
\frac{\Omega_{\text{DM}}}{\Omega_{B}}=\mathcal{O}(10^{2})x^{5/2}e^{-x},~x=\frac{m_{\Phi}}{T^{\ast}},
\ee
which can be about right, ${\Omega_{\text{DM}}}/{\Omega_{B}}\approx5$, for $m_{\Phi}$ in the TeV range.

In this as in other cases of putative DM particles, the problem is to find experimental signals that would not only establish their existence but would allow a clear interpretation of their nature. To this end it is difficult to overestimate the interplay between direct DM searches and the LHC experiments, as we are going to discuss.

\section{Summary of direct detection signals}

In absence of a detailed model, the possible signals in direct detection searches can be discussed by means of effective operators that mediate the interactions between $\Phi$ and the $u,d$ quarks or the photon \cite{Bagnasco:1993bb}.
The $\Phi$ particle can be either a complex scalar or a Dirac fermion.

If $\Phi$ is a scalar, the dominant interactions are described by
\be
O_{1}=\frac{1}{\Lambda^{2}}(\Phi^{\ast}\overleftrightarrow{\partial_{\mu}}\Phi)\sum_{q=u,d}c_{q}(\bar{q}\gamma_{\mu}q),\qquad O_{2}=\frac{ec_{2}}{\Lambda^{2}}\partial_{\mu}F^{\mu\nu}(\Phi^{\ast}\overleftrightarrow
{\partial_{\mu}}\Phi)\,, \l{operators}
\ee
which for $c_{q}\approx c_{2}\approx1$, as expected from NDA, give comparable effects. Taking $O_{2}$ for concreteness, the non-relativistic cross section of $\Phi$ on a nucleus of charge $Z$ and mass $m_{N}\ll m_{\Phi}$ is, up to
form factor effects,
\be
\quad\sigma_{2}=c_{2}^{2}\frac{e^{4}Z^{2}m_{N}^{2}}{\pi\Lambda^{4}}\,.
\ee
For Xenon target this corresponds to the per-nucleon cross section
\be
\frac{\sigma_{2}}{A^{4}}\approx c_{2}^{2}\,1.5\cdot10^{-7}\text{pb}\left(\frac{\text{TeV}}{\Lambda}\right)  ^{4}, \l{eq:scalar}
\ee
to be compared with the XENON100 limit on the coherent spin-independent cross section \cite{Collaboration:2011wb}\footnote{This limit is about $2\div 3$ times stronger than the previous limit set by the CDMS experiment \mbox{$\frac{\sigma_{\text{SI}}}{A^{4}}|_{\text{CDMS}}\lesssim2\cdot10^{-7}\text{pb}\left(\frac{m_{\text{DM}}}{\text{TeV}}\right)$ for $m_{\text{DM}}\gg m_{\text{Ge}}$} \cite{CDMSCollaboration:2009ti}. Even if the old CDMS limit was considered in Ref. \cite{Barbieri:2010p431} the conclusions don't change.}
\be
\frac{\sigma_{\text{SI}}}{A^{4}}|_{\text{XENON100}}\lesssim8\cdot10^{-8}\text{pb}\left(\frac{m_{\text{DM}}}{\text{TeV}}\right)  \qquad(m_{\text{DM}}\gg m_{\text{Xe}}),
\ee
i.e.
\be
c_{2} \lesssim 0.73 \left(  \frac{\Lambda}{\text{TeV}}\right)  ^{2}\left(  \frac{m_{\Phi}}{\text{TeV}}\right)  ^{1/2}. \l{bound}
\ee
For $c_{2}\approx1,$ and $\Lambda\approx m_{\Phi}\approx4\pi v\approx3$ TeV, the expected cross section \eqref{eq:scalar} is about two orders of magnitude below the XENON100 limit.

If instead $\Phi$ is a Dirac fermion, assuming parity invariance (up to anomalies) of the EWSB sector, the dominant operator is a magnetic moment interaction
\be
O_{M}=\frac{iec_{M}}{2\Lambda}(\bar{\Phi}\sigma_{\mu\nu}\Phi)F^{\mu\nu}.
\ee
In Xenon, $O_{M}$ gives rise to the dominant spin-independent cross section due to scattering on the current produced by the nuclear charge:
\be
\frac{d\sigma_{M}}{dE}\approx c_{M}^{2}\frac{e^{4}Z^{2}}{4\pi\Lambda^{2}%
E}(1+\mathcal{O}(E/E_{\text{max}})), \l{eq:fermi-si}%
\ee
where $E$ is the kinetic energy of the recoiling nucleus, ranging from the experimental threshold $\sim10$ keV to $E_{\max}=2v^{2}m_{N}^{2}=\mathcal{O}(50$ keV). The spin-dependent cross section is subleading due to small nuclear spin of Xenon and because of $E_{\max}/E$ enhancement present in Eq.~\eqref{eq:fermi-si} \cite{Bagnasco:1993bb}. A suitable comparison of this cross section with the null XENON100 result gives in this case
\be
c_{M}<10^{-1}\left(  \frac{\Lambda}{\text{TeV}}\right)  \left(  \frac{m_{\Phi
}}{\text{TeV}}\right)  ^{1/2},
\ee
against the NDA estimate $c_{M}\approx1$. Given the uncertainties of these estimates and of the value of the scale $\Lambda$ itself, in no way this bound can be interpreted as ruling out a composite fermionic DM particle. Quite on
the contrary, the message we draw is that a signal in direct DM searches could be around the corner. Yet we find it preferable, at least for reference, to stick in the following to the scalar case.

\section{DM-nucleus interaction mediated by an iso-singlet vector $V$}

Suppose that a positive signal were indeed found in direct DM searches at the level indicated above, in fact not far from the present sensitivity. How would we know that the candidate DM particle is a composite $\Phi$-like particle? As
already mentioned, the LHC should come into play here. However the detection at the LHC of an EW neutral particle of TeV mass that can only be pair produced may not be an easy task. For this reason, we turn the question into a
different but related one. What could mediate the operators in Eq.~\eqref{operators} responsible in the first place for the direct DM signal? We argue that the most likely candidate for this role is a composite vector iso-singlet $V$, the analog of the $\omega$-meson in QCD, strongly coupled to $\Phi$ and mixed with the elementary hypercharge gauge boson $B_{\mu}$, via the diagram of Fig.\ \ref{VB}.

\begin{figure}[ptbh]
\centering
\includegraphics[width=4cm]{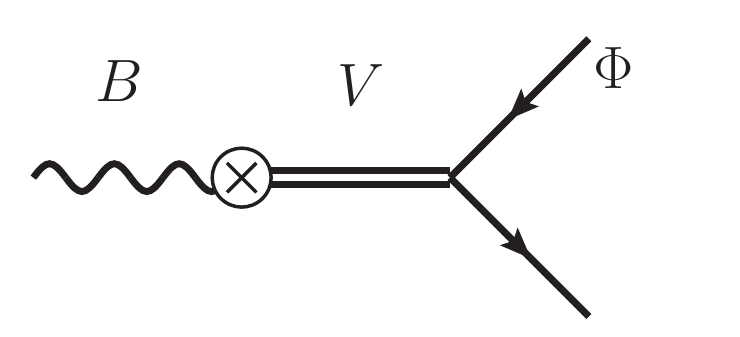} \caption{The diagram which generates $O_{2}$ in Eq.~\eqref{operators} via the $B-V$ mixing.}
\l{VB}
\end{figure}

We base our estimates on the following phenomenological Lagrangian
\be
\La=\La_{V}+\La_{V\Phi} \l{Ltot2}
\ee
where
\be
\La_{V\Phi}=g_{S}V_{\mu}(\Phi^{\ast}\overleftrightarrow{\partial_{\mu }}\Phi),\quad g_{S}=4\pi\frac{M_{V}}{\Lambda}, \l{gS}
\ee
and
\be
\La_{V}=-\frac{1}{4}V_{\mu\nu}^{2}+\frac{1}{2}M_{V}^{2}V_{\mu}^{2}+\frac{g^{\prime}}{4\pi}B_{\mu\nu}V_{\mu\nu}-\frac{i}{8\pi}\epsilon^{\mu\nu\rho\sigma}V_{\mu}\text{tr(}u_{\nu}u_{\rho}u_{\sigma}\text{)}+\frac{g}{4\pi}\epsilon^{\mu\nu\rho\sigma}V_{\mu}\text{tr(}u_{\nu}\hat{W}_{\rho\sigma}\text{)} \l{lagrangian}
\ee
in the same notation for the EWChL used in Chapter~\ref{chap3}. We assume that the couplings proportional to the epsilon tensor, relevant to the following section, are induced, analogously to the QCD case, by a chiral anomaly. The strength of the various couplings are all based on NDA estimates, known to work well for QCD \cite{Klingl:1996ui}, and noticing that the only coupling
in Eq.~\eqref{Ltot2} that corrects at one loop level the $V$ mass $M_{V}$ is $g_{S}$ in Eq.~\eqref{gS}.

From the diagram of Fig.\ \ref{VB} it is straightforward to obtain the operator $O_{2}$ in Eq.~\eqref{operators} with
\be
c_{2}=\frac{2\Lambda}{M_{V}},
\ee
or, from Eq.~\eqref{bound},
\be
M_{V}>2\,\text{TeV}\left(  \frac{\text{TeV}}{\Lambda}\right)  \left(\frac{\text{TeV}}{m_{\Phi}}\right)  ^{1/2},
\ee
which could easily allow, taking $\Lambda\approx m_{\Phi}\approx4\pi v\approx3$ TeV, a $V$ mass as low as 700 GeV. The iso-singlet nature of $V$ makes its exchange innocuous in the EWPT, giving a contribution to the $Y$-parameter \cite{Barbieri:2004p1607} well below the $10^{-4}$ level.

\section{LHC phenomenology of $V$}

The Lagrangian \eqref{lagrangian} allows to calculate the decay widths of $V$. The relevant widths are:

\begin{itemize}
\item From the last term in Eq.~\eqref{lagrangian}, the dominant decay into two standard vector bosons\footnote{This formula corrects a factor 9 error in the RHS of Eq.~(4.7) of Ref.~\cite{Klingl:1996ui}.}
\begin{gather}
\Gamma_{\text{tot}}\approx\Gamma\left(  V\rightarrow W^{+}W^{-},ZZ,Z\gamma \right)  =\frac{g^{2}}{8\pi}\frac{M_{V}^{3}}{(4\pi v)^{2}}\,,\\
\text{BR}\left(  V\rightarrow W^{+}W^{-}\right)  \approx\frac{2}{3} ,\qquad\text{BR}\left(  V\rightarrow ZZ\right)  \approx\frac{\cos^{2} \theta_{W}}{3},\qquad\text{BR}\left(  V\rightarrow Z\gamma\right)
\approx\frac{\sin^{2}\theta_{W}}{3}\,.
\end{gather}

\item From the last but one term in Eq.~\eqref{lagrangian}, the 3-body decay
\be
\Gamma\left(  V\rightarrow W_{L}^{+}W_{L}^{-}Z_{L}\right)  =\frac{\pi}{40}\frac{M_{V}^{7}}{(4\pi v)^{6}}\,.
\ee

\item From the mixing of $V$ with the $B$ boson, the decay into a pair of
standard fermions, e.g.
\be
\Gamma\left(  V\rightarrow e^{+}e^{-}\right)  =\frac{5}{24\pi}\left(\frac{g^{\prime2}}{4\pi}\right)  ^{2}M_{V}\ .
\ee

\end{itemize}

The total width and the subdominant BRs are shown in Fig.\ \ref{width_BR} for $M_{V}$ around 1 TeV. A few features are especially apparent from these figure: the smallness of the $\Gamma/M$ ratio and the strong dominance of the decays into two bosons (among which the $Z\gamma$ channel) over all the other decay modes, in particular the three body $W^{+}W^{-}Z$. \begin{figure}[ptbh]
\begin{minipage}[b]{7.5cm}
\centering
\includegraphics[height=5.1cm]{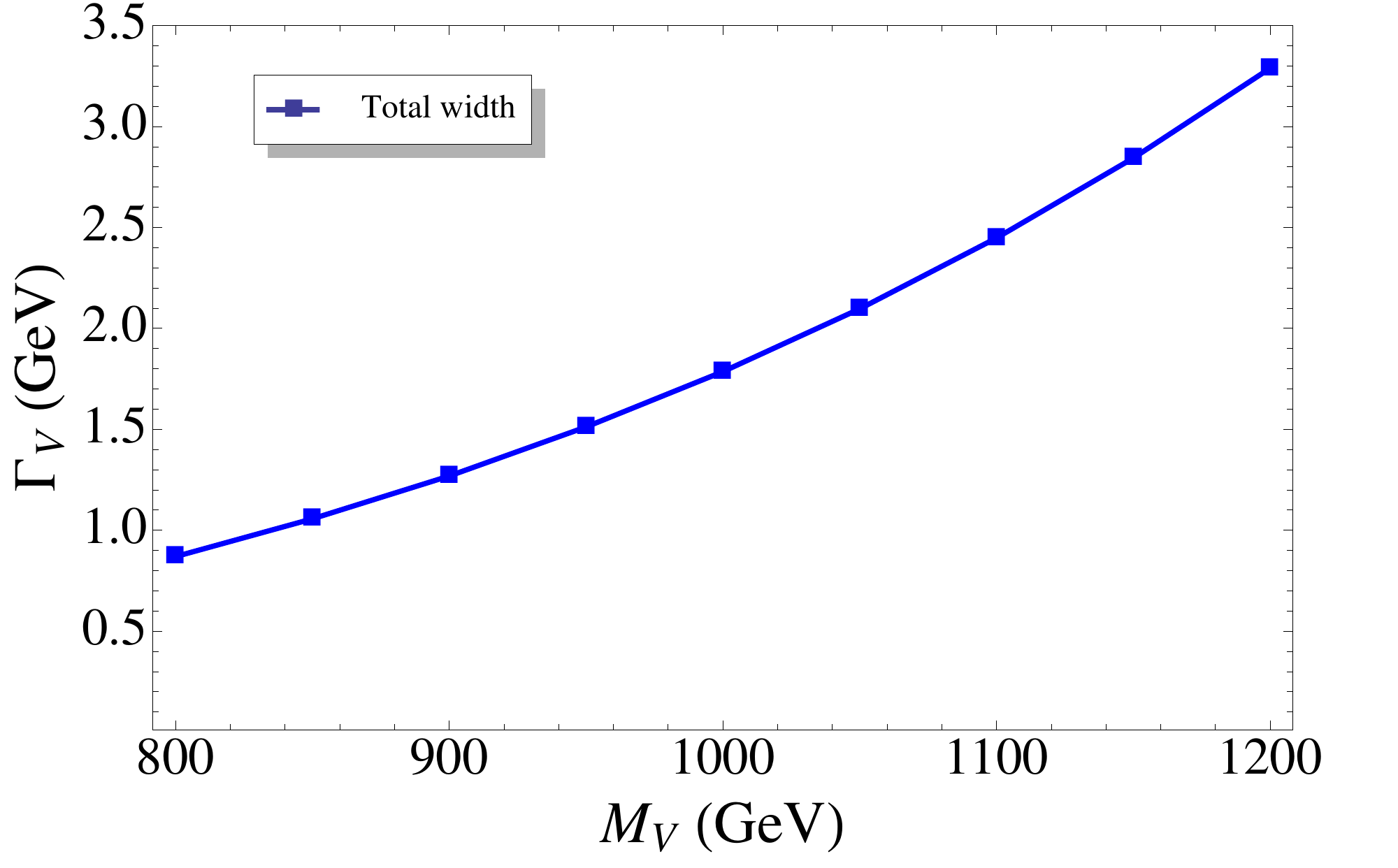}
\end{minipage}
\ \hspace{1mm}\ 
\begin{minipage}[b]{7.5cm}
\centering
\includegraphics[height=5.1cm]{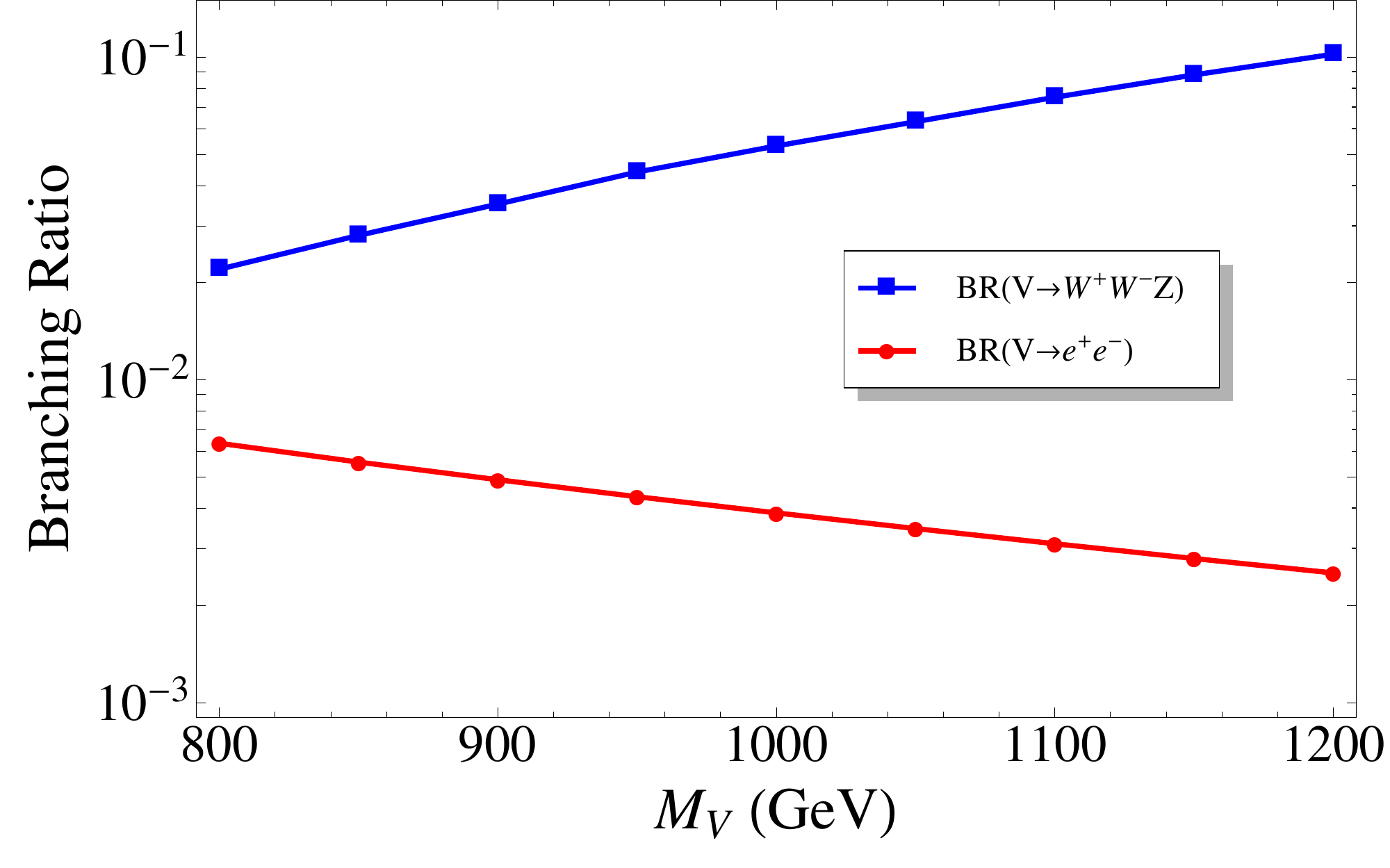}
\end{minipage}
\caption{Left panel: total width of the iso-singlet vector boson $V$ as a function of its mass around $1$ TeV. Right panel: subdominant BRs $BR(V\to e^{+}e^{-})$ and $BR(V\to W^{+}W^{-}Z)$.}
\l{width_BR}
\end{figure}Especially this last feature is at variance with what one might have expected from the analogy with the $\omega$ in QCD. The main reason for this can be traced back to the close degeneracy of the $\omega$ with the $\rho$, as dictated by $SU(3)$, making the decay $\omega\rightarrow3\pi$ dominated by the intermediate $\pi\rho$ state.

The vector $V$ can be produced at the LHC by the DY annihilation or by VBF, as again described by the Lagrangian \eqref{lagrangian}. The corresponding production rates are shown in Fig.~\ref{totalcrosssections}. 
\begin{figure}[ptbh]
\begin{center}
\includegraphics[height=5.1cm]{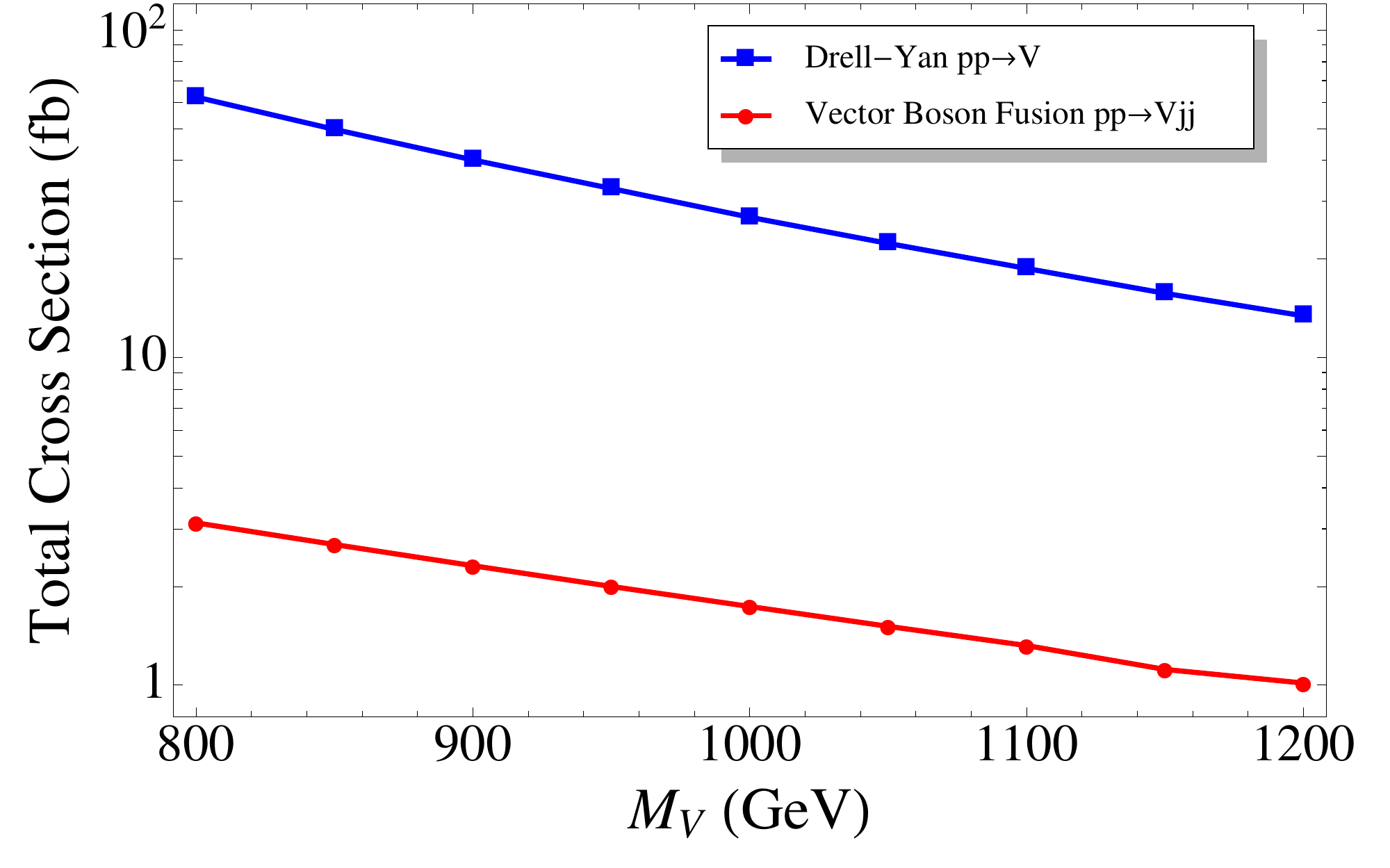}
\end{center}
\caption{Total cross sections for the VBF and the DY iso-singlet vector boson production at the LHC as functions of its mass for $\sqrt{s}=14$ TeV.}
\l{totalcrosssections}
\end{figure}
From the BRs  above, the $Z\gamma$ channel appears most promising. In Fig.\ \ref*{AZ}
we show the number of events expected when the $Z$ decays into $l^{+}l^{-},l=e,\mu$ or into $\nu\bar{\nu}$ respectively at $\sqrt{s}=14$ TeV. The binnings of the events, crucial for discovery, are based on current estimates of the expected resolutions in an advanced phase of the LHC operation\footnote{We are assuming a 1\% energy resolution of the invariant mass of the $l^{+}l^{-}\gamma$ system (comparable to the peak resolution in the $H\to\gamma\gamma$ studies) and a $0.5\%$ resolution of the photon $p_T$. The binnings correspond to $2\sigma$ bins.} The background shown corresponds to the $Z\gamma$ production in the SM. On the basis of these figures, we conclude that the vector $V$ in the TeV mass range could be discovered at the LHC with about 100 fb$^{-1}$ of integrated luminosity.\footnote{See Ref.~\cite{D0Collaboration:2008p115} for a recent D0 search of narrow vector resonances decaying into $Z\gamma$ based on 1 fb${}^{-1}$ of Tevatron data. The resulting limit on $\sigma\times$B.R. is $\sim0.2-0.4$ pb (95\% CL) for $M_V=700-900$ GeV, two orders of magnitude above the values predicted by our model.}


\begin{figure}[ptbh]
\begin{minipage}[b]{7.5cm}
\centering
\includegraphics[height=5.1cm]{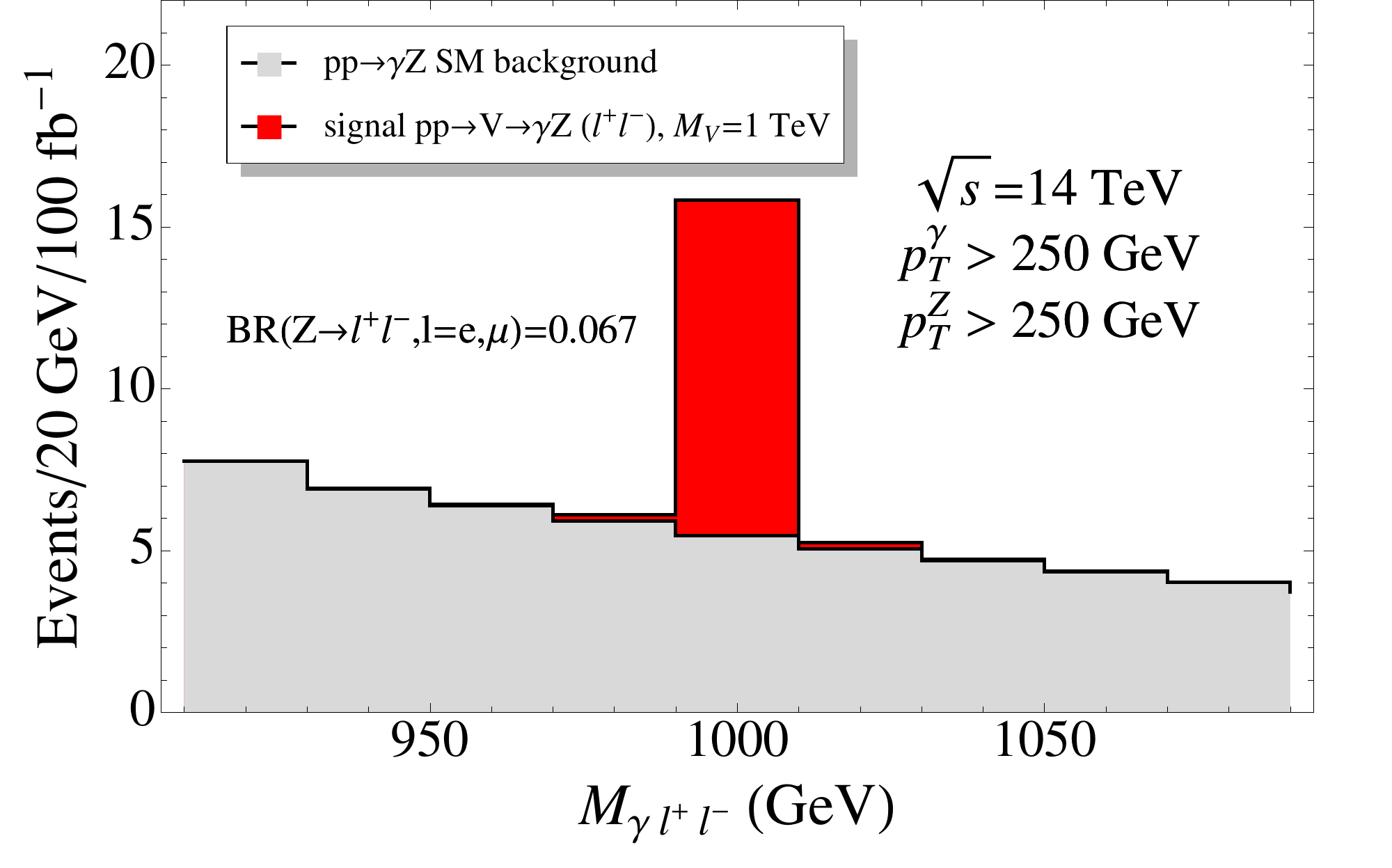}
\end{minipage}
\ \hspace{1mm} \ 
\begin{minipage}[b]{7.5cm}
\centering
\includegraphics[height=5.1cm]{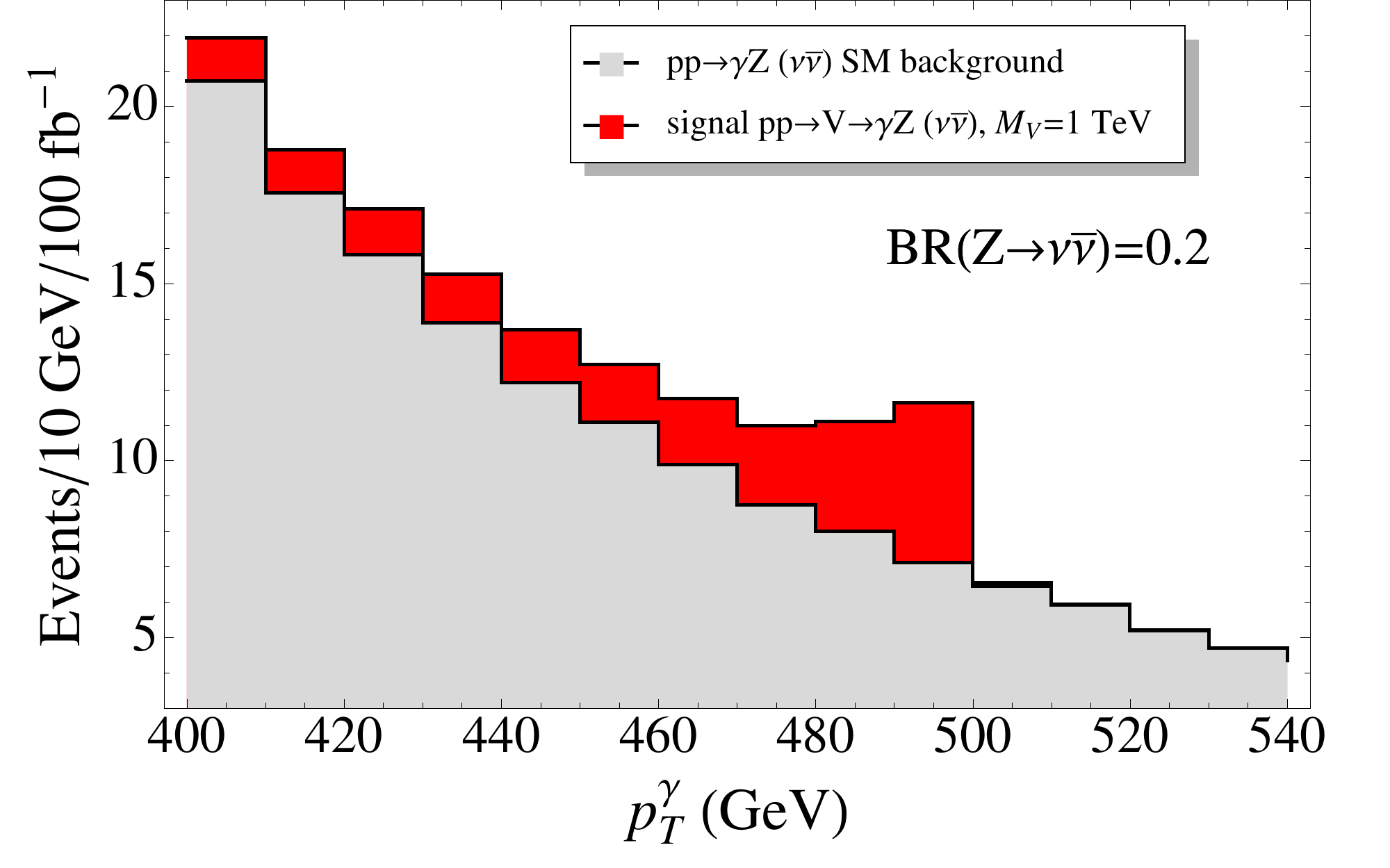}
\end{minipage}
\caption{Left panel: the number of $\gamma l^{+}l^{-}$ events as a function of the total invariant mass. The imposed $p_{T}>250$ GeV cut on the photon and the reconstructed $Z$ boson enhances the $S/B$ ratio. A $\sim5\sigma$ excess from the SM prediction can be seen. Right panel: the number of $\gamma+\slash\!\!\!\!{E}_{T}$ events as a function of the photon transverse momentum. This decay mode will give additional information although not a discovery ($S/\sqrt{B}\sim 2$). Both plots are for 100 fb${}^{-1}$ of data at $\sqrt{s}=14$ TeV. The chosen binnings correspond to twice the expected experimental resolution. }
\l{AZ}
\end{figure}

\section{Summary and conclusions}

In this chapter we have applied the same phenomenological approach of the previous chapter to discuss a possible interplay between a new putative strong dynamics responsible for EWSB and Dark Matter. With this approach we have studied an iso-singlet vector resonances  $V$ (`techni-$\omega$') coupled to the lightest particle $\Phi$ carrying non-vanishing conserved charge (`techni-baryon number'). The $\Phi$ is a candidate DM particle, assumed EW neutral to evade direct detection constraints\footnote{For the alternative possibility of a neutral iso-triplet component, see Ref.~\cite{Frandsen:2009wn}.}.

We have shown that the phenomenology of this sector allows for an interesting interplay between the ongoing direct DM searches and the LHC. Apart from NDA, our main assumption is that the operators describing interactions of $\Phi$
with the SM particles, Eq.~\eqref{operators}, are generated via the cubic $\Phi\Phi V$ coupling in the strong sector, and the mixing between $V$ and the elementary hypercharge gauge field $B,$ Fig.~\ref{VB}. This is the Vector Meson Dominance hypothesis, which is known to work well in QCD not only for the pion \cite{Klingl:1996ui} but also for the nucleons \cite{Lomon:2001ki,Lomon:2002kd}. Two conclusions transpire from our analysis. First, the expected signal in direct detection experiments, estimated already in Ref.~\cite{Bagnasco:1993bb}, is not far from the present experimental bounds. Second, the iso-singlet vector $V$, in its typical mass range and with couplings as in Eq.~\eqref{lagrangian}, appears within reach of the LHC with $O(100$ fb$^{-1})$ of integrated luminosity\footnote{For an old study of the techni-$\omega$ discovery potentials at the SSC see Ref.~\cite{Chivukula:1990jg}. That study correctly identifies the crucial $Z\gamma$ decay mode. Our approach to estimating the production cross section is more direct, in our opinion.}. Its strong sector nature will be easily identifiable due to a characteristic $Z\gamma$ decay mode. The joint observation of nuclear recoil events in direct detection experiments and of a vectorial resonance decaying into $Z\gamma$ at the LHC will then point to a composite DM particle.

\chapter*{Conclusion}\l{chap10}
\addcontentsline{toc}{chapter}{Conclusion}
\markboth{{\scshape Conclusion}}{{\scshape Conclusion}}
\lettrine{T}he possibility that a new strongly interacting sector is responsible for the EWSB should be taken seriously, especially after the null results in the Higgs boson and SUSY searches performed so far, which push the level of fine-tuning of the MSSM in the percent region\footnote{For a recent analysis of the implications of the early LHC searches of supersymmetry see e.g. Ref.~\cite{Strumia:2011dv}.}.
In this thesis we have examined different prospects in the search at the LHC of new composite particles generated by an unknown strong dynamics with masses at the Fermi scale. The main goal of this work has been to apply an effective approach to the discussion of new physics signals in different phases of the LHC without assuming detailed models for the EWSB.

We have organized the discussion in three different parts. In the first part we have performed a detailed analysis of perturbative unitarity and EWPT in different strongly coupled extensions of the SM by adding composite scalars and vectors to the low energy spectrum. 
In particular, making the only assumption that new physics be custodially symmetric, we have studied the implications of the presence in the low energy spectrum of only one vector triplet of different parity, only a scalar singlet or both. We have shown that the presence of both the vector and the scalar can allow for a wider region in the parameter space where perturbative unitarity and EWPT are satisfied. At the same time we have seen that the effective approach always suffers for the presence of many different couplings and a criterion to constrain their values is therefore necessary. For this reason we have always kept in mind the relations with hidden gauge models spontaneously broken by linear or non-linear $\s$ models, in which the different couplings are fixed by gauge invariance. However, we tried to allow for small deviations from minimal gauge models, which, leading to a worst asymptotic behavior of the different scattering amplitudes, can give rise to a strong increase in the related cross sections. With this approach we have studied the phenomenology of the pair productions of scalars and vectors which are sensitive to couplings that are usually quite model dependent and that cannot be measured only looking at the single productions of the same particles. The measurement of this couplings is a leading task in trying to decipher the explicit model which generates the new particles.

In the second part of this thesis we have discussed some possible new physics signals in the first run of the LHC with $\sqrt{s}=7$ TeV and $L=\int \mathcal{L}dt\approx 1-5$ fb$^{-1}$. In particular we have reduced the possible new physics candidates by requiring a large production rate (i.e. the largest luminosity channels) and by considering the lowest possible spin and QCD representations. With a phenomenological approach, based on reasonable effective Lagrangians, we have studied a scalar singlet in the $gg$-channel and a heavy fermion in the $qg$-channel showing that the channels containing at least one photon are probably the most viable at the early LHC. Then we have also studied a charged vector in the $q\bar{q}$-channel, disfavored at the early LHC for parton luminosity reasons, but very weakly constrained from present bounds and therefore with possibly light mass and large couplings. We have seen that in general, the LHC can become competitive with the Tevatron already with some hundreds of inverse picobarns and that it will reach the sensitivity of the Tevatron in almost every channel approaching the inverse femtobarn region\footnote{We remark that at the time of finishing this thesis the LHC had collected slightly more than $1$ fb$^{-1}$ of integrated luminosity.}

Finally, in the third part of this work we have discussed the possible interplay between direct detection DM searches and collider phenomenology in the case of a composite DM candidate. We have seen that the bounds on the spin-independent cross sections for the scattering of DM on nucleus coming from the XENON100 experiment do not rule out a composite DM candidate, either a scalar or a fermion. The interactions between these DM candidates and the SM particles can be mediated by a composite iso-singlet vector analogous to the $\omega$-meson in QCD. This vector can be strongly coupled to the DM and can mix with the elementary hypercharge gauge boson $B_{\mu}$ through a kinetic mixing. It can be produced at the LHC by VBF and DY annihilation and signals of its decay into the $\gamma Z$ final state can appear at the LHC at $\sqrt{s}=14$ TeV with an integrated luminosity of $100$ fb$^{-1}$. Again, the main goal of our analysis has been to show how an effective Lagrangian approach can be used to make predictions which can be soon tested by the LHC.

\chapter*{Aknowledgments}
\markboth{}{}
\addcontentsline{toc}{chapter}{Aknowledgments}
\noindent I'm deeply grateful to Riccardo Barbieri for having guided this work with patience and experience and for having taught me not only a lot of physics, but also the critical thinking and the intellectual honesty which are the base of physics. I'm indebted to Christophe Grojean who made it possible for me to stay a long period at CERN under his supervision, profiting of a very stimulating environment. I'm indebted to Michelangelo Mangano, German Rodrigo, Thomas Gehrmann and Riccardo Rattazzi for their help and support after the end of my PhD and for valuable discussions.\\
I would like to thank all my collaborators and colleagues from whom I've learnt a lot: Gennaro Corcella, Enrico Trincherini, Antonio C\'arcamo, Slava Rychkov, Ennio Salvioni, Duccio Pappadopulo, Roberto Franceschini, Alessandro Strumia, Gino Isidori, Roberto Contino, Paolo Francavilla, Andrea Coccaro, Marco Bonvini, Javi Serra, Andrea Thamm, Sandeepan Gupta. I also thank all my office mates and friends at CERN, EPFL and UZH, especially Agostino Patella, Paolo Panci, Sara Borroni, Ahmad Zein Assi, Jean-Claude Jacot, Ehsan Hatefi, Stefan Prestel, Paolo Torrielli, Andrea De Simone, Alessandro Vichi.\\
I'm very grateful to Andrea Coccaro, Anna Rolando, Paolo Francavilla and Vincent Giangiobbe for their great hospitality in Cessy and Gex and to Damiano Tommasini for the wonderful cohabitation in Z\"urich. I'm indebted to all my colleagues for the innumerable discussions which have contribute to my knowledge of physics.\\
Thanks to Scuola Normale, CERN, UZH and EPFL for their hospitality and to all the secretaries who worked to make me working: Nanie Perrin, Elena Gianolio, Michelle Connor, Jeanne Rostant, Regina Schmid, Anna Maria De Domenico.\\
Thanks to Steve Jobs for the wonderful tools he provided for me.\\
This work has been partially supported by the European Commission under the contract ERC Advanced Grant 226371 {\it MassTeV}, the contract PITN-GA-2009-237920 {\it UNILHC} and the contract PITN-GA-2010-264564 {\it LHCPhenoNet} and by the Swiss National Science Foundation under the Grant n. 200021-125237.

\selectlanguage{italian}
\chapter*{Ringraziamenti}
\markboth{}{}
\addcontentsline{toc}{chapter}{Ringraziamenti}
\noindent Grazie,\\\\
A Francesca, la donna che amo, per la sua testa dura, cos\'i dura da essere riuscita a starmi vicino tutto questo tempo.\\
Alla famiglia: alla mamma che mi ha reso la persona che sono oggi, a nonna Gina che mi ha sempre sostenuto permettendomi di studiare, a nonno Luigi che ha sempre scommesso su di me, a nonna Anna che ha aspettato pazientemente, quasi novant'anni, per avere un pronipote dottore, allo zio Fabry che mi ha aiutato, non solo a costruire la casa, a mia cugina Sara, che mi ha dato due meravigliosi cuginetti, a tutti i Valsusini che pur lontani sono sempre stati vicini.\\
A quelli che non sono qui a condividere con me questi momenti, pap\'a prima di tutti e poi nonna Netta, nonno Rocco, nonno Antonio. A Mario che ha riportato in casa la serenit\'a di cui c'era bisogno.\\
E poi agli amici: a Giulio che con la sua voce meravigliosa ha reso la mia vita una splendida serenata, a zio Paride e alla Turca per le notti nel box e gli indimenticabili compleanni, a Ciccio che \'e sempre come una seconda fidanzata, a Simo con cui ``spesso il male di vivere ho incontrato...'', a Salvo, anzi Sambo, o forse Santo, a Gio che pur di disintossicarsi da me \'e scappato dall'altra parte del mondo, ad Andre che mi ha ospitato, a Matte che non ha pi\'u fumato, a Miche che la corsa ha dominato, a Dani che con me ha ``pi\'u pagato'', a Silvio che la mail non ha pi\'u guardato, al Bersa che il buon umore ha portato, al Baia che per primo ha procreato, a Fede che non ho mai dimenticato.\\
Grazie agli amici che hanno reso la mia permanenza all'estero meno difficile: a Paolo per le serate indimenticabili, a Ennio per avermi sopportato come compagno di ufficio, a Marco per aver offerto, mhh.., la sua amicizia, a Damiano per il pollo arrosto a Zurigo, a Duccio per la partita a tennis che non abbiamo mai giocato ma che prima o poi giocheremo. A tutti gli amici e i colleghi con cui ho condiviso questi anni.
\selectlanguage{english}

\markboth{}{}
\addcontentsline{toc}{chapter}{Bibliography}
\bibliography{00_Thesis}{}
\bibliographystyle{jhep}

\end{document}